\documentclass[preprint2,longabstract]{aastex}

\newcommand{\teff}{\ensuremath{T_{\rm eff}}}
\newcommand{\hy}{\mbox{H}}
\newcommand{\logg}{\ensuremath{\log{g}}}

\shorttitle{limb-darkening}
\shortauthors{Morello et al.}

\usepackage{amsmath,nicefrac}
\usepackage{enumitem}
\usepackage{multirow}

\begin{document}


\title{High-precision stellar limb-darkening in exoplanetary transits}


\author{G. Morello, A. Tsiaras, I. D. Howarth}
\affil{Department of Physics \& Astronomy, University College London, Gower Street, WC1E 6BT, UK}
\email{giuseppe.morello.11@ucl.ac.uk}

\author{D. Homeier}
\affil{Zentrum f\"{u}r Astronomie der Universit\"{a}t Heidelberg, Landessternwarte, K\"{o}nigstuhl 12, D-69117 Heidelberg, Germany}




\begin{abstract}
Characterization of the atmospheres of transiting exoplanets relies on
accurate measurements of the extent of the optically thick area of the
planet at multiple wavelengths with a precision $\lesssim$100 parts
per million (ppm). Next-generation instruments onboard the James Webb
Space Telescope (JWST) are expected to achieve $\sim$10 ppm precision
for several tens of targets. A similar precision can be obtained in
modeling only
if other astrophysical effects, including the stellar limb-darkening,
are properly accounted for. In this paper, we explore the limits on
precision
due to the mathematical formulas currently adopted to approximate the
stellar limb-darkening, and due to the use of limb-darkening coefficients
obtained either from stellar-atmosphere models or empirically. We recommend the use of a
two-coefficient limb-darkening law, named ``power-2'', which
outperforms other two-coefficient laws adopted in the exoplanet literature
in most cases, and particularly for cool stars.
Empirical limb-darkening based on two-coefficient
formulas can be significantly biased, even if the light-curve
residuals are nearly photon-noise limited. We demonstrate an optimal
strategy to fitting for the four-coefficient limb-darkening in the
visible, using prior information on the exoplanet orbital parameters
to break some of the degeneracies that otherwise would prevent the
convergence of the fit. Infrared observations taken with the James
Webb Space Telescope (JWST) will provide accurate measurements of the
exoplanet orbital parameters with unprecedented precision, which can
be used as priors to improve the stellar limb-darkening
characterization, and therefore the inferred exoplanet parameters,
from observations in the visible, such as those taken with Kepler/K2, the
JWST, and other past and future instruments.
\end{abstract}





\section{Introduction}

\begin{figure*}[!t]
\epsscale{0.45}
\plotone{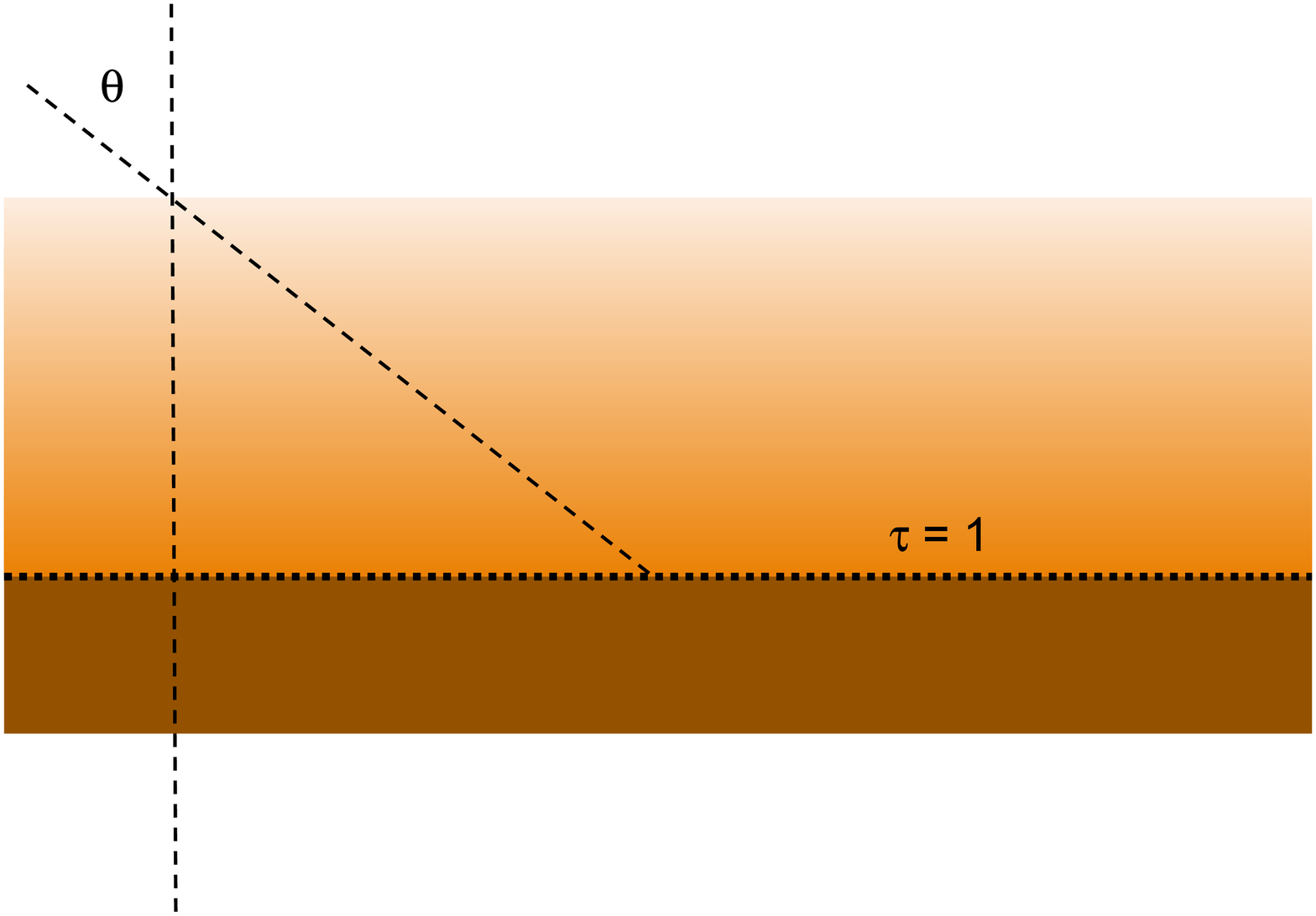}
\plotone{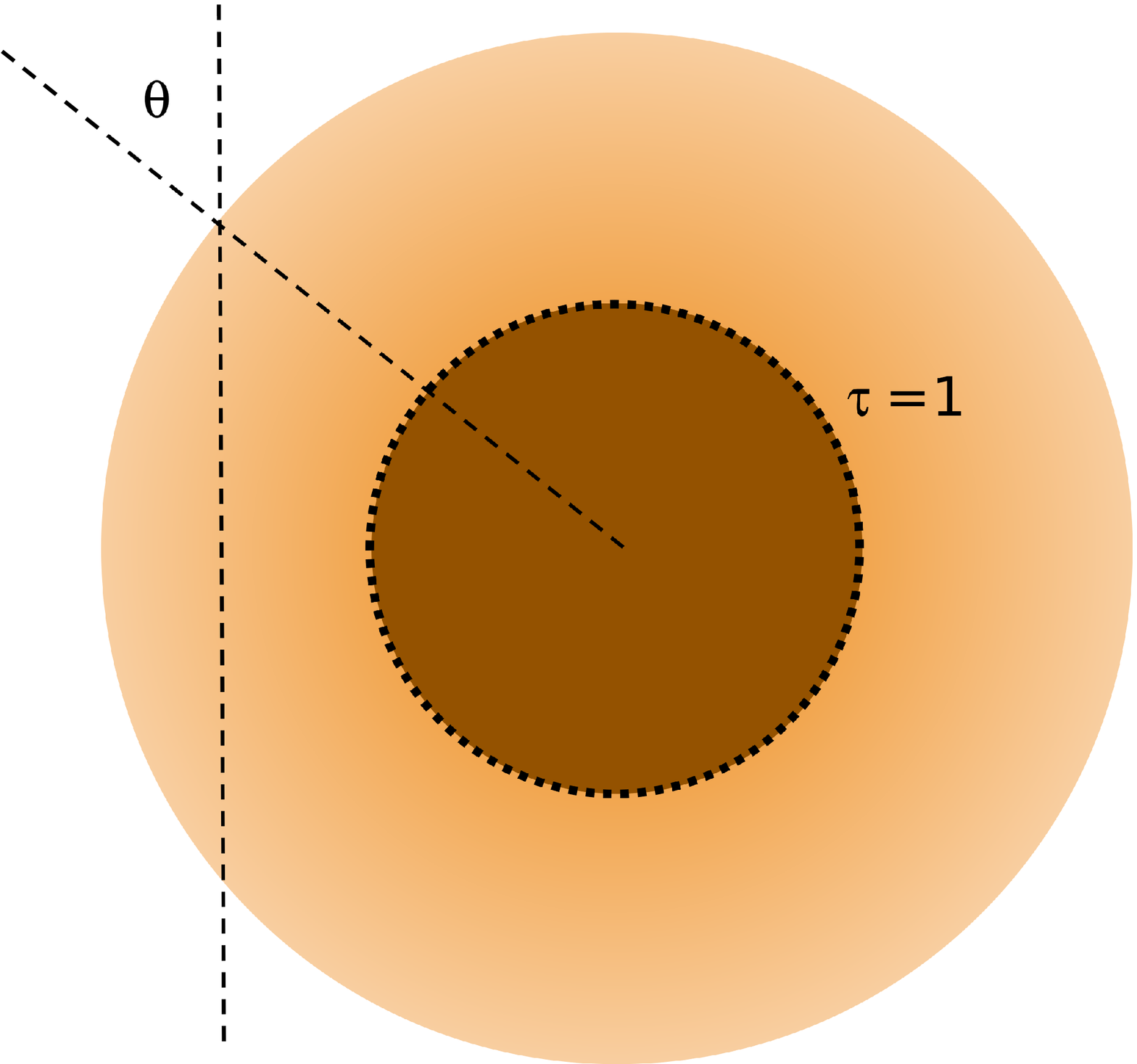}
\caption{Left panel: illustration of a plane-parallel atmosphere. Right panel:  illustration of a spherical-geometry atmosphere (the scale is exaggerated for easier visualization). Note that, differently from the plane-parallel case, the line of sight for the chosen angle $\mu$ does not intersect the shell corresponding to the radial optical depth $\tau = 1$.
\label{fig1}}
\end{figure*}

Observations of transits offer the most accurate means of measuring
exoplanet sizes and orbital inclinations, as well as mean stellar densities
and, if combined with radial-velocity information, system masses. Transits
are revealed through periodic drops in the stellar flux, due to the
partial occultation of the stellar disk by the planet for a portion of
its orbit. The amplitude of the flux decrement is primarily determined
by the size of the planet relative to the star, but also depends on
the location of the occulted area of the stellar disk and the
wavelength observed, because of limb-darkening (the radial decrease in
specific intensity).  Inadequate treatment of limb-darkening may give rise to $\gtrsim$10$\%$ errors in 
exoplanetary radii inferred from transits observed at UV or visible
wavelengths, and accurate modeling is paramount in the study of the
exoplanetary atmospheres, where differences of 10--100 parts per
million (ppm) in transit depths at different wavelengths can be
attributed to the wavelength-dependent optical depth of the external
layers of the planet, rather than to stellar properties.

Stellar-atmosphere models are commonly used to predict the
limb-darkening profiles, but empirical estimates are desirable, both
to test the stellar models and to reduce potential biases
in transit depths due to errors in the theoretical models or to other
second-order effects, such as stellar activity, granulation, gravity
darkening, etc.

Other than for the Sun, the surface of which can be directly observed
in great detail, techniques to map the stellar intensity distributions
rely mainly on optical interferometry (e.g., \citealt{hestroffer97, lane01,
  wittkowski01, aufdenberg05}) and microlensing
(e.g., \citealp{witt95, fields03, dominik04, zub11}).  The former is
useful for only a very limited number of stars with large
angular diameters, while the latter is limited by the low occurrence
rate and non-repeatability of the microlensing events. Eclipsing
binaries offer another route to mapping stellar surfaces, but accurate
modeling of these systems is handicapped by complicating factors
(gravity darkening, reflection effect, tidal distortion$\ldots$), and
a degree of redundancy between limb-darkening and radii. These issues are
much reduced in most star+exoplanet systems, thanks to the smaller mass and size of the planetary companions \citep{wilson71, loeb03, pfahl08}.

In this paper, we explore the potential biases in high-precision exoplanet spectroscopy using approximate stellar limb-darkening parameterizations, with coefficients obtained either from stellar-atmosphere models or empirically.

\subsection{Structure of the paper}

Section~\ref{sec:describing_ld} reviews the limb-darkening
laws most commonly adopted in the exoplanet literature, the proposed power-2 law, and discusses the current approaches to
obtain {\em theoretical} and {\em empirical} limb-darkening
coefficients. Section~\ref{sec:simulations} describes how we 
simulate light-curves from spherical-atmosphere models, and
Section~\ref{sec:results} reports the results of our analyses. In
particular, Section~\ref{sec:1Dvs3D} outlines the main differences
between plane-parallel and spherical stellar-atmosphere models; in
Sections~\ref{sec:accuracy_ld_laws} and~\ref{sec:ldc_fixed} we analyze
the precision with which different limb-darkening laws describe the
intensity profile and the transit morphology, and derive the correct
transit depth. Section~\ref{sec:ldc_free} describes the equivalent
analysis for the case of empirical limb-darkening coefficients, (i.e.,
allowed as free parameters in the light-curve fit).

Section~\ref{sec:WFC3} then focuses on the potential errors in
`narrow-band exoplanet spectroscopy' over
short wavelength ranges, specifically in the context of
\textit{Hubble Space Telescope (HST)}/WFC3 observations.  Section~\ref{sec:lc_fit_emp} examines
the ability to fit a set of transit parameters and limb-darkening
coefficients on transit light-curves, and develops an optimal strategy
to maximize the accuracy in the estimated transit parameters and
limb-darkening coefficients in the visible, if infrared observations
are also available.  Finally, Section~\ref{sec:discussion} discusses
the results of our analysis, with emphasis on the synergies between
the \textit{James Webb Space Telescope (JWST)} and Kepler, and on future surveys.

\begin{figure*}[!t]
\epsscale{0.90}
\plotone{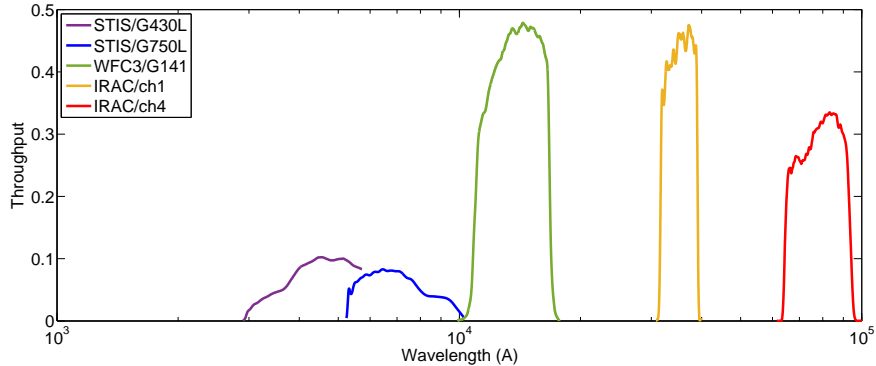}
\caption{Throughputs of the instruments adopted for the
  simulations; from visible to mid-infrared, STIS/G430L
  (purple), STIS/G750L (blue), WFC3/G141 (green), IRAC/ch1 (yellow)
  and IRAC/ch4 (red). 
\label{fig2}}
\end{figure*}

\section{Describing stellar limb-darkening}
\label{sec:describing_ld}

\subsection{Limb-darkening parameterizations}

In exoplanetary studies, 
the stellar limb-darkening profile is typically described by an
analytical function $I_{\lambda}( \mu )$, where $I$ denotes the
specific intensity, $\mu = \cos{ \theta}$, $\theta$ is the angle
between the surface normal and the line of sight, and the $\lambda$
subscript refers to the monochromatic wavelength or effective
wavelength of the passband at which the specific intensities are
given.  For circular symmetry, $\mu = \sqrt{1-r^2}$, where $r$ is the
projected radial co-ordinate normalized to a reference radius.

Numerous functional forms 
to approximate $I_{\lambda}(\mu)$
have been proposed in the literature.  In the study of exoplanetary
transits, the most commonly used of these limb-darkening `laws' are:
\begin{enumerate}
\item the quadratic law \citep{kopal50},
\begin{equation}
\frac{ I_{\lambda}( \mu ) }{ I_{\lambda}( 1 ) } = 1 - u_1 (1-\mu) - u_2 (1-\mu)^2 ;
\label{eqn:law_quadratic}
\end{equation}
\item the square-root law \citep{diaz-cordoves92},
\begin{equation}
\frac{ I_{\lambda}( \mu ) }{ I_{\lambda}( 1 ) } = 1 - v_1 (1- \sqrt{
  \mu }) - v_2 (1-\mu)\text{; and}
\label{eqn:law_sqrt}
\end{equation}
\item the four-coefficient law \citep{claret00},
\begin{equation}
\frac{ I_{\lambda}( \mu ) }{ I_{\lambda}( 1 ) } = 1 - \sum_{n=1}^{4} a_n \left ( 1 - \mu^{ \nicefrac{n}{2}} \right ) ,
\label{eqn:law_claret-4}
\end{equation}
\end{enumerate}
hereinafter referred to as ``claret-4''.
The quadratic, square-root, and claret-4 laws rely on linear
combinations of fixed powers of $\mu$.  In this paper, we advocate an
alternative two-coefficient law incorporating an arbitrary power of
$\mu$ which, to the best of our knowledge, has not previously been
considered in the exoplanet literature (and which we initially
constructed independently):
\begin{enumerate}[resume]
\item the `power-2' law \citep{hestroffer97},
\begin{equation}
\label{eqn:law_power-2}
\frac{ I_{\lambda}( \mu ) }{ I_{\lambda}( 1 ) } = 1 - c \left ( 1-\mu^{\alpha} \right )
\end{equation}
\end{enumerate}
We find that this form offers more flexibility and a better match to
model-atmosphere limb-darkening than do other two-coefficient forms
(Section~\ref{sec:accuracy_ld_laws}).
The claret-4 law can
provide a more accurate approximation to model-atmosphere limb-darkening than other forms, but at the expense of using a larger number of
coefficients.
We note that the quadratic and square-root laws are subsets of the
claret-4 prescription, with $a_1=a_3=0$, $a_2 = u_1+2u_2$, $a_4 =
-u_2$ (quadratic) and $a_3=a_4=0$ (square-root). 
The power-2 form
is a subset only for $\alpha = \nicefrac{1}{2}$,
1, $\nicefrac{3}{2}$, or~2.

\begin{figure*}[!t]
\epsscale{0.90}
\plotone{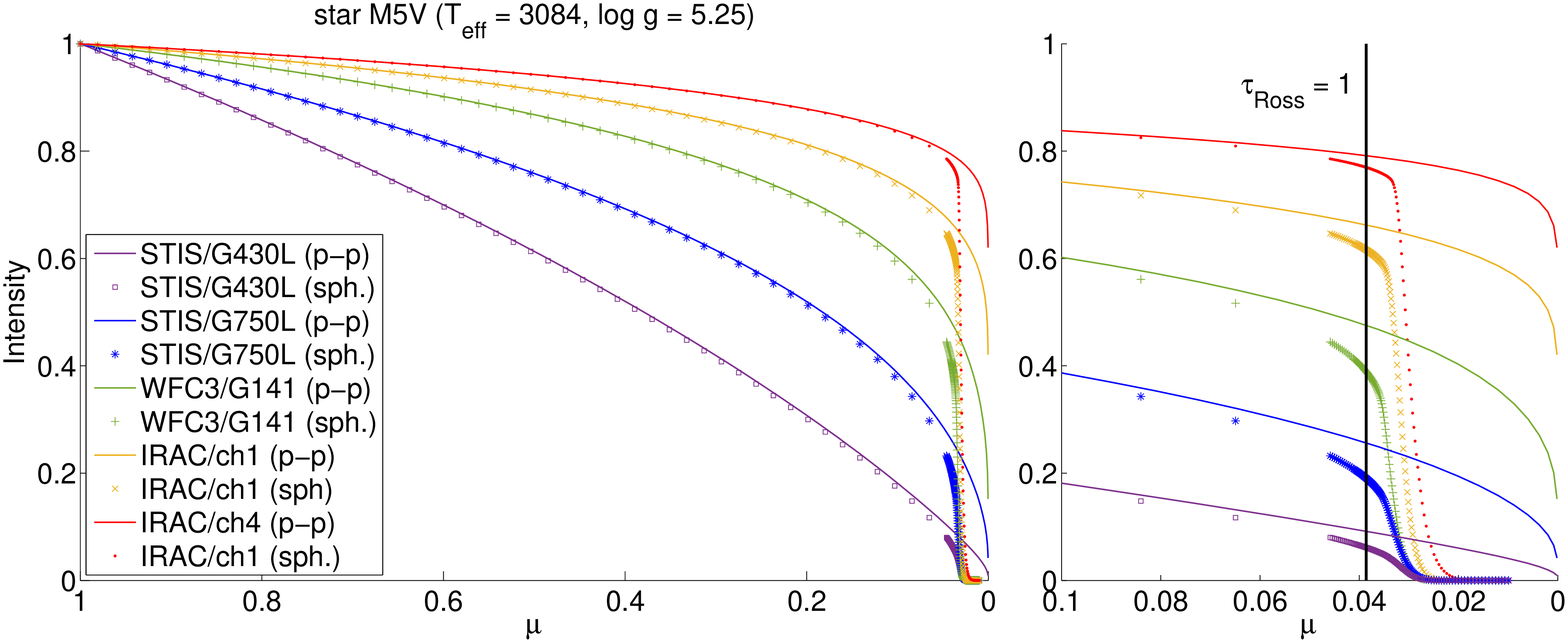}
\plotone{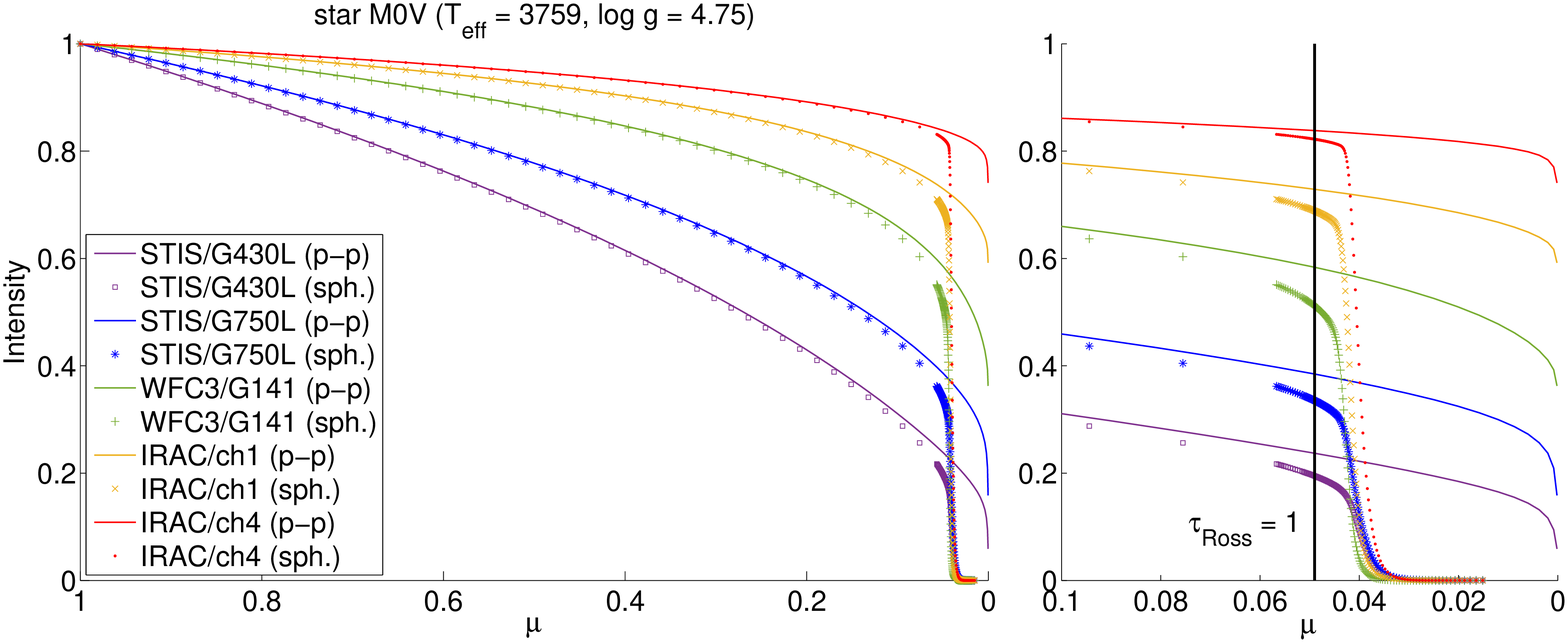}
\plotone{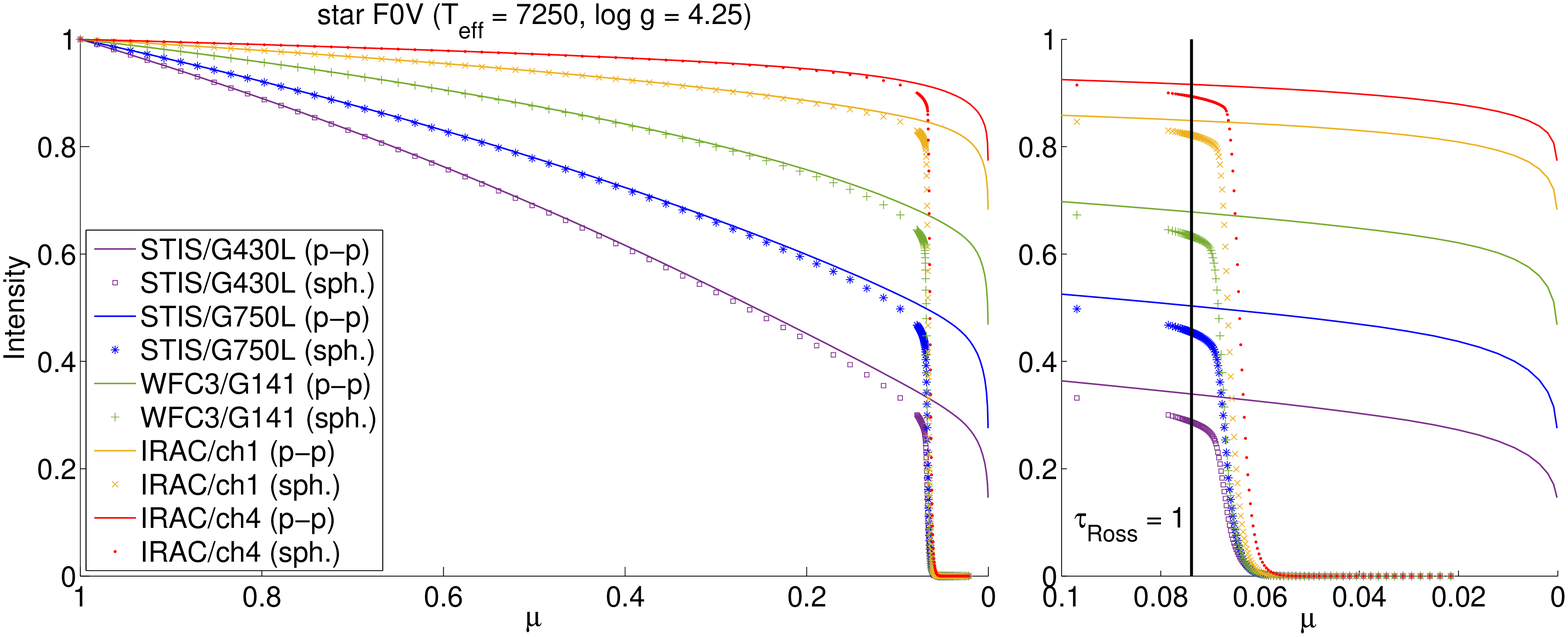}
\caption{Top to bottom: angular intensity distributions for the M5\;V,
  M0\;V, and F0\;V models of Table~\ref{tab1}. Symbols represent
  spherical-geometry, passband-integrated intensities for STIS/G430L
  (purple squares), STIS/G750L (blue ``*''), WFC3/G141 (green ``+''),
  IRAC/ch1 (yellow ``x'') and IRAC/ch4 (red dots); the corresponding
  plane-parallel intensities are shown as continuous lines of the same
  colors. The right-hand panels show the limb region (0.1$\ge \mu
  \ge$0.0) at a larger scale and the $\mu$ angle corresponding to $\tau_\mathrm{Ross}(r) = 1$.}
\label{fig3}
\end{figure*}


\subsection{Intensity distributions: plane-parallel vs.\ spherical}
\label{sec:1d2d}

\begin{figure*}[!t]
\epsscale{1.0}
\plotone{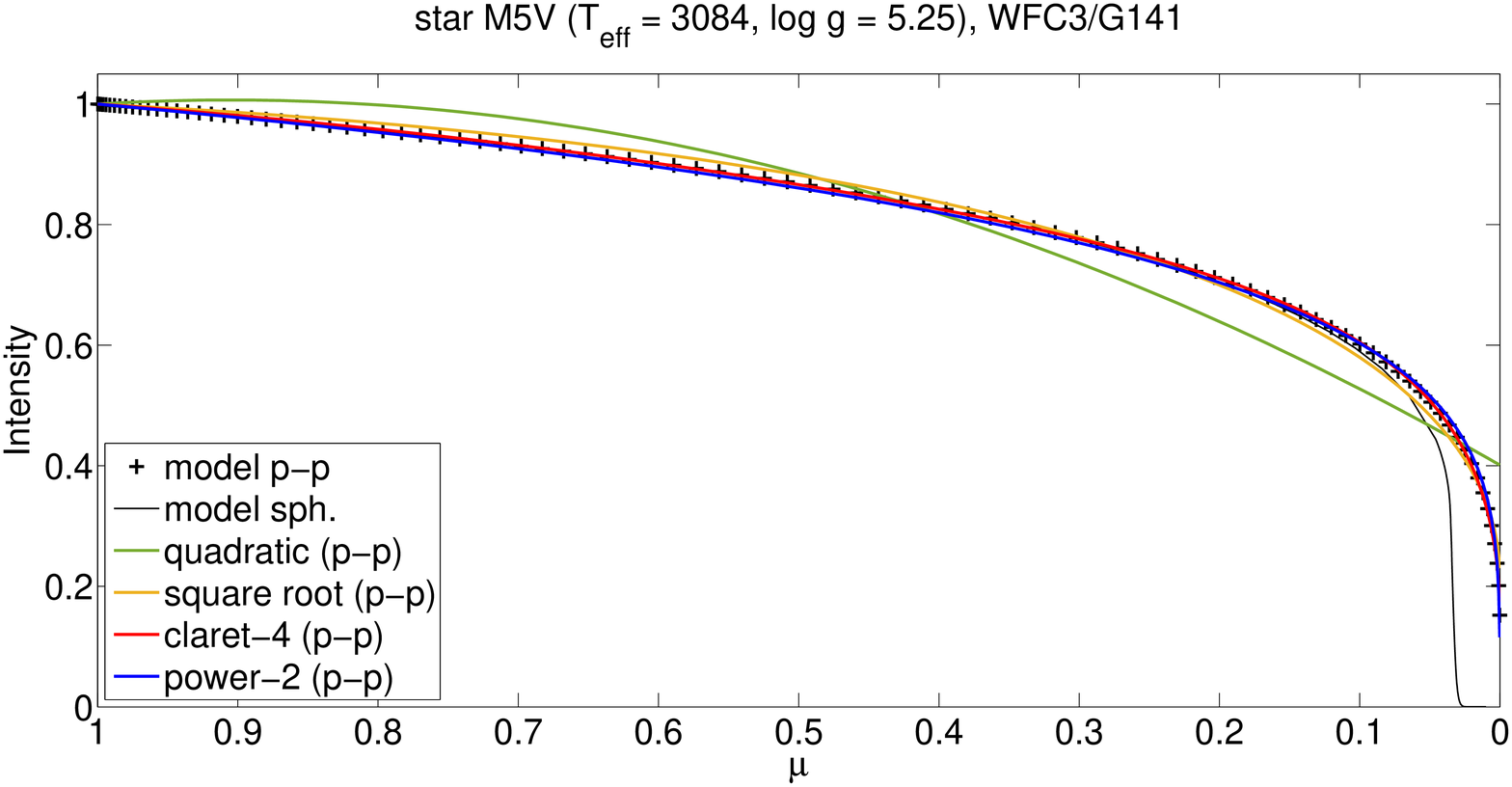}\\
\plotone{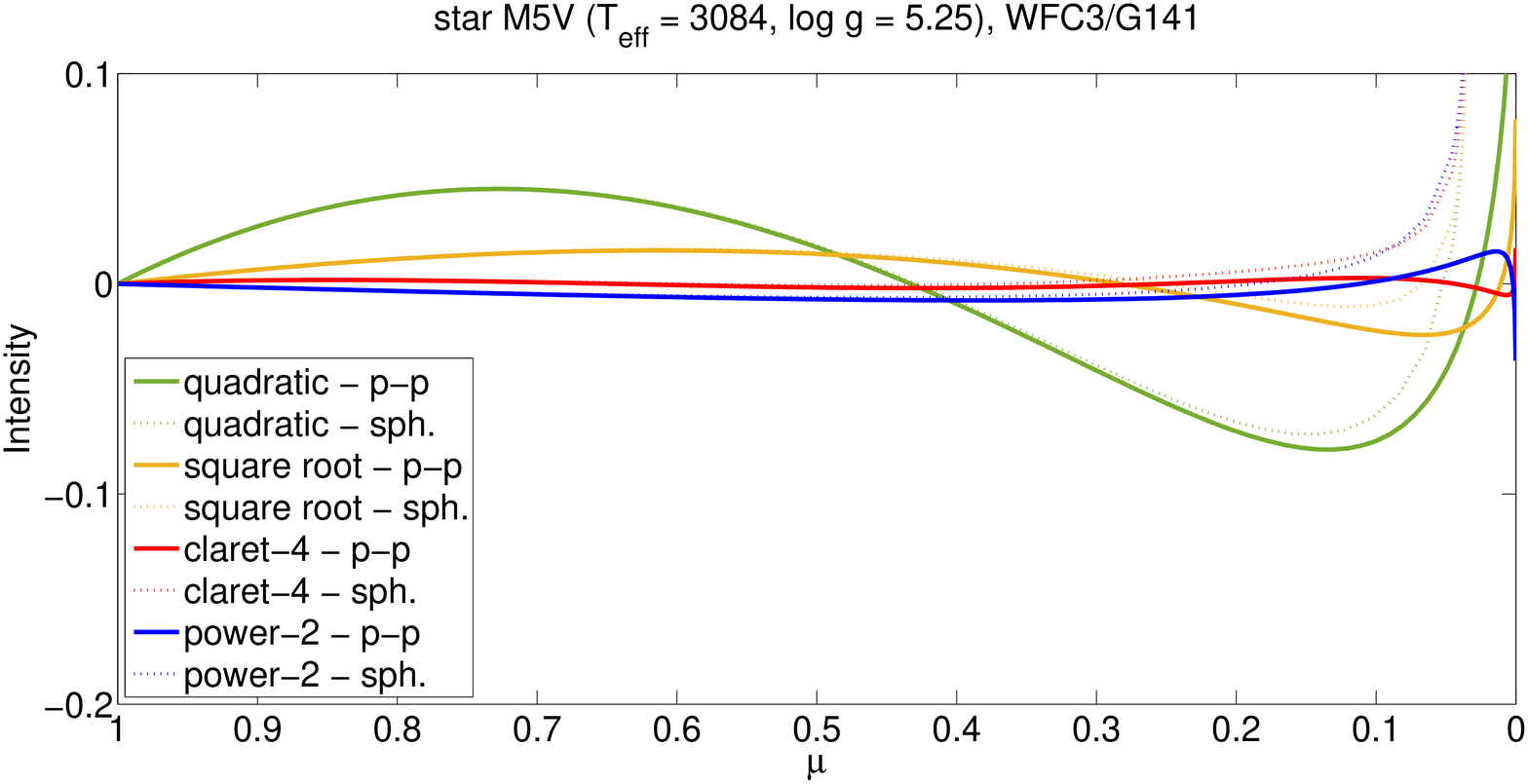}
\caption{Top panel: Plane-parallel (black ``+'') and spherical (black
  line) model-atmosphere intensities vs.\ $\mu$ for the M5\;V star in the WFC3/G141 passband. Parametric limb-darkening functions fitted to plane-parallel intensities (theoretical models) are
  quadratic (green), square-root (yellow), claret-4 (red), and power-2
  (blue) laws. Bottom panel: large-scale plots of residuals of
  parametric limb-darkening laws for model-atmosphere
  intensities in the plane-parallel and
 spherical geometries (continuous and dashed
  lines, respectively).\label{fig4}}
\end{figure*}

{\em Theoretical} limb-darkening coefficients can be obtained from
stellar-atmosphere models, by fitting a parametric law
(such as Equations~\ref{eqn:law_quadratic}--\ref{eqn:law_power-2}) to
detailed numerical evaluations of $I_{\lambda}(\mu)$
using some suitable numerical technique  -- typically least squares,
though
detailed numerical results depend on both the method chosen
and the data sampling (e.g., \citealp{heyrovsky07, claret08, howarth11}). 
Tables of theoretical limb-darkening coefficients as a function of
stellar parameters (usually the effective temperature, gravity, and
metallicity) have been published by several authors for various
photometric passbands. Most calculations are based on plane-parallel
atmosphere models \citep{claret00, claret04, claret08, sing10, howarth11b, claret12, claret13,  reeve16}, but some authors have
considered spherical geometry, claiming, in some cases, noteworthy
improvements compared to the use of plane-parallel models
\citep{claret11, hayek12, neilson13a, neilson13b, claret14, magic15}.


These spherical models show a characteristic steep drop-off in intensity at small, but finite $\mu$ (see, e.g., Figure~\ref{fig3}). The explanation for this drop-off is straightforward (Figure~\ref{fig1}).

In a plane-parallel atmosphere, the optical depth 
\textit{always} reaches unity somewhere along the line of sight, even at grazing incidence. The
limb of the star is, consequently, geometrically well-defined (and
wavelength independent), and the intensity at the limb is comparable 
to the intensity at the center of the disk, to within a
factor of a few.

In spherical geometry, in contrast, there are no constant angles $\mu$ at which the characteristic rays intersect the shells; instead they vary as a function of radius. For technical reasons the emergent intensities are usually specified as functions of the angle as measured at the outer boundary of the model atmosphere, which is originally set by the modeller at an arbitrary physical radius or reference optical depth, subject only to the condition that it has a suitably small opacity and emissivity even at the cores of strong lines.
The outermost layer of the model atmosphere, corresponding to $\mu = 0$ in this reference frame, is therefore always optically thin and does not correspond to what would be observed as `the' stellar radius in investigations involving interferometric
imaging, lunar occultation, or exoplanetary transits.
Furthermore, the rapid changes in $I_{\lambda}(\mu)$ that arise at small $\mu$ in spherical models are, inevitably, not well approximated by any of the standard parametric laws developed to represent results of plane-parallel models, because the intensity does not converge to zero at any given radius (see, e.g., Figure 1 of \citealp{neilson17}).

\cite{wittkowski04} and \cite{espinoza15} therefore suggested to re-define $\mu = 0$ to a radius outside of which almost no flux is observed, i.e. at the inflection point of the intensity profile so that the tail-like extension originating from the optically thin outer layers is excluded from fitting the limb-darkening profile.
The radius is reasonably well defined by the point at which the gradient $dI(\mu)/d\mu$, or almost equivalently $|dI(r)/dr|$, reaches a maximum. \cite{wittkowski04} found that this radius corresponded closely to the Rosseland radius, defined by a Rosseland optical depth along the normal $\tau_\mathrm{Ross}(r) = 1$, for the M giant models they presented. Close to the limb, however, one can expect to observe significant emission as long as the {\em line-of-sight} optical depth is at least of order unity. Additionally, in the context of modeling exoplanetary transits, the projected radius for which the total line-of-sight optical depth within the observed wavelength band becomes one should be considered. In \cite{wittkowski04} the differences between radial and line-of-sight optical depths at a given physical depth were relatively modest because the giant-star atmosphere they modeled has a large radial extension with corresponding large angles $\mu=0.3$. Furthermore, they studied the emergent intensities in the $K$ band, which has relatively smaller mean opacity than the Rosseland mean, thus partly cancelling the effect of the off-normal incidence.

In this paper, we test empirically the choice of stellar radius, based on the ratios between best-fit and input transit depths (see Section~\ref{sec:1Dvs3D}), finding different values for the best renormalization radius in different wavelength bands (corresponding to the different band-mean opacities). But these radii are larger throughout than the respective $\tau_\mathrm{Ross}(r) = 1$, which correspond to $\mu = $ 0.0386, 0.049, and 0.0738 for the three models displayed in Figure~\ref{fig3}, confirming the effect of the off-normal incidence on the emission near the limb.

\subsection{Limitations on empirical limb-darkening coefficients}

{\em Empirical} limb-darkening coefficients can be inferred by fitting a
parametric model to an observed transit light-curve \citep{mandel02,
  southworth04, pal08}
Two-coefficient laws are typically used for this purpose
(e.g. \citealp{southworth08, claret09, kipping11, kipping11b,
  hebrard13, muller13}), as parameter degeneracies hamper convergence
when fitting higher-order models (e.g., the claret-4
characterization).

Measuring empirical limb-darkening coefficients is important to test
the validity of the stellar-atmosphere models and, if results are
sufficiently accurate, to select the best theoretical
models. Furthermore, fixing limb-darkening coefficients at incorrect
theoretical values can significantly bias other fitted transit
parameters, leading to incorrect inferences about planetary sizes and 
masses, or confusing the spectral signature of a
planetary atmosphere \citep{csizmadia13}. In active stars, the
presence of dark or bright spots on the surface can 
change the `effective' limb-darkening coefficients relative to the
unperturbed case, as well as the inferred stellar parameters adopted
to compute the theoretical coefficients \citep{ballerini12,
  csizmadia13}. Other light-curve distortions may arise from gravity
darkening in fast-rotating stars \citep{vonzeipel24, barnes09,
  claret12}, stellar oscillations \citep{broomhall09, kjeldsen11},
granulation \citep{chiavassa16}, beaming \citep{zucker07, shporer12},
ellipsoidal variations \citep{pfahl08, welsh10}, reflected light
\citep{borucki09, snellen09}, planetary thermal emission
\citep{kipping10}, or exomoons \citep{sartoretti99, kipping09,
  kipping09b}. The photometric amplitudes of such distortions can be 
up to $\sim$100~ppm.

\section{Simulated transit light-curves}
\label{sec:simulations}

In order to investigate the consequences of various approximations to
limb-darkening, we calculated `exact' synthetic transit photometry as
a reference, using new
model-atmosphere intensities coupled to an accurate numerical
integration scheme for the light-curves.

\subsection{Stellar models}
We generated three representative model atmospheres using the \textsc{Phoenix}
simulator \citep{allard12}; input parameters are summarized in
Table~\ref{tab1}. These models are
intended to bracket the range in effective
temperature of known exoplanet host stars, and embrace $\sim$98$\%$ 
of those listed in the \texttt{exoplanet.eu} database
\citep{schneider11} as of 2016 December 12.

\begin{table}
\begin{center}
\caption{Input parameters (effective
  temperature, gravity) for solar-abundance \textsc{Phoenix}
  stellar-atmosphere models adopted for the simulations.}
\begin{tabular}{cccc}
\tableline\tableline
Sp. type & \teff\ & \logg\ & $[M/\hy]$\\
\tableline
M5\;V & 3084 & 5.25 & 0.0 \\
M0\;V & 3759 & 4.75 & 0.0 \\
F0\;V & 7250 & 4.25 & 0.0 \\
\tableline
\end{tabular}
\end{center}
\tablecomments{Default values
  for other parameters are specified in
  \citet{allard12}. The corresponding spectral types are based on
the calibration reported in \citet{gray09book}.}
\label{tab1}
\end{table}


For each stellar model, $I_\lambda(\mu)$ profiles were calculated
in both plane-parallel and spherical geometries.  In the former case,
intensities were calculated at 96 values of $\mu$, chosen as the
anchor points for a Gaussian-quadrature integration; the intervals
$\Delta \mu_i = | \mu_{i+1} - \mu_{i} |$ vary in the range
$7\times 10^{-4}$--$1.6 \times 10^{-2}$ and are smallest for $\mu
\sim$0 and $\sim$1. In spherical geometry, the $\mu$ values
were determined by the properties of the model atmosphere, and the number of
grid points is model-dependent (169--177 in the cases
considered here); the limb is the most finely sampled region, down to $\Delta
\mu \sim 6 \times 10^{-5}$. Passband-integrated intensities were
calculated for five instruments which have been widely used in the
field of exoplanet spectroscopy, from the visible to mid-infrared
wavelengths: the STIS/G430L, STIS/G750L and WFC3/G141 gratings onboard
\textit{HST},\footnote{The Kepler passband is
similar to the combined STIS passbands, and
the results for the
STIS passbands are therefore a good proxy for Kepler.} and the IRAC photometric channels 1 and 4 onboard
\textit{Spitzer}. The throughputs of these instruments are shown in
Figure~\ref{fig2}, and the corresponding plane-parallel and spherical-model
intensities are shown in Figure~\ref{fig3}. 



\subsection{Computing transit light-curves from spherical model atmospheres}
\label{sec:ideal}
We generated two sets of `exact' transit light-curves for the
exoplanet-system parameters reported in Table~\ref{tab2}. Each set
consists of fifteen transit light-curves, one for each stellar model
and instrument passband, using the spherical-geometry intensities;
the sets differ only in the impact parameter (or, equivalently, the
orbital inclination).
Each
light-curve contains 2001 data points with 8.4 s sampling time, over a
$\sim$4.7 hr interval centered on the mid-transit (the total
duration of the transits is $\sim$2 hr, with the central transit
being $\sim$10 min longer).

\begin{table}[t]
\begin{center}
\caption{Input transit parameters adopted in simulations \label{tab2}}
\begin{tabular}{cccccc}
\tableline\tableline
$p$ & $a_R$ & $i$ ($^\circ$) & $e$ & $b$ & $P$ (days) \\
\tableline
0.15 & 9.0 & 90.0 & 0 & 0.0 & 2.218573 \\
0.15 & 9.0 & 86.81526146 & 0 & 0.5 & 2.218573 \\
\tableline
\end{tabular}
\end{center}
\tablecomments{Transit parameters are similar to those of HD189733b: $p = R_p/R_*$ is the ratio of planet-to-star radii, $a_R = a/R_*$ the
  orbital semimajor axis in units of the stellar radius, $i$ the
  orbital inclination, $e$ the eccentricity, $b (= a_R
  \cos{i})$ the impact parameter, and $P$ the orbital
  period.}
\end{table}

The orbital parameters determine $z$, the sky-projected star--planet
separation in units of the stellar radius at any given time; for
a circular orbit
\begin{equation}
z(t) = a_R \sqrt{1 - \cos^2
\left({ 
\frac{ 2 \pi (t - t_0) }{P}
}\right)
\sin^2(i) },
\end{equation}
where $a_R$ is the semimajor axis in units of the stellar radius, $P$
is the orbital period, $i$ is the inclination relative to the sky, and
$t_0$ is the time of conjunction.  

The fraction of stellar light occulted by the planet is, for a given
intensity profile, a function $F(p,z(t))$, where $p$ is the ratio of
planet-to-star radii. Instead of using an analytical function
(requiring a numerical approximation to the intensity profile), we computed the light-curve by direct
integration of the occulted stellar flux, using our purpose-built
`tlc' algorithm. The algorithm:
\begin{enumerate}
\item divides the sky-projected stellar disk into a user-defined
  number of annuli, $n$, with uniform radial separation, $dr = 1/n$;
\item evaluates the intensity at the central radius of each annulus,
  $I(r_i)$, where $r_i = (0.5 + i)/n$ for $i = 0 \dots n-1$,
  interpolating in $\mu$ from the input stellar-intensity profile
  (and $r_i = \sqrt{1-\mu_i^2}$);
\item evaluates the flux from each annulus, $F_i = I(r_i) \times 2 \pi
  r_i dr$, and hence the total stellar flux, $F_* =
  \sum_{i=0}^{n-1}F_i$.
\item
 The occulted flux  is then calculated as 
$F_{\rm occ}(p,z) = \sum_{i=0}^{n-1} F_i
  f_{z,p}(r_i)$, where $f_{z,p}(r_i)$ is the fraction of circumference
  of each annulus covered by the planet,  given by
\begin{equation}
f_{p,z}(r_i) = 
\begin{cases}
\frac{1}{ \pi} \arccos{ \frac{r_i^2 + z^2 - p^2}{2zr_i}}  & |z-p| < r_i < z+p \\
0 & r_i \le z-p \quad\text{or}\quad r_i \ge z+p, \\
1 &  r_i \le p-z
\end{cases}
\end{equation}
\item whence the normalized flux is $F(p,z) = 1 - F_{\rm occ}/F_*$.
\end{enumerate}
Before calculating the actual transit light-curves from the spherical model
intensities, we tested the accuracy of the algorithm using a wide range
of parametric intensity profiles as input, with the same grid of $\mu$ values as
the spherical models, comparing the resulting
light-curves to those from analytical
calculations. We found that, with $n=100\,000$ annuli, the maximum
differences between tlc and analytical
light-curves were $<5\times 10^{-7}$ in the worst-case
scenarios -- negligible compared to the minimum uncertainties
that can be obtained with any current or forthcoming instrument.


\section{modeling transit light-curves}
\label{sec:results}

\begin{figure*}[!t]
\epsscale{0.95}
\plotone{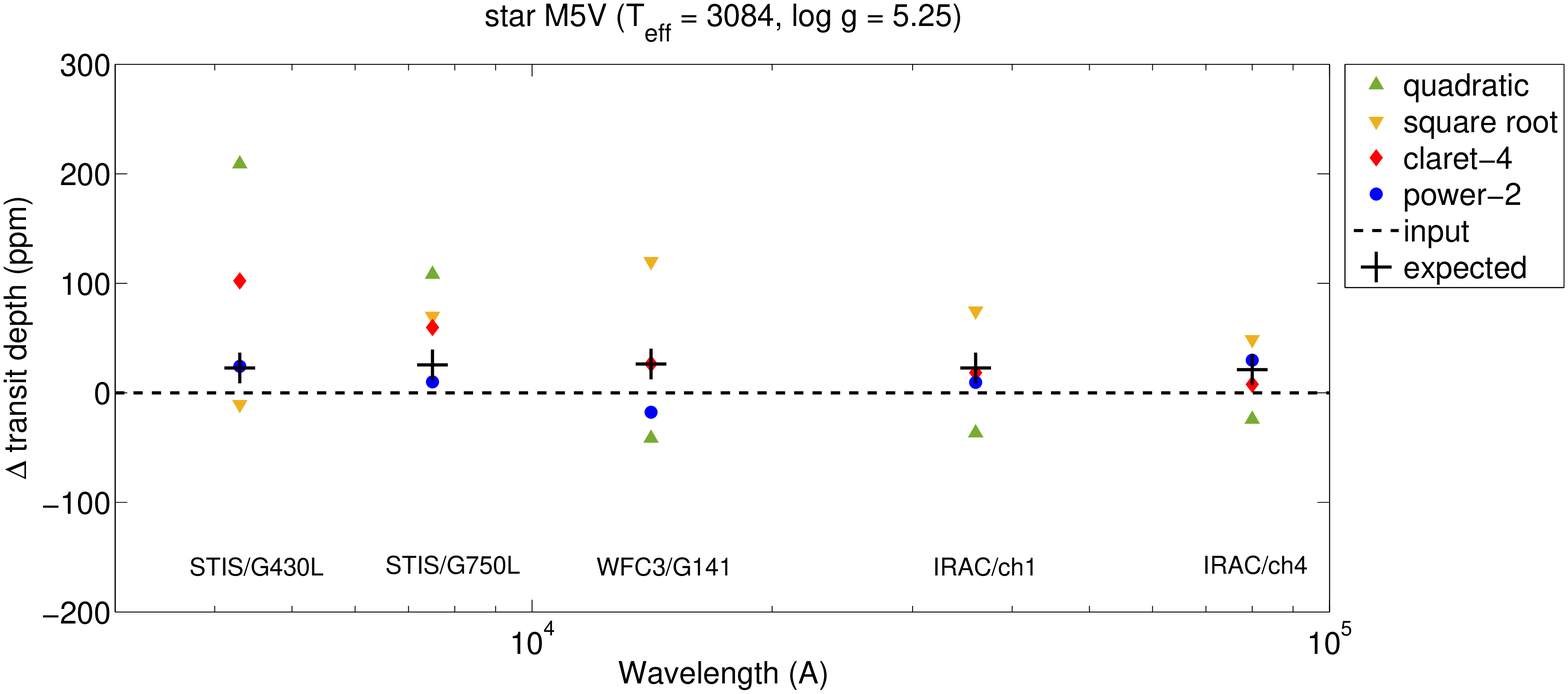}
\plotone{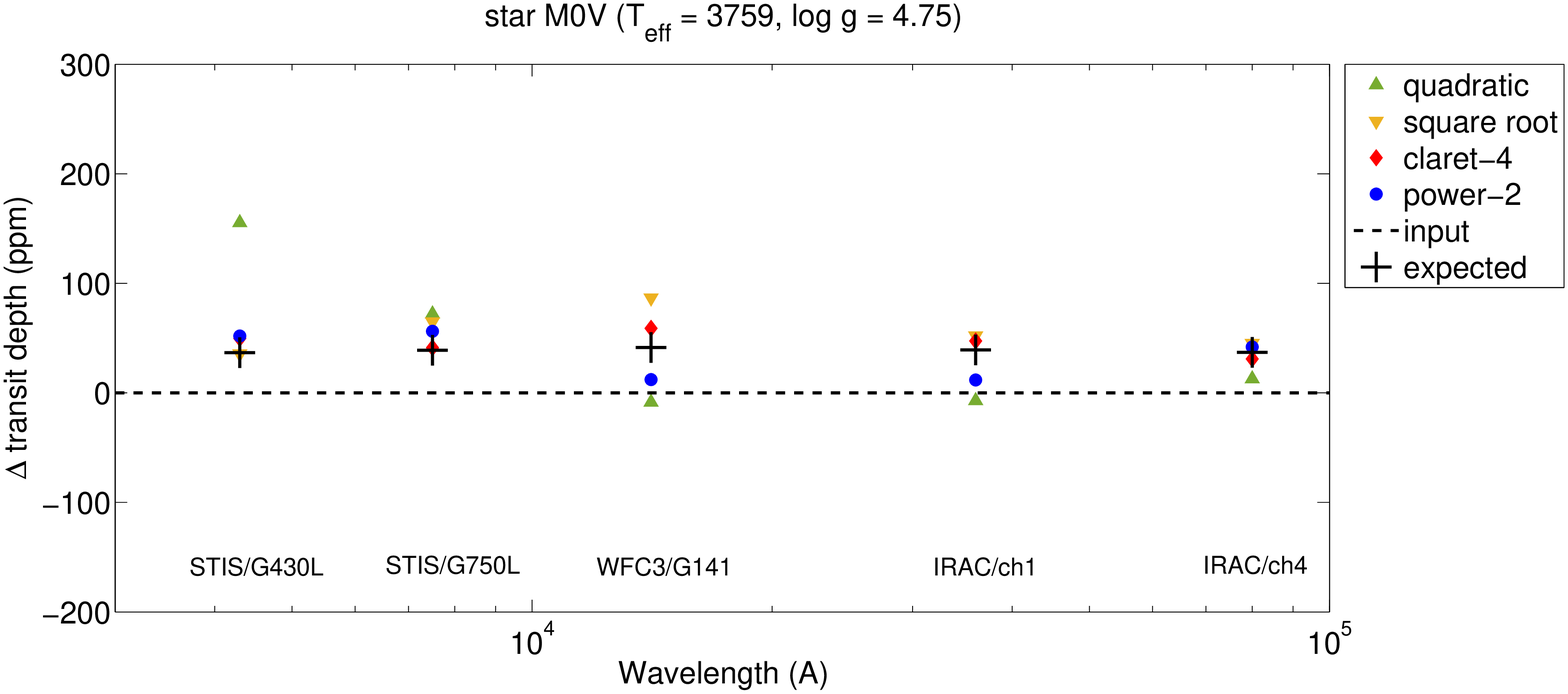}
\plotone{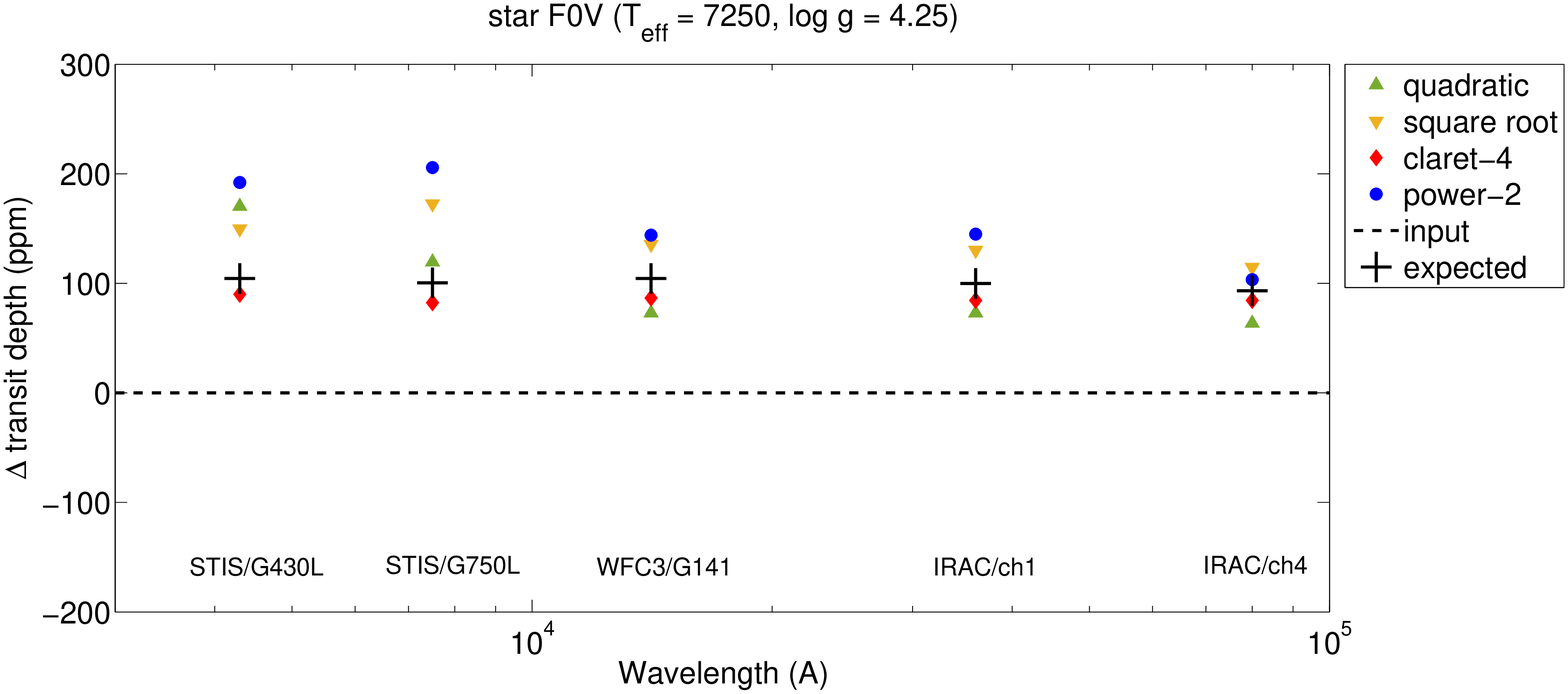}
\caption{Top panel: differences between the best-fit and the input
  transit depths for the edge-on transits in front of the M5\;V model, 
using \textit{theoretical} quadratic (green,
  upward triangles), square-root (yellow, downward triangles),
  claret-4 (red diamonds), and power-2 (blue circles) limb-darkening
  coefficients; the expected values 
(Equations~\ref{eqn:p2_expected})
are indicated with black ``+''. Middle, bottom panels: the same for M0\;V,
and 
  F0\;V models. 
\label{fig5}}
\end{figure*}

\begin{figure*}[!t]
\epsscale{1.0}
\plotone{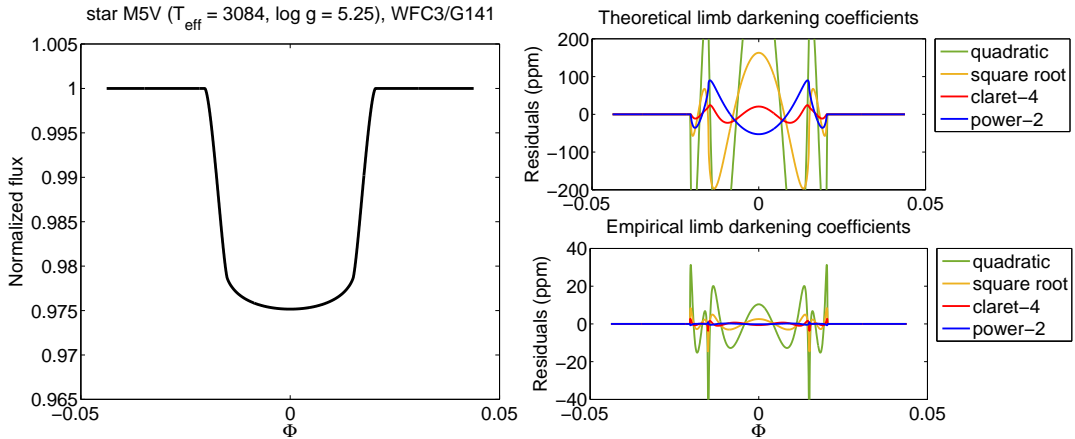}
\caption{Left panel: exact transit light-curves obtained with $b=0$ for the M5\;V star in the WFC3/G141 passband. Right, top panel: residuals for the best-fit transit models using theoretical quadratic (green), square-root (yellow), claret-4 (red), and power-2 (blue) limb-darkening coefficients. Right, bottom panel: residuals obtained with the empirical limb-darkening coefficients. \label{fig6}}
\end{figure*}

\begin{figure*}[!t]
\epsscale{0.95}
\plotone{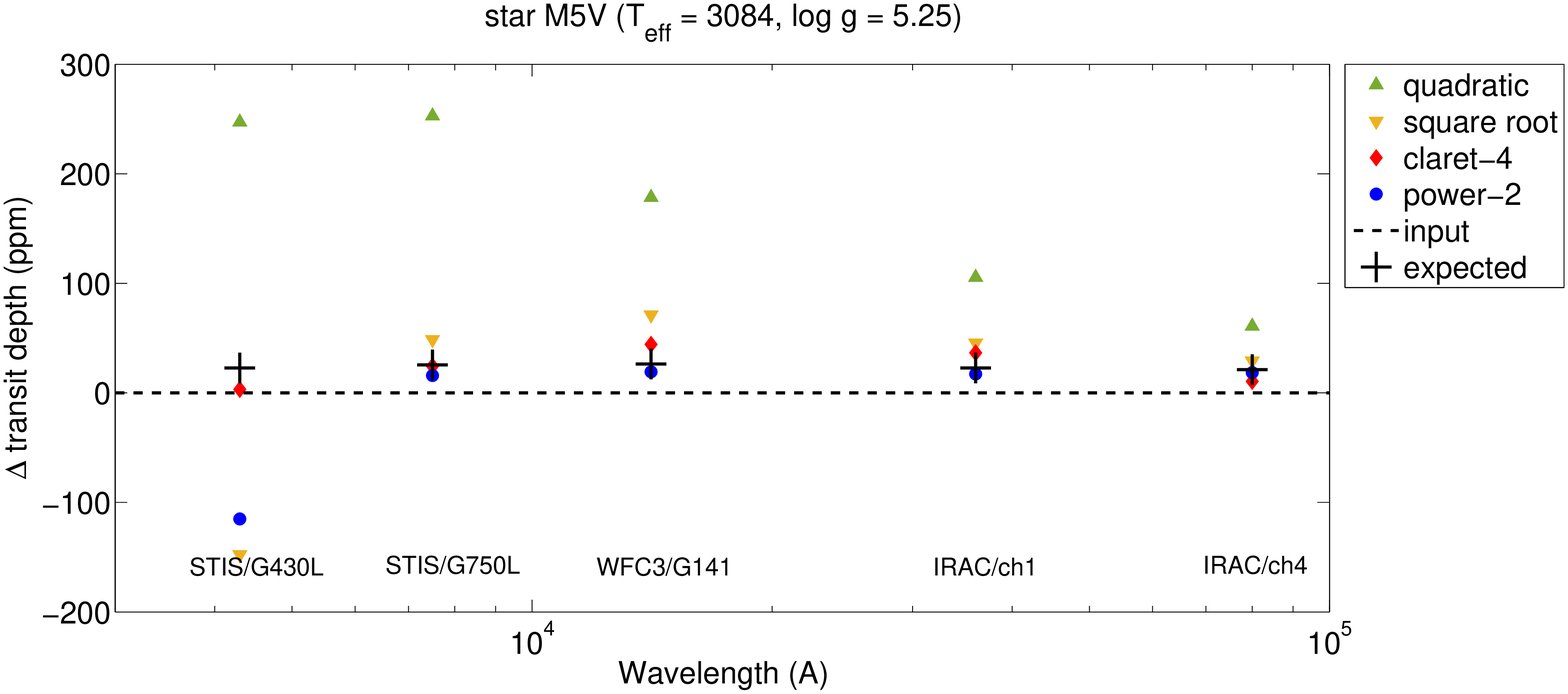}
\plotone{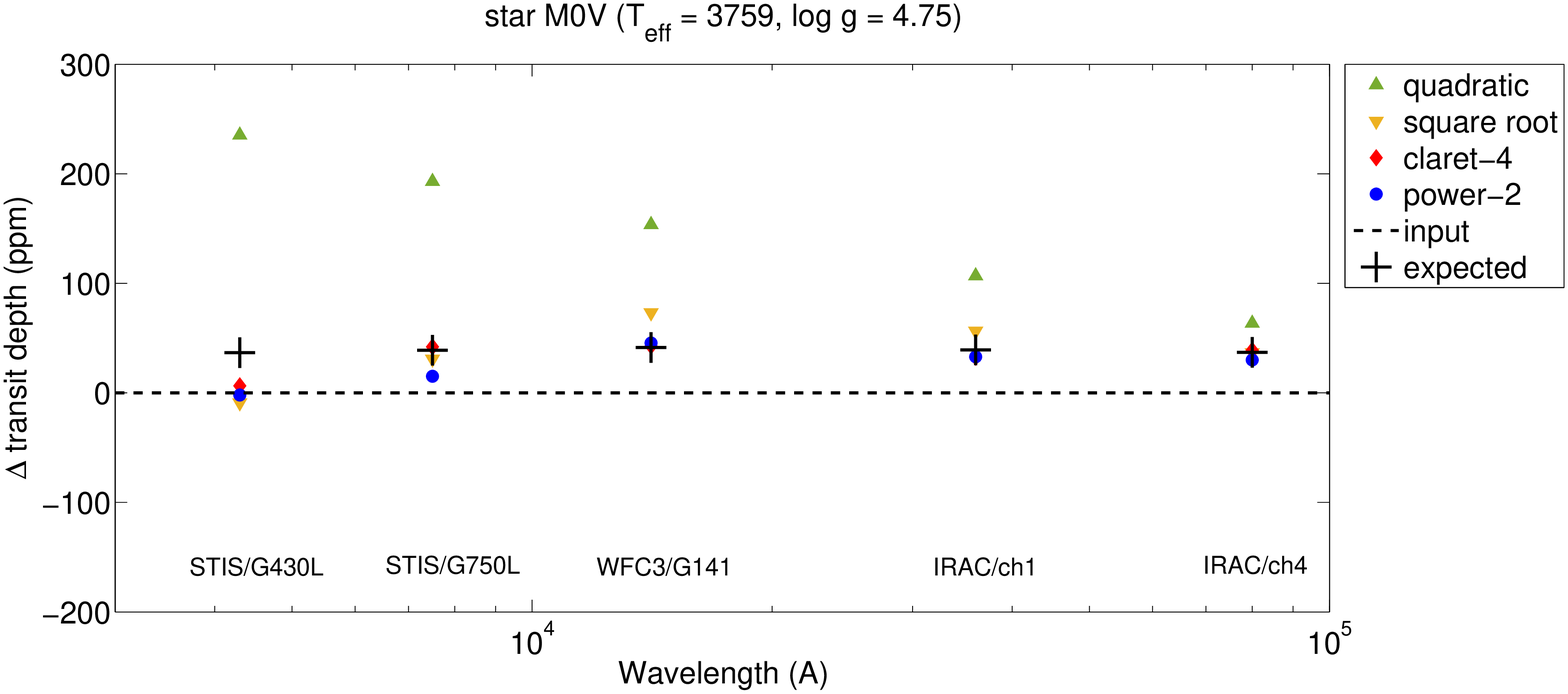}
\plotone{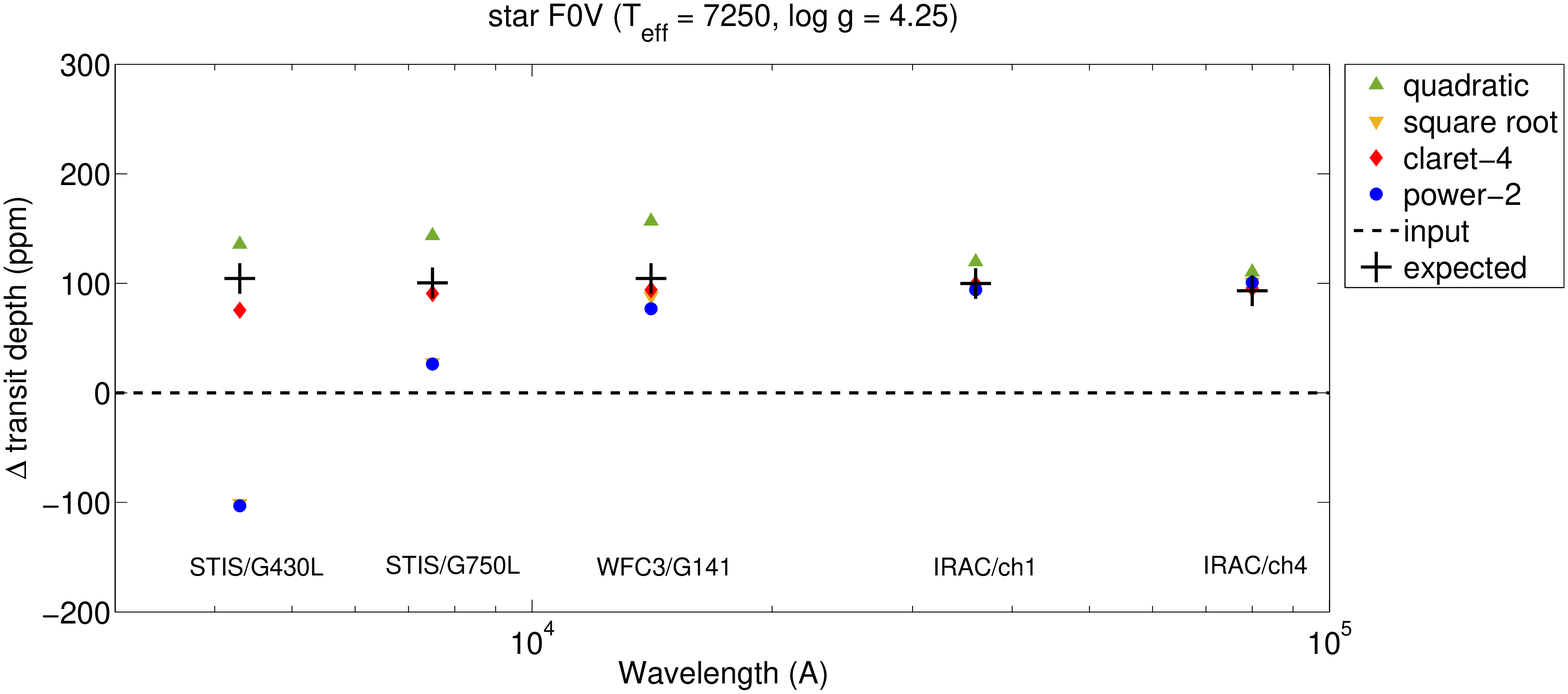}
\caption{Top panel: differences between the best-fit and the input transit depths for the edge-on transits in front of the M5\;V model, using empirical quadratic (green, upward triangles), square-root (yellow, downward triangles), claret-4 (red diamonds), and power-2 (blue circles)  limb-darkening coefficients; the expected values (Equations~\ref{eqn:p2_expected}) are indicated with black ``+''. Middle, bottom panels: the same for M0\;V and F0\;V models.
\label{fig7}}
\end{figure*}

\begin{figure*}[!t]
\epsscale{1.0}
\plotone{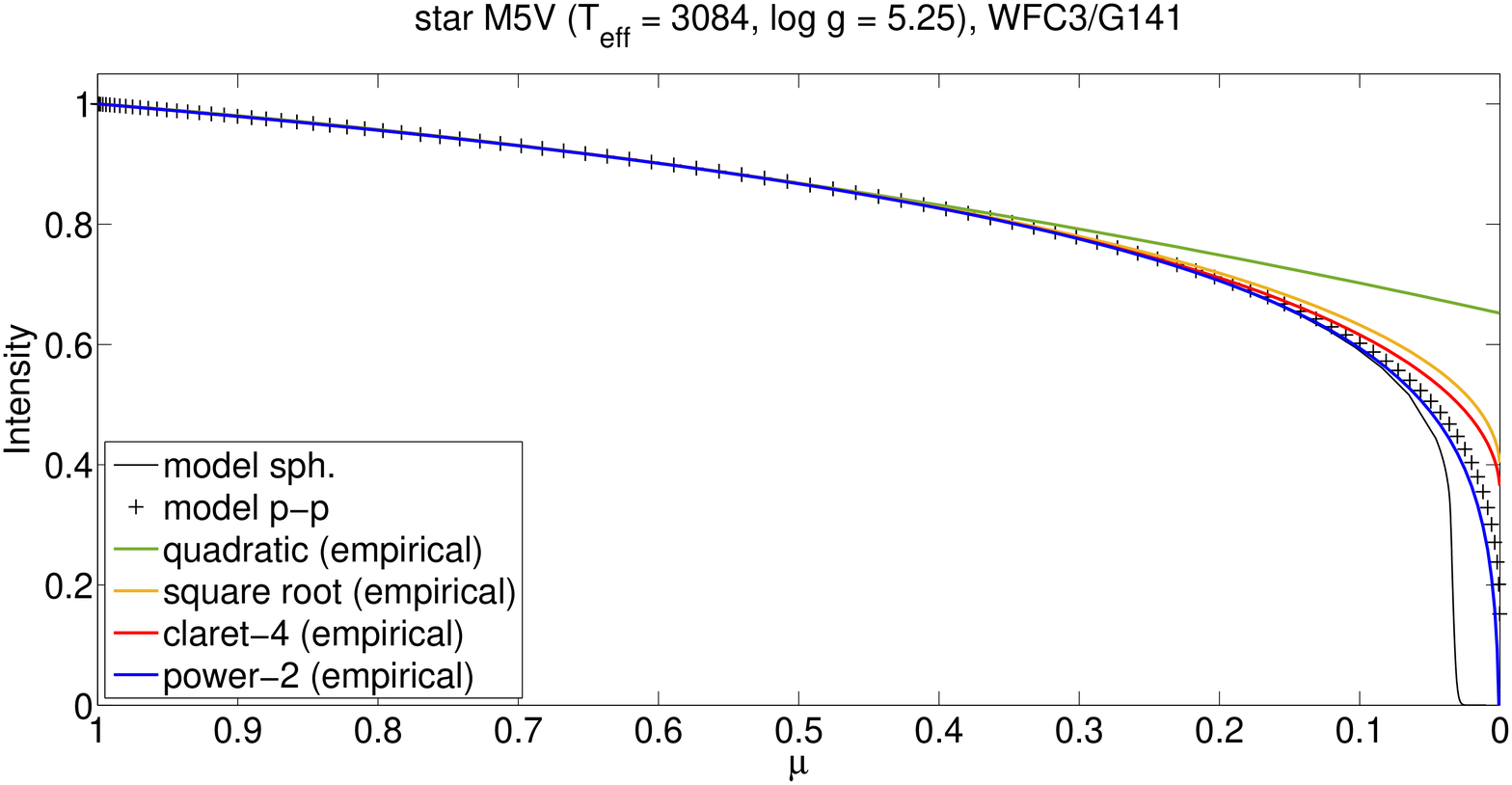}\\
\plotone{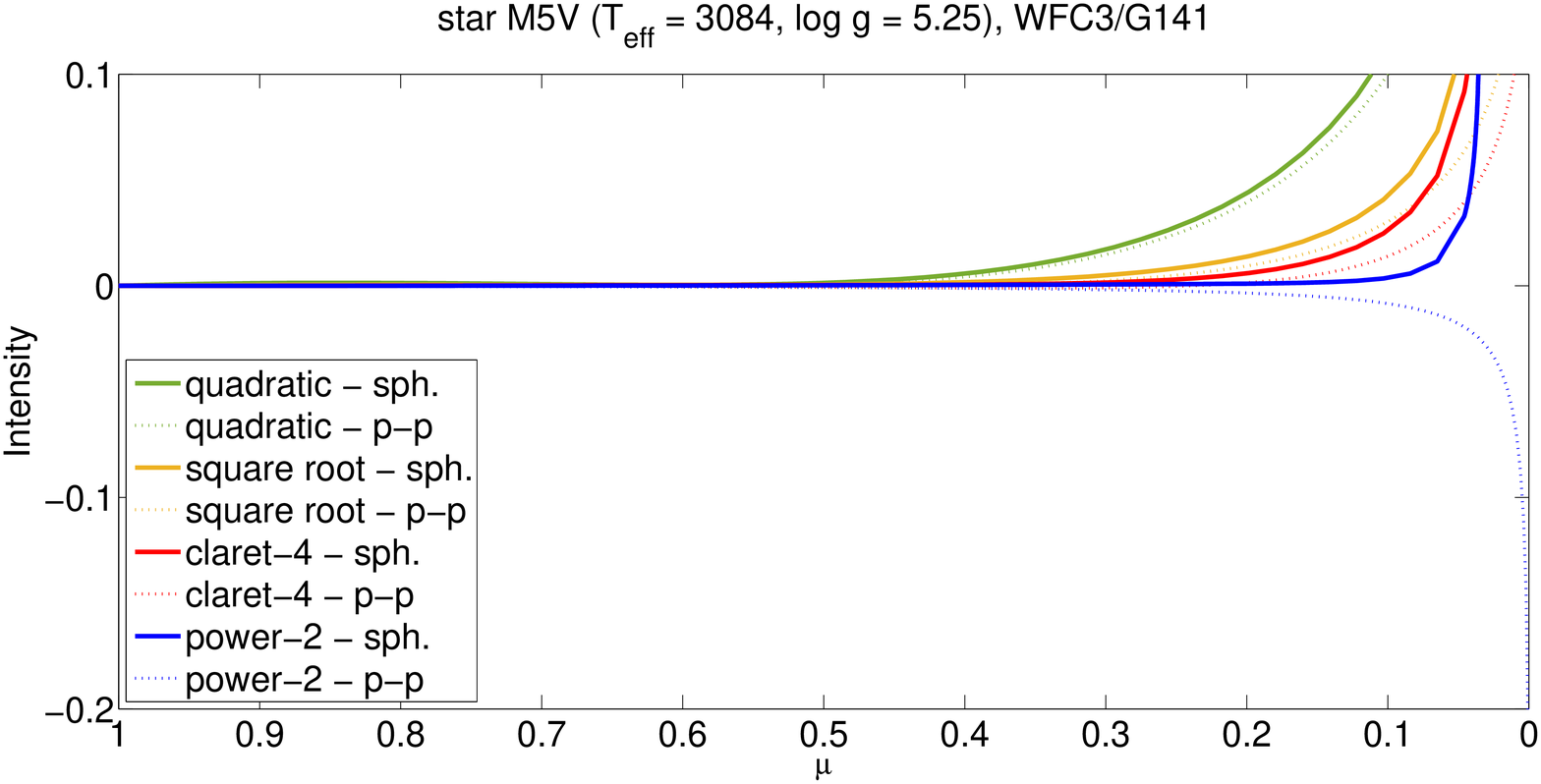}
\caption{Top panel: Plane-parallel (black ``+'') and spherical (black
  line) model-atmosphere intensities vs.\ $\mu$ for the M5\;V star in the WFC3/G141 passband. Parametric limb-darkening functions inferred from the transit light-curves (empirical models) are
  quadratic (green), square-root (yellow), claret-4 (red), and power-2
  (blue) laws. Bottom panel: large-scale plots of residuals of
  empirical limb-darkening laws for model-atmosphere
  intensities in the spherical and plane-parallel geometries (continuous and dashed
  lines, respectively).\label{fig8}}
\end{figure*}

\begin{figure*}[!t]
\epsscale{0.95}
\plotone{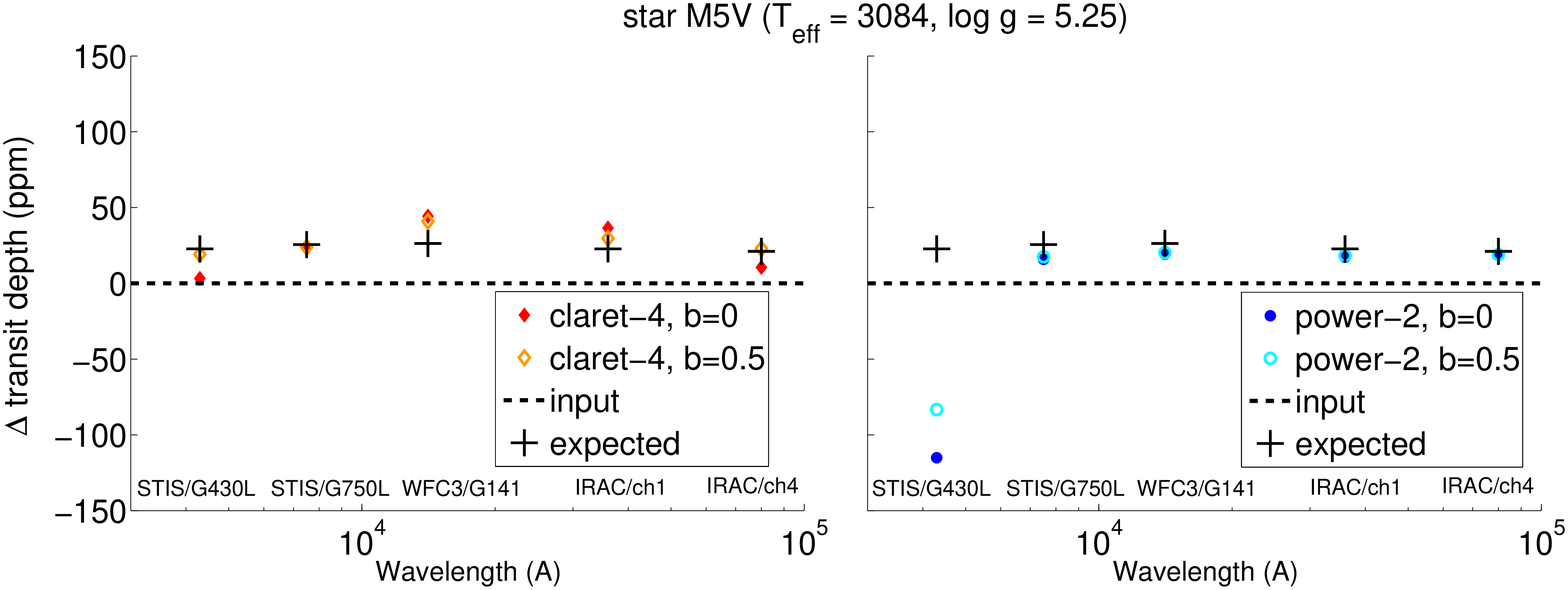}
\plotone{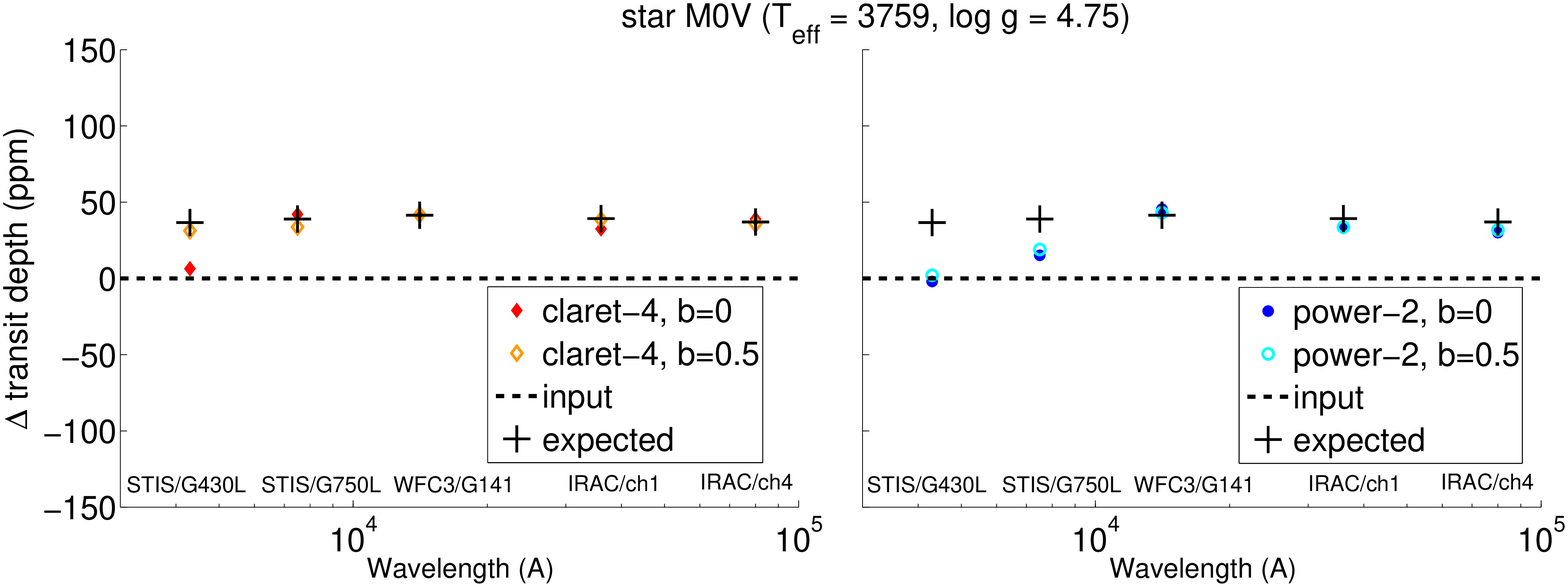}
\plotone{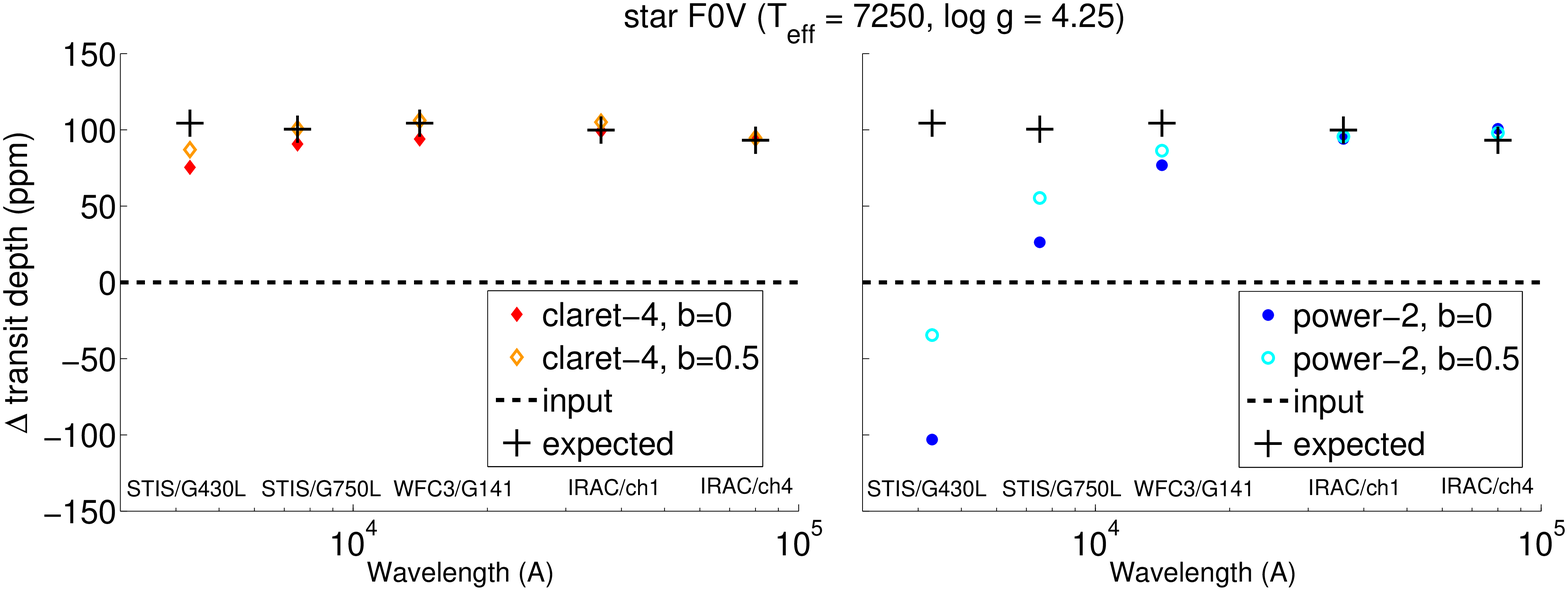}
\caption{Left panels: differences between the best-fit and input
  transit depths for the three stellar models, with $b=0$ (red, full
  diamonds), and 0.5 (orange, empty diamonds), using empirical
  claret-4 limb-darkening coefficients; the expected values are
  denoted by black ``+''.  Right panels: the same, but with empirical
  power-2 limb-darkening coefficients, $b=0$ (blue, full circles), and
  0.5 (cyan, empty circles). \label{fig9}}
\end{figure*}

\begin{figure*}[!t]
\epsscale{1.0}
\plotone{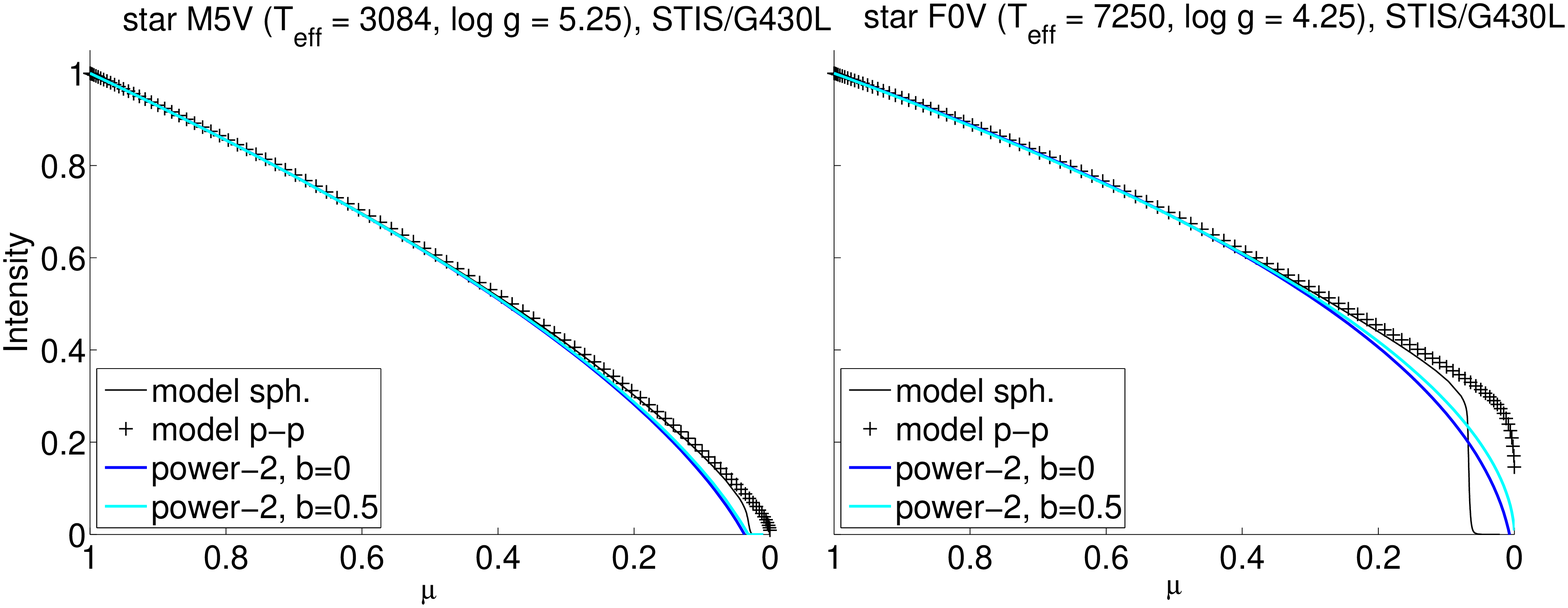}
\plotone{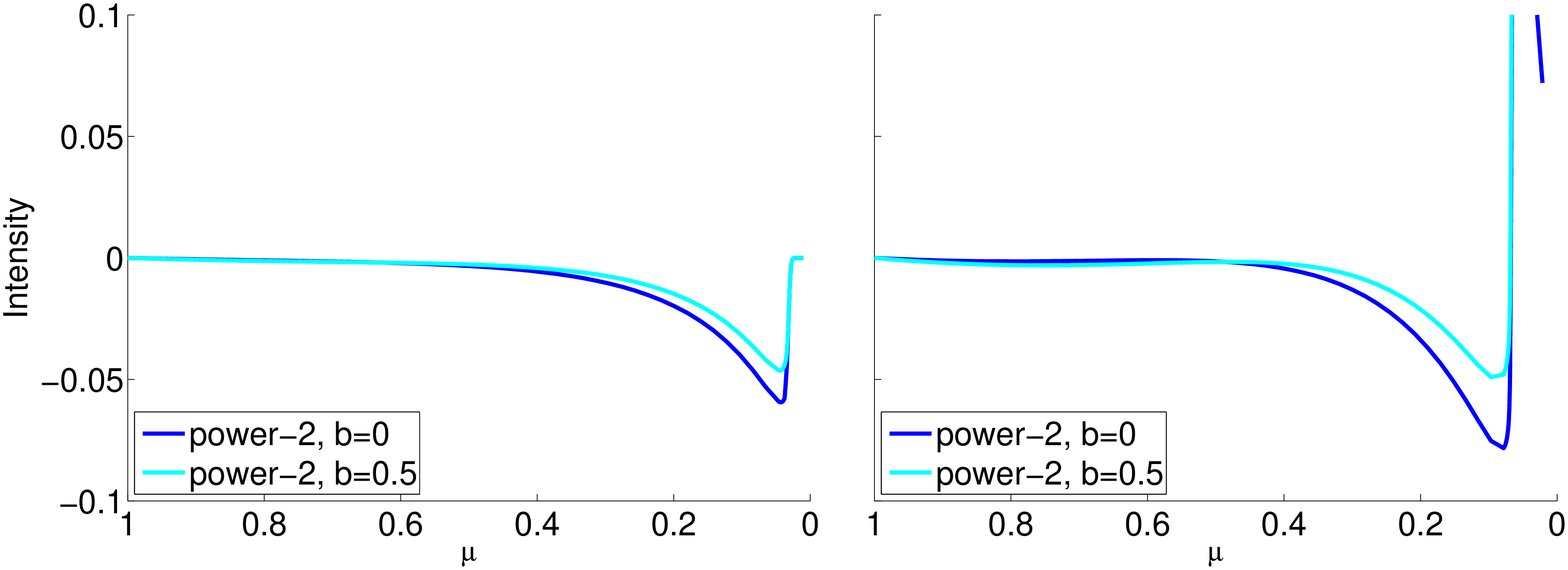}
\caption{Top, left panel: plane-parallel (black ``+'') and spherical (black line) angular intensity vs.\ $\mu$ for the M5\;V model in the STIS/G430L passband; parametric limb-darkening with empirical power-2 coefficients fitted to the transit light-curves with $b=0$ (blue) and 0.5 (cyan). Bottom, left panel: residuals between the parametric models and the spherical intensities. Right panels: the same for the F0\;V model in the STIS/G430L passband. \label{fig10}}
\end{figure*}

In empirical studies, it is generally convenient to analyse observed
transit light-curves using parameterized models 
in order to fit for
the unknown transit parameters and/or limb-darkening coefficients.  To
mimic this observational approach we employed \textsc{pylightcurve},\footnote{\texttt{https://github.com/ucl-exoplanets/pylightcurve}} our
pipeline dedicated to the fast computation of model transit
light-curves with a parametric limb-darkening profile. The
power-2 parameterization (Equation~\ref{eqn:law_power-2}) was
implemented in the code for this work. Based on our proposal, the power-2 law has been implemented also in the \textsc{batman} code \citep{kreidberg15b}.

\subsection{Plane-parallel vs.\ spherical limb-darkening models}
\label{sec:1Dvs3D}

As outlined in Section~\ref{sec:1d2d}, discrepancies between the
plane-parallel and spherical limb-darkening models are larger at
smaller $\mu$ (i.e., closer to the stellar limb), solely because of
the manner in which $\mu$ is defined, at least in the first step, for the spherical models. 
The spherical models present a steep drop-off in the normalized
intensity $I(\mu)/I(1)$, approaching zero at some $\mu >$0, while
for the plane-parallel models the intensity is significantly
greater than zero for all $\mu$ (see Figure~\ref{fig3}). 

It is reasonable to suppose that the `photometric' stellar radius
relevant to transit studies is better represented by the projected
radius of the intensity drop-off than by $\mu = 0$, the arbitrary uppermost
layer in the atmospheric model.
As a pragmatic approach, we assign this projected photometric radius
to the point in the intensity distribution at which the gradient
$\text{d}I(\mu)/{\text d}\mu$ reaches a maximum (estimated as the mean
$\mu$ value between the two consecutive $\mu$ values in the model with
the maximum difference quotient, $| I( \mu_{i+1} ) - I( \mu_i ) | / |
\mu_{i+1} - \mu_i |$). 
The corresponding radius, $r_0 = \sqrt{1-\mu_0^2}$, hereinafter
called the `apparent' radius,
is the ratio between the stellar photometric radius and the radius of
the outermost layer in the model.
Our approach is similar to what is suggested by \cite{wittkowski04} and \cite{espinoza15}, but here we compute the photometric radius for each passband, while they calculate one wavelength-averaged photometric radius.

With this working definition, the best-fit model parameters for any
of our simulated transit light-curves are expected to deviate from
their input values according to:
\begin{align}
p_{\rm expected}^2 &\simeq \left ( \frac{ p_{\rm input} }{ r_0 } \right )^2 
\label{eqn:p2_expected}\\
a_{R,{\rm expected}} &\simeq \frac{ a_{R,{\rm input}} }{ r_0 } 
\label{eqn:aR_expected}\\
i_{\rm expected} &\simeq i_{\rm input}.
\label{eqn:i_expected}
\end{align}
Table~\ref{tab3} reports the ranges of $r_0$ over the five instrument
passbands for the given stellar model, the corresponding percentage variation in
transit depth, $p^2 =  (R_p/R_*)^2$, and the absolute variation evaluated at $p_{\rm input} = 0.15$.
\begin{table}[t]
\begin{center}
\caption{Measured ranges of apparent stellar radius, $r_0$, over five
  instrument passbands for the three stellar models; corresponding
  percentage variations in transit depth, $p^2 =  (R_p/R_*)^2$; and absolute variations, evaluated
  at 
  $p_{\rm input}
  =0.15$. \label{tab3}}
\begin{tabular}{cccc}
\tableline\tableline
Sp. type & \multirow{2}{*}{$r_0$ range} & \multirow{2}{*}{$\frac{p_{\rm exp}^2 - p_{\rm input}^2}{p_{\rm input}^2}$} & $p_{\rm exp}^2 - p_{\rm input}^2$ \\
(\teff, \logg) & & & $|_{p =0.15}$\\
\tableline
M5\;V, & 0.99942--0.99953 & 0.09--0.12$\%$ & 21--26~ppm \\
(3084, 5.25) & & & \\
M0\;V, & 0.99908--0.99919 & 0.16--0.18$\%$ & 37--41~ppm \\
(3759, 4.75) & & & \\
F0\;V, & 0.99769--0.99794 & 0.41--0.46$\%$ & 93--104~ppm \\
(7250, 4.25) & & & \\
\tableline
\end{tabular}
\end{center}
\end{table}
In the analytical approximations represented by
Equations~\ref{eqn:p2_expected}--\ref{eqn:aR_expected},
 we find the apparent stellar radius to
be systematically smaller than the radius of the uppermost layer of
these models by 0.05--0.1$\%$ for the M dwarfs, and up to
$\sim$0.2$\%$ for the F0-star model; the corresponding percentage errors in
transit depths are about twice as large. For the case of a transiting
hot Jupiter ($p=0.15$), 
the discrepancies in transit depth
are at the level of
$\sim$20, 40, and 100~ppm for the two M dwarfs and for the F0 model,
respectively. The discrepancies measured for the F0 model currently
represent practical upper limits for exoplanet host stars, given that
$\sim$99\%\ of the current population is cooler, and hence have
less extended atmospheres (for given \logg). The wavelength-dependence
of the apparent radius is negligible over the parameter space explored
here, with a peak-to-peak amplitude 
of 11~ppm, in transit depth, from visible to
mid-infrared wavelengths in the worst-case scenario (see Table~\ref{tab3}).

\subsection{Accuracy of the theoretical limb-darkening laws}
\label{sec:accuracy_ld_laws}

We fitted the limb-darkening laws
to the plane-parallel intensity profiles by adopting a simple least-squares method in the fits. We checked, both by using subsets of the precalculated intensity grids and interpolating at different angles, that similar results would be obtained using a uniform sampling in $\mu$. Figure~\ref{fig4} shows the corresponding best-fit models, hereinafter referred as ``theoretical'' limb-darkening models, and their residuals, for the case of the M5\;V observed in the WFC3 passband. The full list of models and the relevant residuals are reported in Figure~\ref{fig4app} (Appendix~\ref{app1}).
The power-2 law
(Equation~\ref{eqn:law_power-2}) outperforms the other
two-coefficient laws at describing the stellar limb-darkening of all
stars observed at near to mid-infrared wavelengths with the
\textit{HST}/WFC3 and \textit{Spitzer}/IRAC instruments;  in some
cases, the power-2 model outperforms even the corresponding claret-4
one. At visible wavelengths, the square-root and power-2 models have
comparable success, while the claret-4 models fit best. The
average errors in specific intensity predicted by the power-2 models
are in the range 0.1--1.0\%, with a maximum error up to $\sim$5--7\%\
for the F0\;V model in the visible passbands. The claret-4 models are
more uniformly robust among all the configurations, with average
errors in the range 0.05--0.6\%\ and maximum errors $<$4\%. The
quadratic models are the least accurate of those tested, with
average errors in the range 1--6\%\ and maximum errors of up to 25\%\
(for the M0\;V model in the WFC3 passband).

\subsection{Transit models with theoretical limb-darkening coefficients}
\label{sec:ldc_fixed}

We measured the potential biases in the model transit depths by fixing
the limb-darkening coefficients at the theoretical values obtained
from the plane-parallel stellar-atmosphere models and fitting
the
exact light-curves described in Section~\ref{sec:ideal}.
The free parameters
in the fit were $p$, the ratio of planet-to-star radii, $a_R$, the
orbital semimajor axis in units of the stellar radius, and $i$, the
orbital inclination. We used a Nelder--Mead minimization algorithm to
find the values of these parameters which minimize the residuals
between the model fits and the exact light-curves. We then carried out
Markov-chain Monte-Carlo (MCMC) runs with 300,000 iterations to assess
the robustness of the point estimates.  Unlike previous
investigations reported in the literature
(e.g. \citealp{csizmadia13}), we seek to isolate the potential biases
arising from the analysis method, and particularly the use of
simplified geometry and a parameterization to characterize the stellar
limb-darkening. No other astrophysical sources of error are considered
in this study.

Figure~\ref{fig5} illustrates the differences between the best-fit
transit depths and input values for $i=90^\circ$; the expected values
(from Equations~\ref{eqn:p2_expected}--\ref{eqn:aR_expected}) are also
indicated. For all stellar models, the results are less dependent on
the parametric law at longer wavelengths; this is to be expected,
since the limb-darkening is smaller at longer wavelengths. In
particular, the transit depths obtained at 8$\mu$m (IRAC channel 4)
are all within 45~ppm of expected values, or within 13~ppm if
adopting the power-2 or claret-4 coefficients. Overall,
the transit depths obtained using the claret-4 coefficients deviate by
less than $\sim$20~ppm from expected values, other than for the
M5\;V model in the visible passbands, where the discrepancy reaches
34~and 80~ppm for the STIS/G750L and STIS/G430L passbands. The
peak-to-peak amplitudes
in best-fit transit depths over the five passbands
are 94, 28, and 8~ppm, going from the coolest to the
hottest model. The results obtained with the power-2
coefficients are more robust for the
cooler stars, and are within 44~ppm of expected values,
except for the F0\;V model in the visible passbands, where
the inferred transit depths are 105 and 88~ppm larger for 
the STIS/G750L and STIS/G430L passbands. The peak-to-peak amplitudes
in best-fit transit depths over the five passbands
are 47, 44, and 102~ppm, again from the coolest to the hottest model.
The quadratic-law coefficients have the largest scatter in
the best-fit transit depth across the different passbands for all
models, with peak-to-peak amplitudes of 250, 164, and 107~ppm.

Even though the true value of the transit depth is not known in a
`real-world' scenario, the presence of biases can be revealed by
time-correlated noise in the light-curve residuals. Figure~\ref{fig6}
shows the residuals between the exact light-curve and the best-fit
parametric model for the M5\;V star in the WFC3 passband. The full list of light-curve residuals is reported in Figure~\ref{fig6app}. The amplitudes of the time-correlated residuals
(maximum discrepancies from zero) are in the ranges 97--456,
8--105, and
11--75~ppm with quadratic, power-2, and claret-4
models respectively. Residuals at infrared wavelengths are typically
smaller than in the visible, as expected. \cite{neilson17} report similar amplitudes for the residuals between the exact light-curves, computed with their CLIV stellar-atmosphere models, and parametric light-curve models.
For comparison, residuals with $\sim$10 ppm root mean square (rms) amplitude have been obtained from the phase-folded Kepler photometry of several targets (e.g., \citealp{barnes11, muller13}), and $\sim$50--200 ppm rms amplitude is typically obtained for the white light-curves observed with the \textit{HST}/WFC3 (e.g., \citealp{deming13, tsiaras16, tsiaras16b}).

It is possible that better results would be obtained if the limb-darkening coefficients were fitted adopting a different sampling in $\mu$ (e.g., uniform in $r$ rather than in $\mu$),  a different method (e.g., imposing flux-conservation), and/or using spherical intensities (e.g., \citealp{sing10, claret11, howarth11, espinoza15}). A detailed study of the different approaches is beyond the scope of this paper, but the analysis in Section~\ref{sec:ldc_free} provides some clear indications.




\subsection{Transit models with \textit{empirical} limb-darkening coefficients}
\label{sec:ldc_free}
\subsubsection{Edge-on transits}
\label{sec:papild_b0}
We repeated the fits to the exact light-curves with the limb-darkening
coefficients as free parameters (in addition to $p$, $a_R$, and
$i$). The increased flexibility allows parametric models that better
match the transits, as shown by the smaller time-correlated residuals
in Figures~\ref{fig6} and \ref{fig6app} (Appendix~\ref{app1}). The residual amplitudes are in the range 10--63,
0--31, and 0--4~ppm with quadratic, power-2, and claret-4 models respectively.
The time-correlated residuals due to
imperfections in the transit models with {\em empirical} limb-darkening coefficients are hardly detectable with current instruments.

Figure~\ref{fig7} shows the corresponding best-fit transit
depths. Despite the very small light-curve residuals, the inferred
transit depth can be significantly biased. The bias obtained with
quadratic limb-darkening is roughly linear in the logarithm of
wavelength for the two M dwarfs, ranging from 27--40~ppm at 8$\mu$m
to 200--225~ppm at 0.4$\mu$m; there is no evident trend for the F0\;V
model, but the transit depth is again systematically over-estimated, by
17--52~ppm. The square-root and power-2 laws have similar
performances, with deviations from the expected values smaller than 
45~ppm, except for three `bad' points: the STIS/G430L passband for
the M5\;V model, and the STIS/G430L and STIS/G750L passbands for the
F0\;V model, for which the transit depth estimates are smaller than
the apparent values by 170--138, 207--205, and 74--73~ppm for the two
laws. The transit depth estimates obtained when fitting
the claret-4 coefficients are the most accurate, with deviations from
the expected values being smaller than 30~ppm (largest in the STIS/G430L
passband) and peak-to-peak amplitudes of 41, 36, and 24~ppm from the
coolest to the hottest model.

Figure~\ref{fig8} shows the empirical limb-darkening models and their
residuals for the case of the M5\;V observed in the WFC3 passband. The full list of models and the relevant residuals are reported in Figure~\ref{fig8app} (Appendix~\ref{app1}). It appears that the
empirical limb-darkening models are better constrained at larger $\mu$
values, corresponding to the inner part of the disk. We also found that the empirical coefficients can be obtained from the stellar-atmosphere models if a uniform sampling in $r$ rather than in $\mu$ is used. However, if a functional
form is not able to reproduce the intensity profile, the empirical
model will be particularly discrepant at the limb, causing larger biases in the best-fit transit depth compared to the case of theoretical coefficients with a uniform sampling in $\mu$. Quadratic models,
especially, always overpredict the intensities at the limb, so that an
apparently larger planet would be needed to occult the extra stellar flux, in
agreement with the larger transit depth estimates.


\begin{figure*}[!t]
\epsscale{1.0}
\plotone{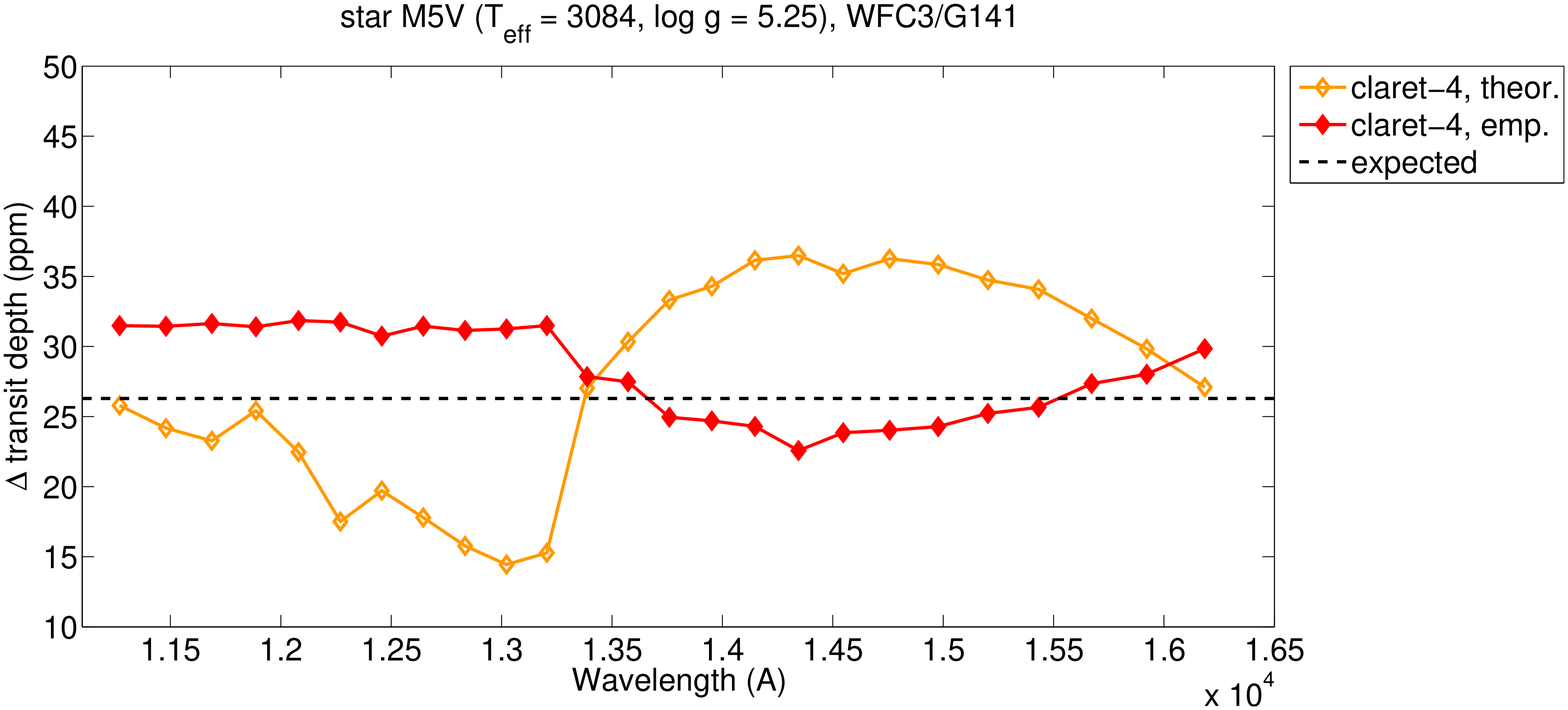}
\plotone{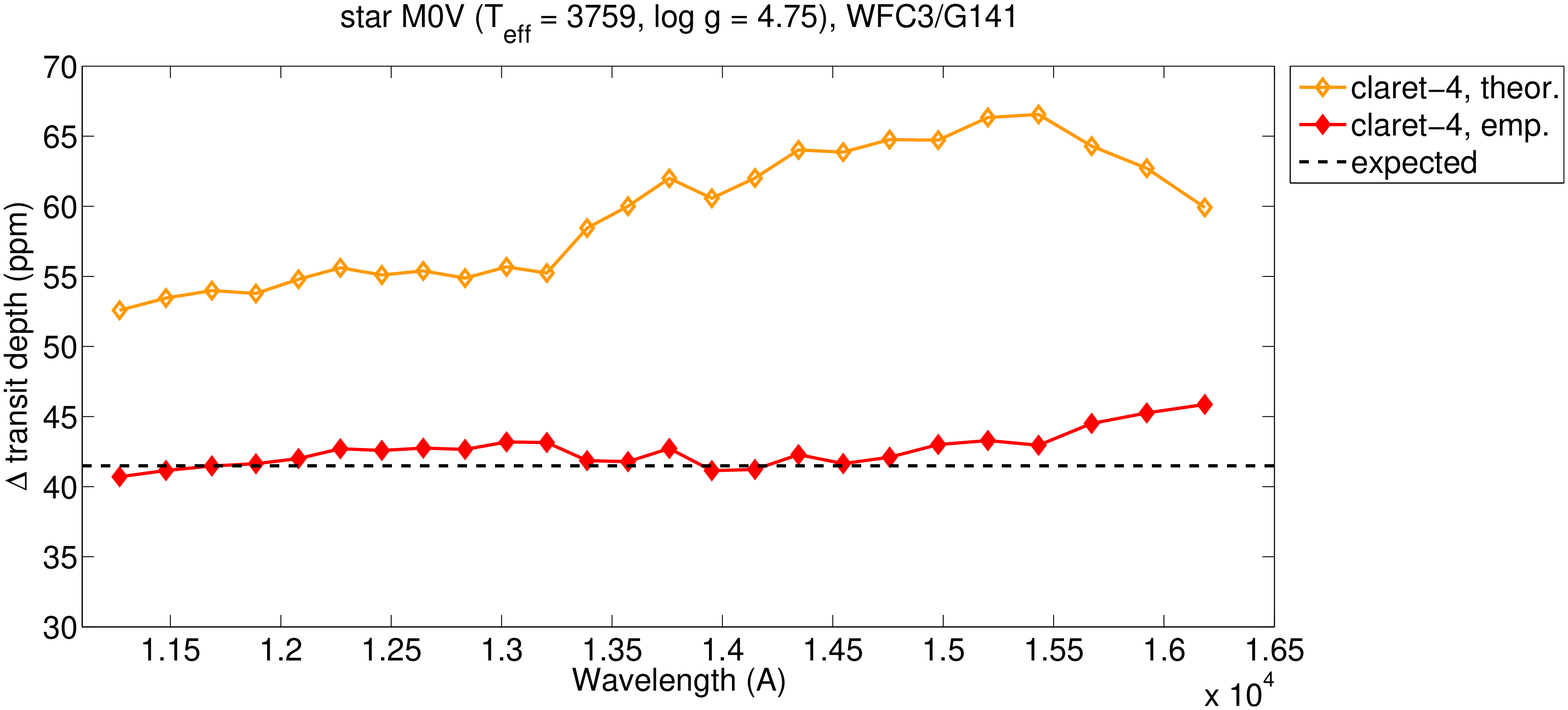}
\plotone{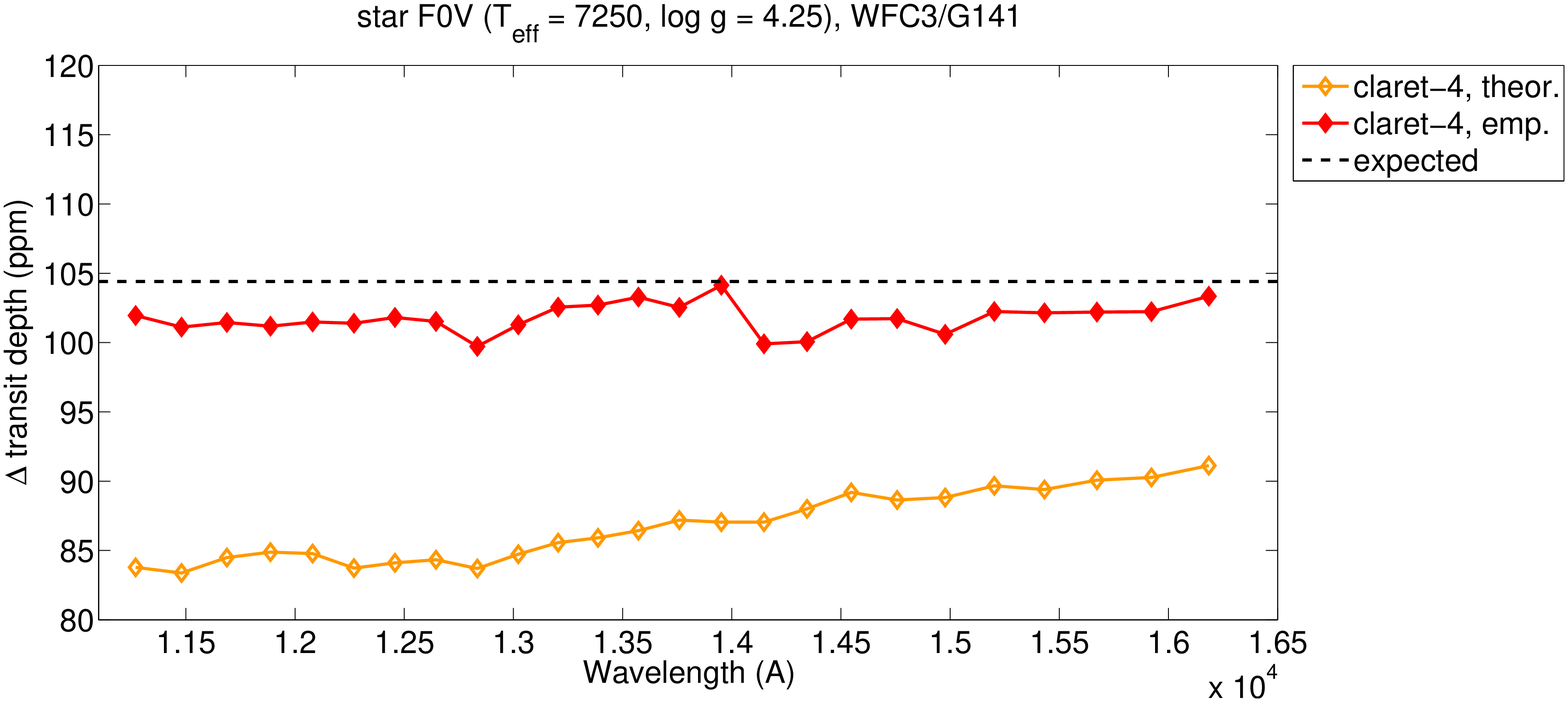}
\caption{Top panel: differences between the best-fit and the input
  transit depths for the M5\;V model over twenty-five wavelength bins
  in the WFC3/G141 passband, with $b=0$, using theoretical (orange, empty diamonds)  and empirical (red, full
  diamonds) claret-4
  coefficients. \label{fig11}}
\end{figure*}

\subsubsection{Inclined transits}
For randomly orientated orbits, the inclinations $i$ are distributed
such that the probability density of $\cos{i}$ is uniform between 0
and 1.  For circular orbits, therefore, the impact parameter, $b = a_R
\cos{i}$, is uniformly distributed between 0 and $a_R$, the semimajor
axis in units of stellar radius.  An exoplanet transits if and only if
$0 \le b < 1 + p$.  We tested the ability to constrain the
stellar limb-darkening profile and to measure the correct transit depth
changes for the case $b=0.5$. This study was conducted for the
claret-4 and power-2 laws, as they led to more robust results than
the other parameterizations. In this configuration, the area of the
stellar disk with $r < b-p = 0.35$, or $\mu \gtrsim 0.94$, is never
occulted by the transiting exoplanet.

Figure~\ref{fig9} shows the comparison between the transit depths
estimated for the cases with $b=0$ and 0.5, using the claret-4 and power-2 laws. In most cases, there are no significant differences in transit depth obtained for the cases with $b=0$ and 0.5. The largest discrepancies (29--68~ppm) are registered for the three `bad' points of the power-2 law, which are highlighted in Section~\ref{sec:papild_b0}.
The empirical limb-darkening profiles are also very
similar. Figure~\ref{fig10} shows the difference for the two most
discrepant cases. The parametric models obtained from the transits
with $b=0.5$ approximate the intensities at the limb slightly better
than those obtained from the transits with $b=0$, but the bias is
also significant. In general, it appears that, if a parametric law
does not allow a good approximation of the limb-darkening profile, the
empirical model is optimized toward the center of the disk and
significantly deviates at the limb (see also the quadratic fits in
Figure~\ref{fig8}). In these cases, inclined transits automatically
attribute slightly higher weights to the intensities at the limb, as
the planet spends more time occulting areas further from the
center. Even if the innermost region of the stellar disk is not sampled
during the transit, this is not often a problem, at least if $b
\lesssim0.5$, because the intensity gradient does not vary
significantly near the center.  The examples discussed here may
suggest that inclined transits can lead to smaller biases in the
inferred transit parameters and limb-darkening profiles than edge-on
transits, but the improvements appear to be quite small and, more
importantly, the error bars have not yet been considered.


\subsection{Narrow-band (WFC3-like) exoplanet spectroscopy}
\label{sec:WFC3}
The results discussed in Section~\ref{sec:papild_b0} may suggest that
it is difficult to model the depth of a hot-Jupiter transit with an
absolute precision better than $\sim$10--30~ppm, because of the
intrinsic limitations of the stellar limb-darkening parameterizations
(claret-4 being the most accurate among those currently
used). This potential bias will be particularly important when
analyzing visible to mid-infrared exoplanet spectra measured with
the JWST, as it is comparable with the
instrumental precision limit \citep{beichman14}.

In this section, we investigate the potential errors in relative
transit depth over multiple narrow bands within a limited wavelength
range, so-called ``narrow-band exoplanet spectroscopy''.  This kind of
measurement has been performed with \textit{HST}/STIS
\citep{charbonneau02, vidal03, vidal04}, \textit{Spitzer}/IRS
\citep{grillmair07, grillmair08, richardson07}, \textit{HST}/NICMOS
\citep{swain08, swain09, swain09b, tinetti10}, \textit{HST}/WFC3
\citep{deming13, crouzet14, fraine14, knutson14, knutson14b,
  kreidberg14, kreidberg14b, kreidberg15, mccullough14, stevenson14,
  line16, tsiaras16, tsiaras16b}, and other space- and ground-based
spectrographs (e.g. \citealp{redfield08, snellen08, swain10,
  danielski14}), leading to the discovery of a long
list of atomic, ionic, and molecular species in the atmospheres of
exoplanets. 

Since the detection of the chemical species relies on their
spectral features, it is not affected by a constant offset in transit
depth; hence only the errors in  transit depth differences at multiple
wavelengths, referred to as relative error, are important.
Here, we study the case of exoplanet spectroscopy with
\textit{HST}/WFC3, for which the narrow-band spectra reported in the
literature often have $\sim$20--40~ppm error bars. 

To fix the ideas, we considered 25 wavelength bins, identical
to those ones adopted in \cite{tsiaras16b}, to generate one set of
exact light-curves (as in Section \ref{sec:ideal}) for each stellar
model, and calculate the theoretical limb-darkening coefficients. We
modeled each exact light-curve using the two approaches outlined in
Sections~\ref{sec:ldc_fixed} and \ref{sec:ldc_free}, i.e., with the
theoretical and empirical limb-darkening coefficients,
respectively. 

Figure~\ref{fig11} shows the spectra obtained by using the
most accurate claret-4 law. The spectra calculated with the
theoretical limb-darkening coefficients are offset by $+1$, $+18$,
and $-18$~ppm, on average, in excellent agreement with the measured
biases for the broadband light-curves reported in
Section~\ref{sec:ldc_fixed}. The relative errors, as measured by the
peak-to-peak amplitudes, are 22, 14, and 8~ppm, from the coolest to
the hottest model. The use of empirical limb-darkening
coefficients reduces the spectral offsets, in these cases, to less
than 3~ppm, and also reduces the peak-to-peak amplitudes down to 9, 5,
and 4~ppm, respectively. Note that, in all configurations, the
peak-to-peak amplitudes across the WFC3 narrow bands are smaller than
the respective amplitudes for the broadband photometry from visible to
mid-infrared reported in Section~\ref{sec:ldc_fixed} and
\ref{sec:ldc_free}, and they also decrease with the increasing model
temperature.

We remind the reader that the results discussed up to this section
focus on the potential biases due to the approximate stellar
limb-darkening parameterizations, in the absence of noise. The limits of
the actual parameter fitting for light-curves with a low, but
realistic, noise level are discussed in Section \ref{sec:lc_fit_emp}. An additional
complication is the presence of temporal gaps in the transit
light-curves observed with instruments onboard the \textit{HST}, and from
satellites operating on low orbits in general. The impact of such
gaps in the retrieval of the transit parameters will be discussed in a
separate paper (Karpouzas et al. 2017, in preparation).



\begin{figure*}[!t]
\epsscale{0.9}
\plotone{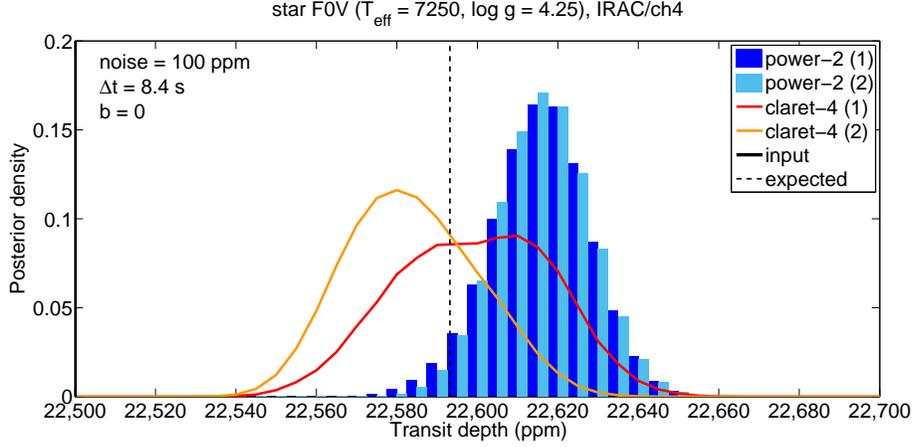}
\caption{MCMC-sampled posterior distributions of modeled transit
  depth for the edge-on transit, F0\;V model, IRAC/ch4 passband, with
  100~ppm gaussian noise (fitting $p$, $a_R$, $i$, normalization factor, and limb-darkening
  coefficients). The histogram channels
  (blue and light-blue) are for two chains with 1\,500\,000
  iterations, using the power-2 law; the channels are half-thick
  and shifted to improve their visualization. The red and orange lines
  denote the analogous posterior distributions when using the claret-4
  law. The input and expected transit depths are also indicated (black
  vertical lines, continuous and dashed, respectively). \label{fig12}}
\end{figure*}
\begin{figure*}
\epsscale{0.82}
\plotone{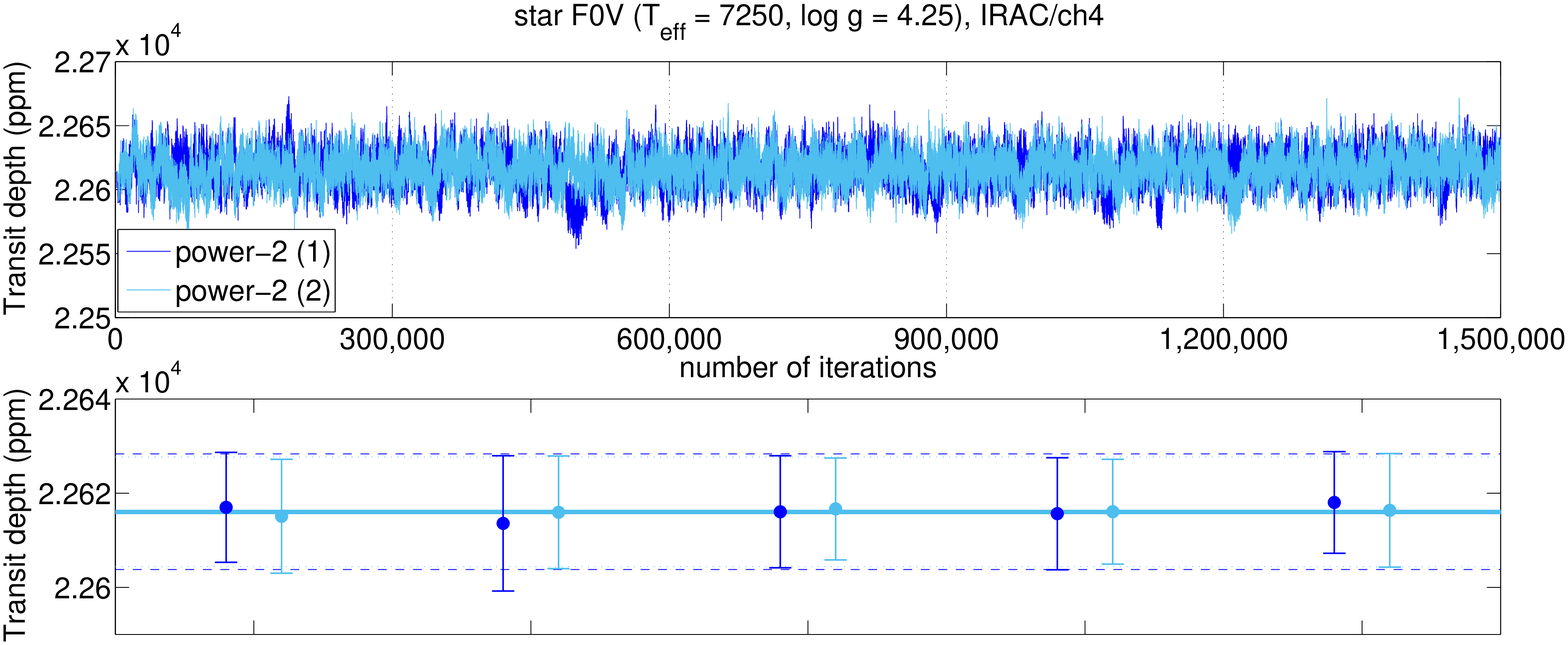}
\plotone{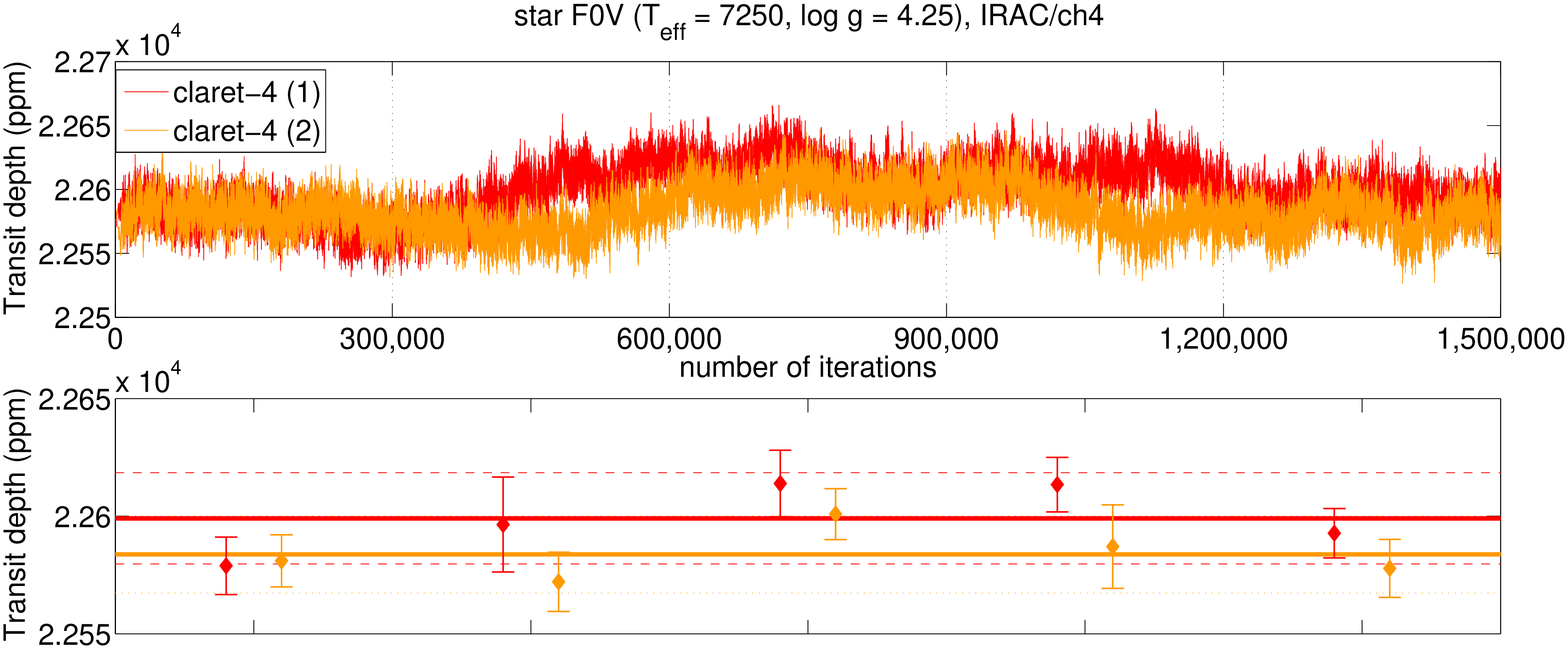}
\caption{Top panels: transit depth chains for the edge-on transit,
  F0\;V model, IRAC/ch4 passband, with 100~ppm
  gaussian noise, fitting $p$, $a_R$, $i$, normalization factor, and power-2 (blue and light-blue) or claret-4 (red and orange) limb-darkening
  coefficients. Bottom panels: mean
  values and standard deviations calculated over fractional chains
  with 300\,000 iterations; the horizontal lines indicate the mean
  values calculated over the full chains (continuous lines), and the
  mean values plus or minus the standard deviations (dashed
  lines). \label{fig13}}
\end{figure*}

\begin{figure*}[!t]
\epsscale{1.0}
\plotone{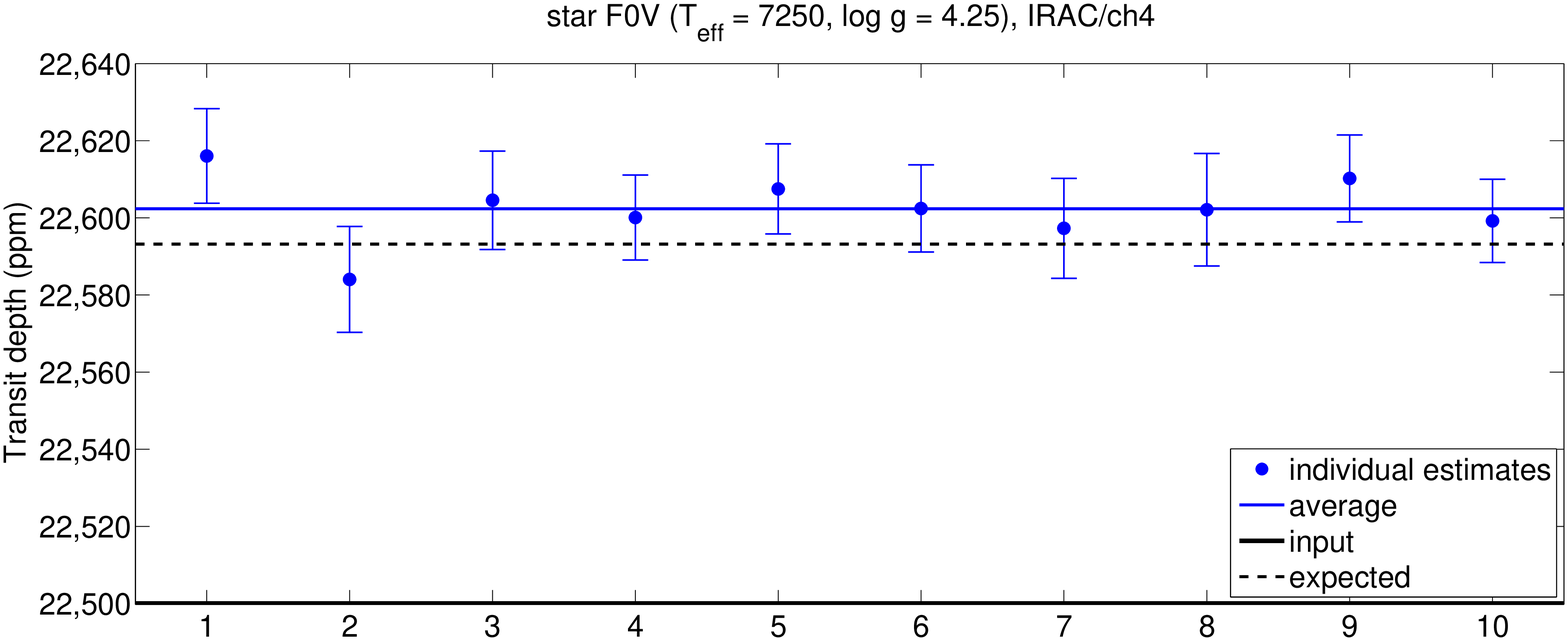}
\caption{Transit depth estimates for the edge-on transit in front of
  the F0\;V model, IRAC/ch4 passband, with 100~ppm gaussian noise
  (different noise time series, blue); fitting $p$, $a_R$, $i$, normalization
  factor, and
  power-2 limb-darkening coefficients. Average over the ten light-curves (blue, continuous line),
  input (black, continuous line) and expected (black, dashed line)
  values. \label{fig14}}
\end{figure*}

\section{Light-curve fitting with empirical limb-darkening coefficients}
\label{sec:lc_fit_emp}

Sections~\ref{sec:ldc_fixed} and~\ref{sec:ldc_free} discussed the
intrinsic biases due to the use of parametric models with theoretical
(plane-parallel) or empirical limb-darkening coefficients. We now
consider fitting the transit models with empirical limb-darkening
coefficients to more realistic light-curves with noise. Those
light-curves are obtained by adding gaussian time series to the exact
light-curves (see Section~\ref{sec:ideal}). The standard
deviation for the noise time series was set at $\sim$100~ppm, which is
similar to the best photon-noise limit possible for a short-cadence
Kepler frame \citep{kih09}, or for a single \textit{HST}/WFC3 scan
\citep{deming13, tsiaras16b}, taking into account the different
integration times.  We focused on four cases: two transits with
$b=0$ and 0.5 across the hottest model star, at 8 $\mu$m (F0\;V
model, IRAC/ch4 passband), and across the coolest model
star, at $\sim$430 nm (M5\;V model, STIS/G430L passband). The cases
considered correspond to those for which the limb-darkening effects
are weakest (mid-infrared) and strongest (visible).

\subsection{F0\;V model, IRAC/ch4 passband}
\label{sec:noisy_F0V_irac4}
Figure~\ref{fig12} shows the transit depth posterior distributions for
the $b=0$ transit  and F0\;V model in the IRAC/ch4
passband, MCMC sampled with 1\,500\,000 iterations. Figure~\ref{fig13}
reports the relevant chains. Similar parameter chains are computed, in
parallel, for $a_R$, $i$, the limb-darkening coefficients, a
normalization factor, and the likelihood's variance. In two cases, the
limb-darkening coefficients for the power-2 law are fitted, in the other
two cases the claret-4 ones. 
The sampled posterior distributions of the transit depth, using the
power-2 limb-darkening law, are almost identical and, in particular,
the mean values and standard deviations differ by less than
1~ppm. Even considering subsets of the chains with 300\,000 samples,
the mean values and standard deviations are stable to better than
5~ppm. It may appear from Figure~\ref{fig12} that the results are
biased, given that the peak of the posterior distribution is more than
1$\sigma$ away from the expected value. However, by repeating the
analysis with different noise realizations, the best-fit transit depth is
within 1$\sigma$ of the expected value in 7 cases out of 10,
consistent with the expectations for gaussian noise (see
Figure~\ref{fig14}). Interestingly, the average of the individual
best-fit transit depths is 9~ppm above the expected value, which is
very close to the 7~ppm bias measured in the noiseless case
(Section~\ref{sec:ldc_free}).  The chains calculated with the claret-4
law present significant long-term modulations, resulting in wider
posterior distributions; there are also moderately large differences
between the two repetitions, indicating that they have not converged. The lack of convergence when fitting the claret-4 coefficients is not a surprise, and it is due to strong correlations between the coefficients and with the other transit parameters.

Figures~\ref{fig15app} and \ref{fig16app} (Appendix~\ref{app2}) show the posterior distribution
and chains obtained for the inclined transit, with $b=0.5$, using the
power-2 limb-darkening law. The posterior distributions are wider than
those obtained for the edge-on transit, with $1\sigma \approx
20$~ppm rather than 12~ppm. The estimates from the partial chains with
300\,000 samples are also less robust, with the corresponding mean values
scattered over a 12~ppm interval.

The accuracy and precision of the empirical limb-darkening profiles are discussed in Appendix~\ref{app3}.


\subsection{M5\;V model, STIS/G430L passband}
\label{sec:noisy_M5V_STIS_G430L}
\begin{figure*}[!t]
\epsscale{0.9}
\plotone{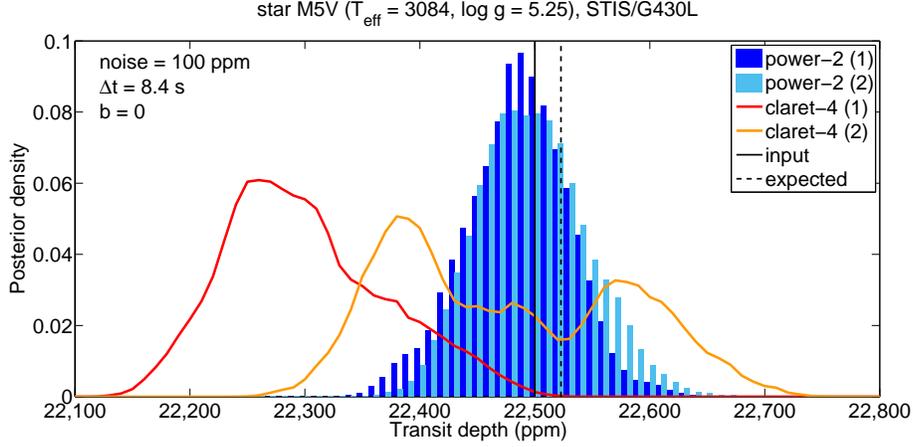}
\caption{MCMC-sampled posterior distributions of modeled transit
  depth for the edge-on transit, M5\;V model, STIS/G430L passband, with
  100~ppm gaussian noise (fitting $p$, $a_R$, $i$, normalization factor, and limb-darkening
  coefficients). The histogram channels
  (blue and light-blue) are for two chains with 1\,500\,000
  iterations, using the power-2 law; the channels are half-thick
  and shifted to improve their visualization. The red and orange lines
  denote the analogous posterior distributions when using the claret-4
  law. The input and expected transit depths are also indicated (black
  vertical lines, continuous and dashed, respectively). \label{fig19}}
\end{figure*}

\begin{figure*}
\epsscale{0.82}
\plotone{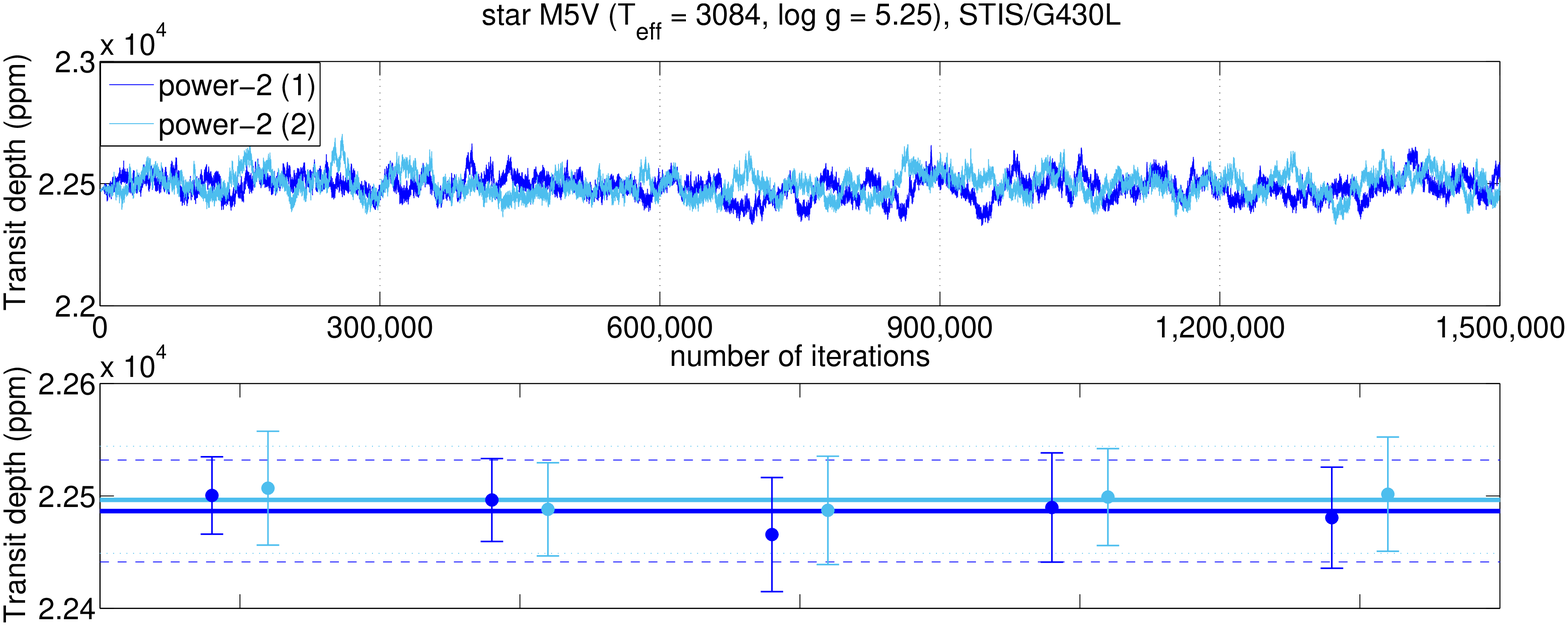}
\plotone{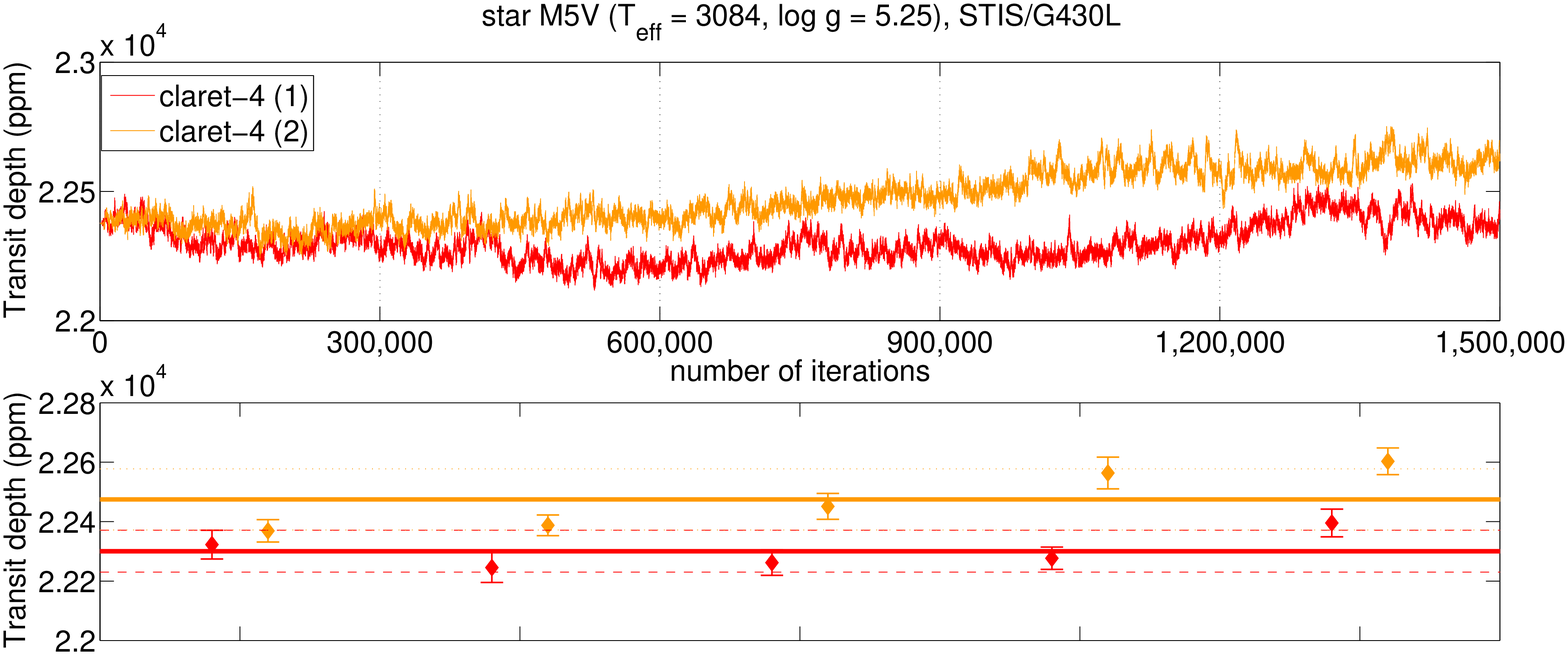}
\caption{Top panels: transit depth chains for the edge-on transit,
  M5\;V model, STIS/G430L passband, with 100~ppm
  gaussian noise, fitting $p$, $a_R$, $i$, normalization factor and power-2 (blue and light-blue) or claret-4 (red and orange) limb-darkening
  coefficients. Bottom panels: mean
  values and standard deviations calculated over fractional chains
  with 300\,000 iterations; the horizontal lines indicate the mean
  values calculated over the full chains (continuous lines), and the
  mean values plus or minus the standard deviations (dashed
  lines). \label{fig20}}
\end{figure*}

\begin{figure*}
\epsscale{1.0}
\plotone{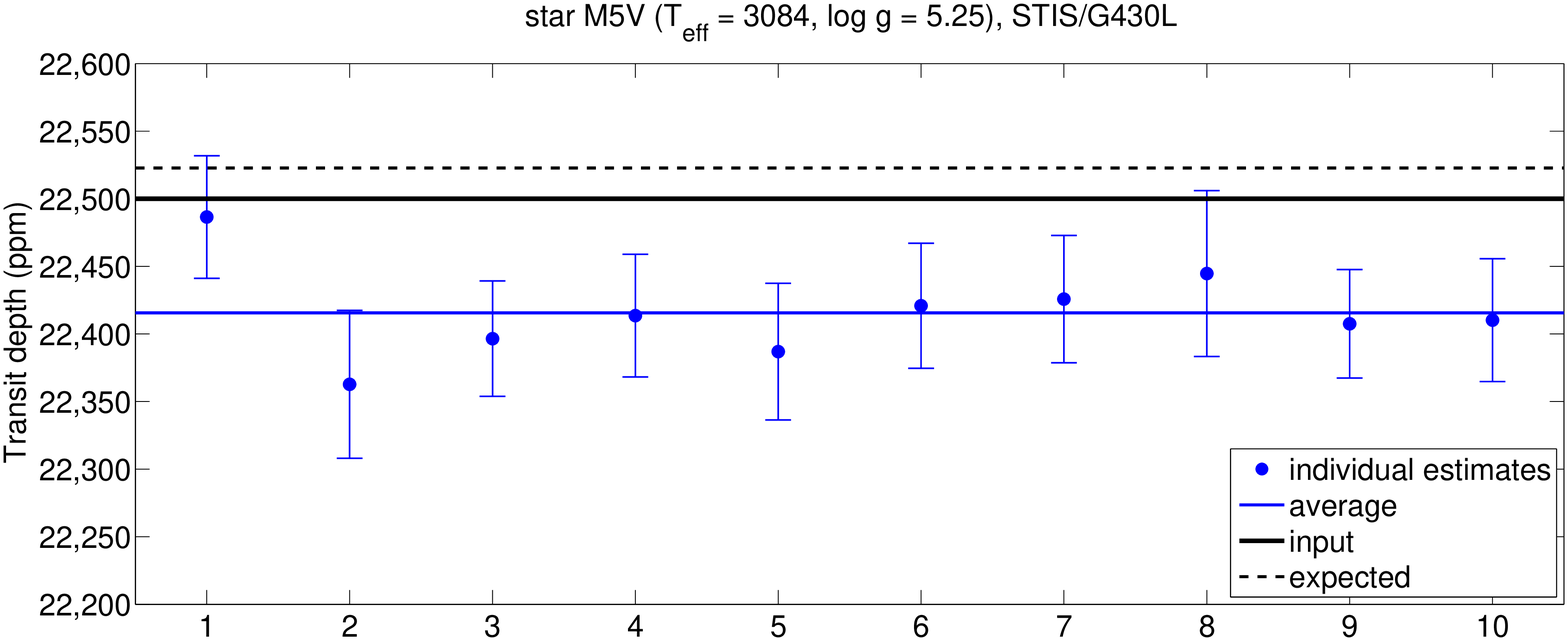}
\caption{Transit depth estimates for the edge-on transit in front of the M5\;V model, STIS/G430L passband, with 100~ppm gaussian noise (different noise time series, blue); fitting $p$, $a_R$, $i$, normalization factor, and power-2 limb-darkening coefficients. Average over the ten light-curves (blue, continuous line), input (black, continuous line) and expected (black, dashed line) values. \label{fig21}}
\end{figure*}

We conducted corresponding studies for two transits in front of the
M5\;V model in the STIS/G430L passband.  Figure~\ref{fig19} shows the
transit depth posterior distributions for the edge-on
transit, and Figure~\ref{fig20} reports the relevant chains. The sampled
posterior distributions of the transit depth, using the power-2
limb-darkening law, are in good agreement and, in particular, the
mean values differ by 10~ppm, with standard deviations of 45 and
48~ppm, respectively. As expected, the error bars are larger than those
obtained for the less limb-darkened case (with identical noise).
The estimates from the fractional chains with 300\,000
samples may differ by up to 35~ppm, and the relevant standard deviations
are in the range 34--51~ppm. Even in this case, the MCMC process failed to
converge when fitting the claret-4 coefficients.  Figure~\ref{fig21}
reports the transit depth estimates obtained with 10 different noise realizations,
using the power-2 law. Note that their average is
significantly biased in the same direction as the bias obtained in
absence of noise (see Section~\ref{sec:ldc_free}), and the 1-$\sigma$
error bars are smaller than the bias.

Figures~\ref{fig22app} and \ref{fig23app} (Appendix~\ref{app2}) show the posterior distribution and chains obtained for the inclined transit, with $b=0.5$, using the power-2 limb-darkening law. The posterior distributions are wider than those obtained for the edge-on transit, i.e. 1$\sigma  \approx$ 115~ppm rather than 45~ppm.

The accuracy and precision of the empirical limb-darkening profiles are discussed in Appendix~\ref{app3}.


\subsection{The benefits of using prior information}
\label{sec:w_prior}
The examples discussed in Section~\ref{sec:noisy_F0V_irac4} and
\ref{sec:noisy_M5V_STIS_G430L} show that:
\begin{enumerate}
\item if using the power-2 law, the empirical limb-darkening
  coefficients can be fitted with the other transit parameters ($p$,
  $a_R$, $i$; see Section~\ref{sec:ideal}) and a normalization
  factor, but the results can be significantly biased, depending on
  the stellar model and passband;
\item the analogous fits, using the claret-4 law, fail to converge (at
  least, using our MCMC routine with up to 1\,500\,000 iterations).
\end{enumerate}
Unfortunately, all the two-coefficients laws are biased for some
stellar types and wavelengths (see Section~\ref{sec:ldc_fixed} and
\ref{sec:ldc_free}), but most of them are sufficiently accurate in the
infrared wavelengths. Some authors proposed to fitting one or two coefficients of the claret-4 law, while keeping the other coefficients fixed (e.g., \citealp{howarth17}). We found that the validity of this approach relies on a good choice of the fixed coefficients, and thus is not fully empirical.

\begin{figure*}[!t]
\epsscale{0.95}
\plotone{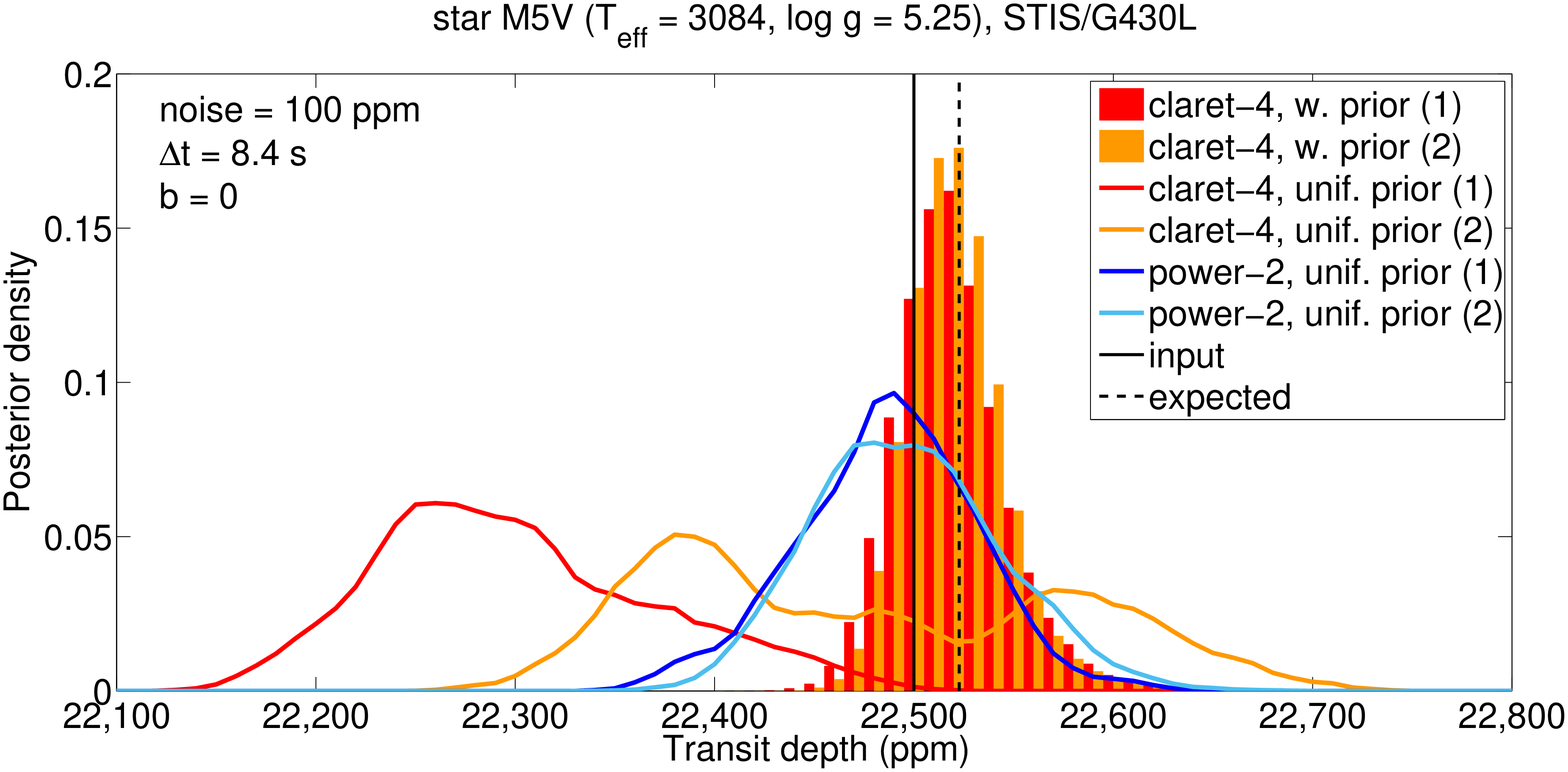}
\plotone{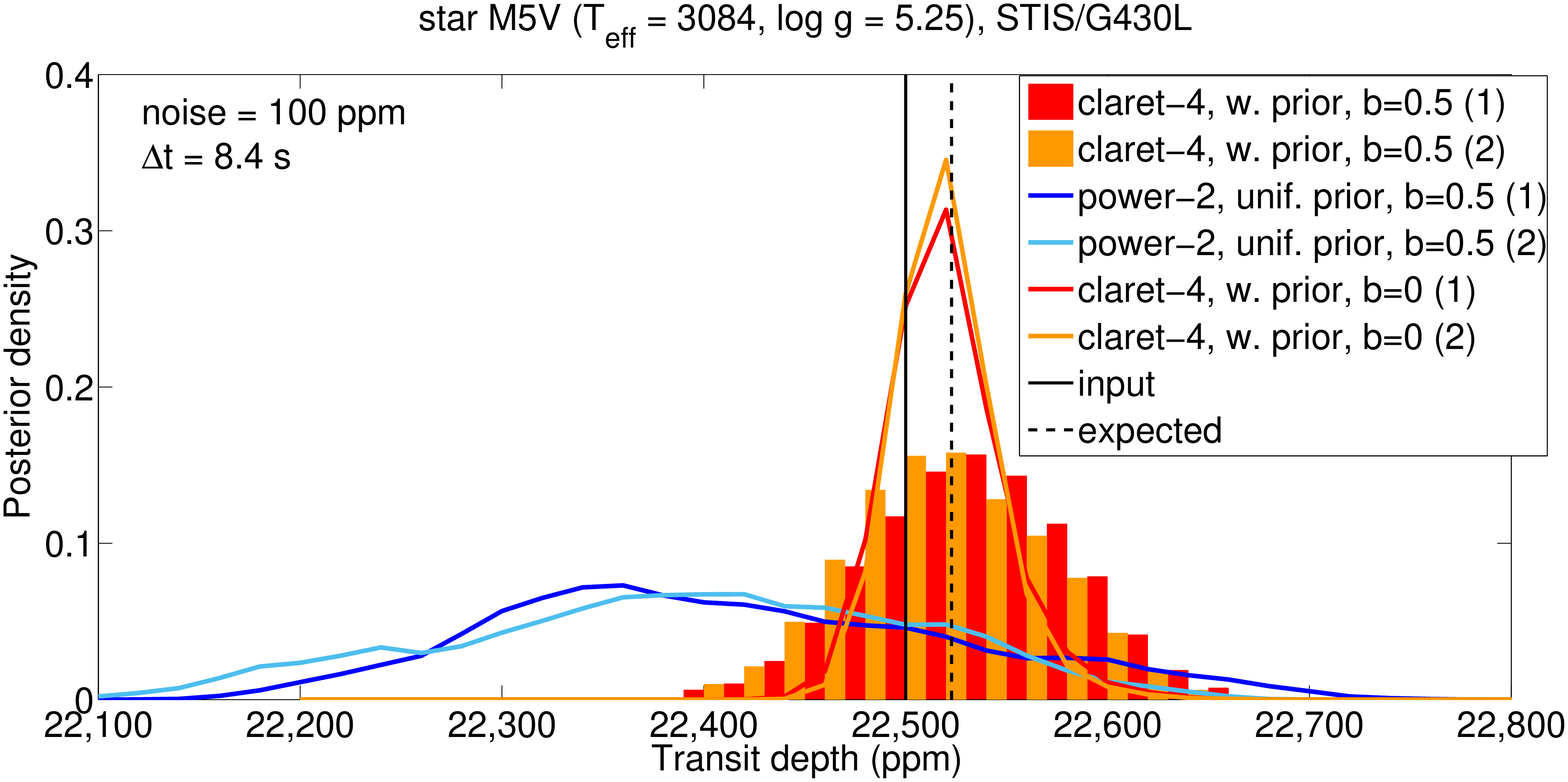}
\caption{Top panel: MCMC-sampled posterior distributions of modeled transit depth for the edge-on transit in front of the M5\;V model, STIS/G430L passband, with 100~ppm gaussian noise; fitting $p$, $a_R$, $i$, normalization factor and limb-darkening coefficients. The histogram channels (red and orange) are for two chains with 1\,500\,000 iterations, using claret-4 coefficients and gaussian priors on $a_R$ and $i$; the channels are half-thick and shifted to improve their visualization. 
Other lines denote the posterior distributions obtained with non-informative priors for all parameters and claret-4 (red and orange) or power-2 (blue and light-blue) limb-darkening coefficients. The input and expected transit depths are also indicated (black vertical lines, continuous and dashed, respectively).
Bottom panel: analogous for the inclined transit ($b=0.5$), except the red and orange lines report the histograms from the top panel for comparison. 
\label{fig26}}
\end{figure*}

\begin{figure*}[!t]
\epsscale{1.0}
\plotone{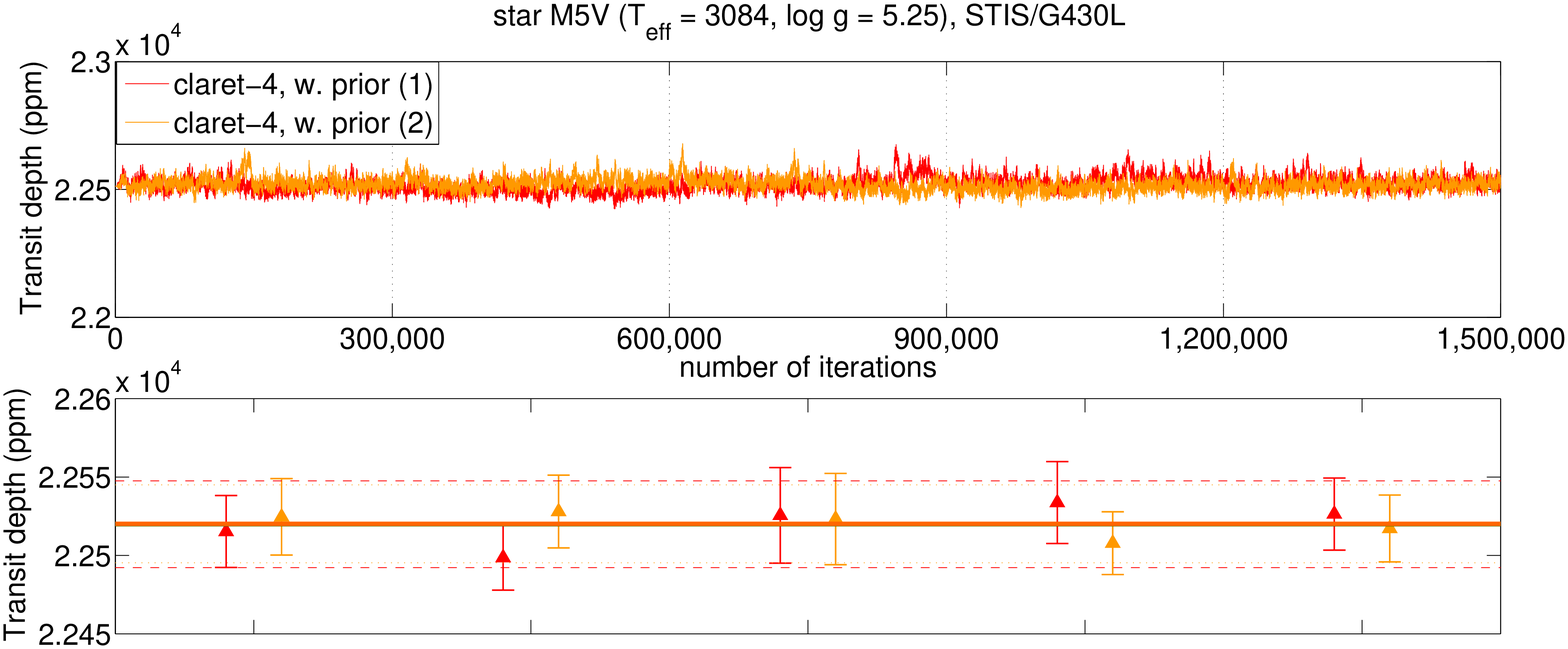}
\plotone{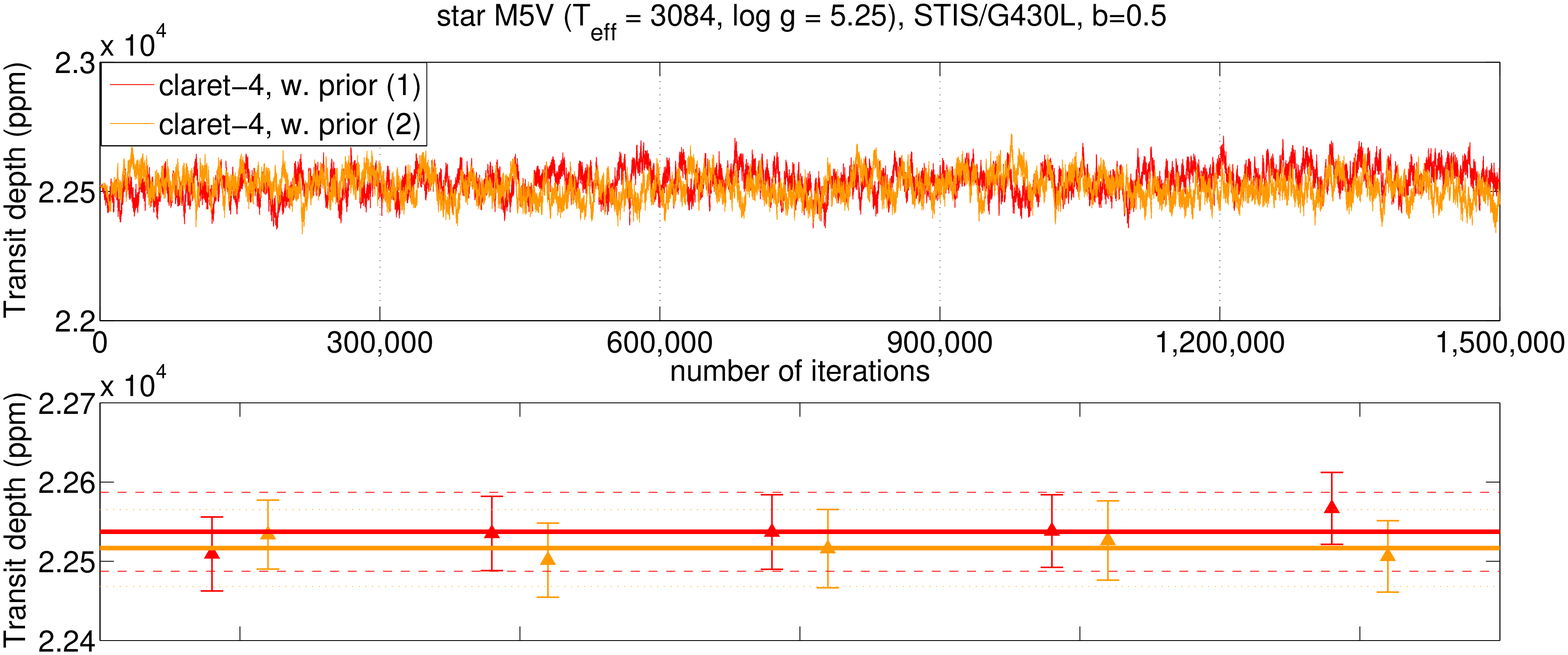}
\caption{Top panels: transit depth chains for the edge-on transit,
  M5\;V model, STIS/G430L passband, with 100~ppm
  gaussian noise, fitting $p$, $a_R$, $i$, normalization factor and claret-4 limb-darkening
  coefficients, adopting gaussian priors on $a_R$ and $i$. Bottom panels: mean
  values and standard deviations calculated over fractional chains
  with 300\,000 iterations; the horizontal lines indicate the mean
  values calculated over the full chains (continuous lines), and the
  mean values plus or minus the standard deviations (dashed
  lines).  \label{fig27}}
\end{figure*}

\begin{figure*}
\epsscale{1.0}
\plotone{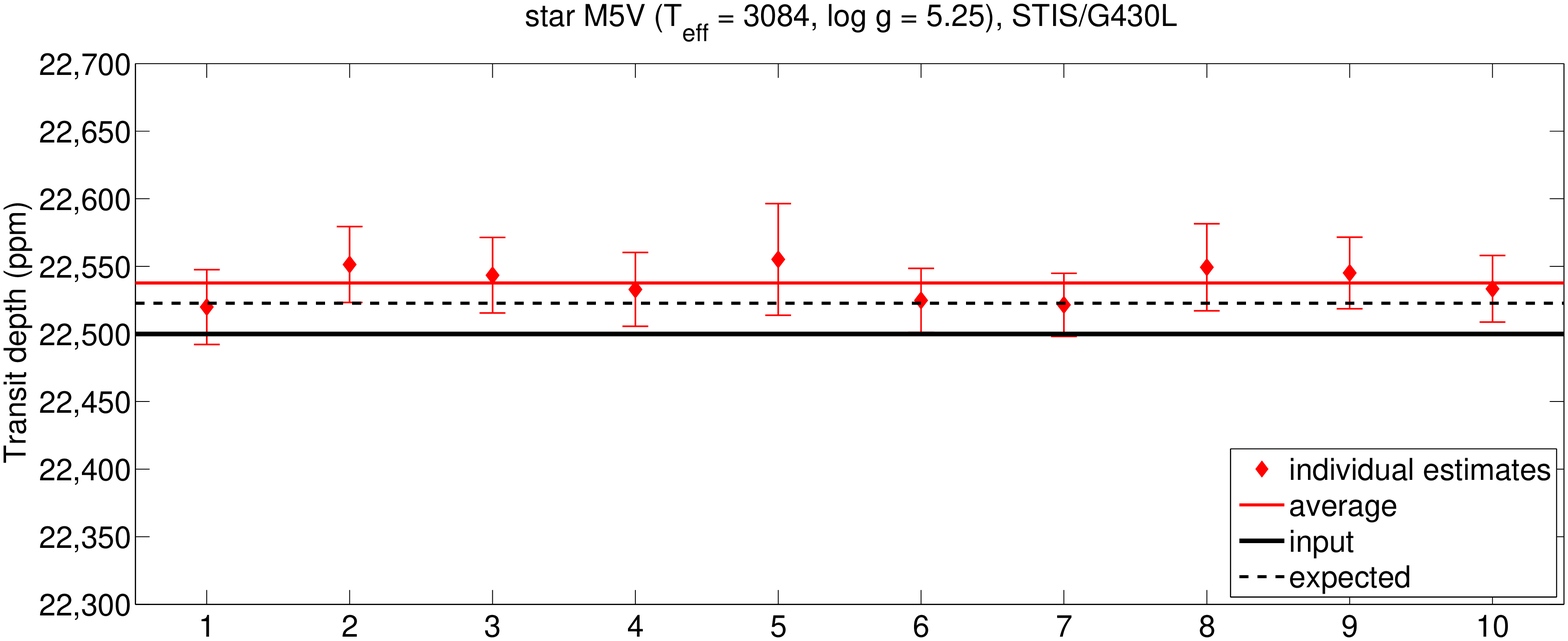}
\plotone{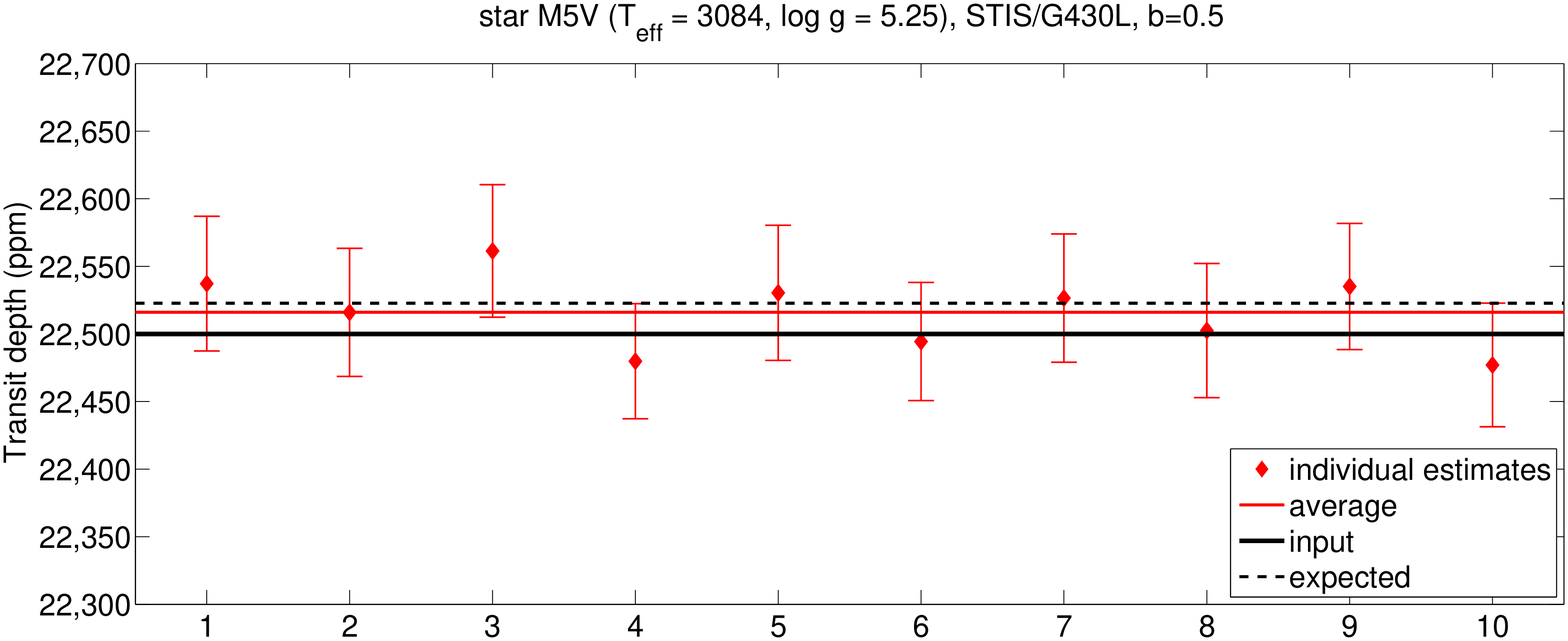}
\caption{Top panel: transit depth estimates for the edge-on transit in front of the M5\;V model, STIS/G430L passband, with 100~ppm gaussian noise (different noise time series, red); fitting $p$, $a_R$, $i$, normalization factor and claret-4 limb-darkening coefficients, adopting gaussian priors on $a_R$ and $i$. Average over the ten light-curves (red, continuous line), input (black, continuous line) and expected (black, dashed line) values. Bottom panel: the same, for the inclined transit ($b=0.5$). \label{fig28}}
\end{figure*}

Our proposal is that, if the `geometric' parameters,
$a_R$ and $i$, are measured in the infrared, the results can be
implemented as an informative prior when fitting at shorter
wavelengths, thanks to their small or negligible wavelength-dependence,
based on Equations~\ref{eqn:aR_expected}, \ref{eqn:i_expected} and
Table~\ref{tab3}.  We tested fitting for $p$, $a_R$, $i$, the claret-4
limb-darkening coefficients, and a normalization factor, on the
transit light-curves obtained for the M5\;V model in the STIS/G430L
passband, adopting gaussian priors on $a_R$ and $i$. The parameters of
the gaussian priors are reported in Table~\ref{tab4}.
\begin{table}[t]
\begin{center}
\caption{Parameters of the gaussian priors adopted in Section~\ref{sec:w_prior}. \label{tab4}}
\begin{tabular}{ccccc}
\tableline\tableline
 & $\mu (a_R)$ & $\sigma (a_R)$ & $\mu (i)$ & $\sigma (i)$ \\
\tableline
$b=0$ & 9.0042 & 0.004 & 90 & 0.18 \\
$b=0.5$ & 9.0042 & 0.006 & 86.81526146 & 0.01 \\
\tableline
\end{tabular}
\end{center}
\tablecomments{$\mu (a_R)$ and $\mu (i)$ are the set to the expected transit parameter values, $\sigma (a_R)$ and $\sigma (i)$ are the error bars obtained from the corresponding light-curves in the IRAC/ch4 passband, with 100~ppm noise level, 8.4 s sampling time, when fitting for $p$, $a_R$, $i$, normalization factor, and power-2 limb-darkening coefficients.}
\end{table}
Figure~\ref{fig26} shows the transit depth posterior distributions for
the edge-on and the inclined transits, obtained with 1\,500\,000 MCMC
samples. Figure~\ref{fig27} shows the relevant chains.
The use of gaussian priors on $a_R$ and $i$ leads to convergence of
the MCMC fits with claret-4 coefficients. The error bars in transit
depth are significantly smaller than those estimated with power-2
limb-darkening coefficients and non-informative priors on $a_R$ and
$i$, falling to $\sim$25 and $\sim$50 (compared with
$\sim$45 and $\sim$115~ppm), for the edge-on and inclined
transits, respectively.  The biases, averaged over 10 light-curves
with different noise levels, are also smaller (+15 and
-7~ppm; Figure~\ref{fig28}).

\begin{figure*}
\epsscale{1.0}
\plotone{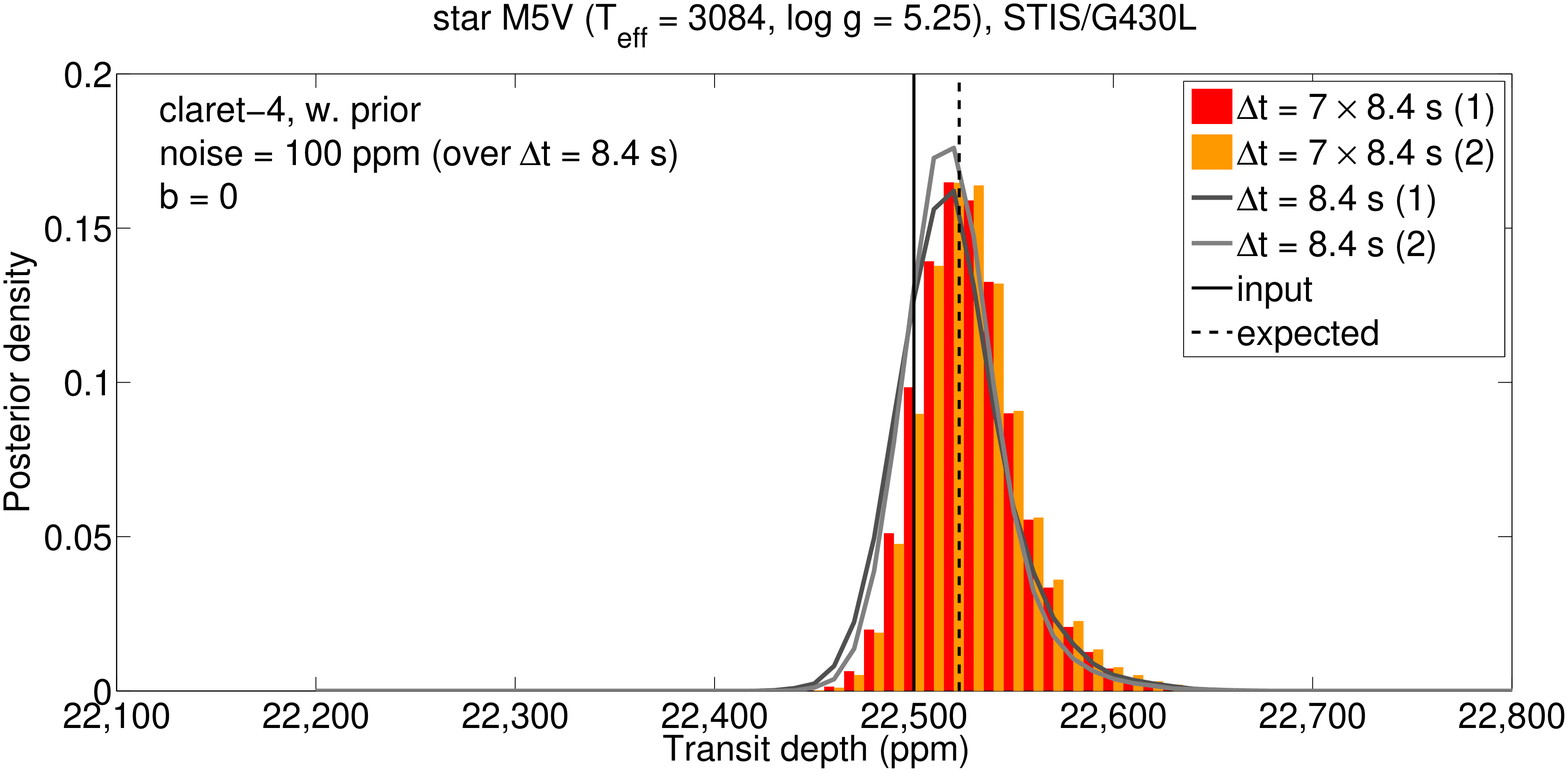}
\plotone{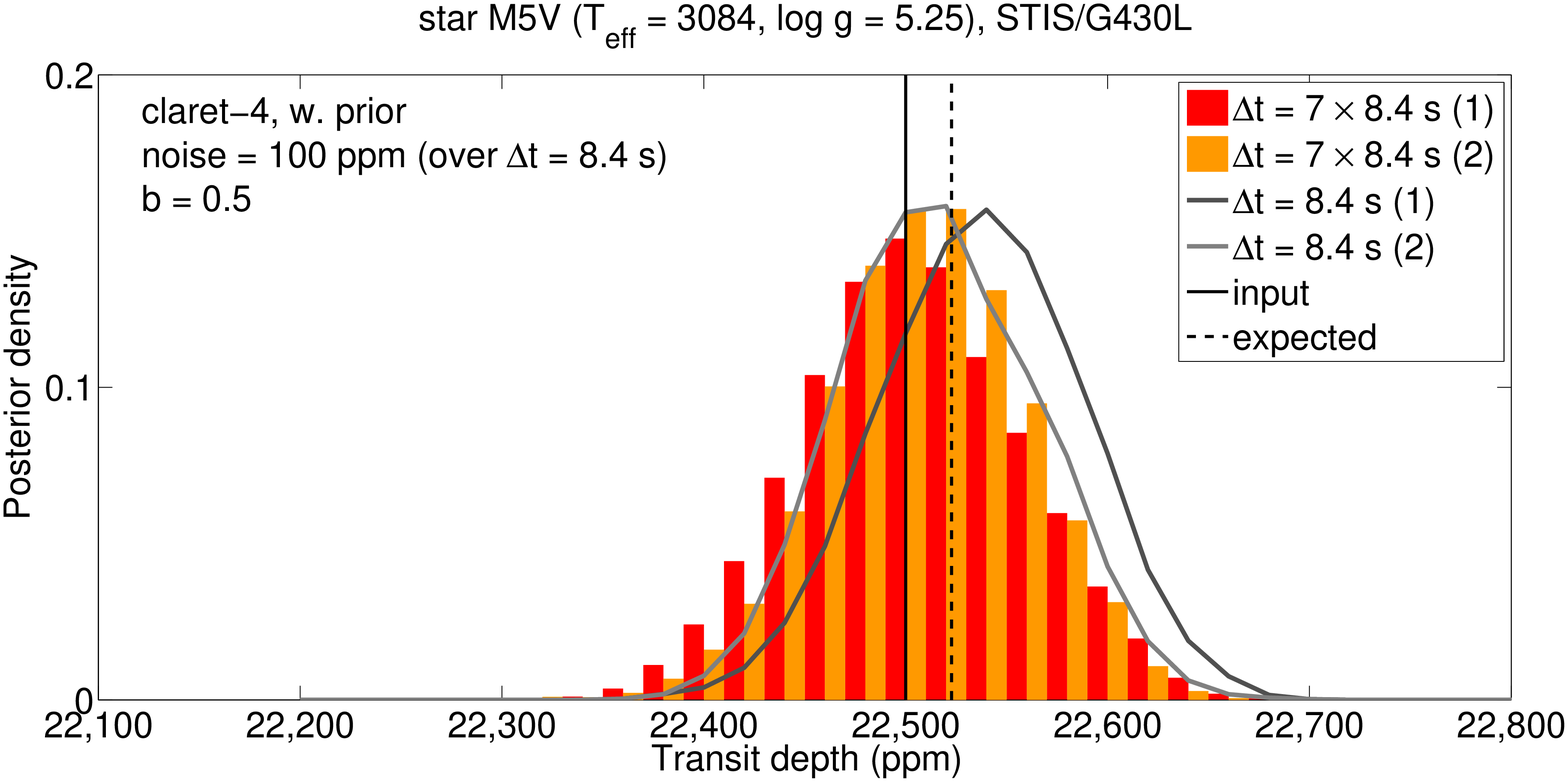}
\caption{Top panel: MCMC-sampled posterior distributions of modeled transit depth for the edge-on transit in front of the M5\;V model, STIS/G430L passband, with 100~ppm gaussian noise, then binned over 7 points; fitting $p$, $a_R$, $i$, normalization factor and claret-4 limb-darkening coefficients, adopting gaussian priors on $a_R$ and $i$. The histogram channels (red and orange) are for two chains with 1\,500\,000 iterations, using claret-4 coefficients and gaussian priors on $a_R$ and $i$; the channels are half-thick and shifted to improve their visualization. The grey lines denote the analogous posterior distributions without binning.
The input and expected transit depths are also indicated (black vertical lines, continuous and dashed, respectively).
Bottom panel: the same, for the inclined transit ($b=0.5$). \label{fig29}}
\end{figure*}

As a final test, we investigated the effect of having a longer
integration time, similar to that of the Kepler short-cadence
frame. The longer integration time is simulated by binning the transit
light-curves over 7 points ($7\times 8.4=58.8$~s). The relevant
transit depth posterior distributions are shown in
Figure~\ref{fig29}; they are almost identical to the
non-binned ones.
We conclude that an integration time of $\sim$1 min, as for the
Kepler short-cadence frames, does not affect the accuracy (error bar)
of the fitted transit depth, compared to shorter integration times.


\section{Discussion}
\label{sec:discussion}

\subsection{Synergies between JWST and Kepler, K2, TESS}
Empirical limb-darkening coefficients determined from exoplanetary
transit light-curves are desirable, not only to validate the
stellar-atmosphere models, but also to improve both the absolute and
relative precision of inferred exoplanetary spectra. No
two-coefficient limb-darkening law is accurate for all stellar types
and/or wavelengths, but can still give near-perfect fits to the
transit light-curves, albeit with significantly biased transit
parameters and limb-darkening coefficients. To overcome this issue,
fitting for the claret-4 limb-darkening coefficients is necessary, but
some prior knowledge of the orbital parameters $a_R$ and $i$ is
required to enable convergence of the fitting algorithms.

Such knowledge can be obtained from the infrared observations, for
which the effect of limb-darkening is smaller, and simple
two-coefficient laws may be sufficiently accurate. The MIRI instrument
onboard JWST will provide suitable observations for tens of exoplanets. A
re-analysis of the Kepler and K2 targets, with the approach developed
in Section~\ref{sec:w_prior}, can address some of the controversies
reported in the literature (e.g. \citealp{southworth08, claret09, kipping11, muller13}), if the only problem was
the use of inadequate two-parameter limb-darkening laws. The same
approach should be used for new observations that will be
obtained, in the visible, by TESS and/or other JWST instruments.

\section{Conclusions}
We studied the potential biases in transit depth due to the use of
theoretical stellar limb-darkening coefficients obtained from
plane-parallel model atmospheres, and when fitting for empirical
limb-darkening coefficients, over a range of model temperatures and
instrumental passbands. We propose the use of a two-coefficient law, named
``power-2'', which outperforms the most common two-coefficient laws
adopted in the exoplanet literature, especially for the M-dwarf
models. Nevertheless, the Claret four-coefficient law is
significantly more robust than any simpler one, especially at visible
wavelengths. Our results indicate that an absolute precision of $\lesssim$30~ppm can be achieved in the modeled transit depth 
at visible and infrared wavelegths, with $\lesssim$10~ppm relative precision over the
\textit{HST}/WFC3 passband, depending on the stellar type. The
intrinsic bias due to the use of theoretical limb-darkening
coefficients obtained from plane-parallel models is also
$\lesssim$30~ppm for most exoplanet host stars (F--M spectral types),
but this estimate does not take into account the uncertainties in the
stellar models and in the measured stellar parameters, or the effect
of stellar activity and other second-order effects.

Finally, we developed an optimal strategy to fitting for the
four-coefficient limb-darkening in the visible, using prior
information on the exoplanet orbital parameters to break some of the
degeneracies. This novel approach could solve some of the
controversial results reported in the literature, which relies on
empirical estimates of quadratic limb-darkening coefficients. The
forthcoming JWST mission will provide accurate information on the orbital
parameters of transiting exoplanets through observations performed by
MIRI, enabling wide application of the approach developed in this paper.

\acknowledgments
This work was supported by STFC (ST/K502406/1) and the ERC project ExoLights (617119). D.H. is supported by Sonderforschungsbereich SFB 881 ``The Milky Way System'' (subproject A4)
of the German Research Foundation (DFG).

\clearpage

\appendix

\section{Supplemental figures: best-fit models and residuals}
\label{app1}

\begin{figure}[!b]
\epsscale{0.29}
\hspace{-1cm}
\plotone{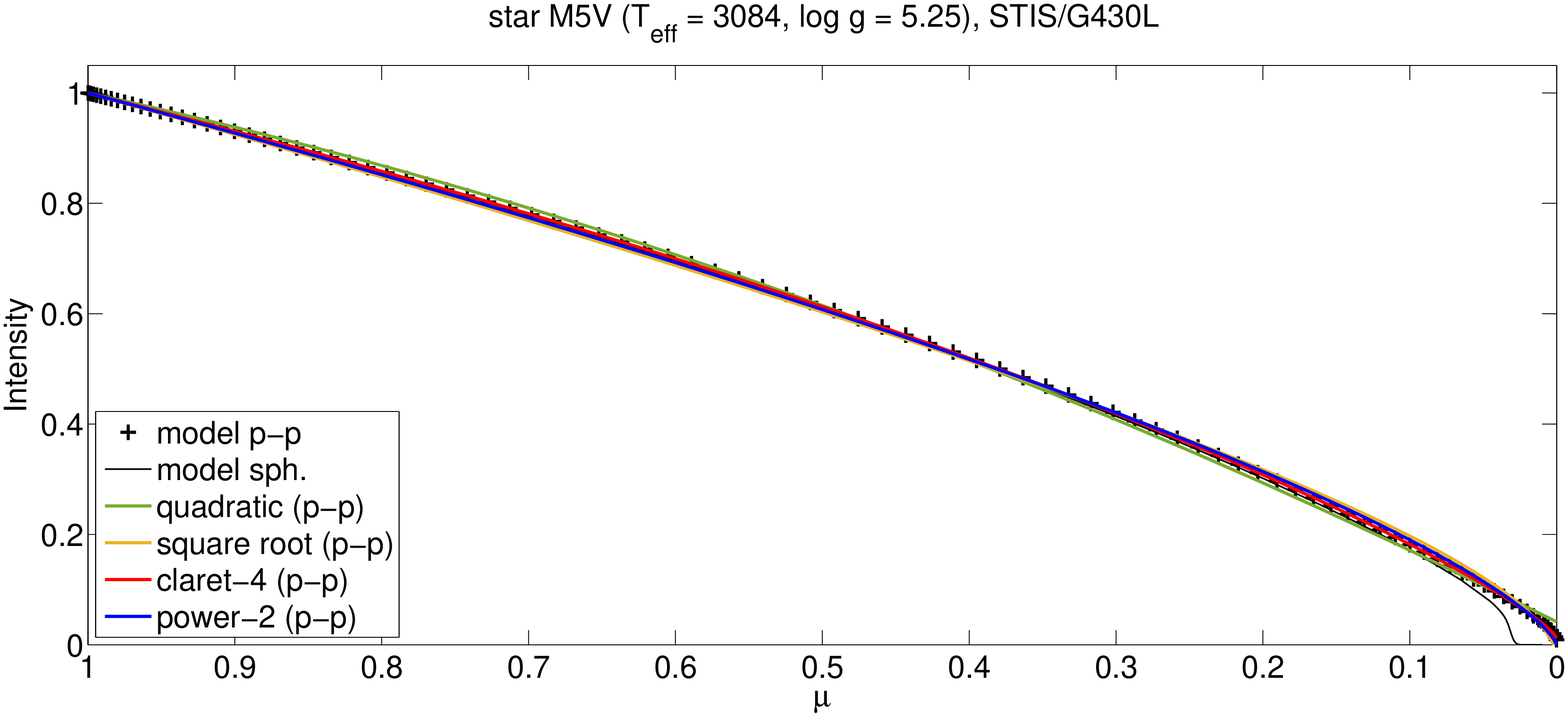}
\plotone{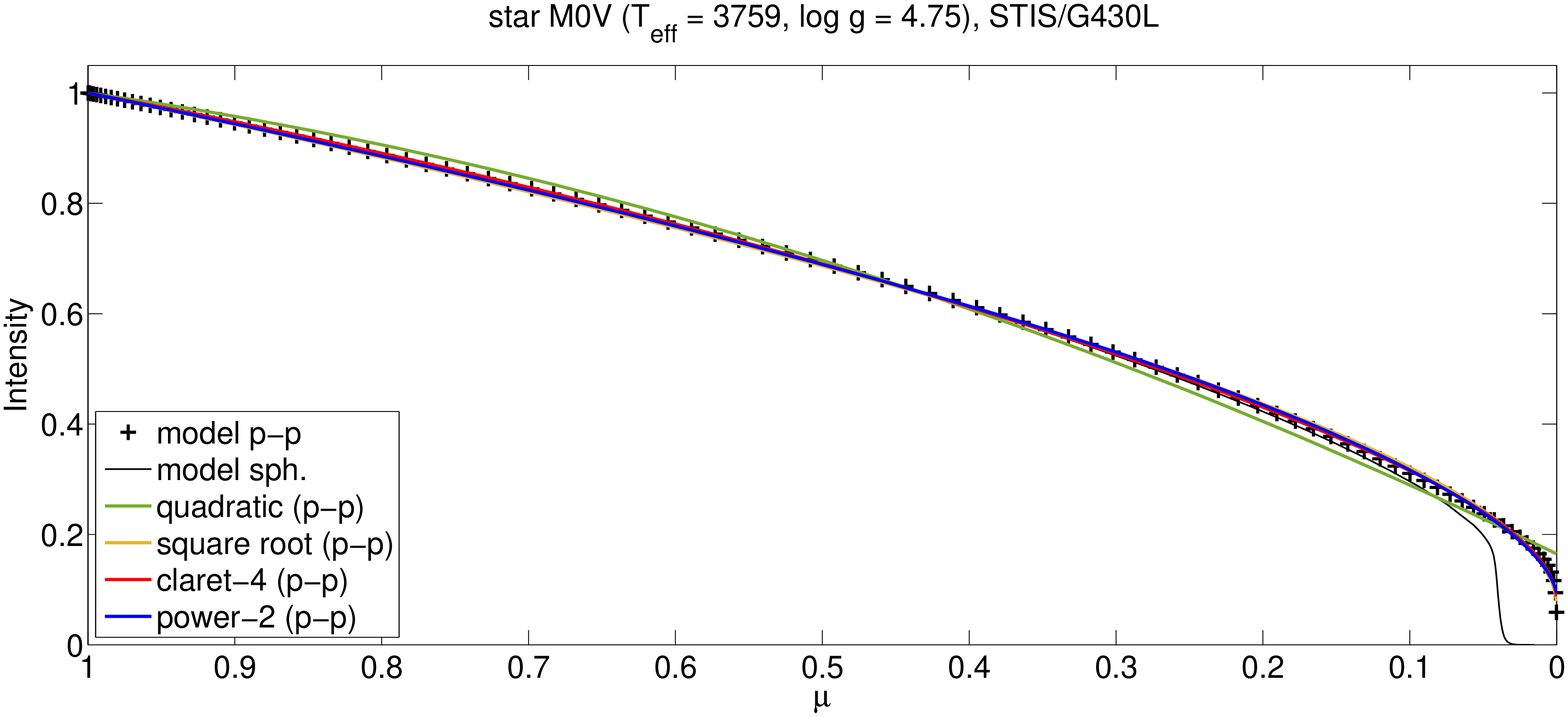}
\plotone{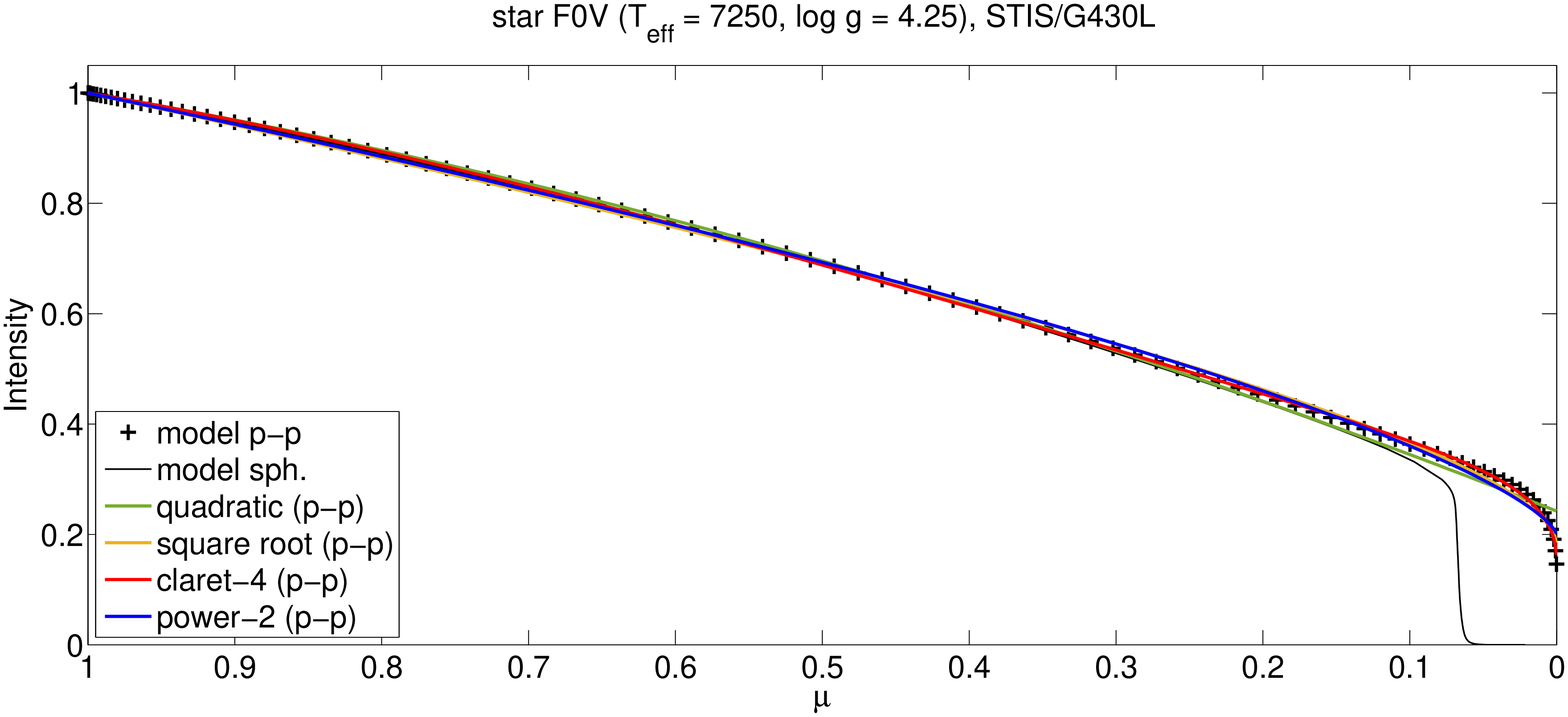}
\\
\hspace{-1cm}
\plotone{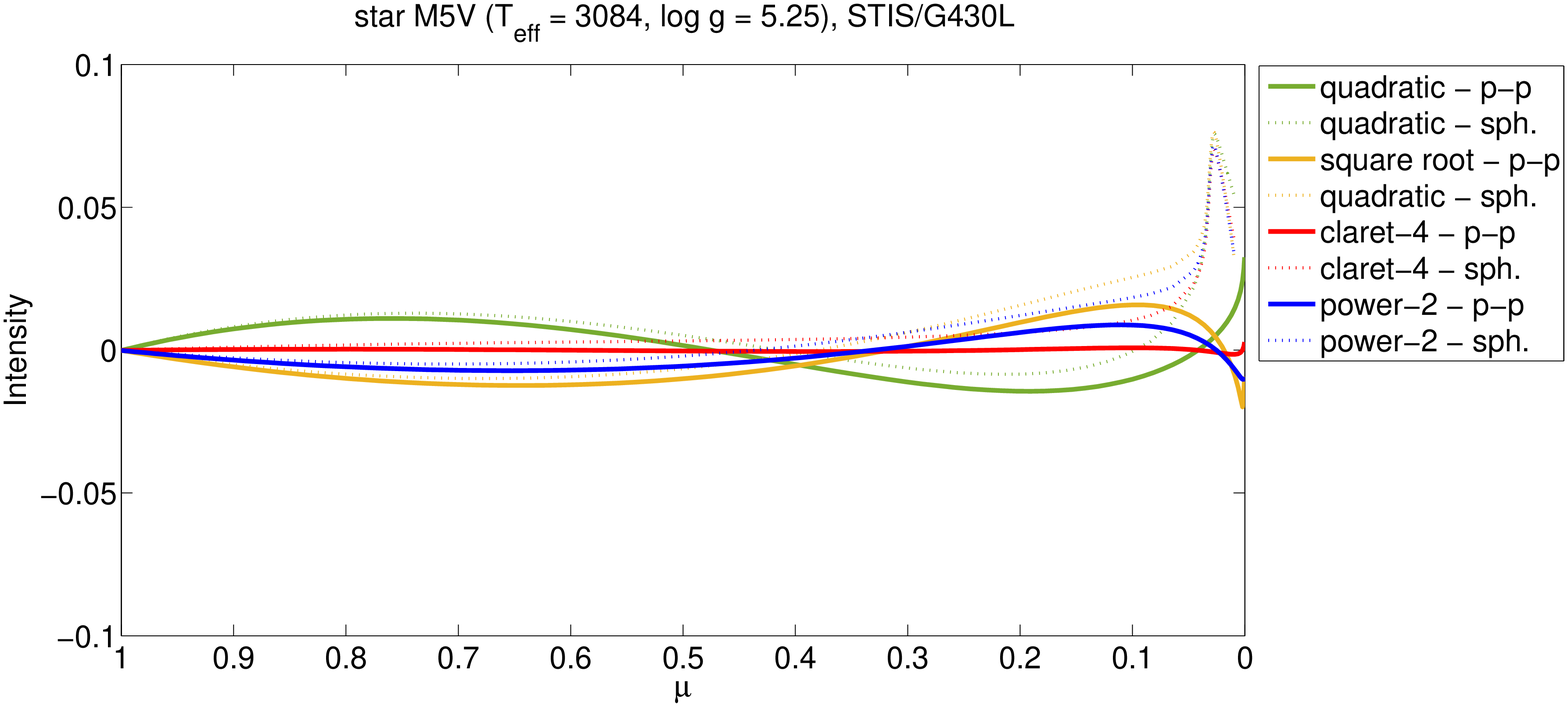}
\plotone{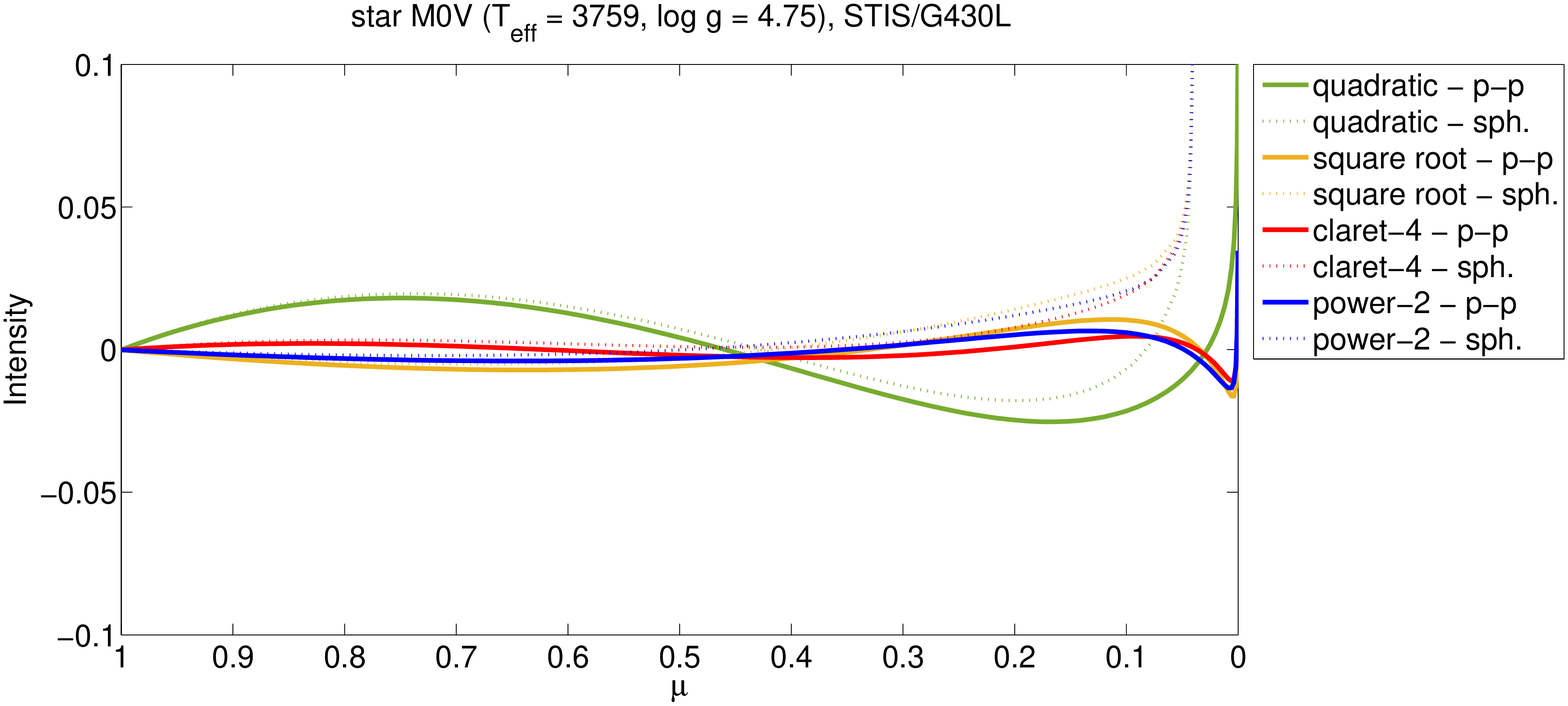}
\plotone{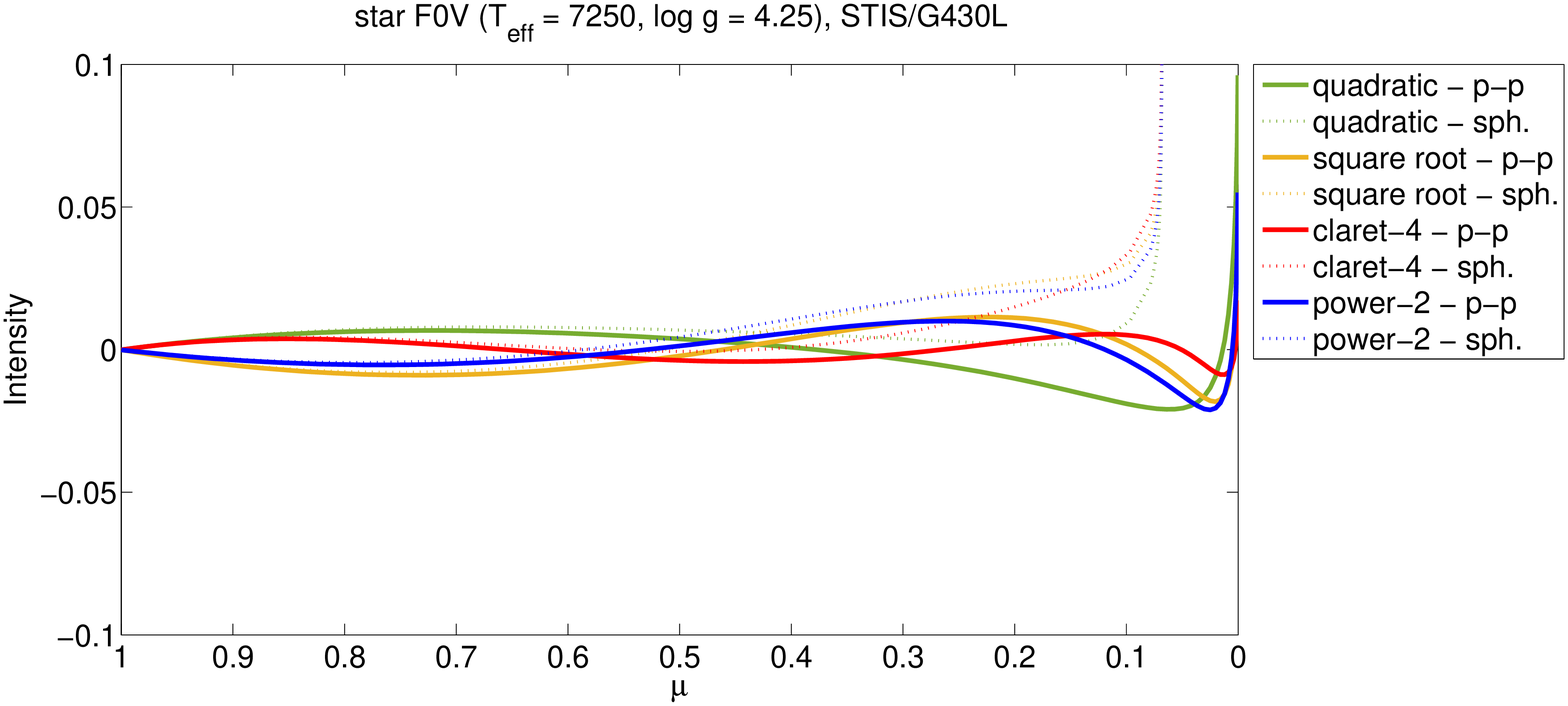}
\\
\hspace{-1cm}
\plotone{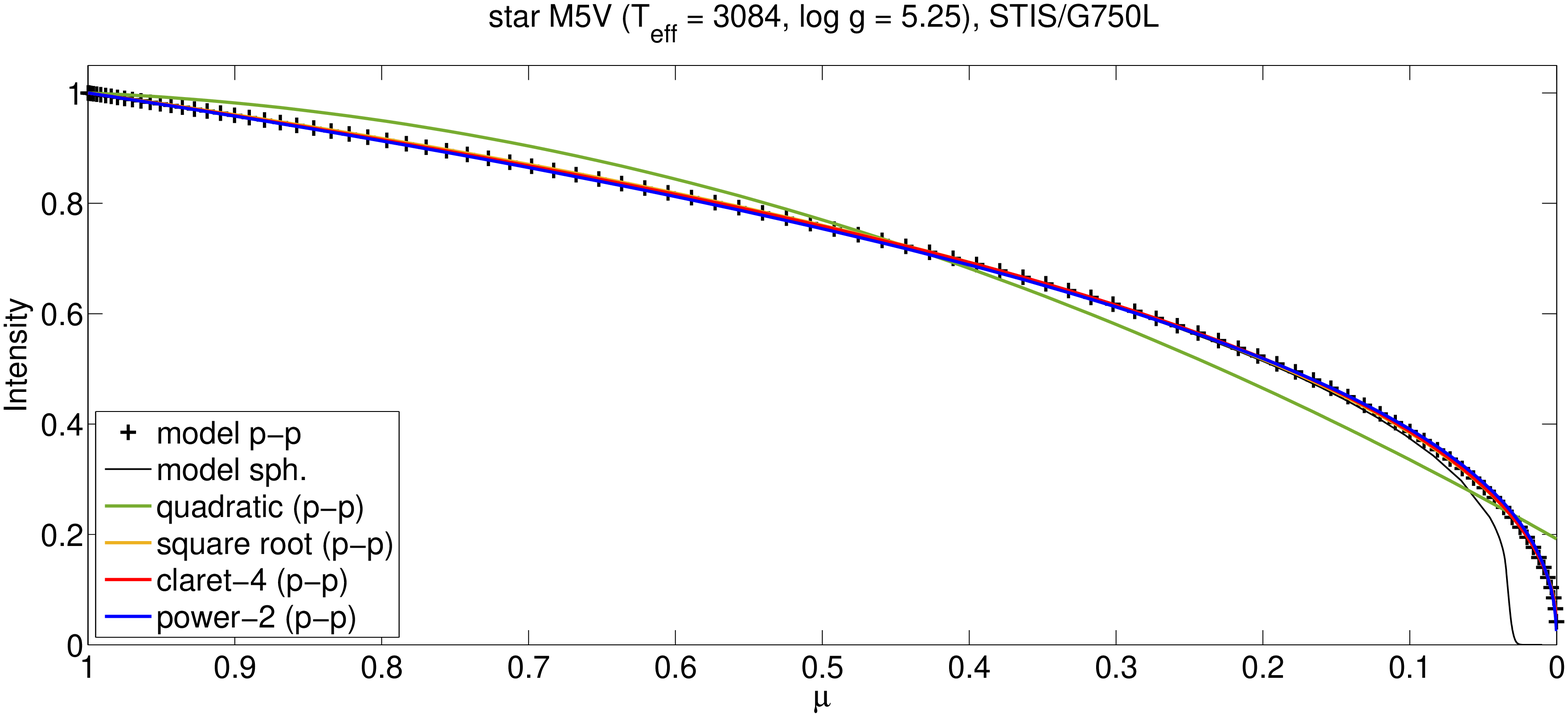}
\plotone{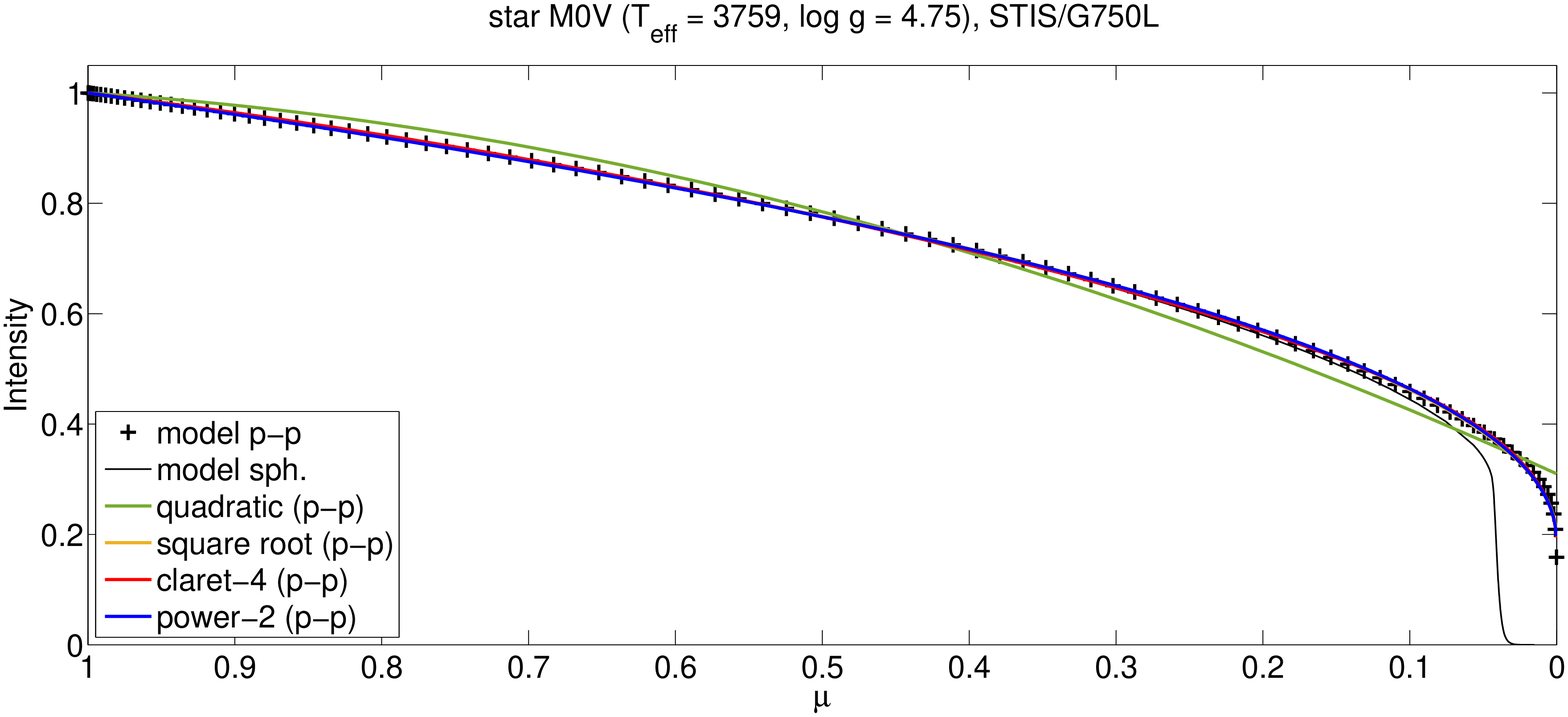}
\plotone{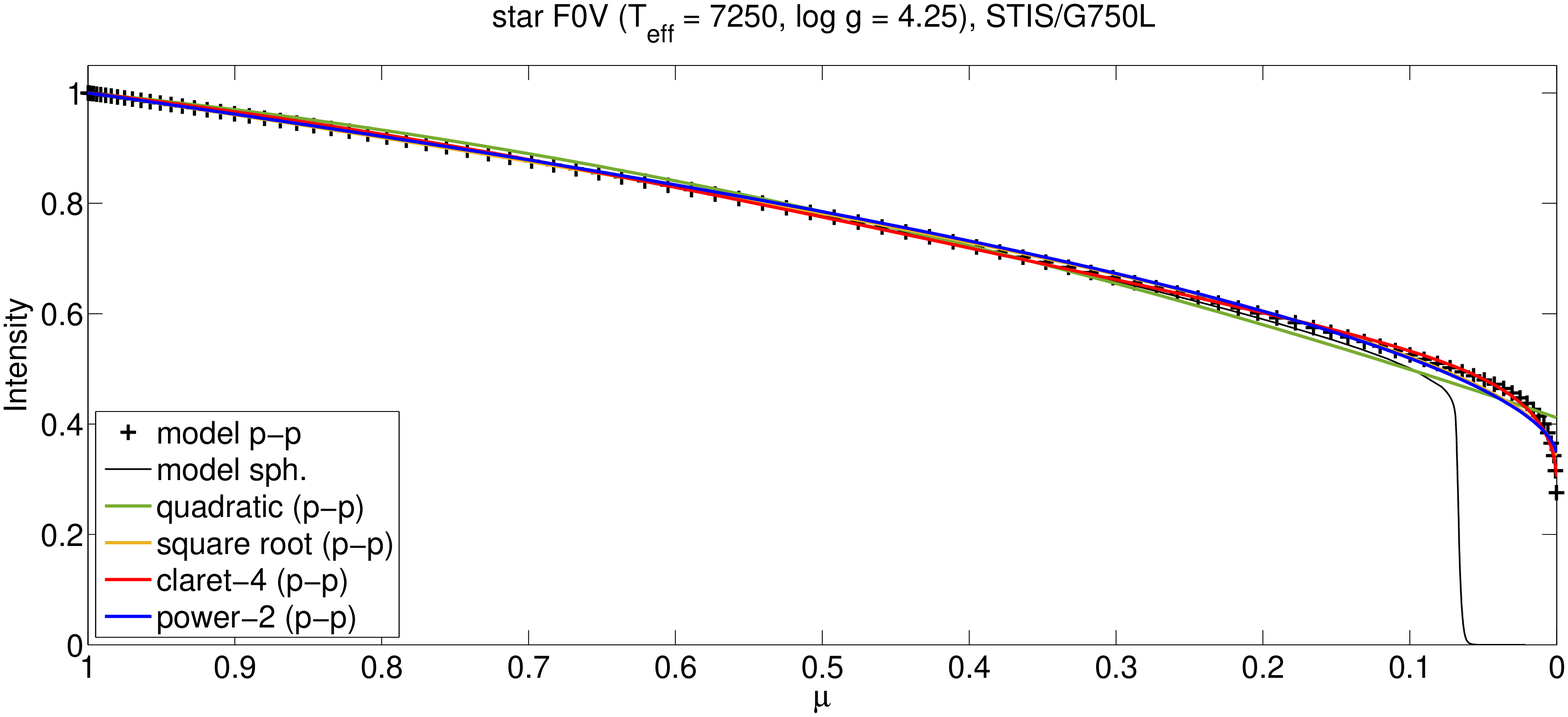}
\\
\hspace{-1cm}
\plotone{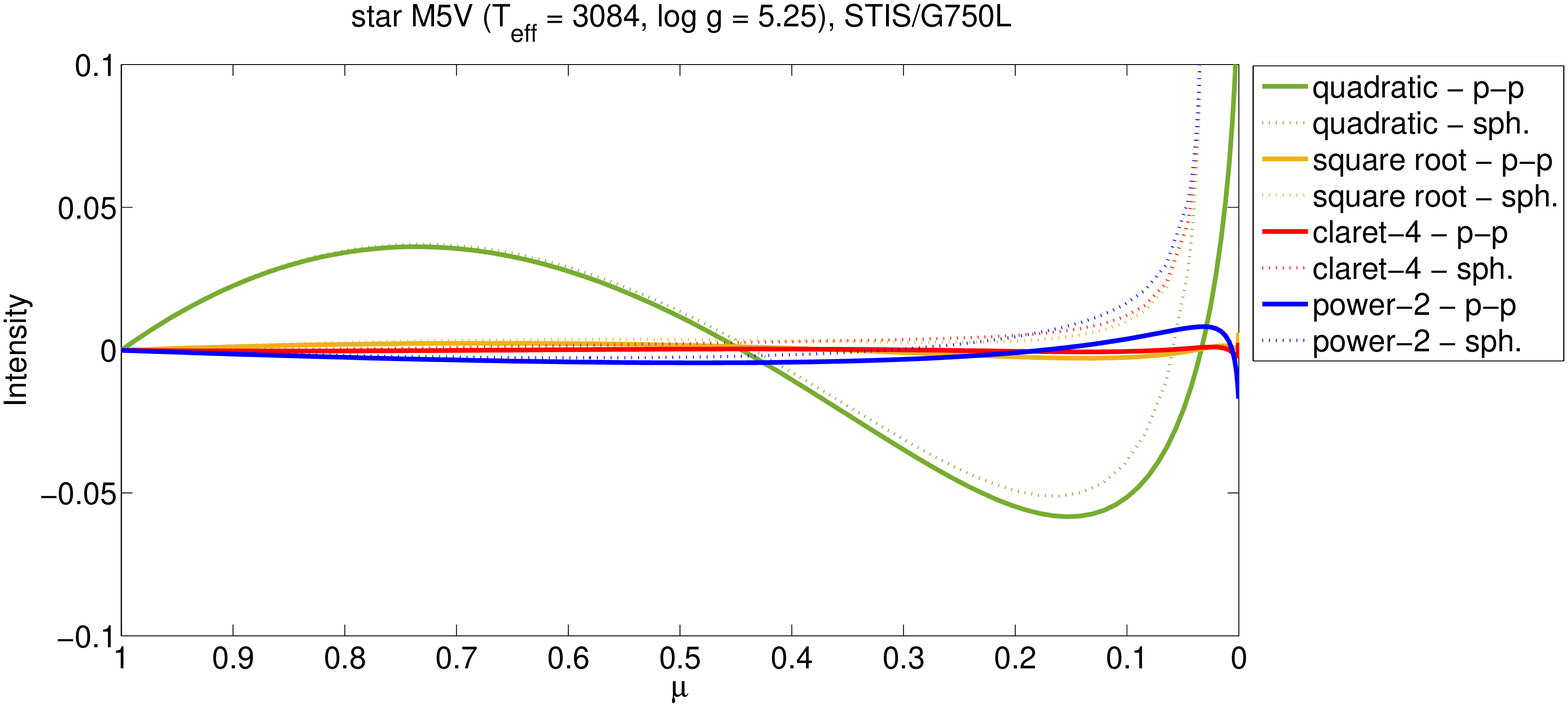}
\plotone{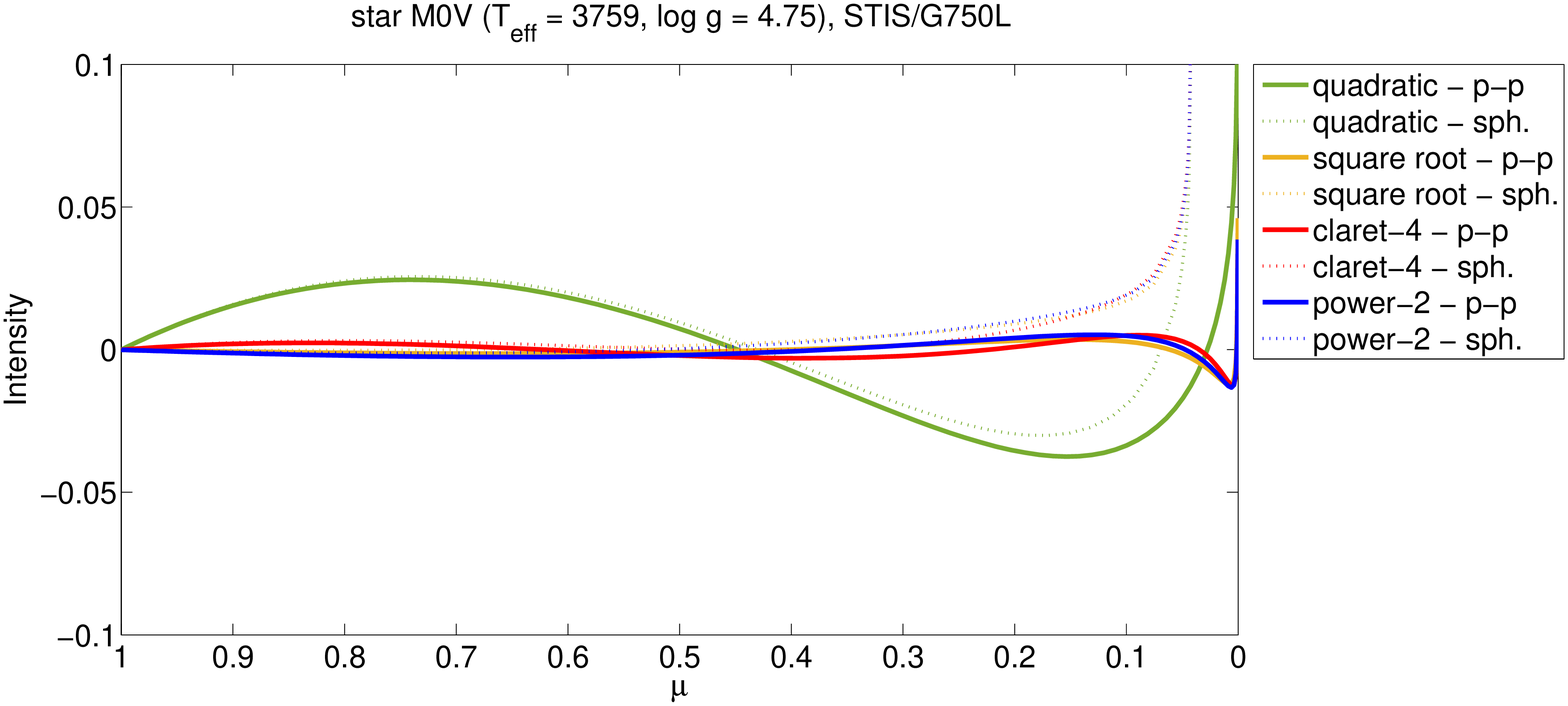}
\plotone{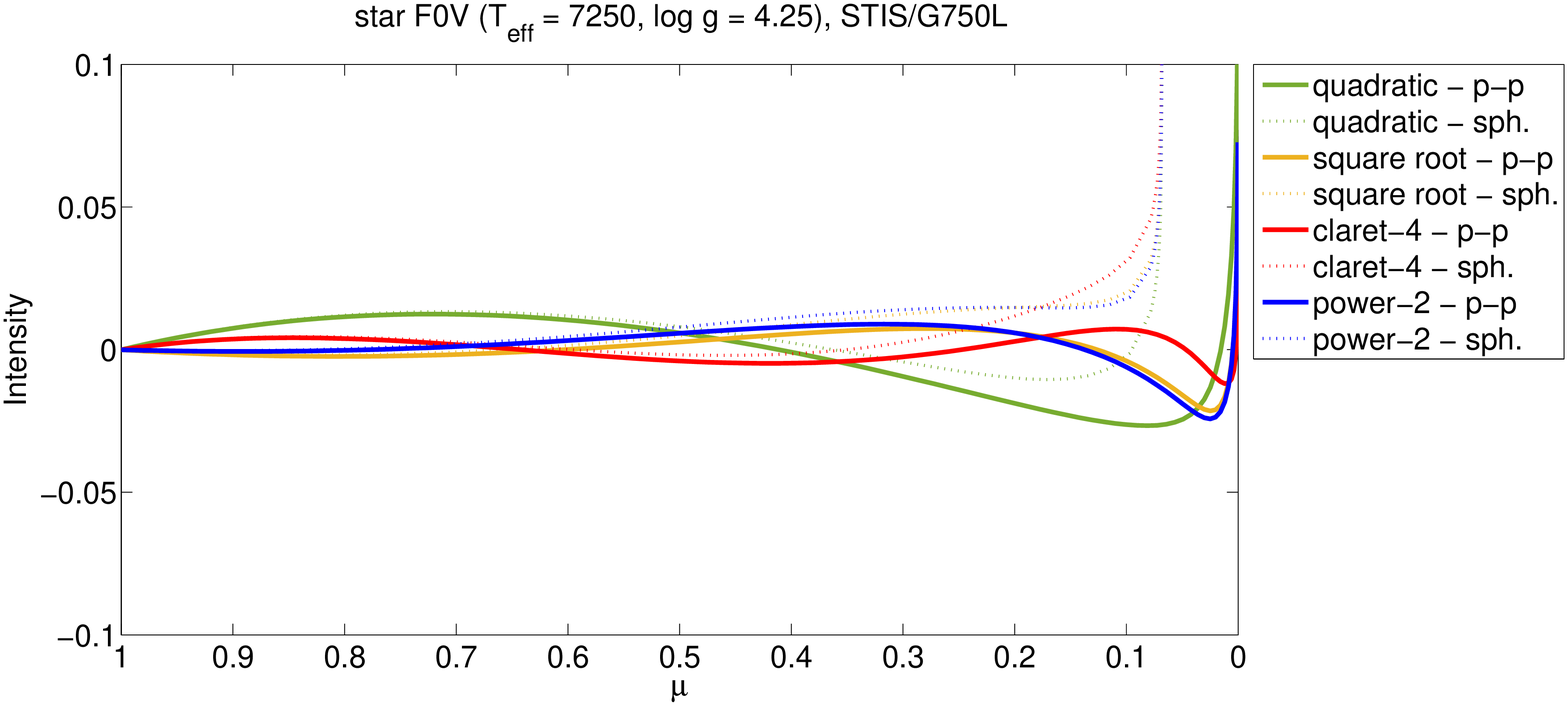}
\\
\hspace{-1cm}
\plotone{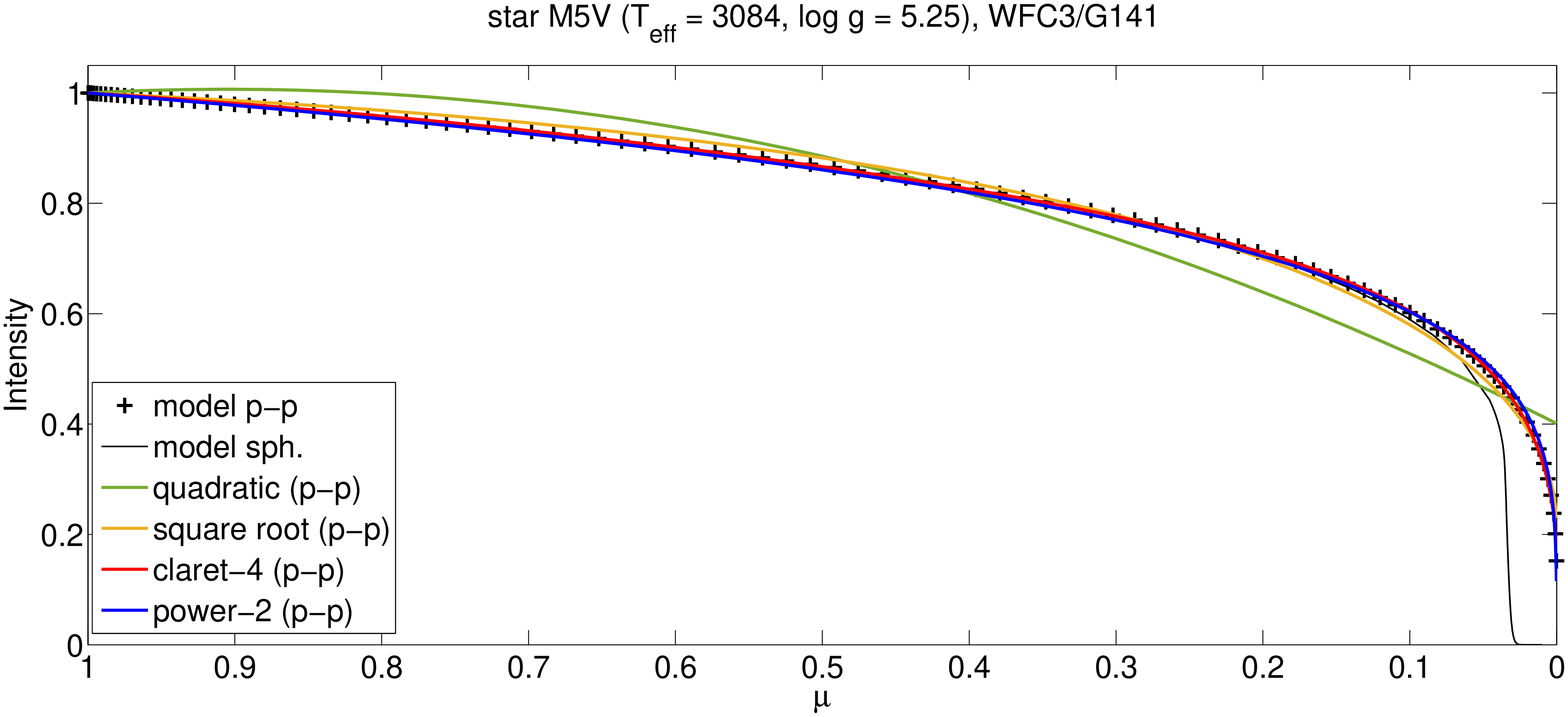}
\plotone{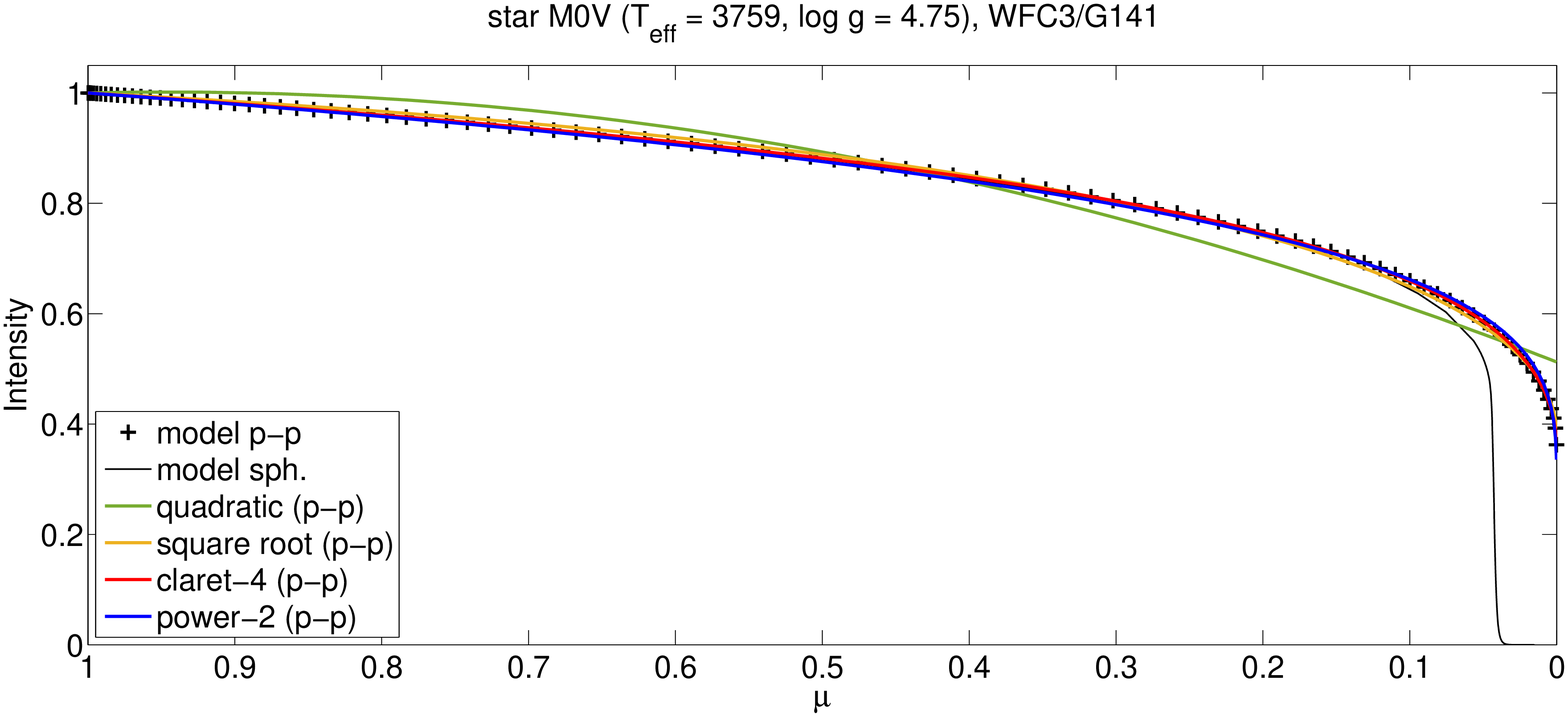}
\plotone{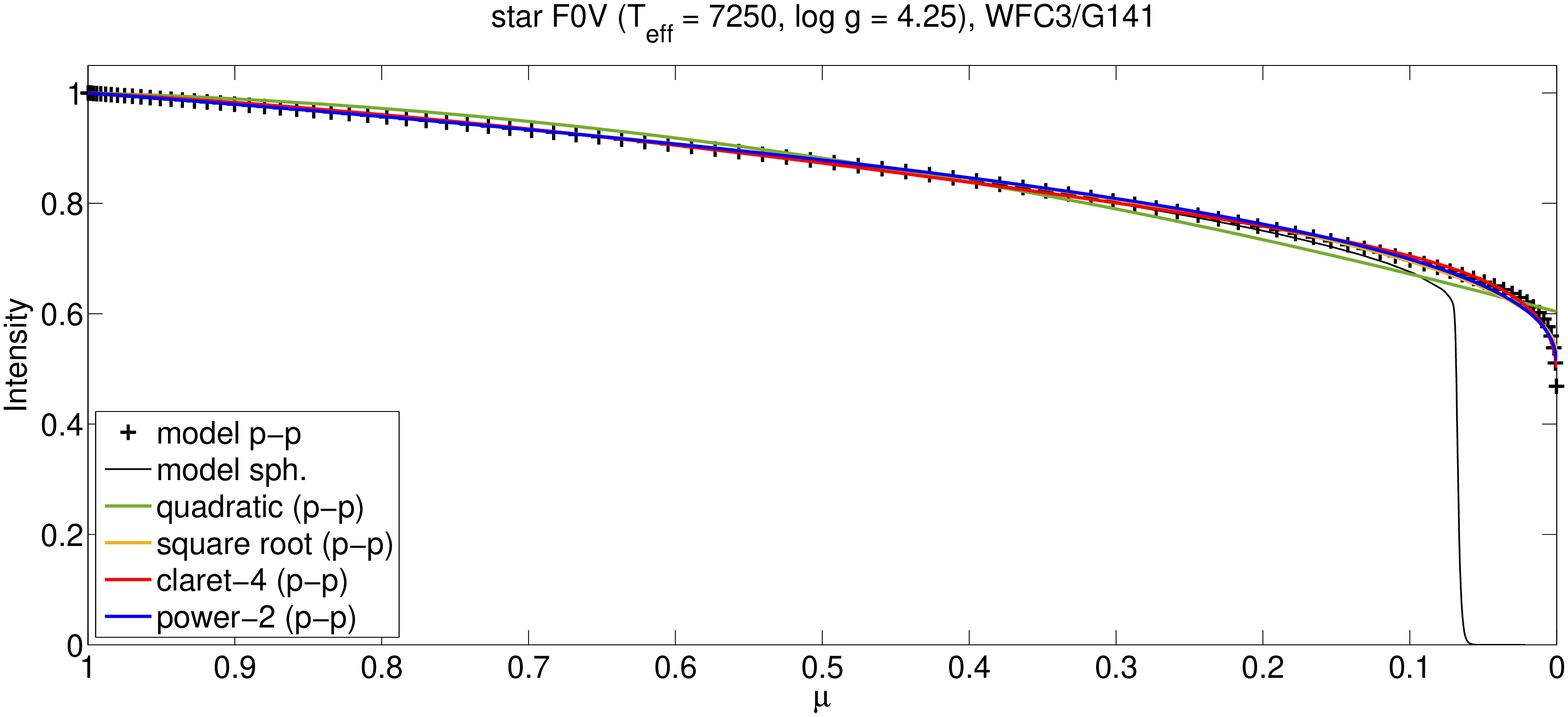}
\\
\hspace{-1cm}
\plotone{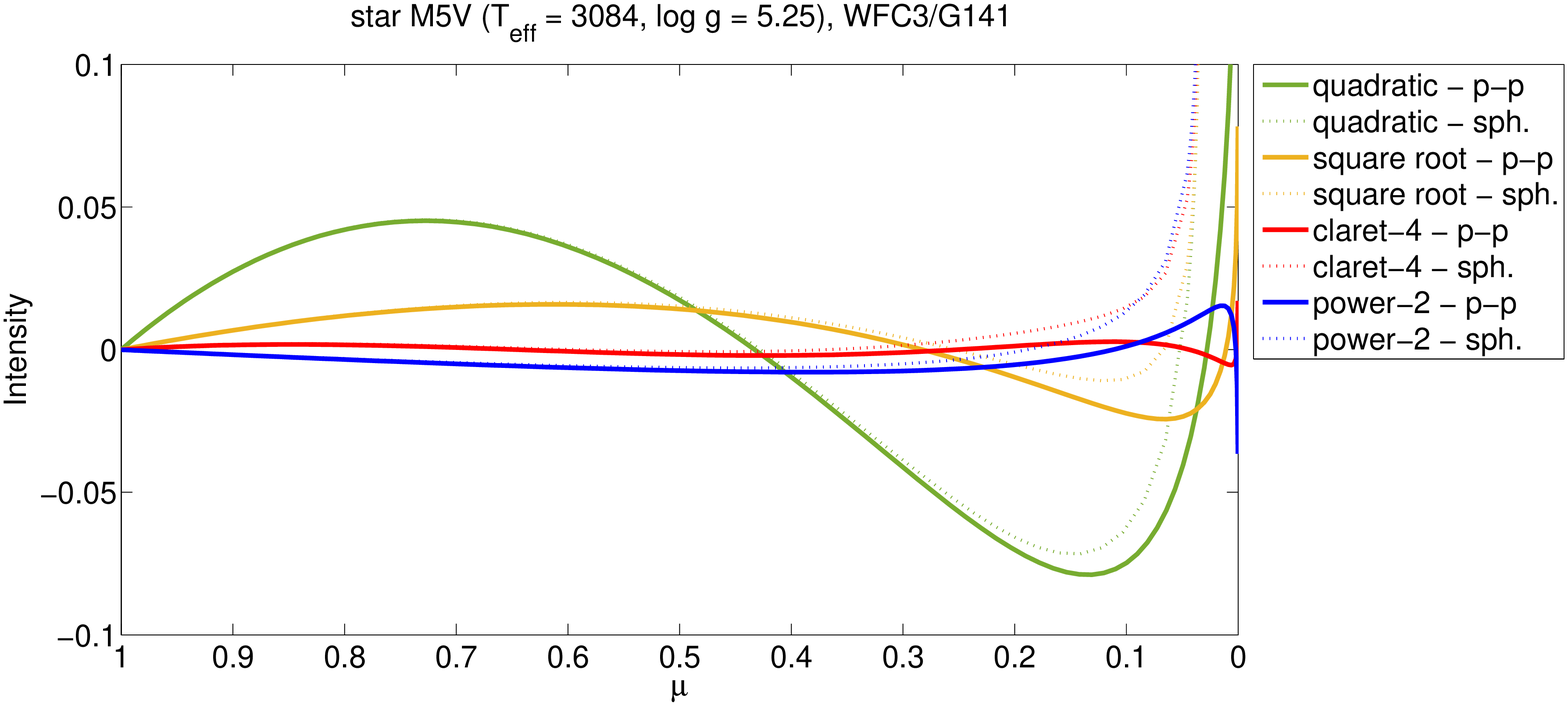}
\plotone{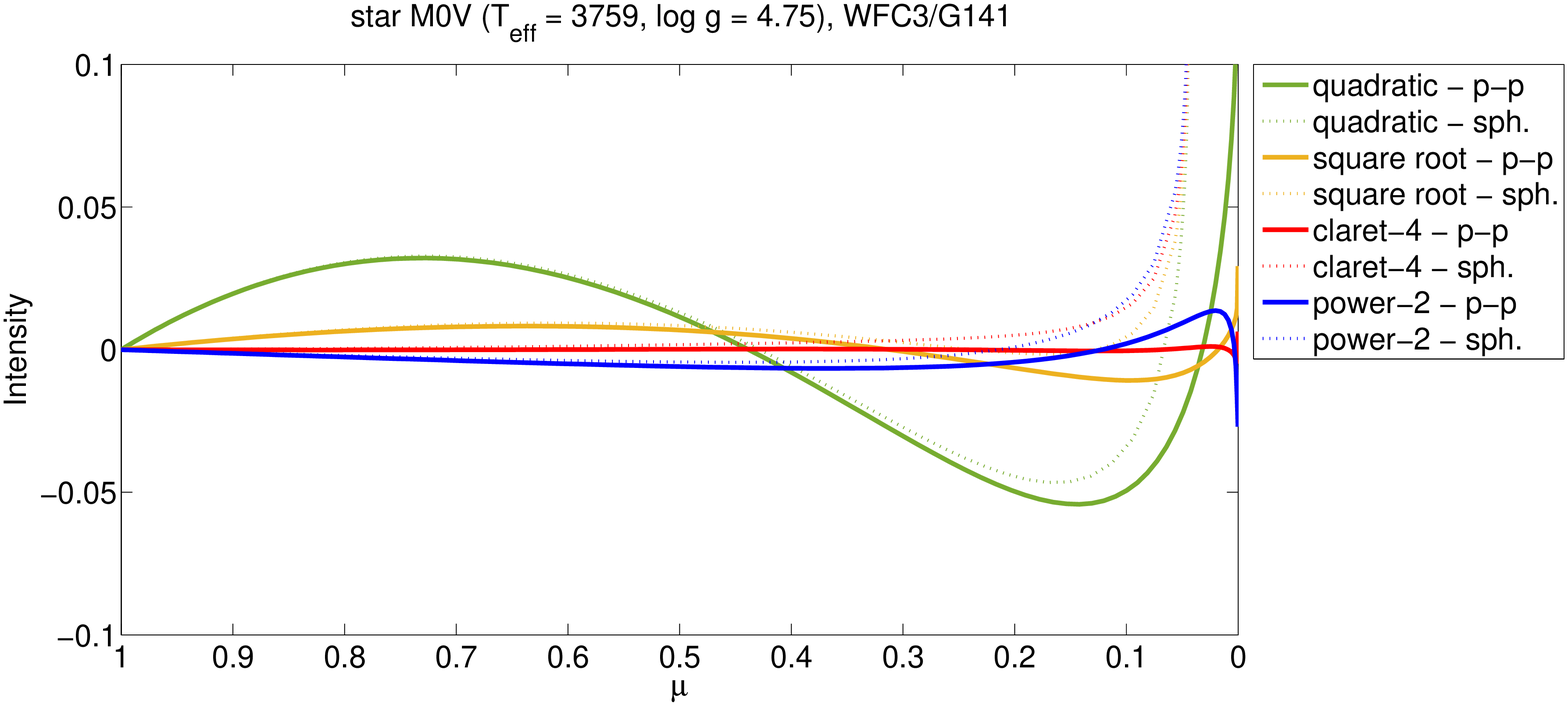}
\plotone{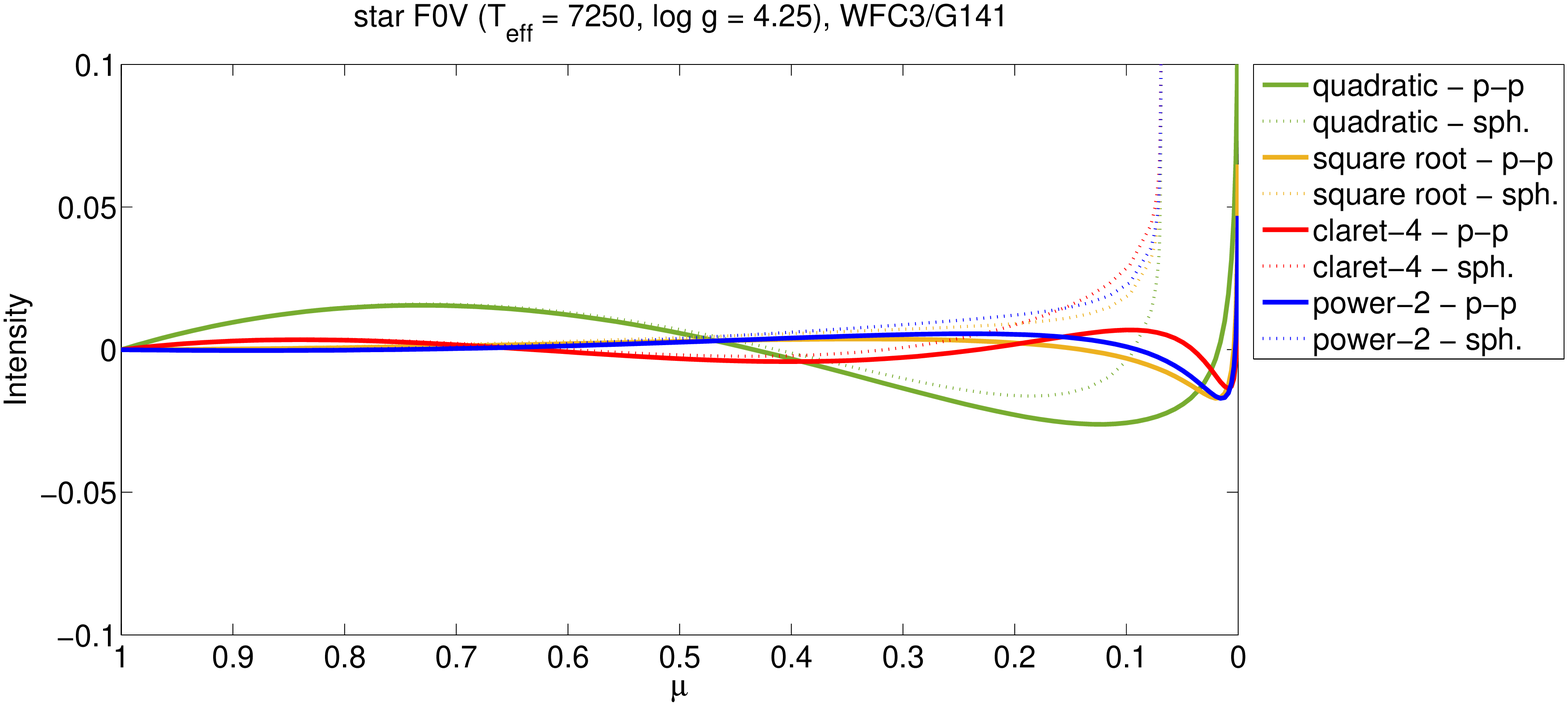}
\\
\hspace{-1cm}
\plotone{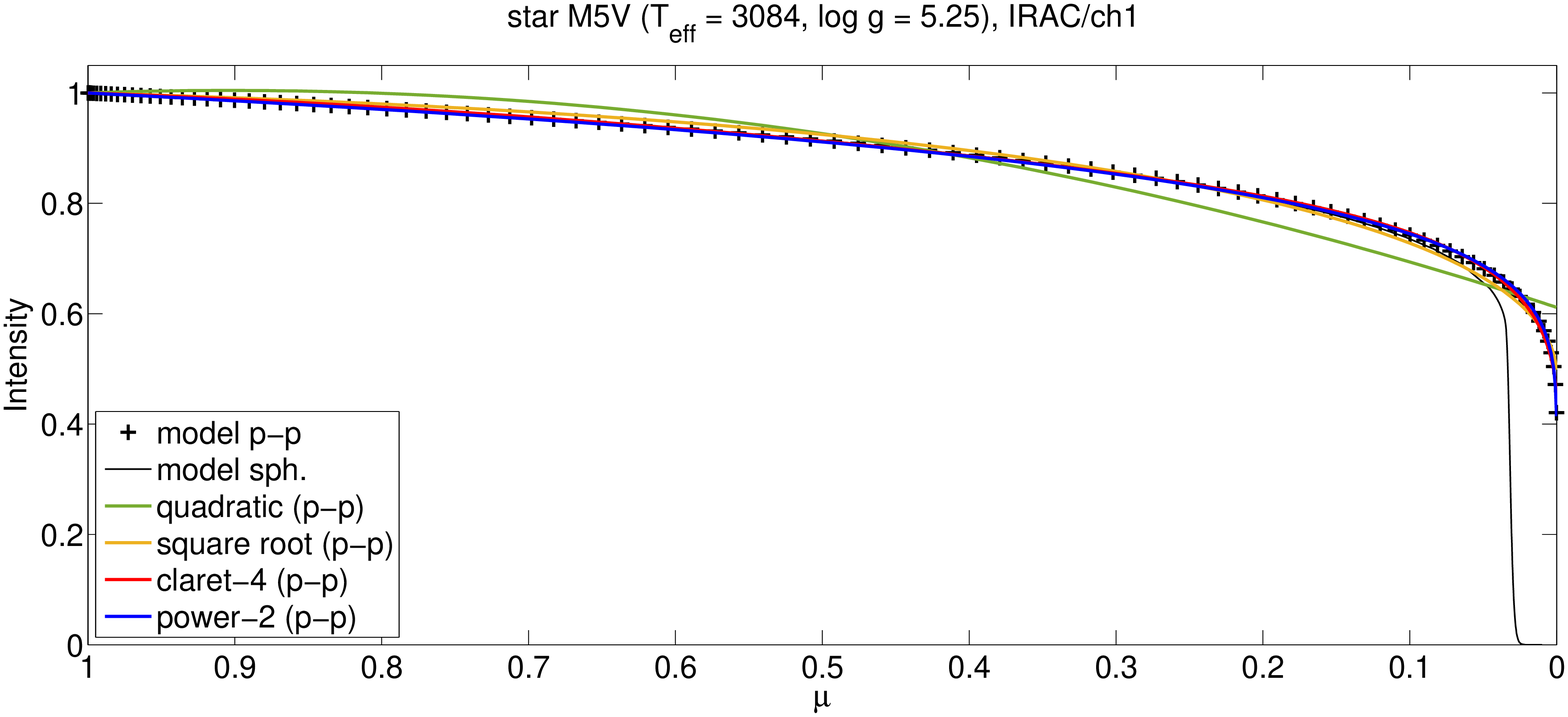}
\plotone{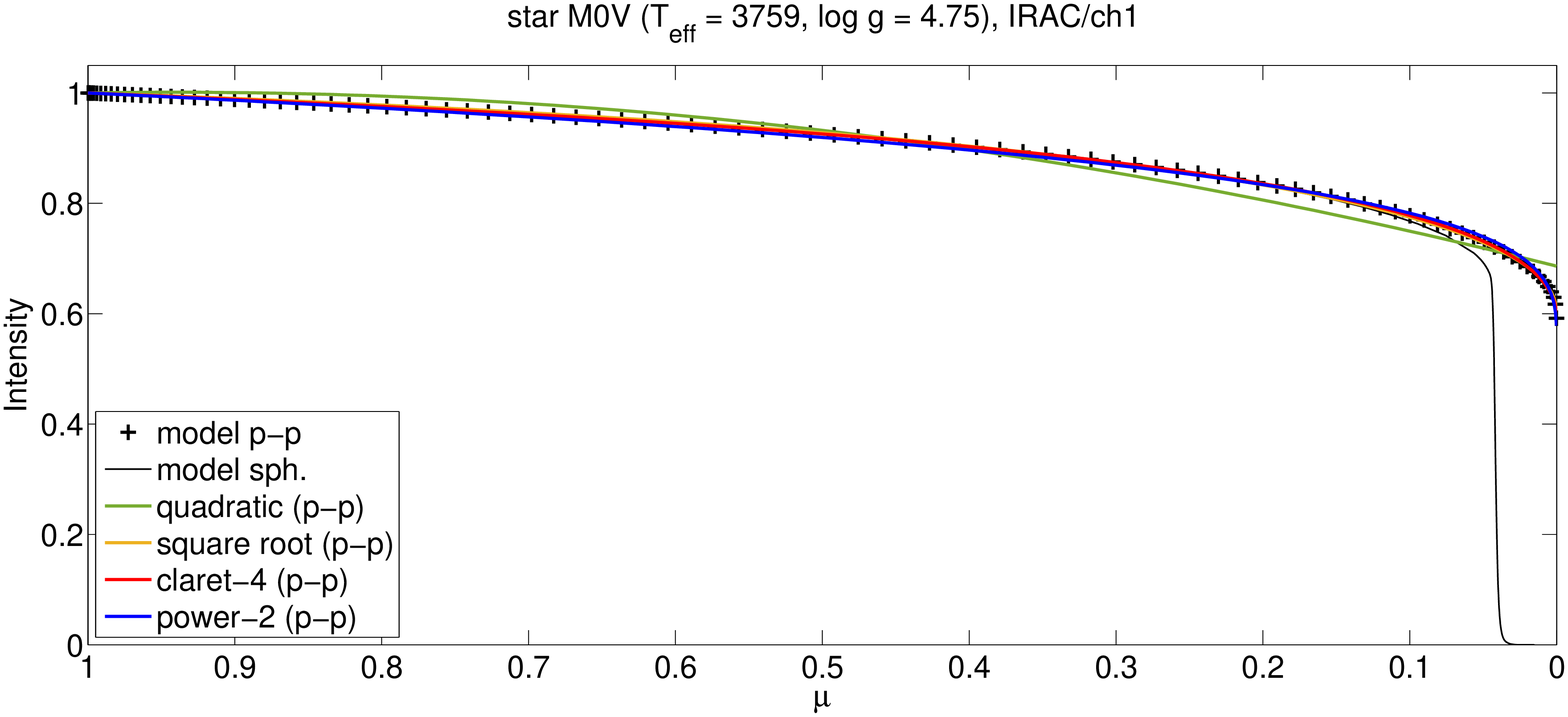}
\plotone{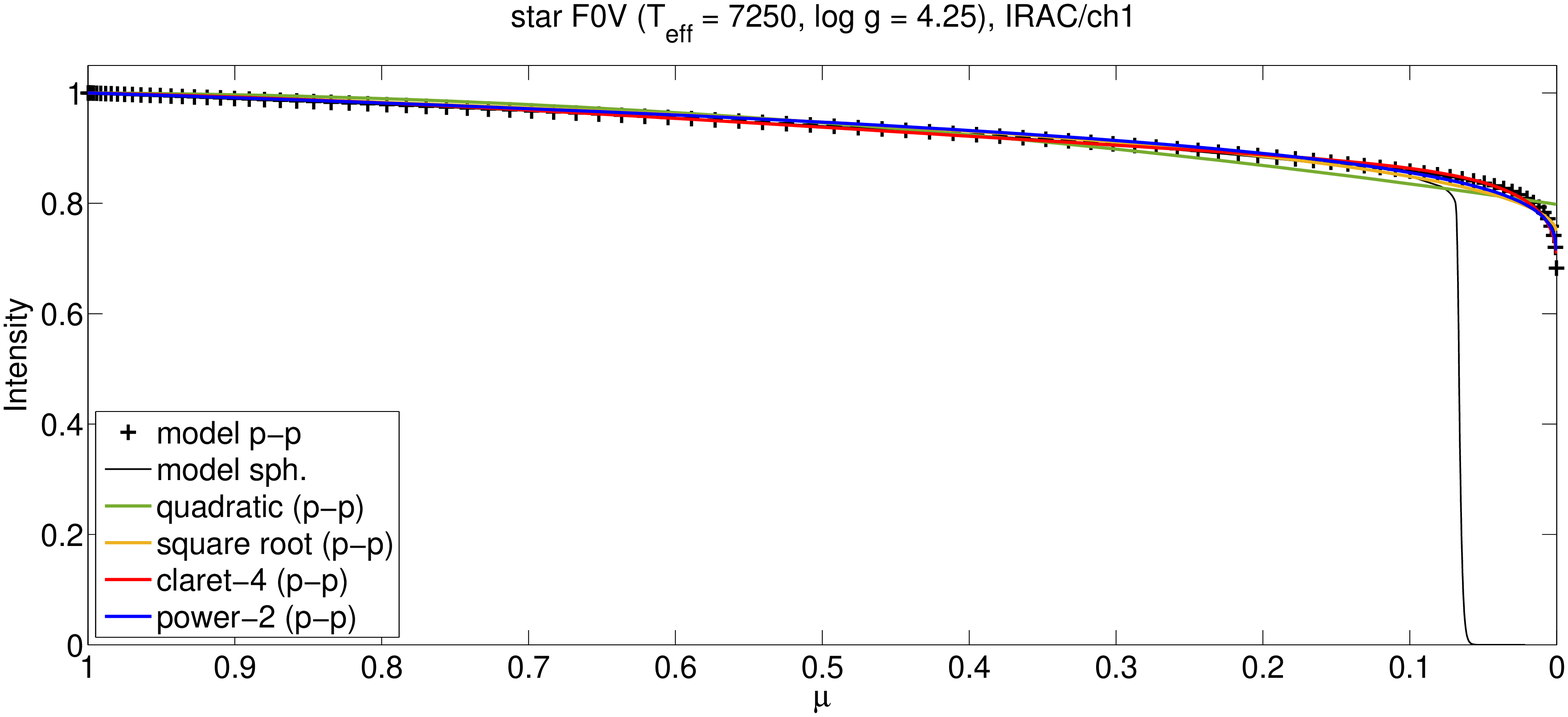}
\\
\hspace{-1cm}
\plotone{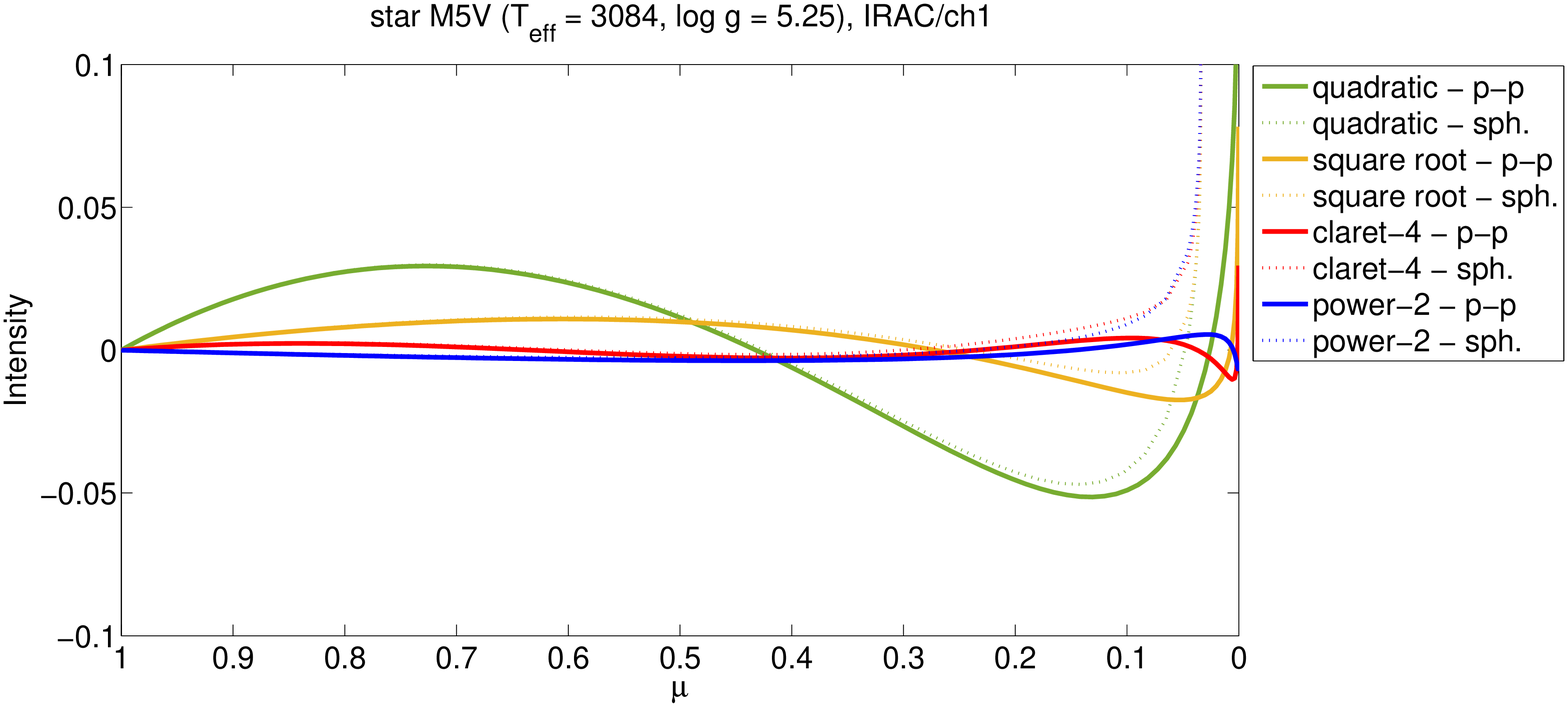}
\plotone{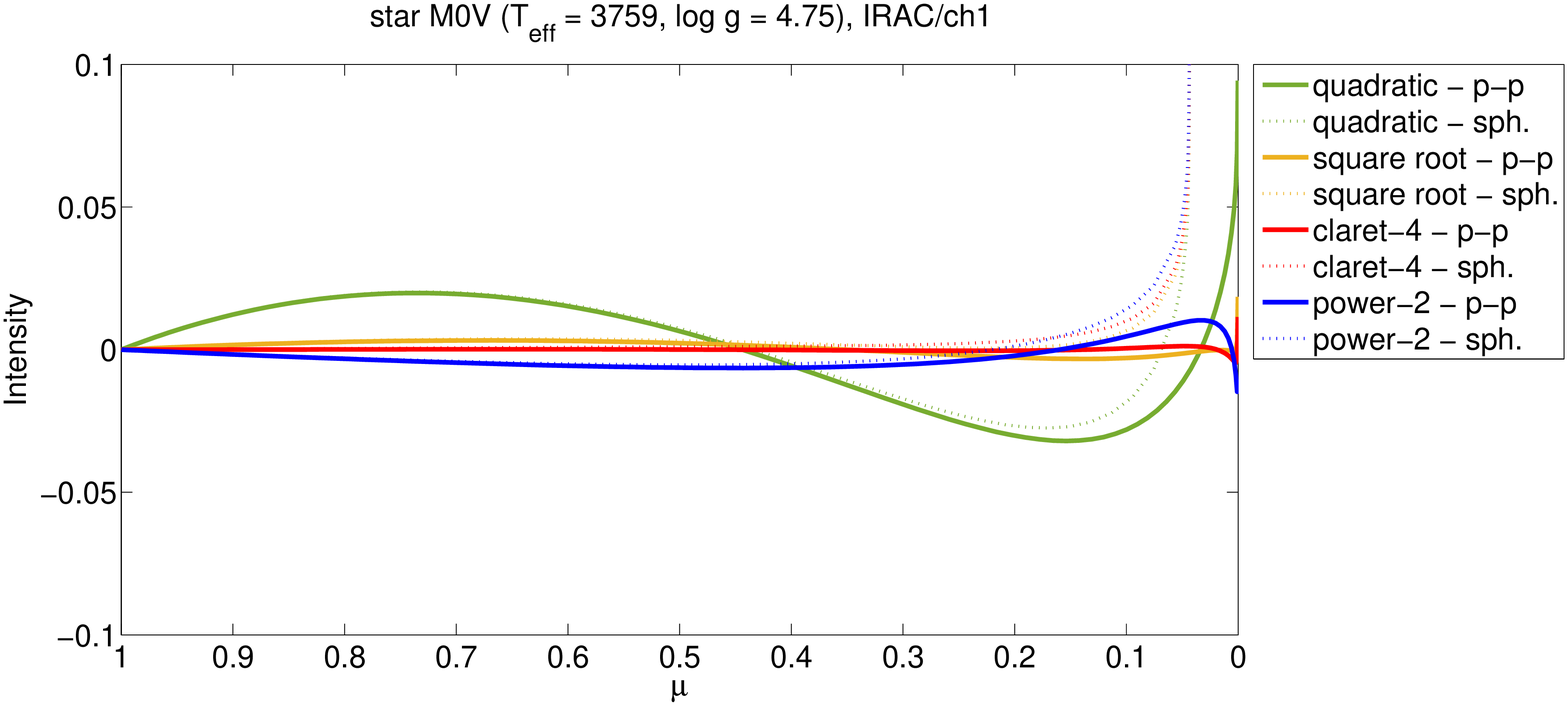}
\plotone{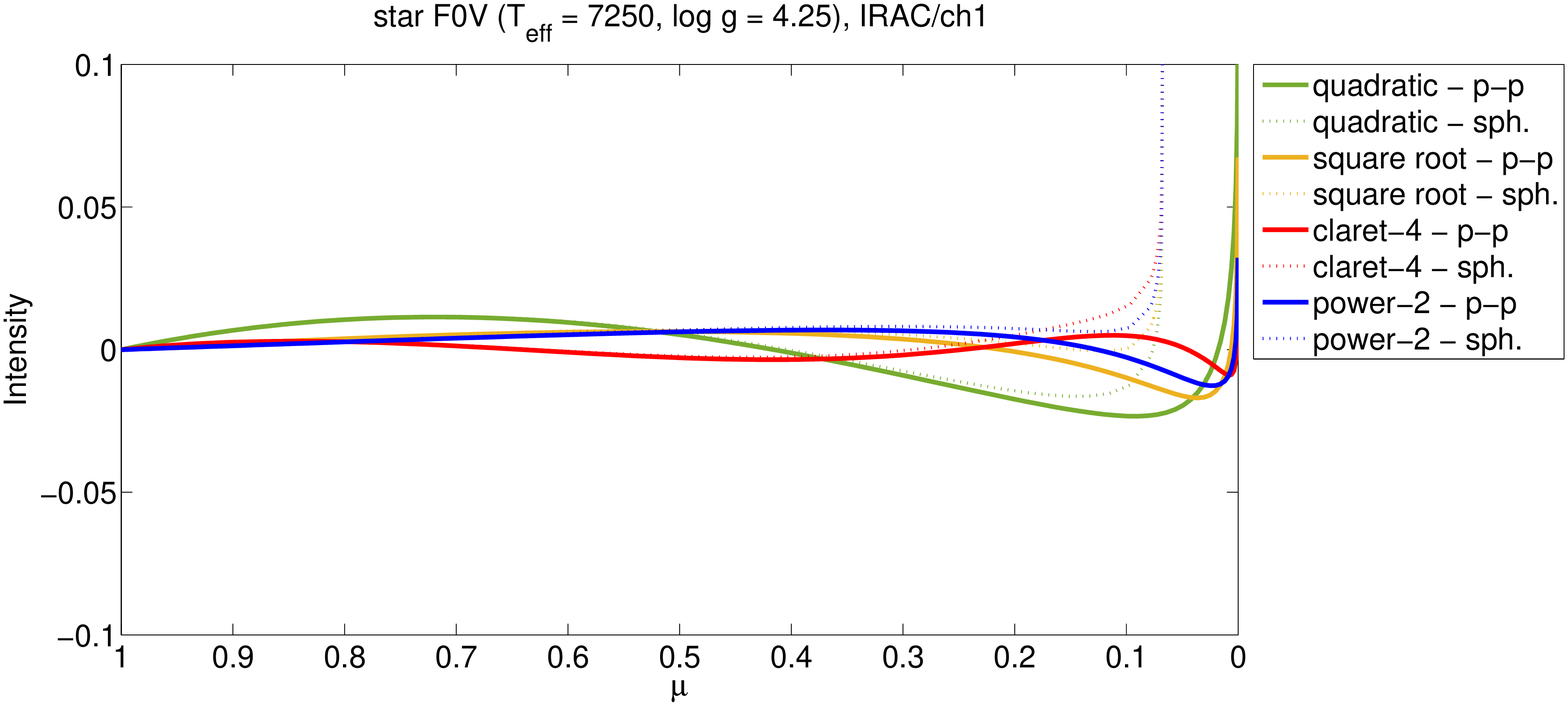}
\\
\hspace{-1cm}
\plotone{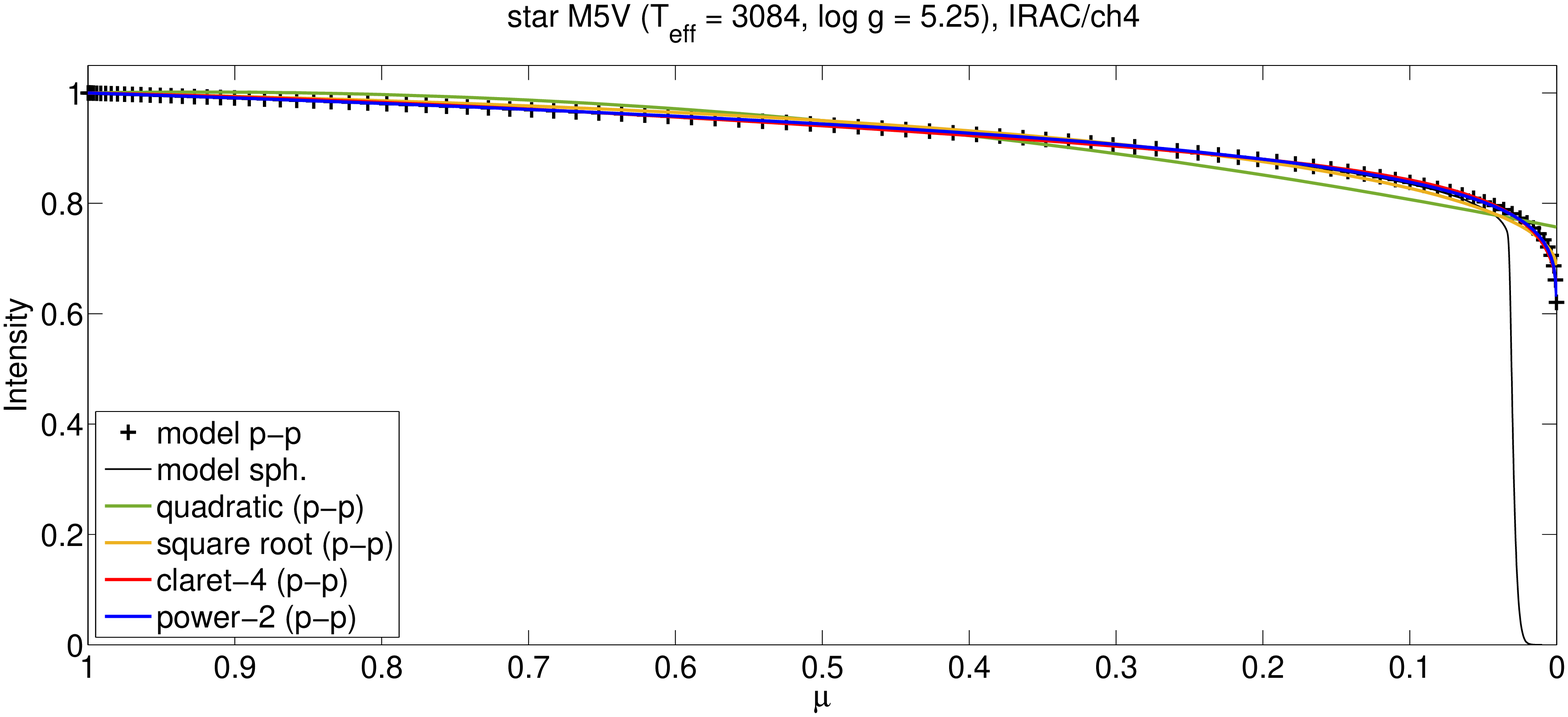}
\plotone{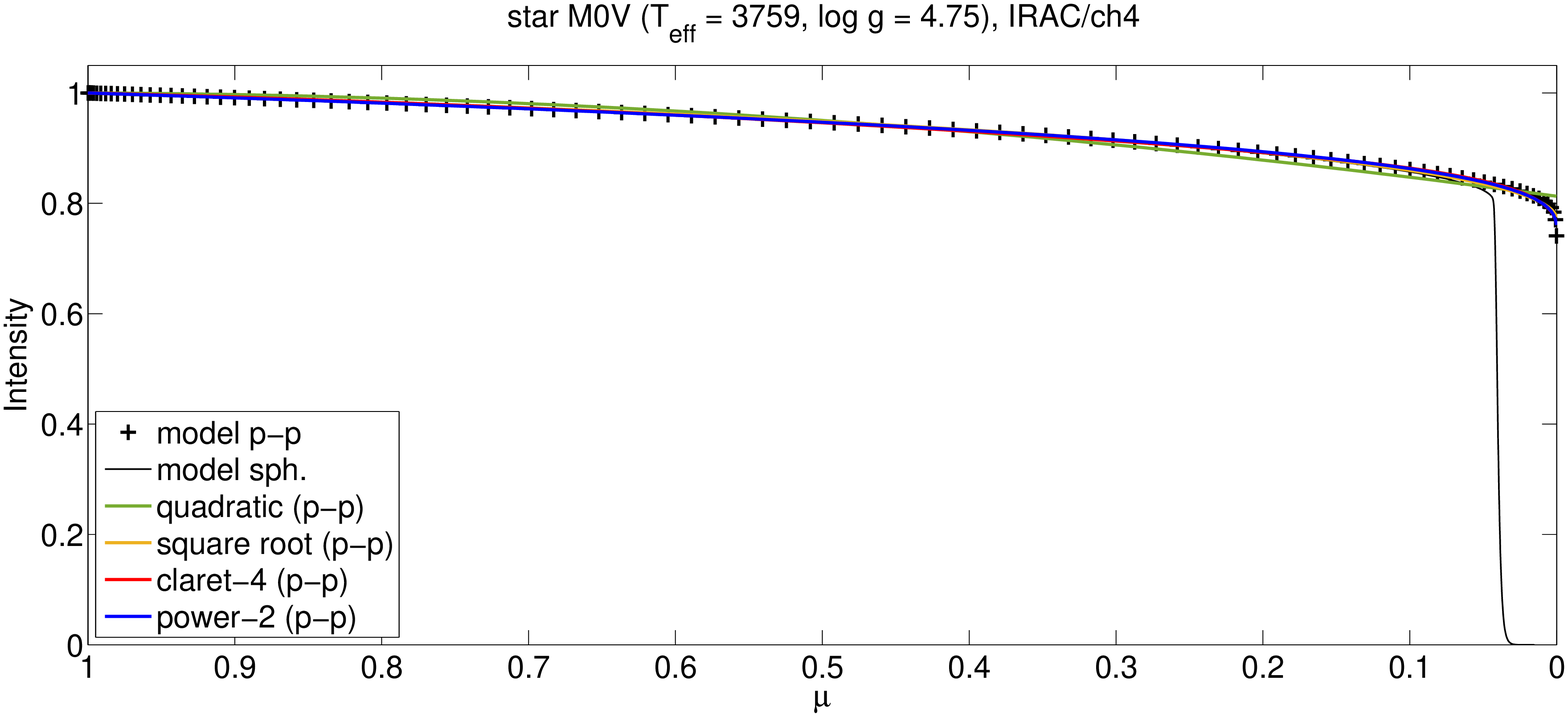}
\plotone{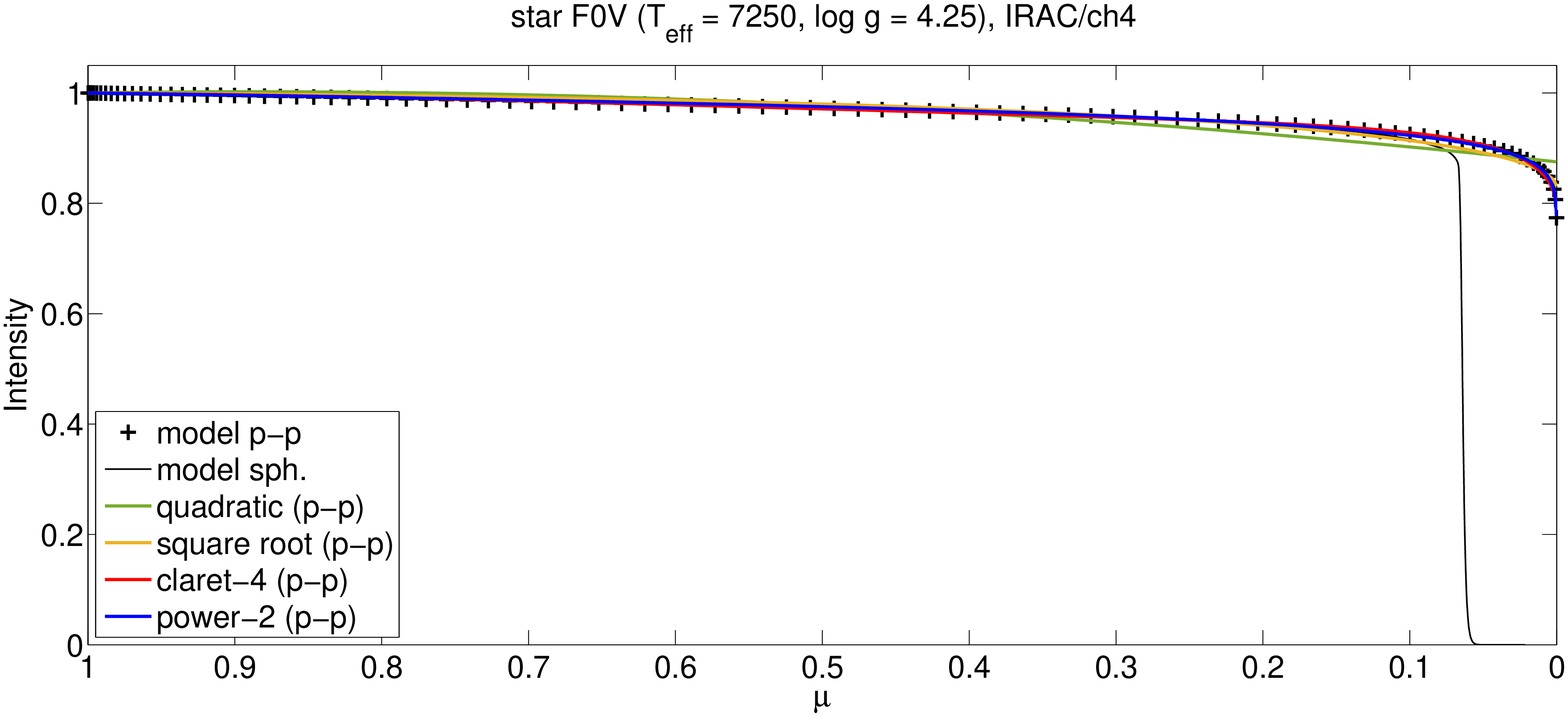}
\\
\hspace{-1cm}
\plotone{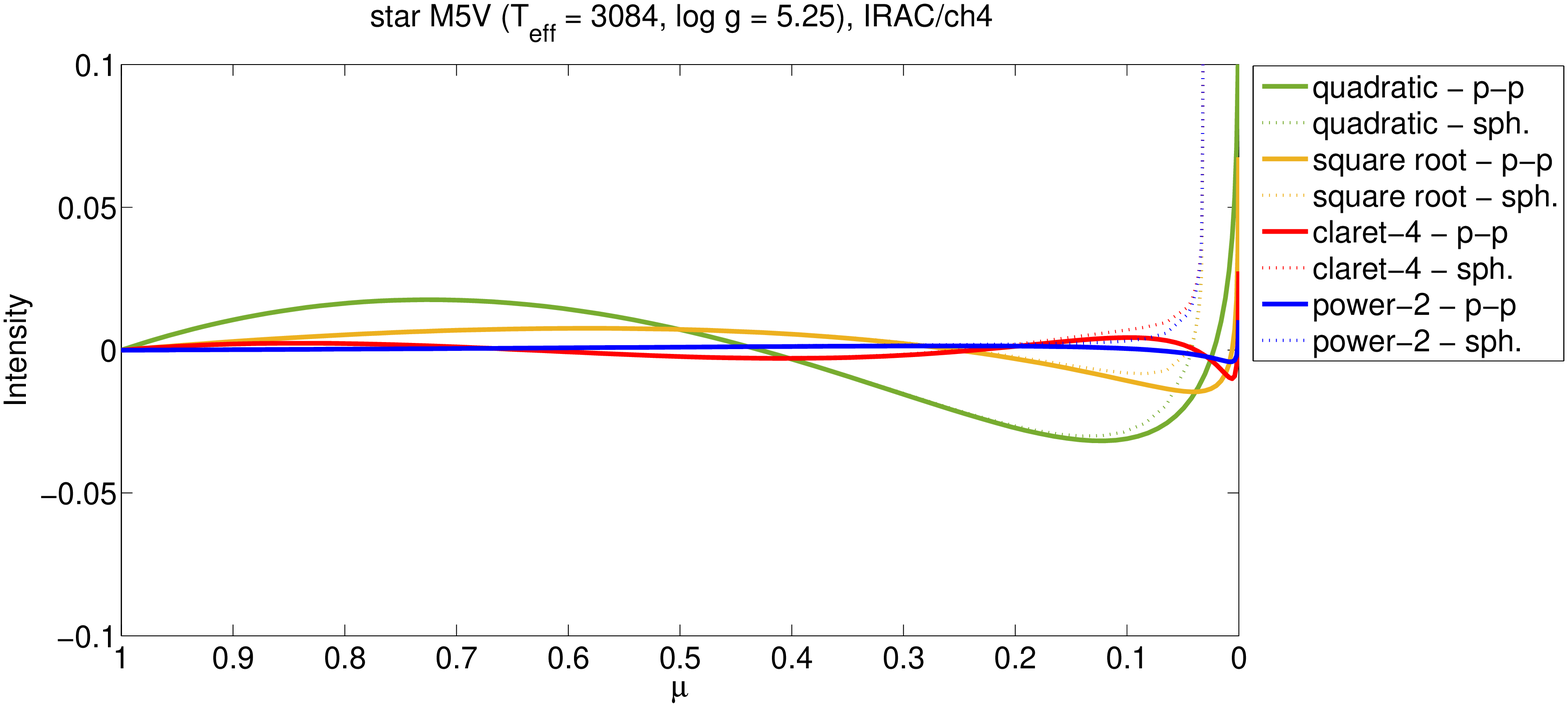}
\plotone{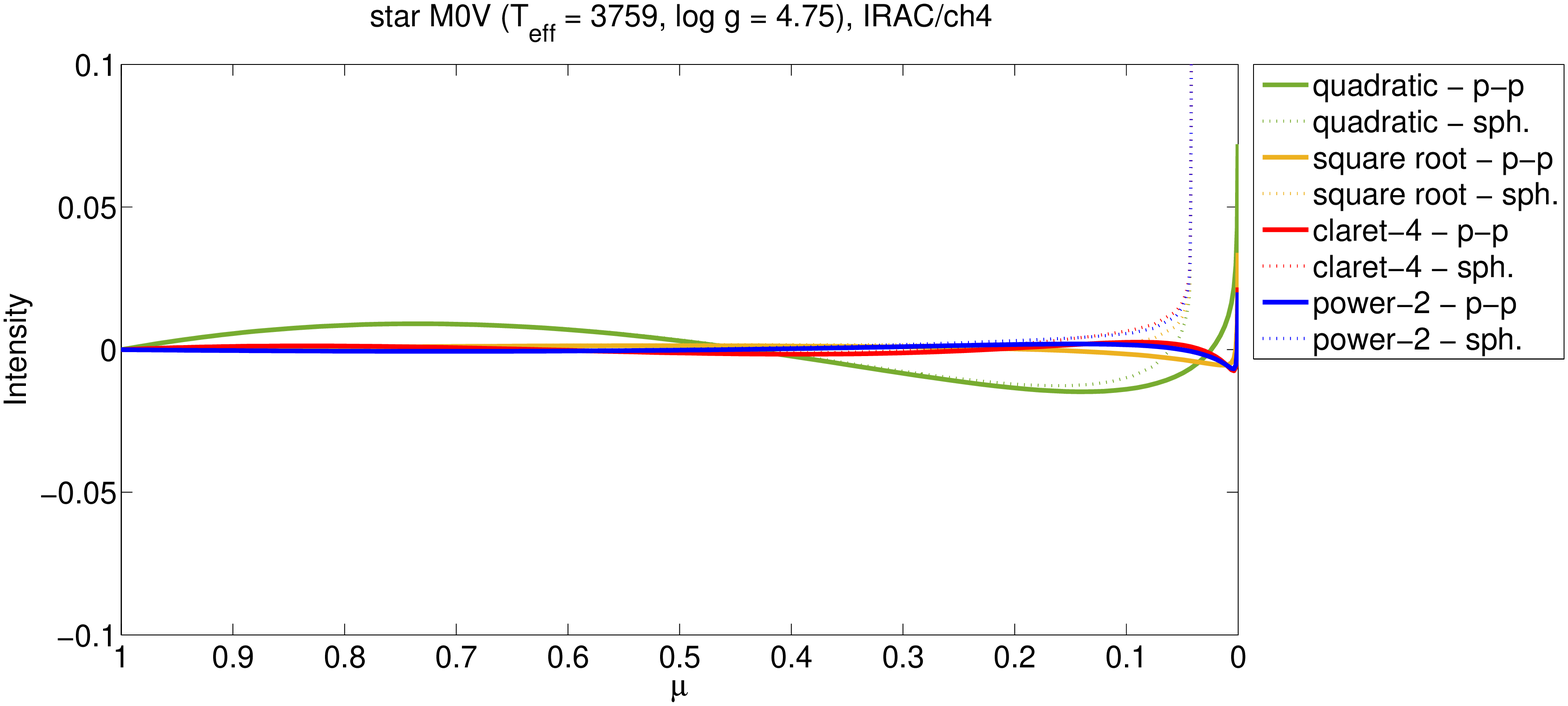}
\plotone{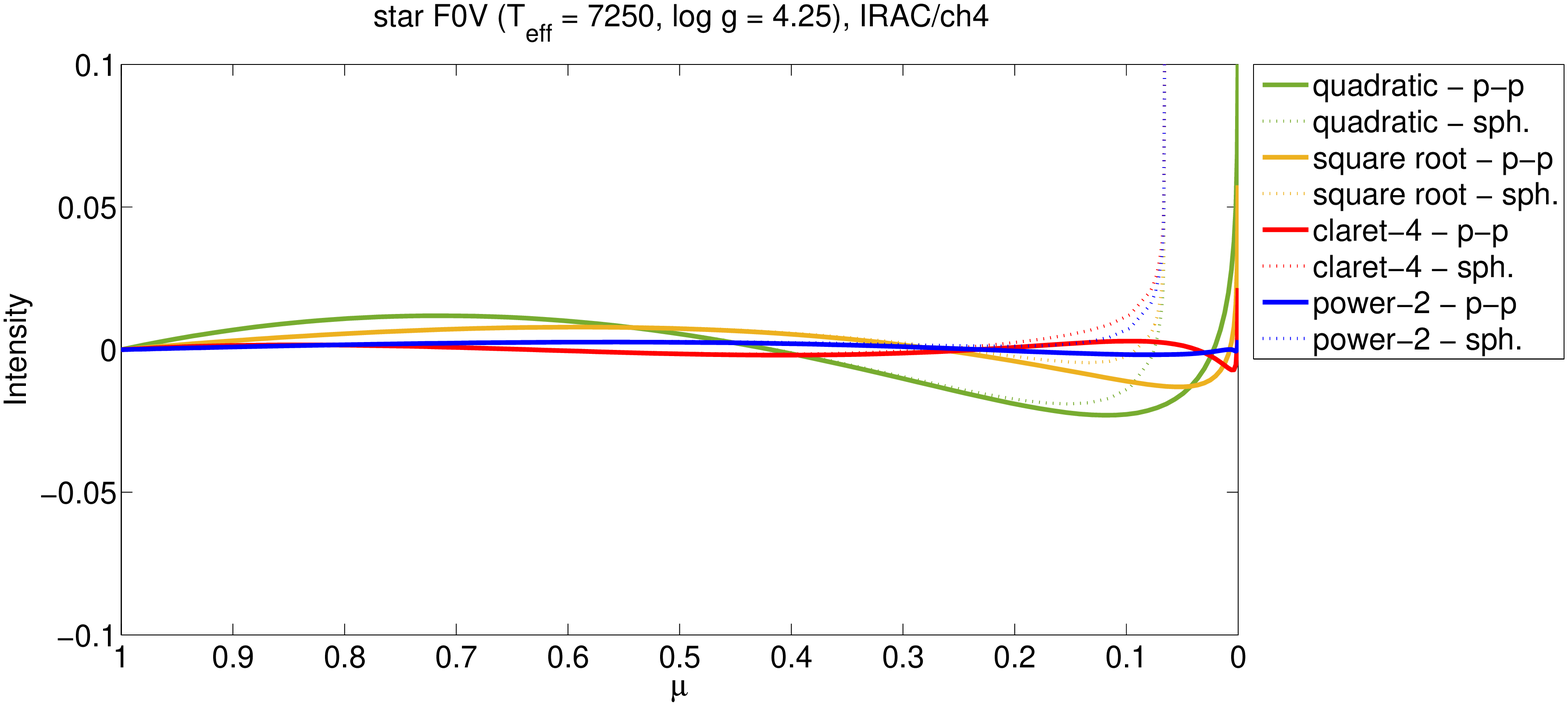}
\caption{Top panels: Plane-parallel (black `+') and spherical (black
  line) model-atmosphere intensities vs.\ $\mu$. Parametric limb-
  darkening functions fitted to plane-parallel intensities are
  quadratic (green), square-root (yellow), claret-4 (red), and power-2
  (blue) laws.\newline Bottom panels: large-scale plots of residuals of
  parametric limb-darkening laws for model-atmosphere
  intensities in the plane-parallel and
 spherical geometries (continuos and dashed
  lines, respectively).\label{fig4app}}
\end{figure}

\begin{figure}
\epsscale{0.32}
\plotone{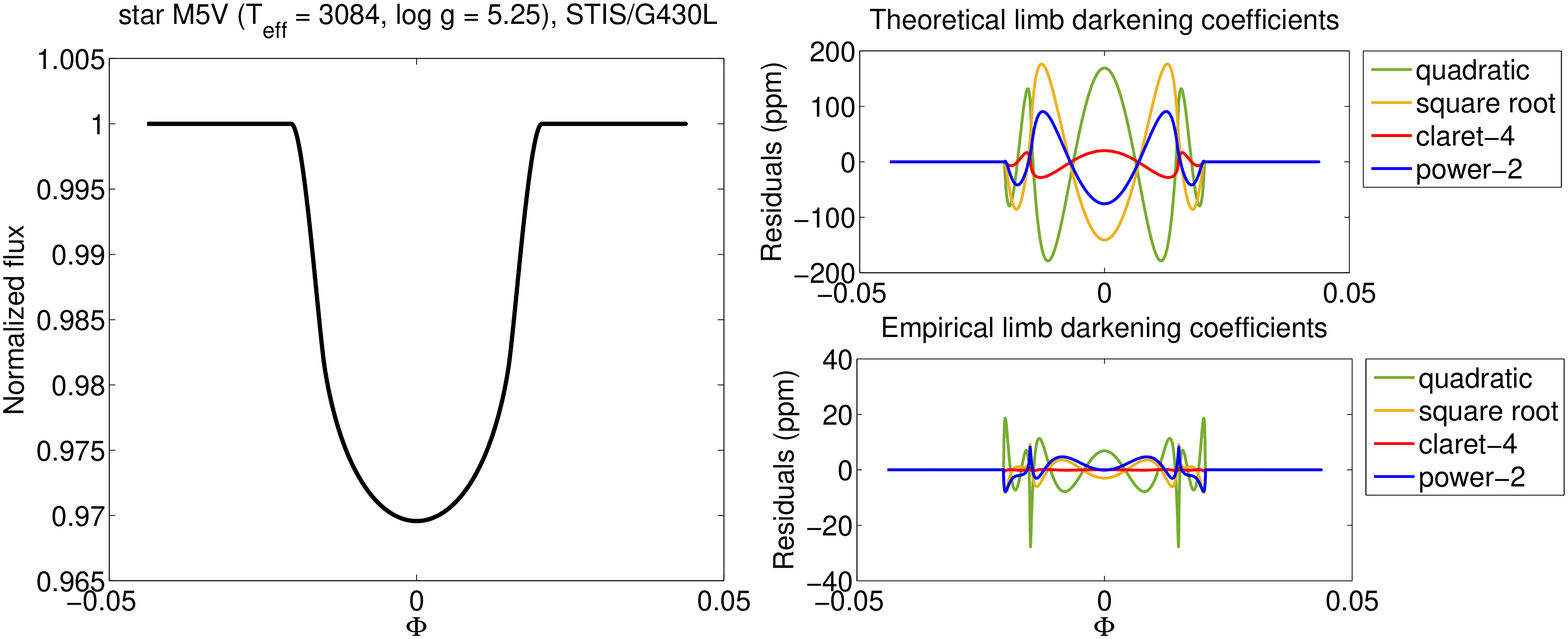}
\plotone{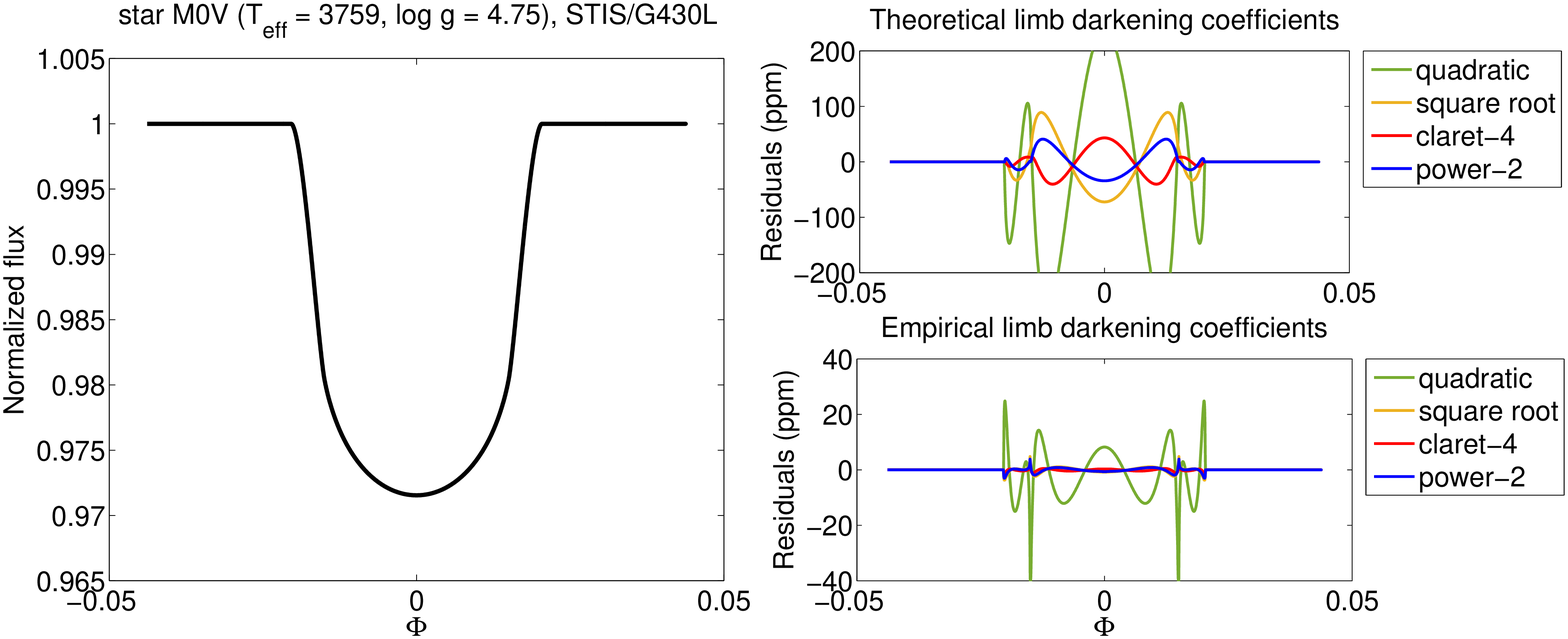}
\plotone{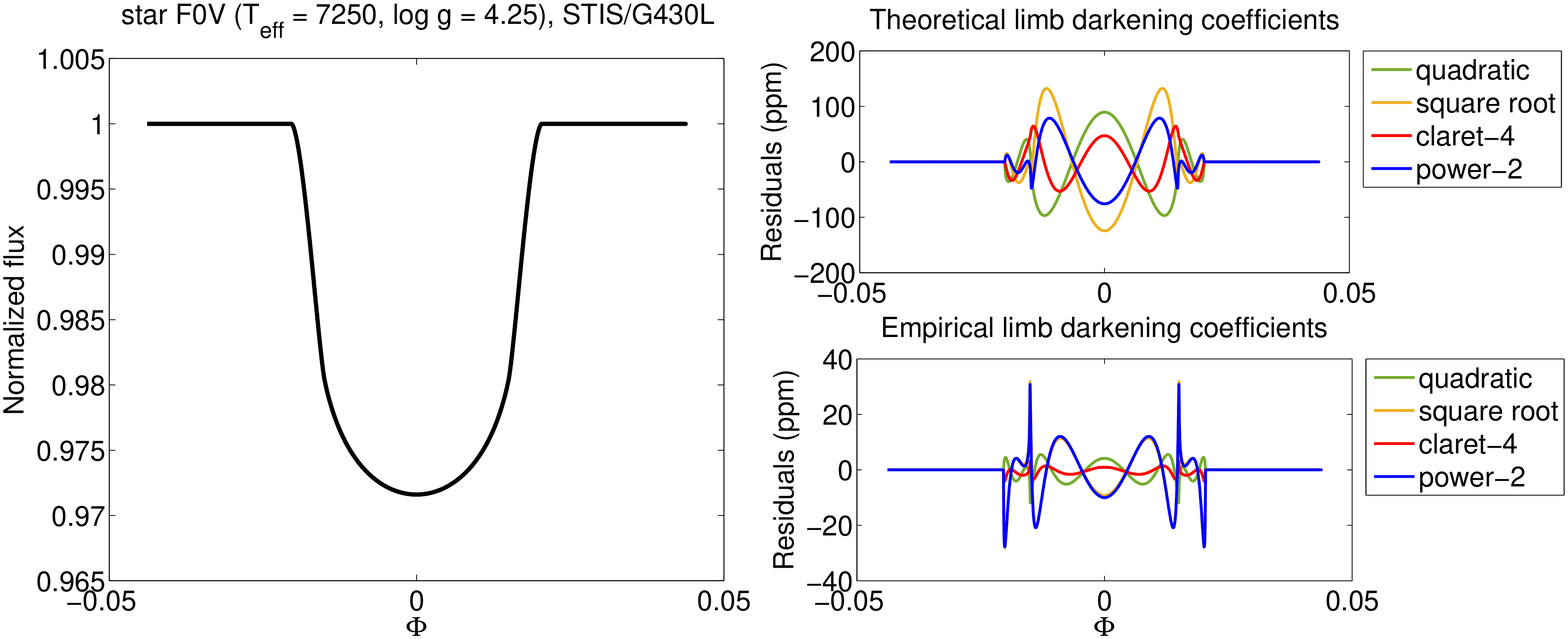}
\plotone{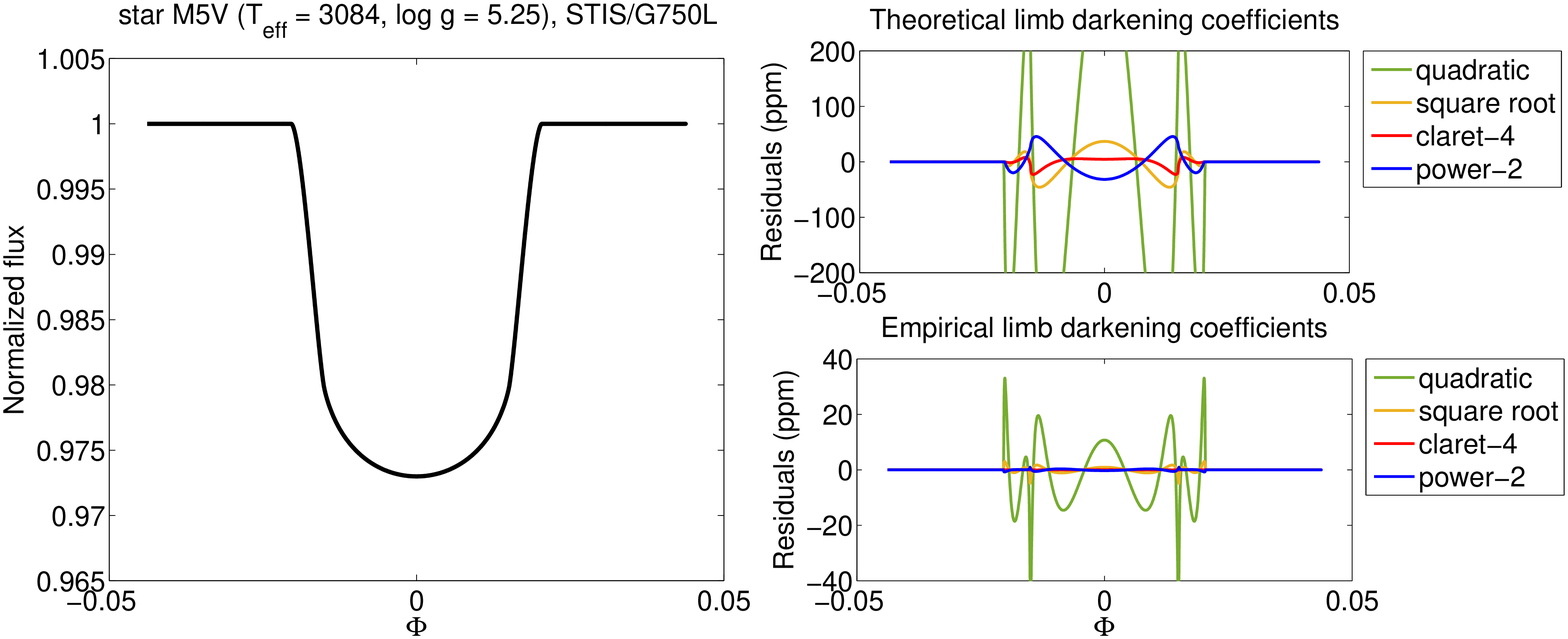}
\plotone{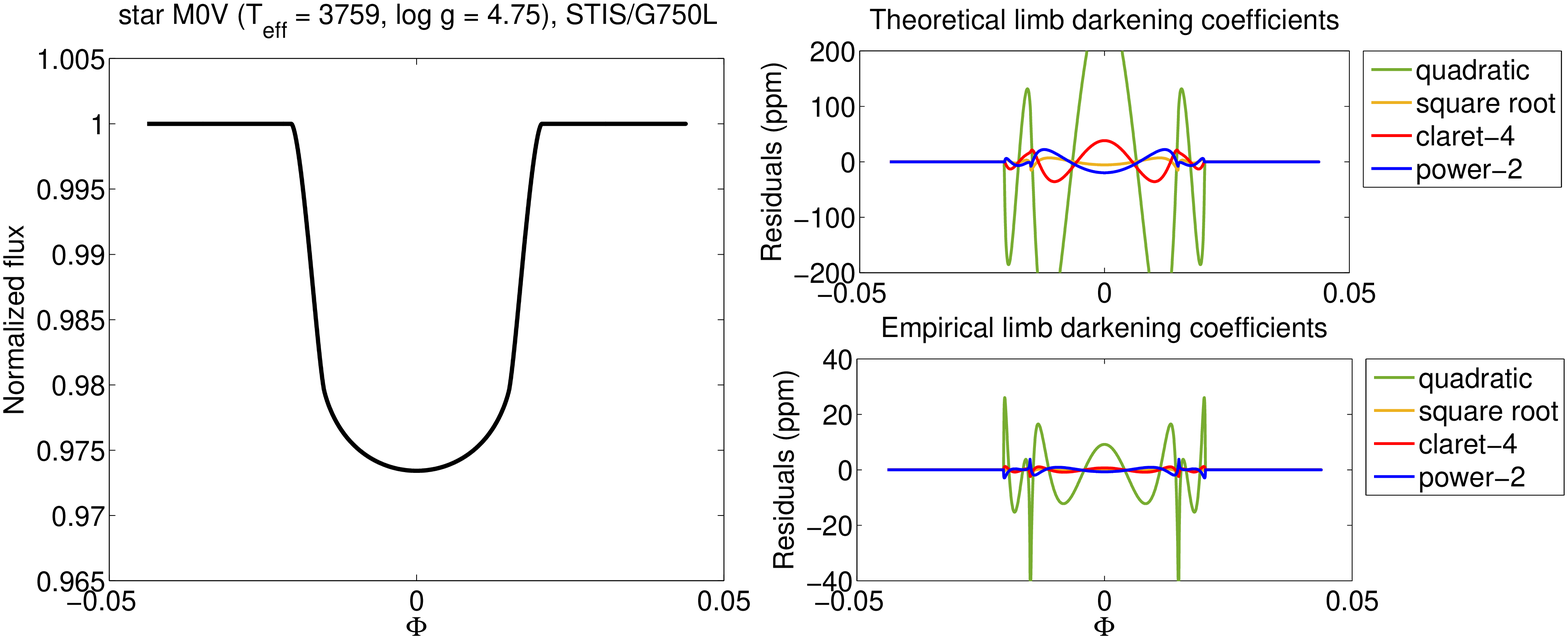}
\plotone{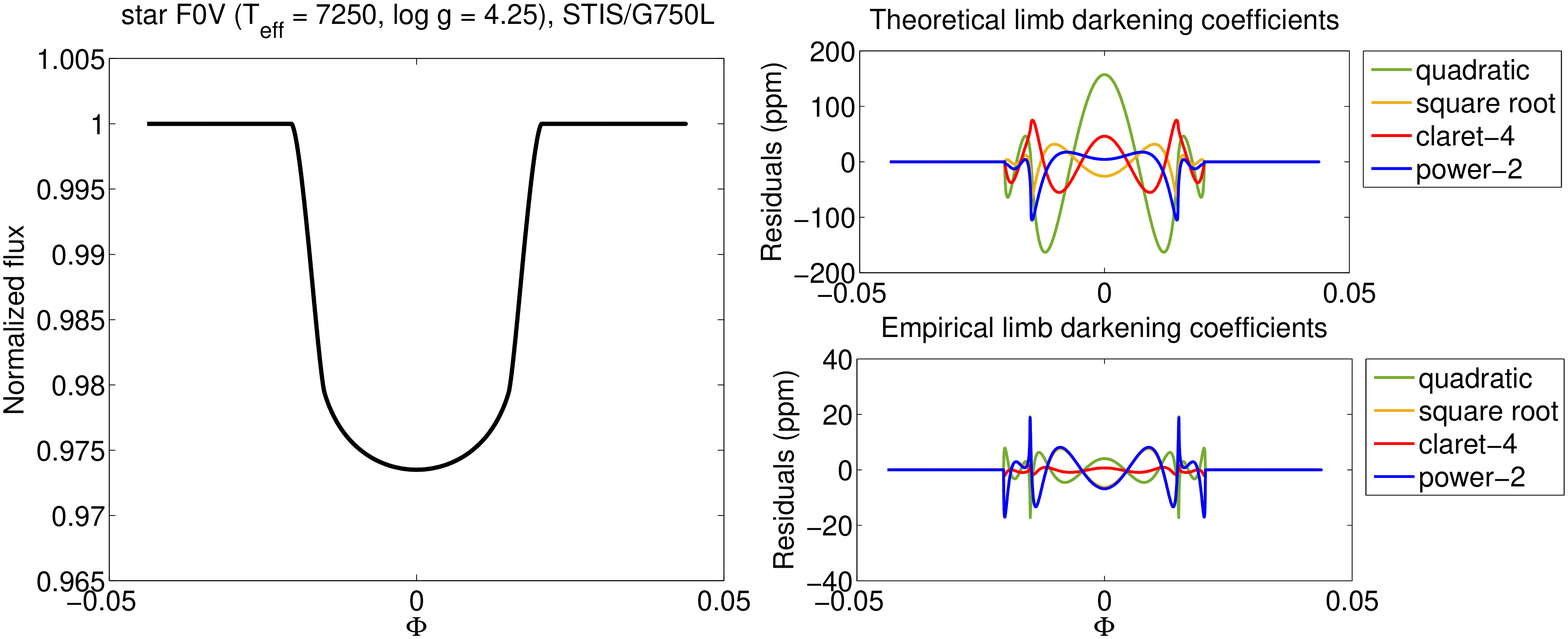}
\plotone{f6g.eps}
\plotone{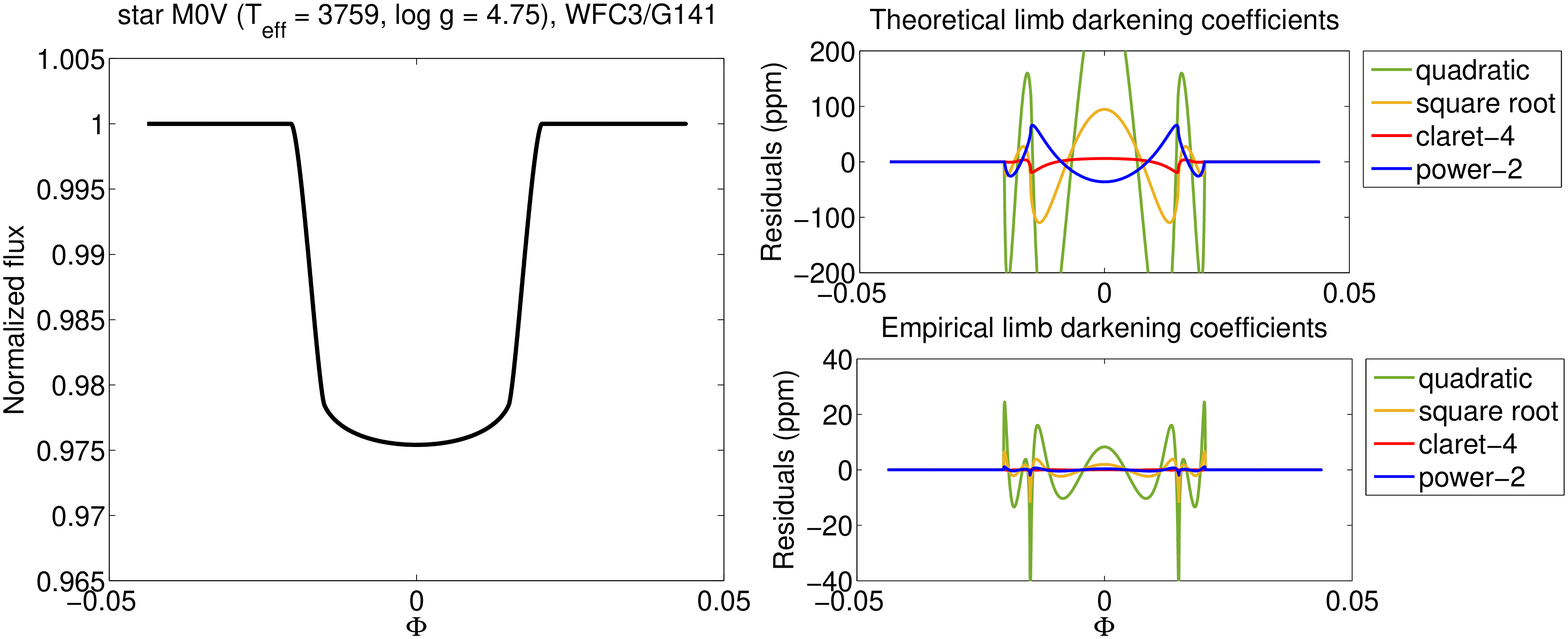}
\plotone{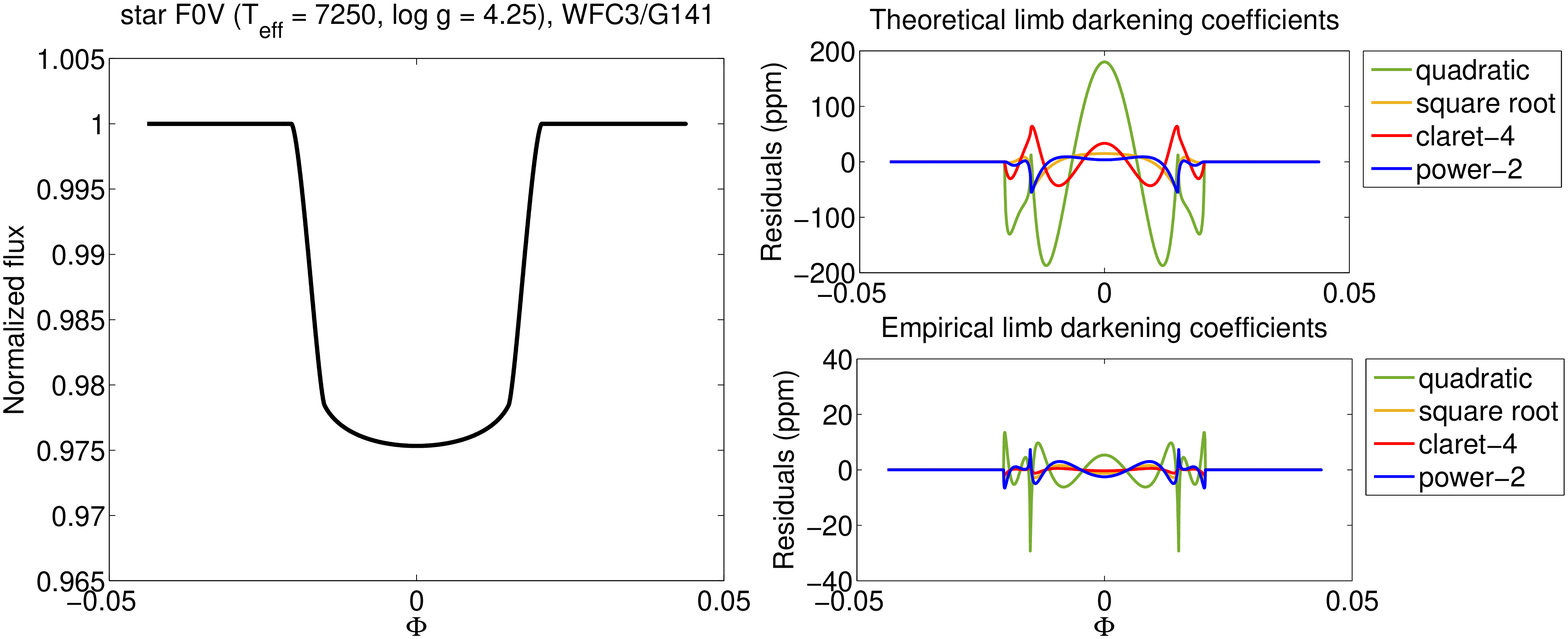}
\plotone{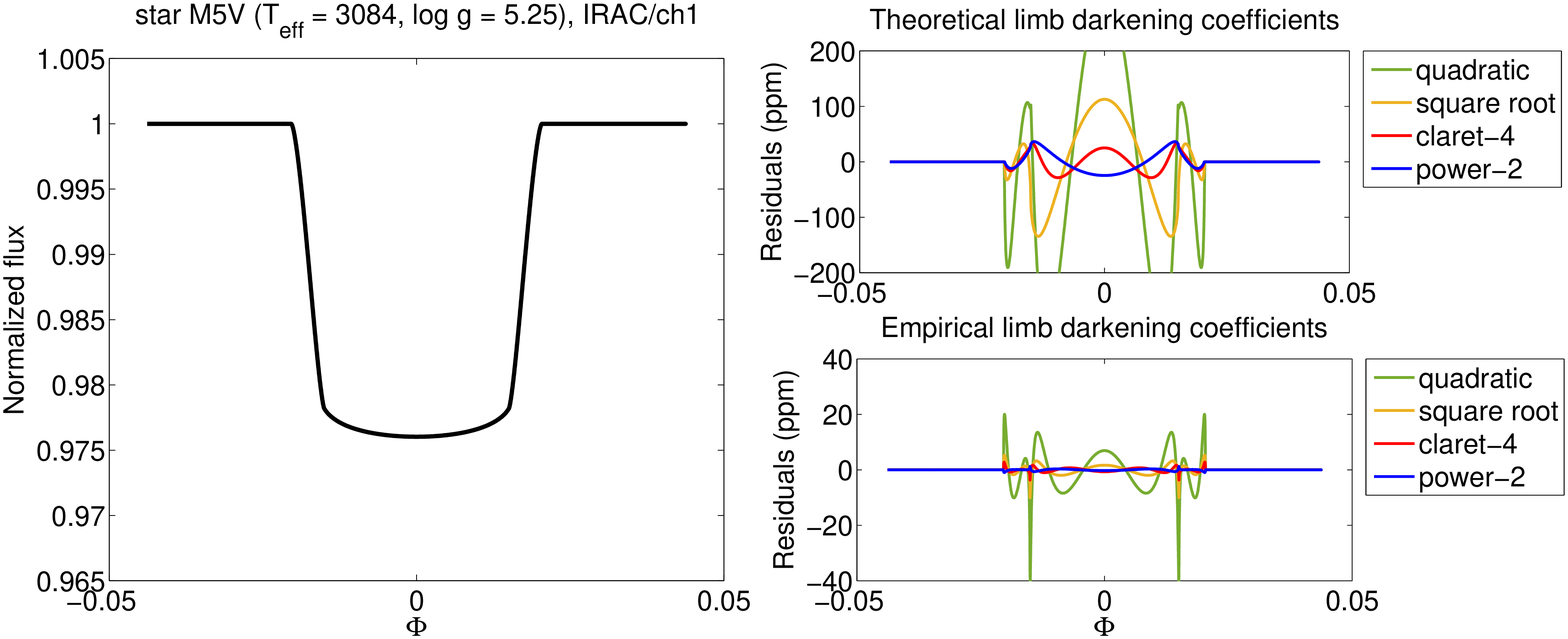}
\plotone{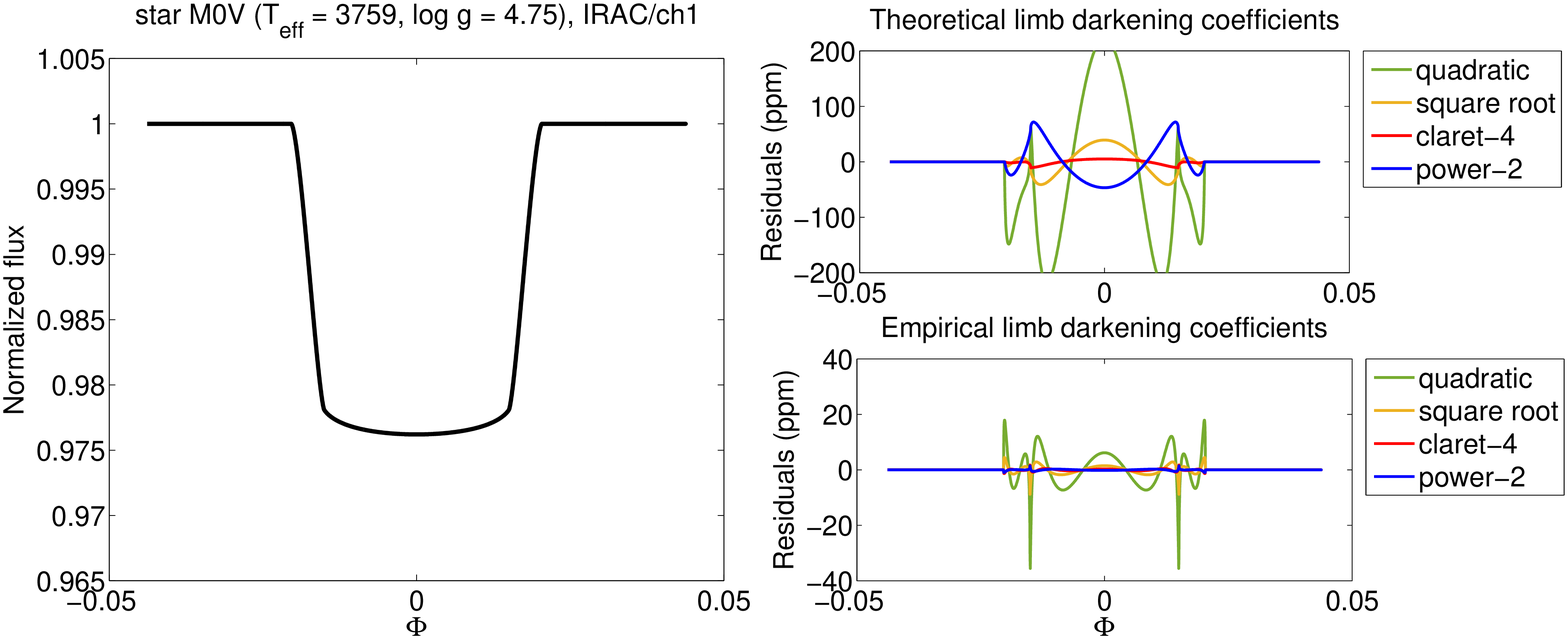}
\plotone{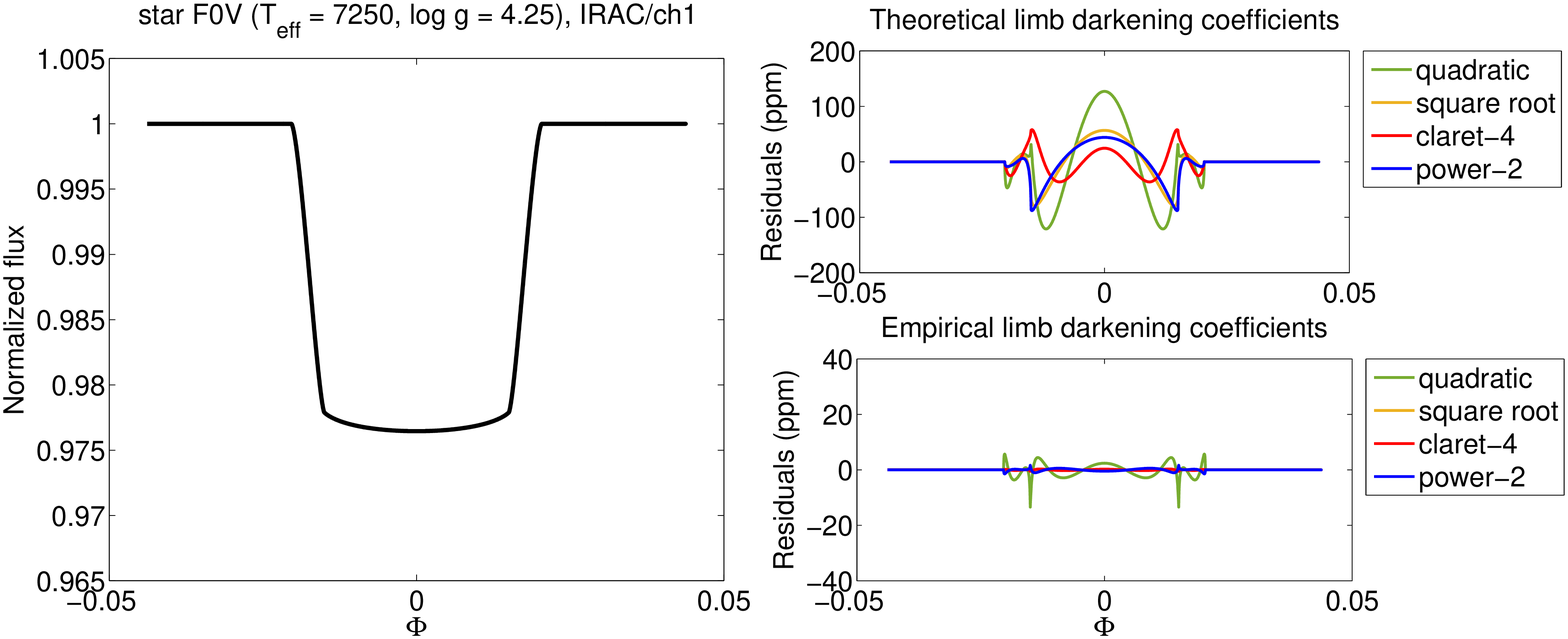}
\plotone{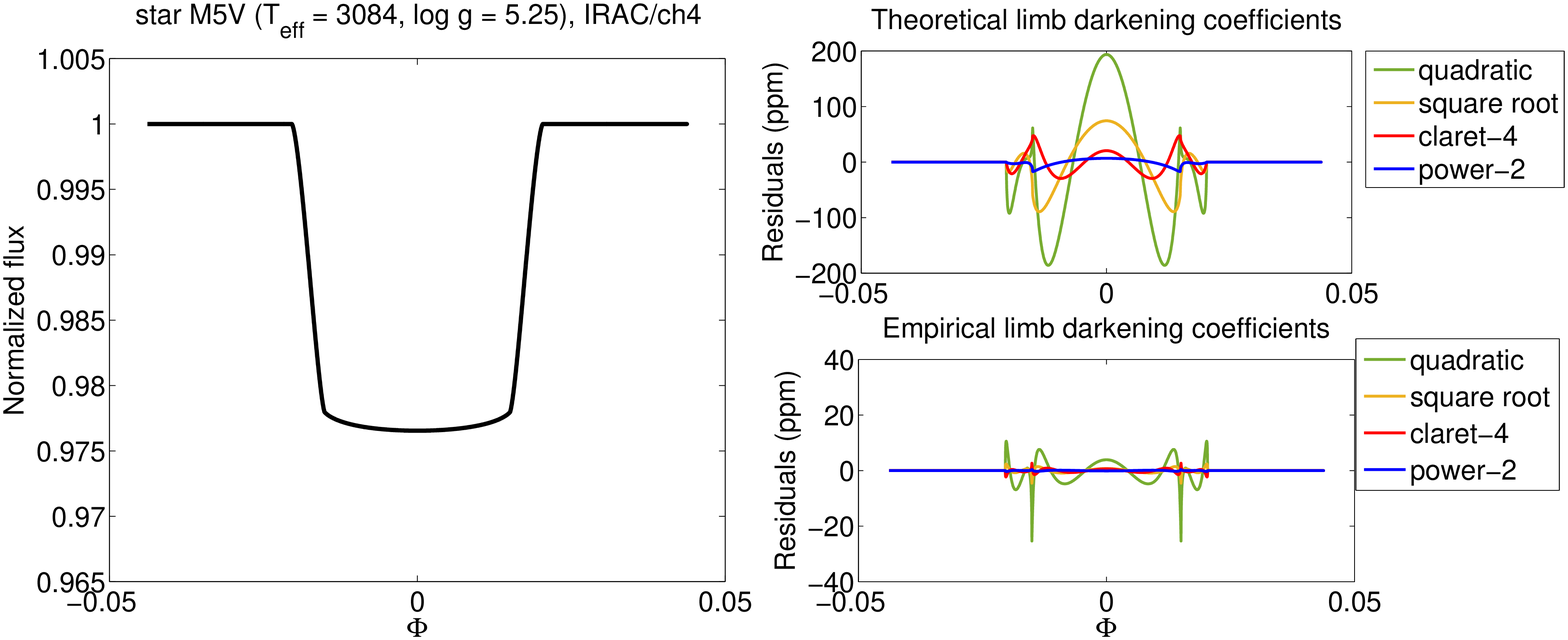}
\plotone{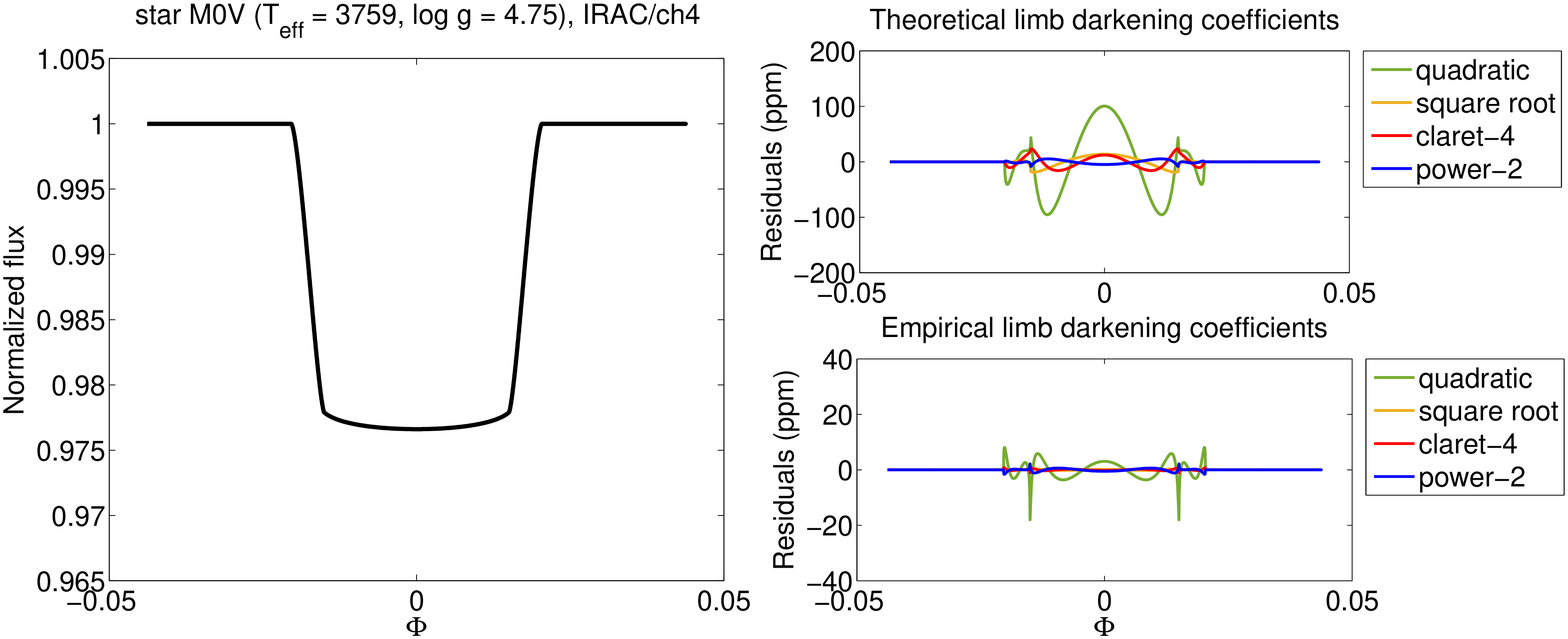}
\plotone{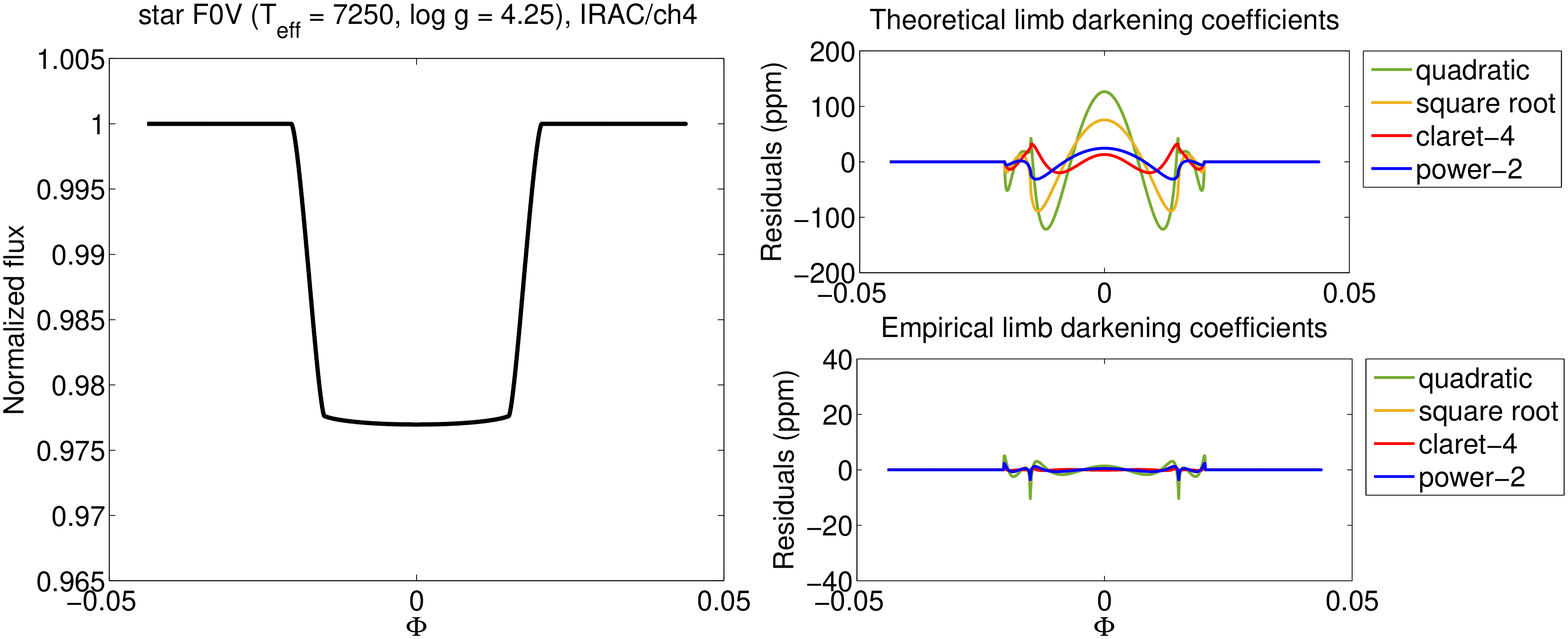}
\caption{Left panels: exact transit light-curves obtained with $b=0$. Right, top panels: residuals for the best-fit transit models using fixed quadratic (green), square-root (yellow), claret-4 (red), and power-2 (blue) limb-darkening coefficients. The coefficients are fitted to the plane-parallel angular intensities. Right, bottom panels: residuals obtained with the empirical limb-darkening coefficients. \label{fig6app}}
\end{figure}

\begin{figure}
\epsscale{0.3}
\hspace{-1cm}
\plotone{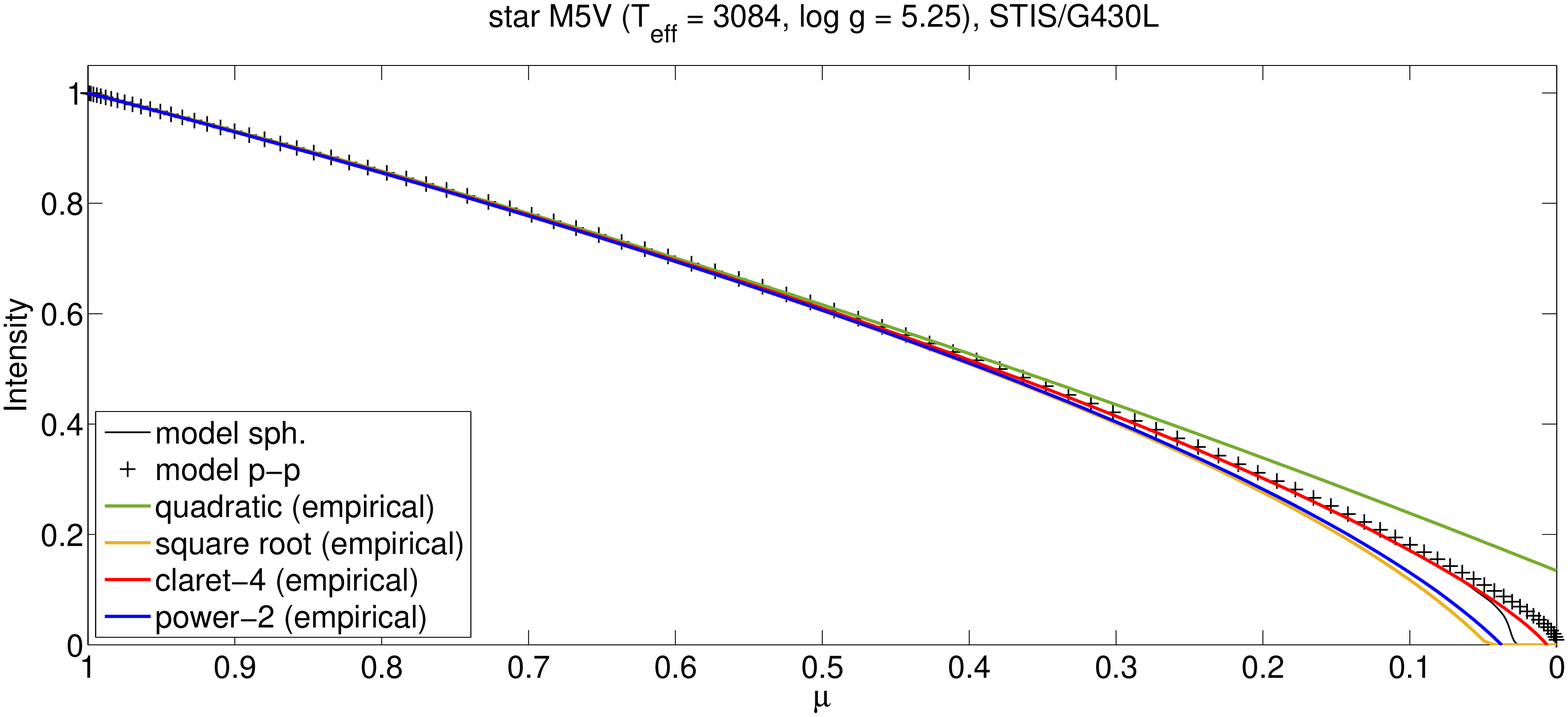}
\plotone{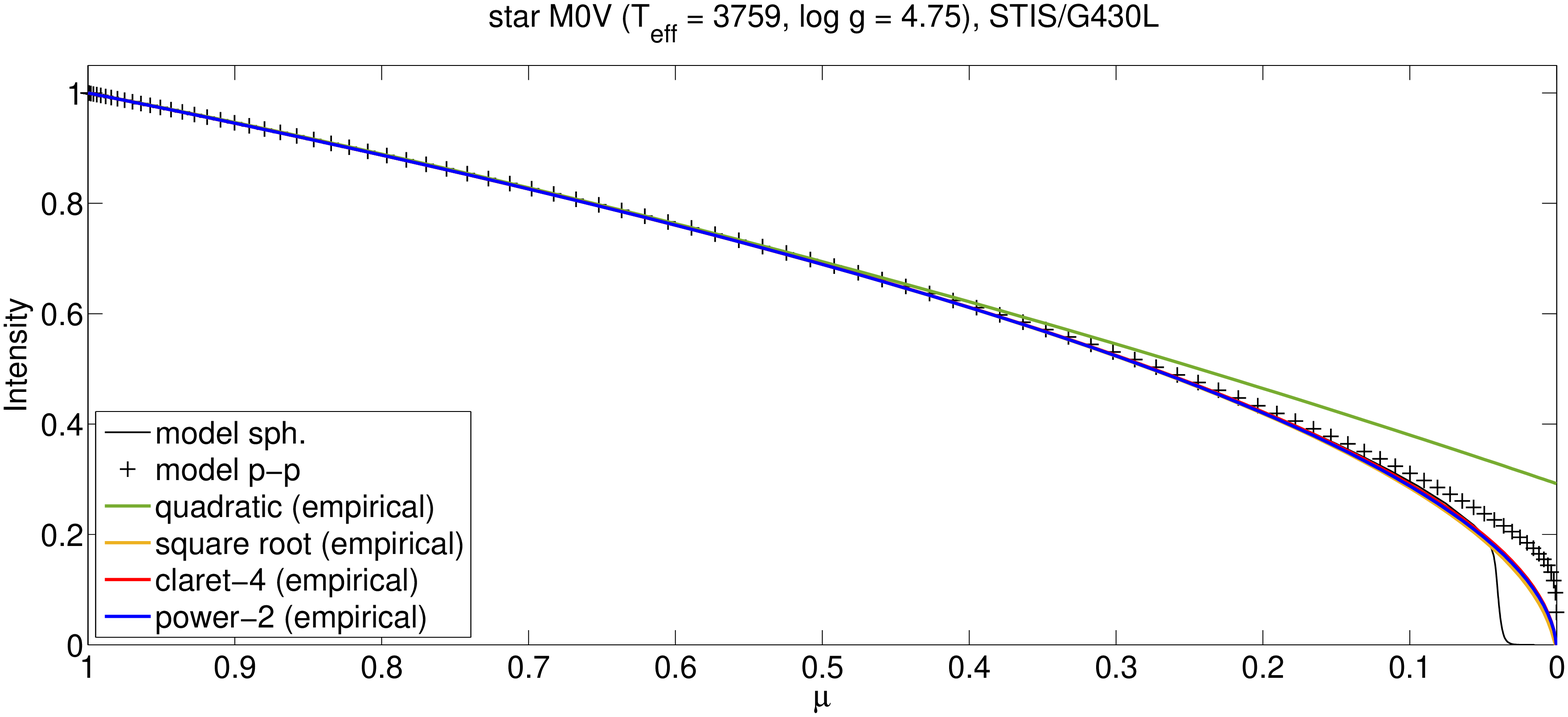}
\plotone{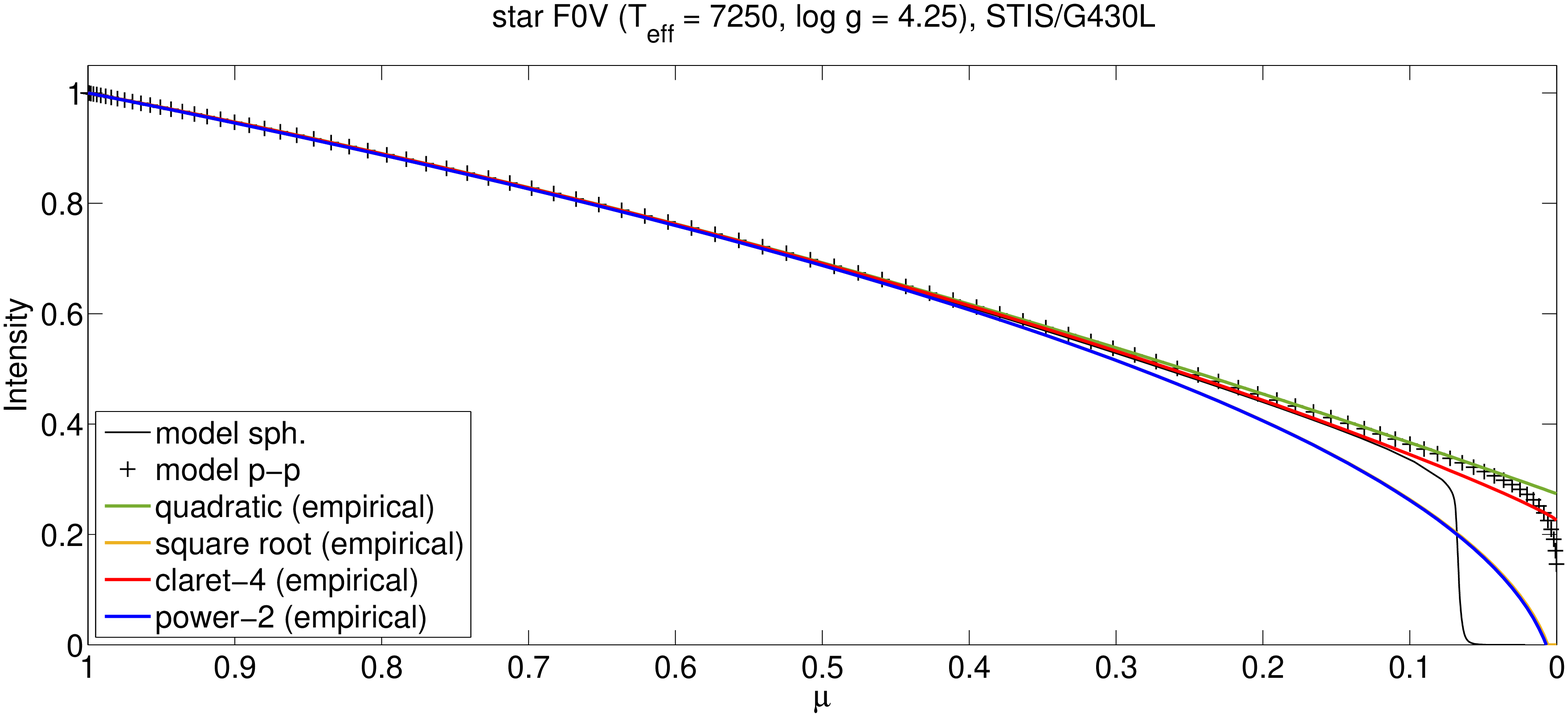}
\\
\hspace{-1cm}
\plotone{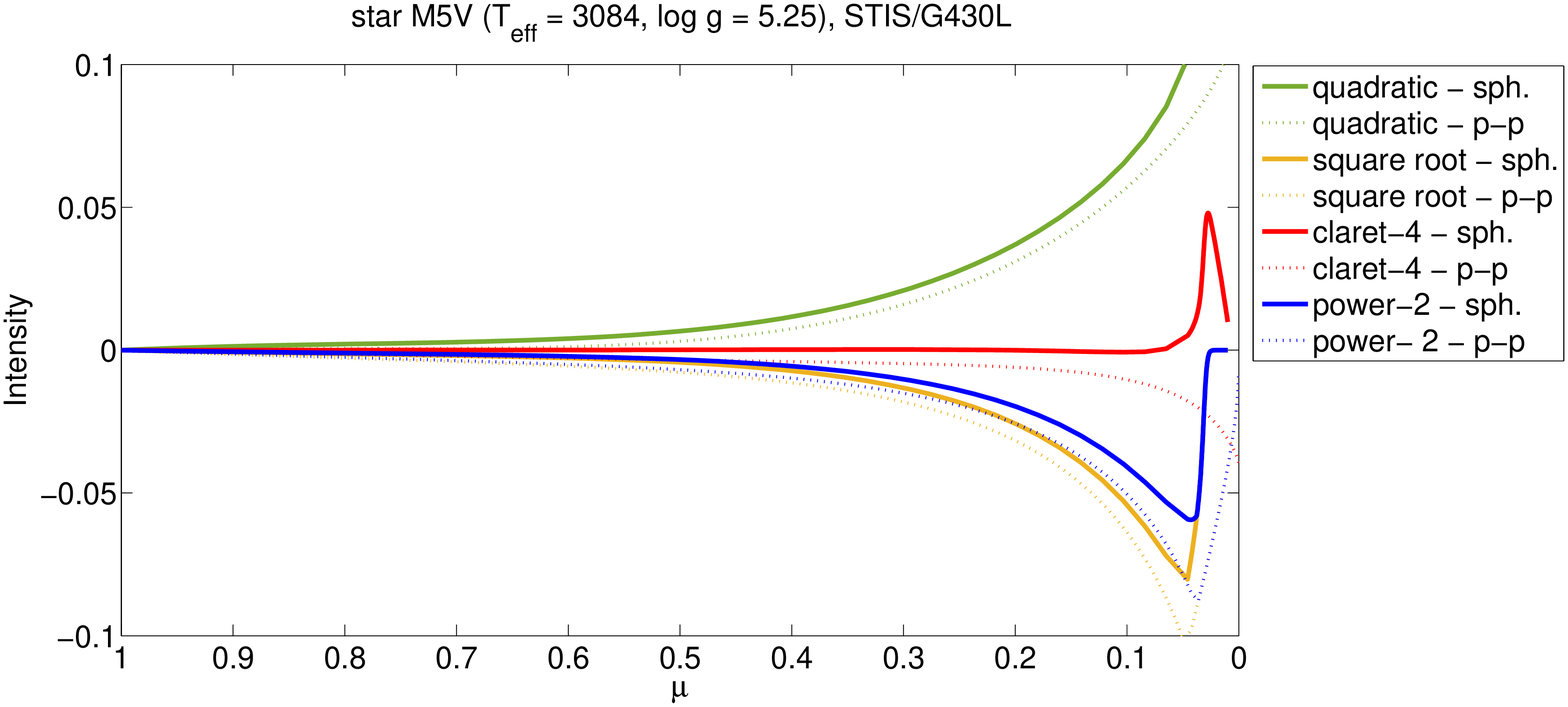}
\plotone{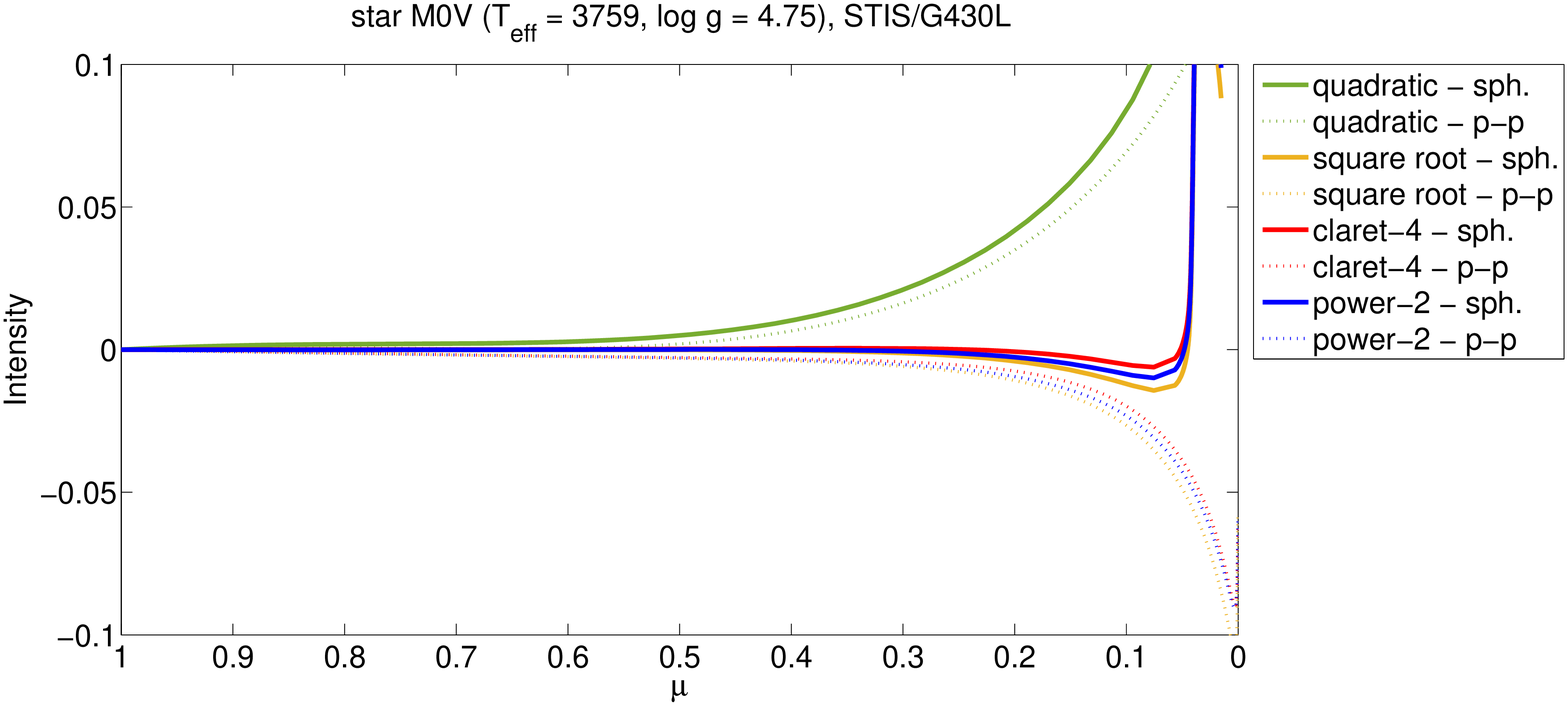}
\plotone{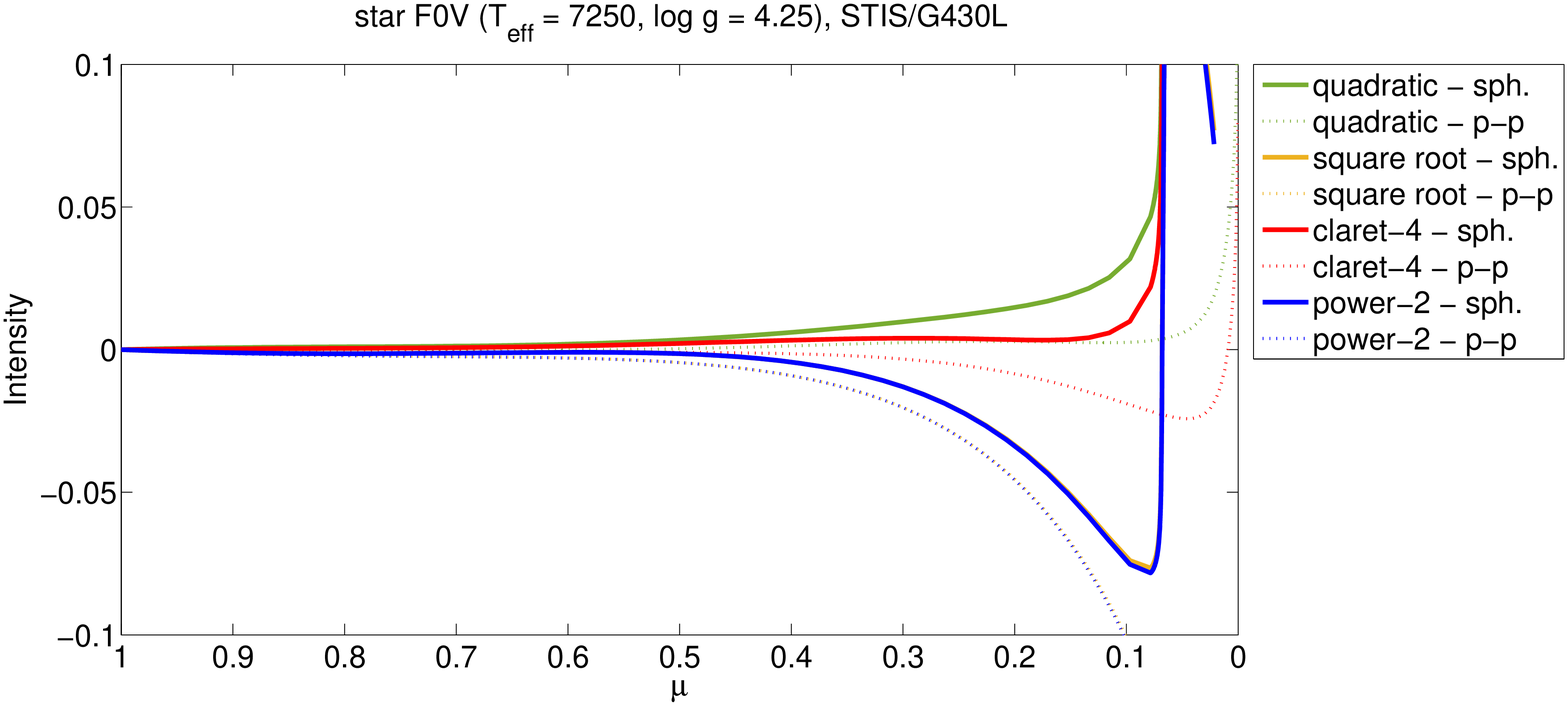}
\\
\hspace{-1cm}
\plotone{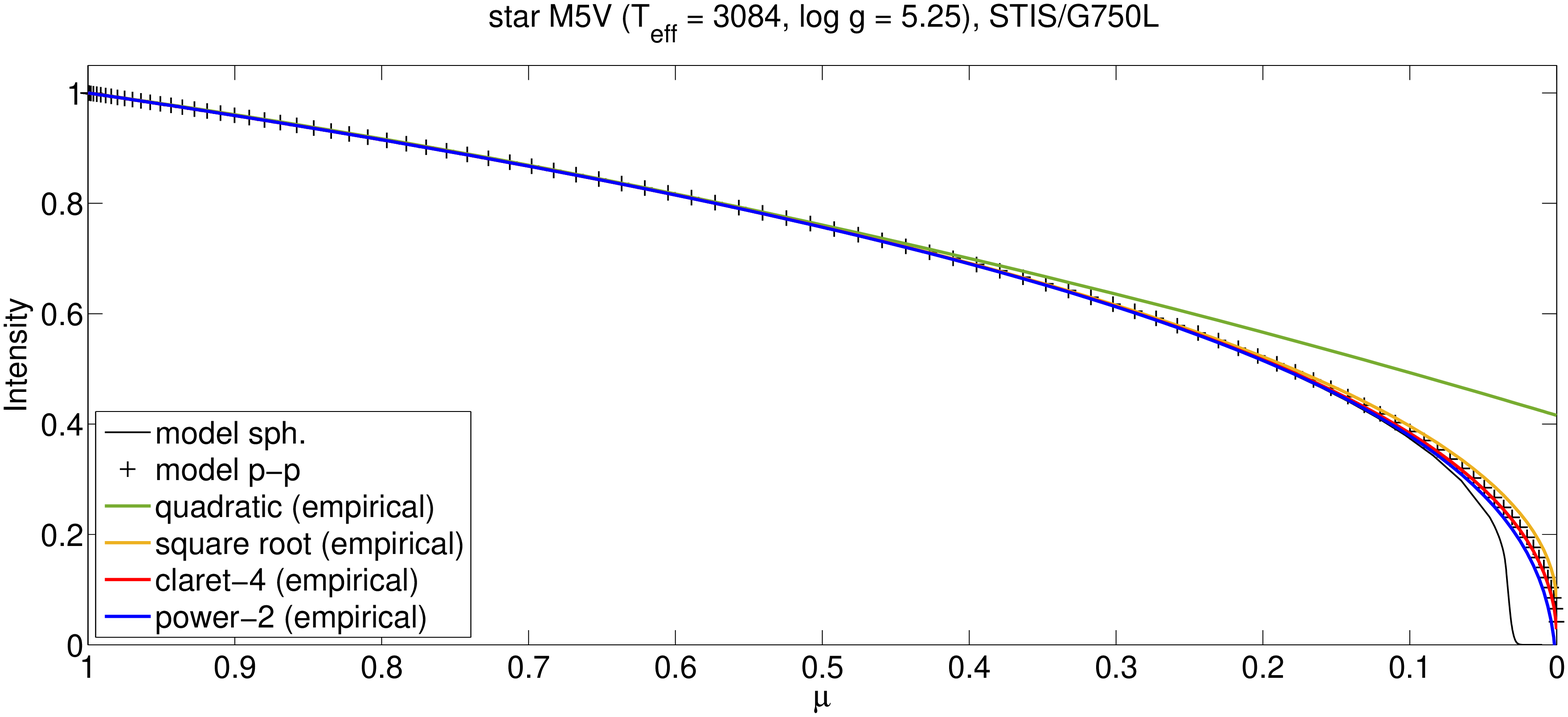}
\plotone{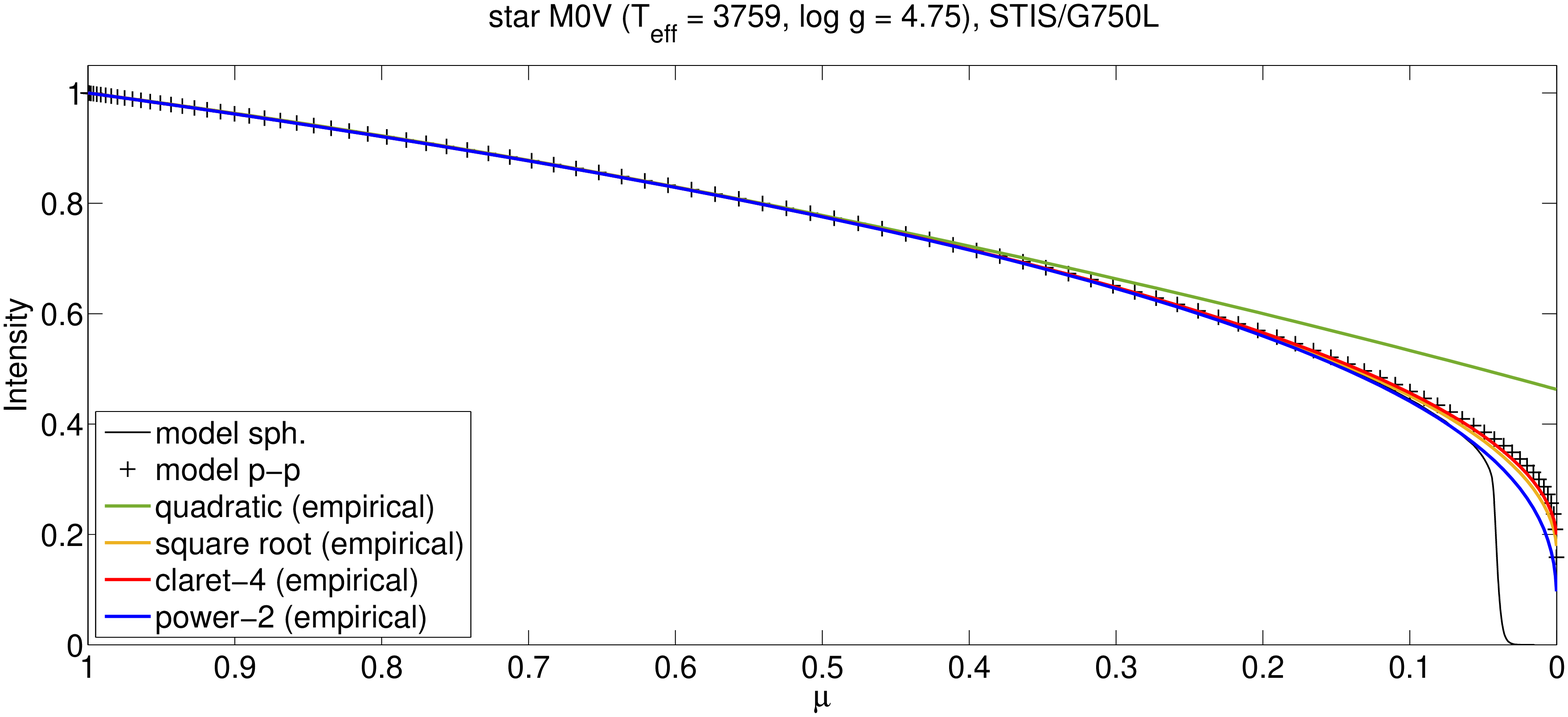}
\plotone{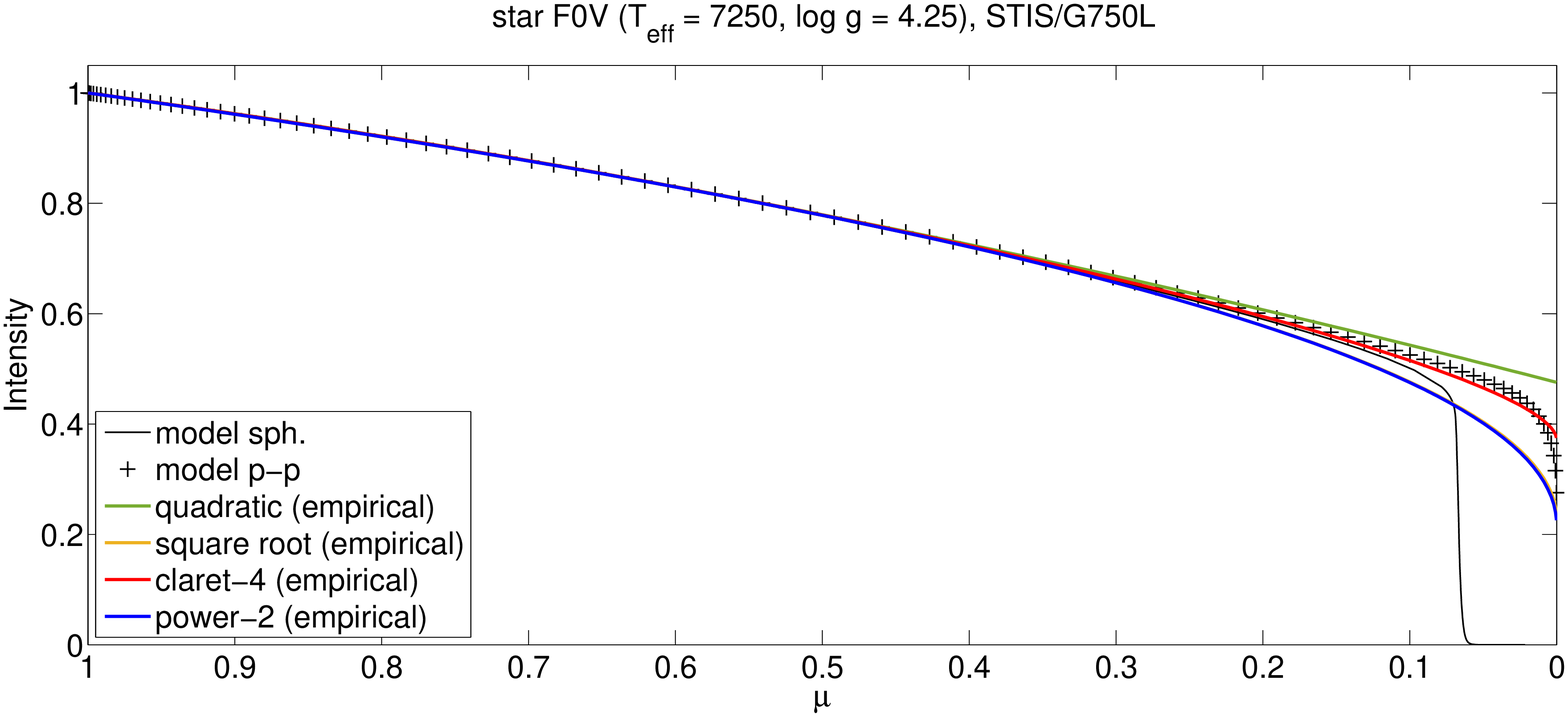}
\\
\hspace{-1cm}
\plotone{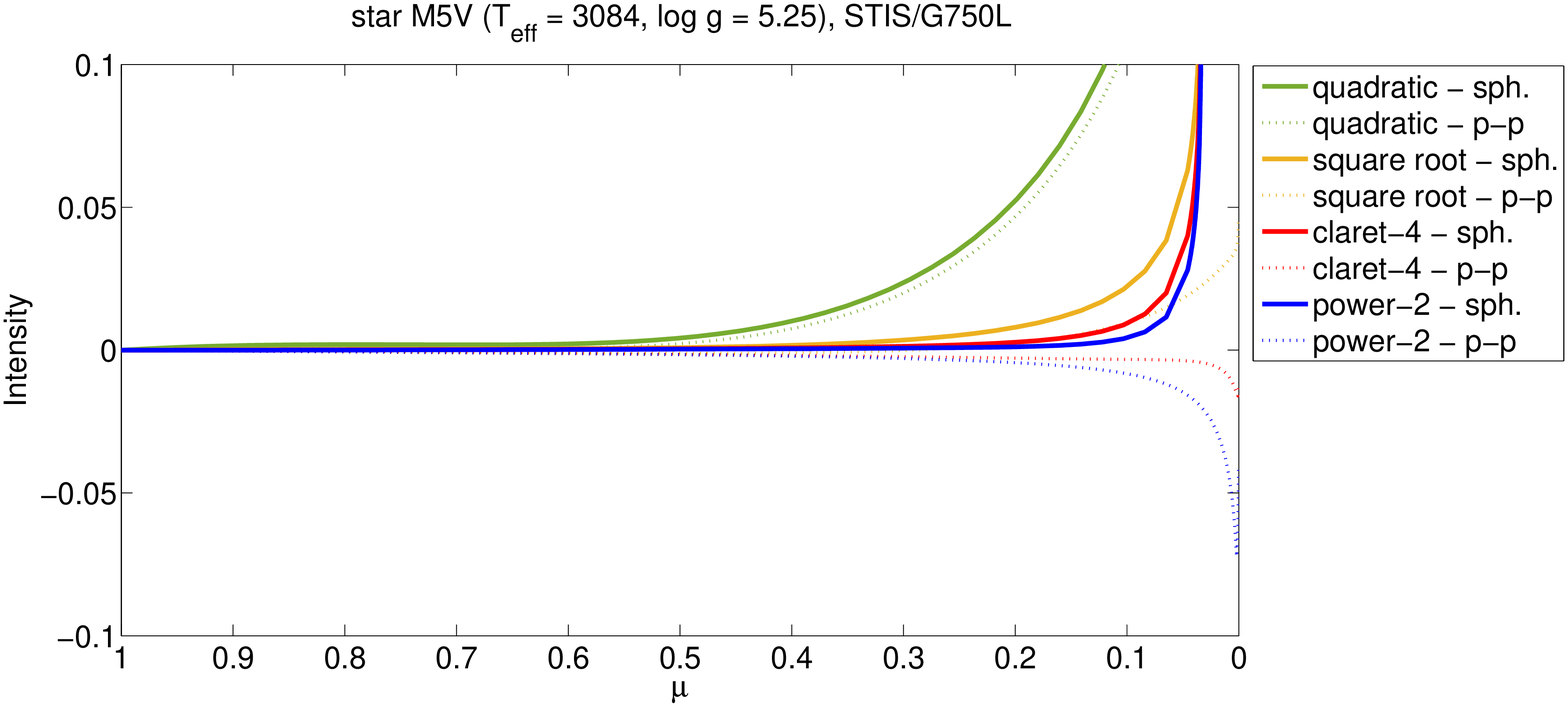}
\plotone{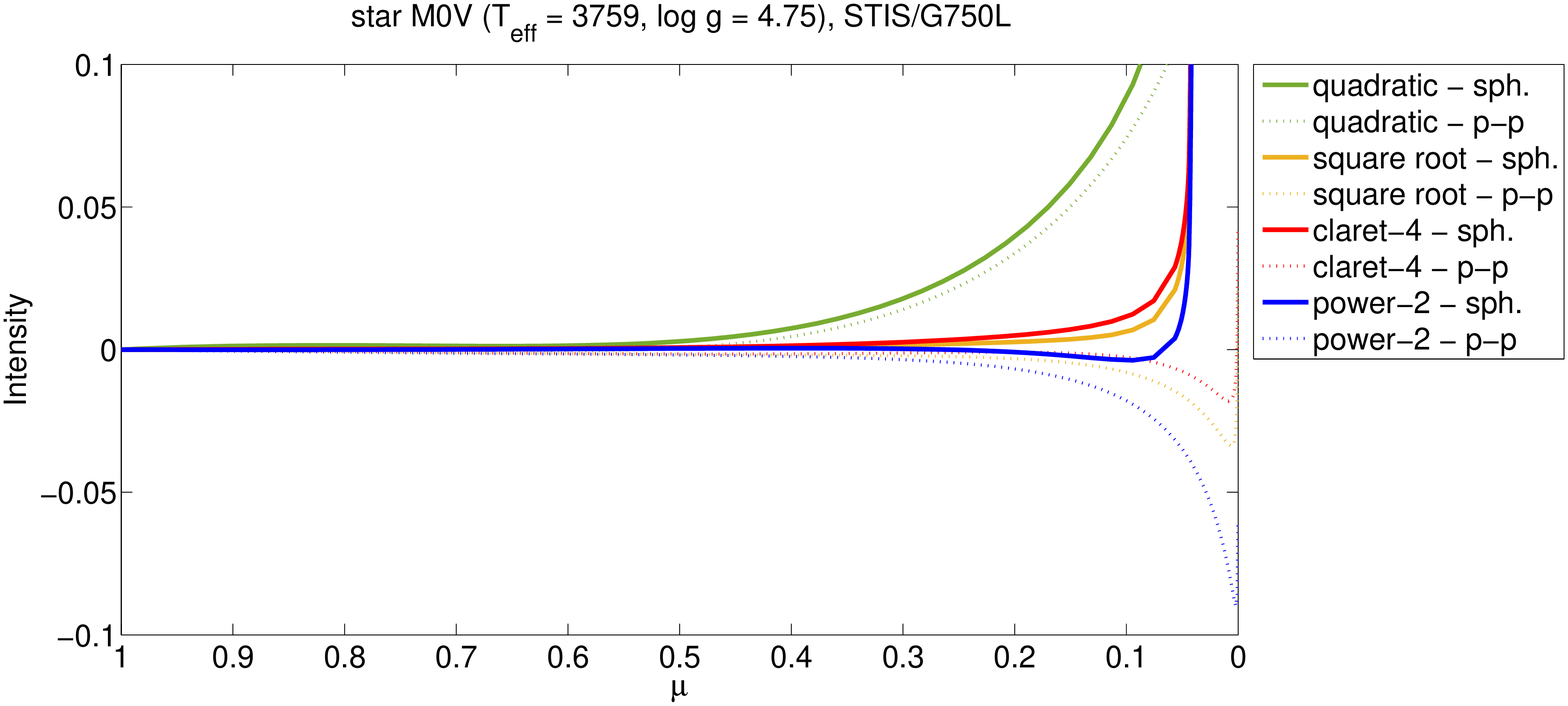}
\plotone{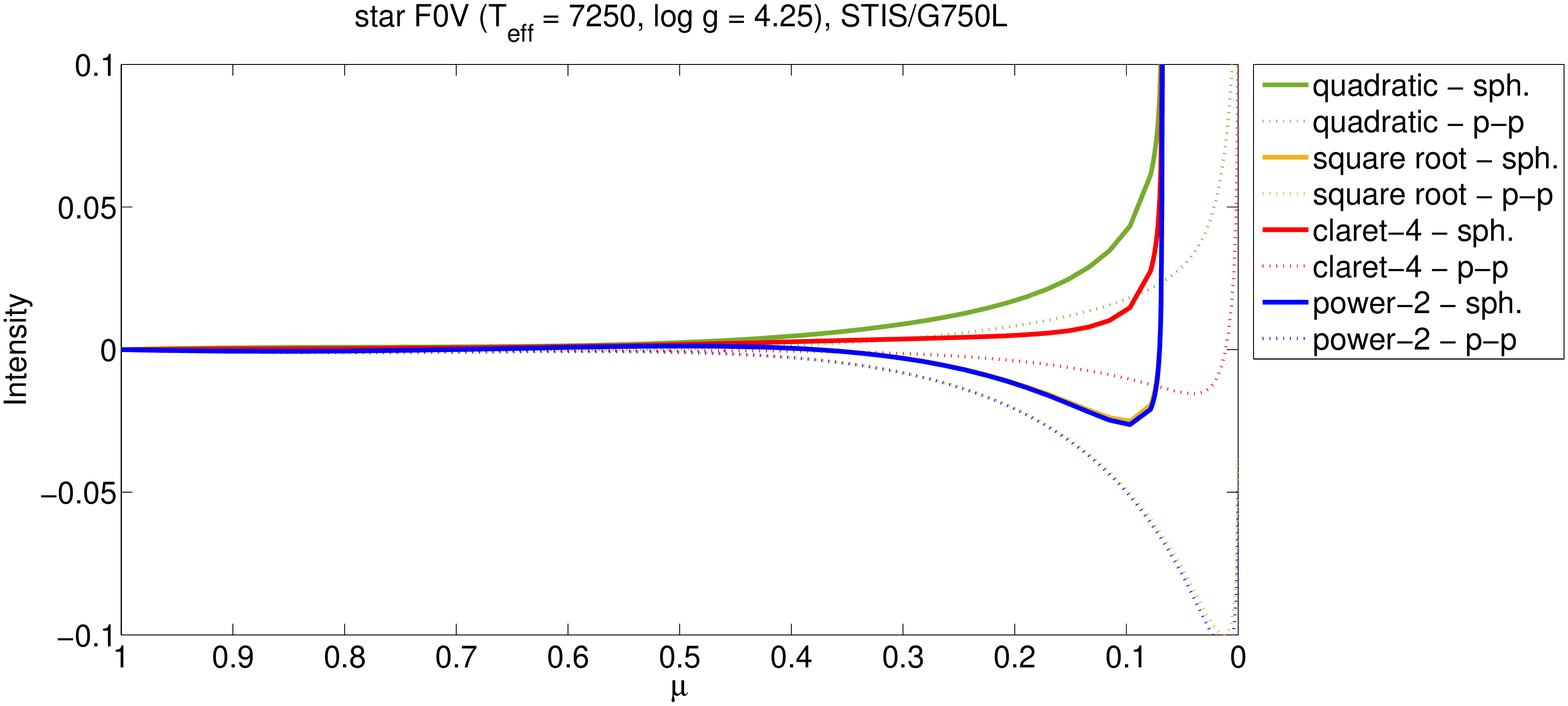}
\\
\hspace{-1cm}
\plotone{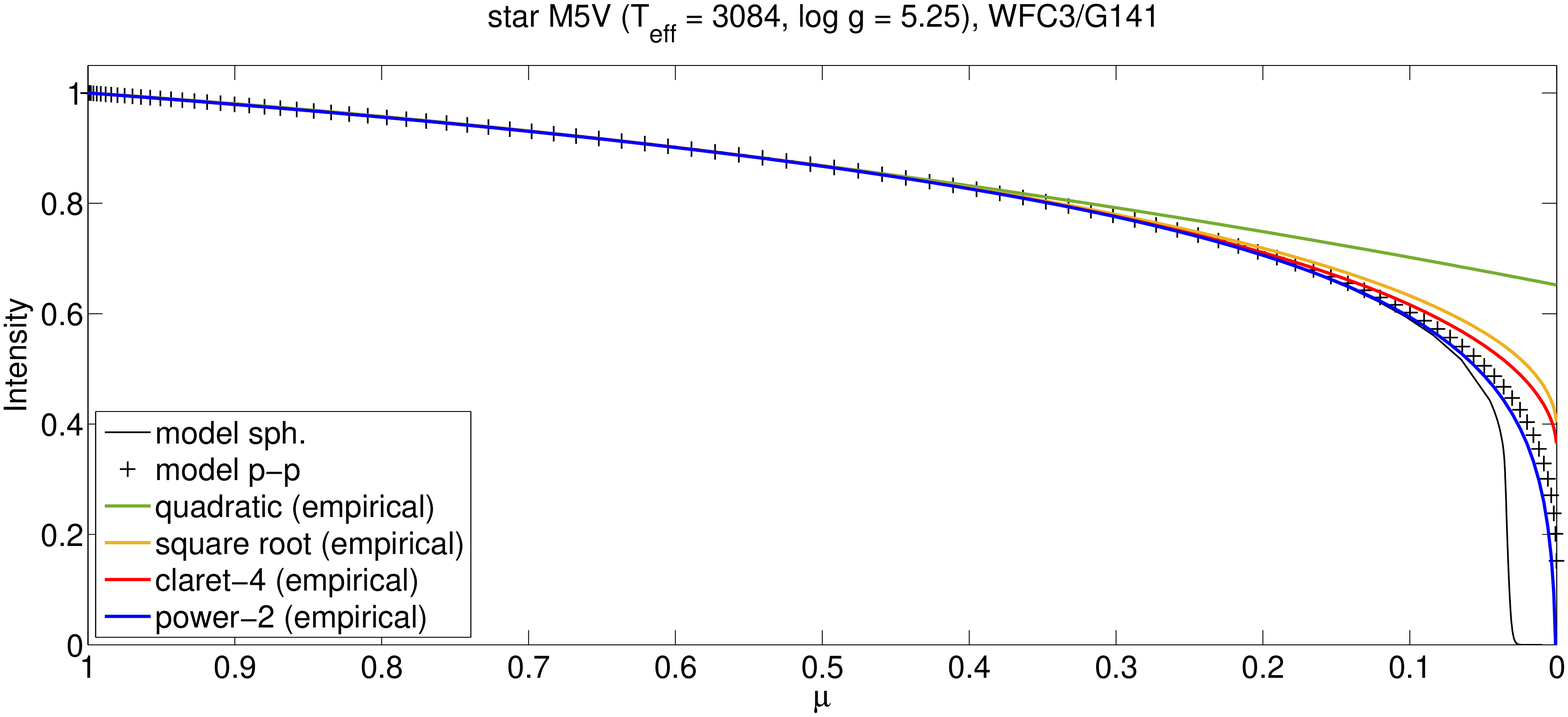}
\plotone{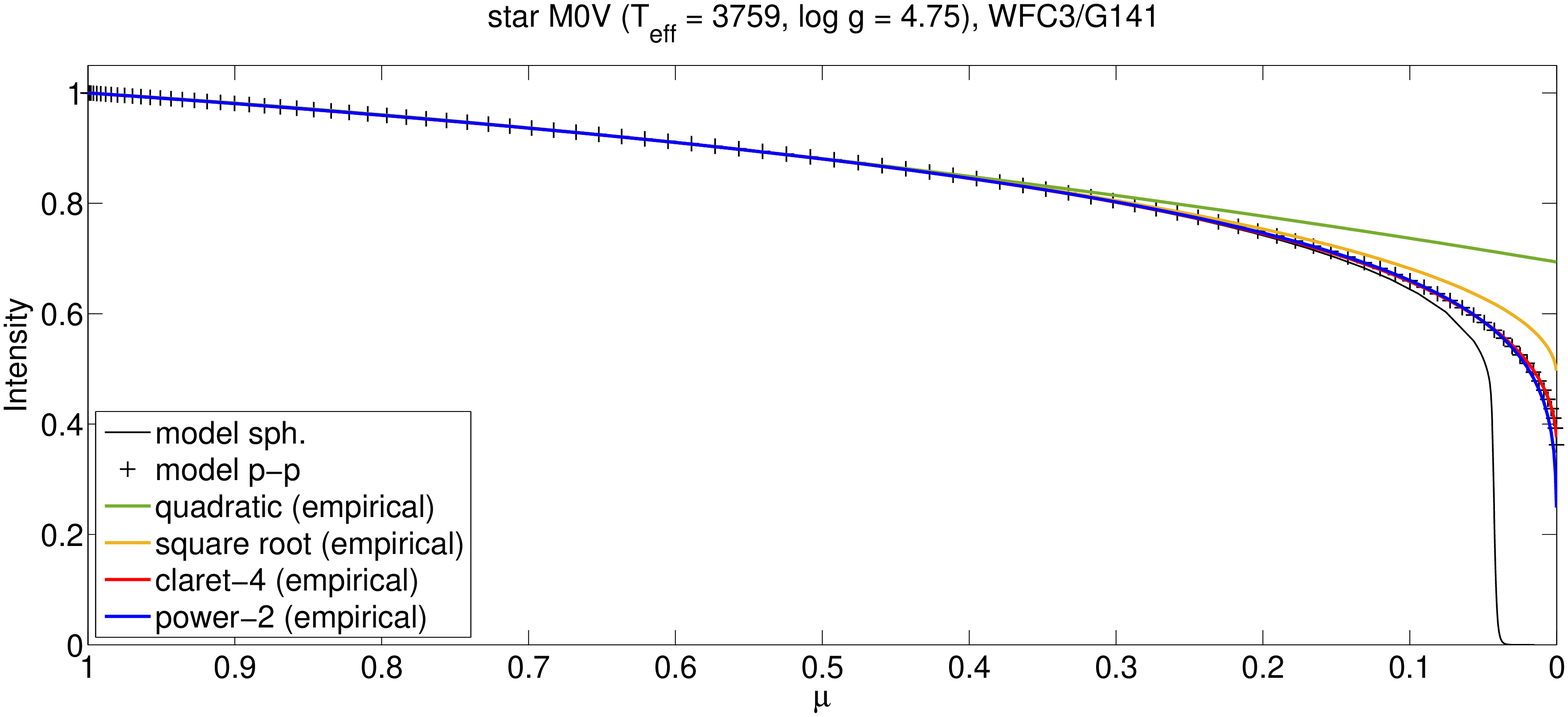}
\plotone{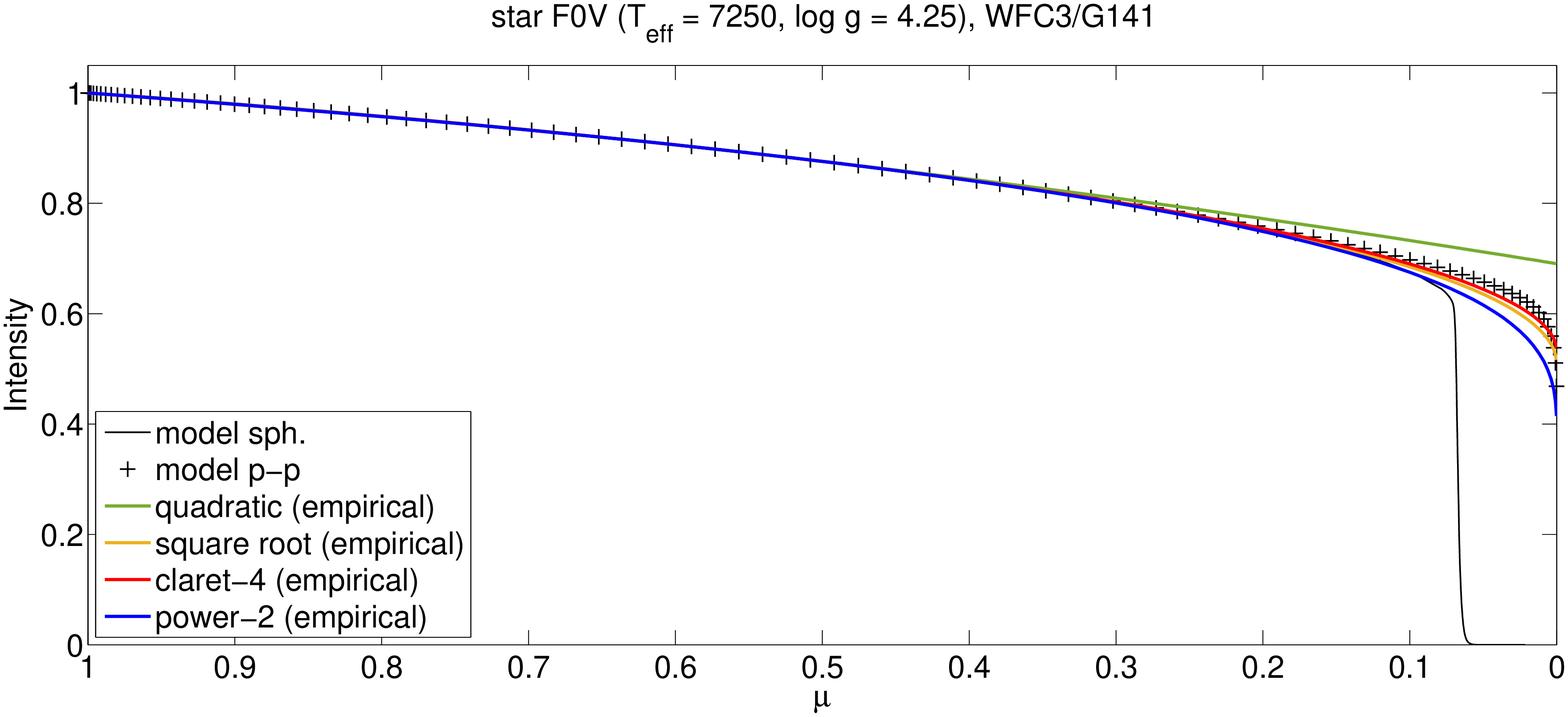}
\\
\hspace{-1cm}
\plotone{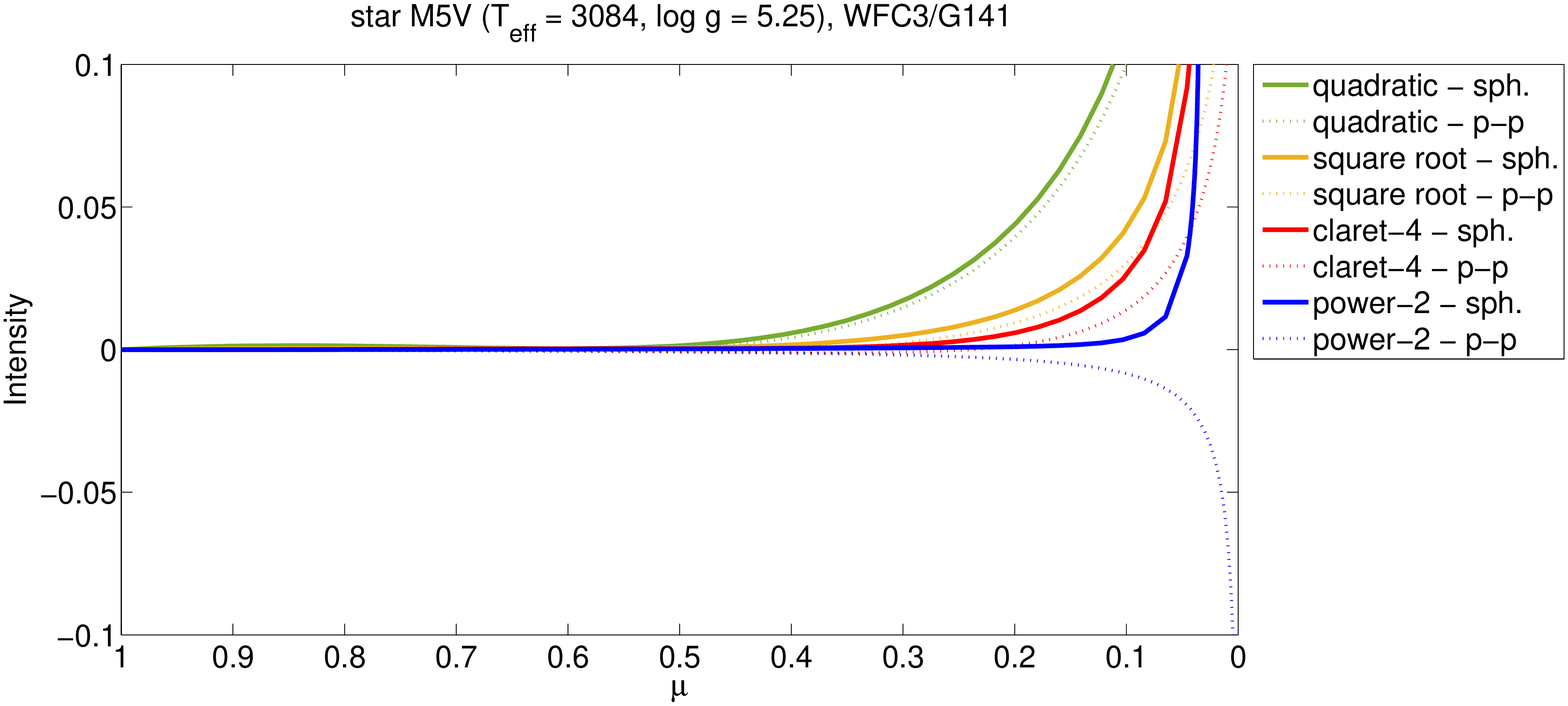}
\plotone{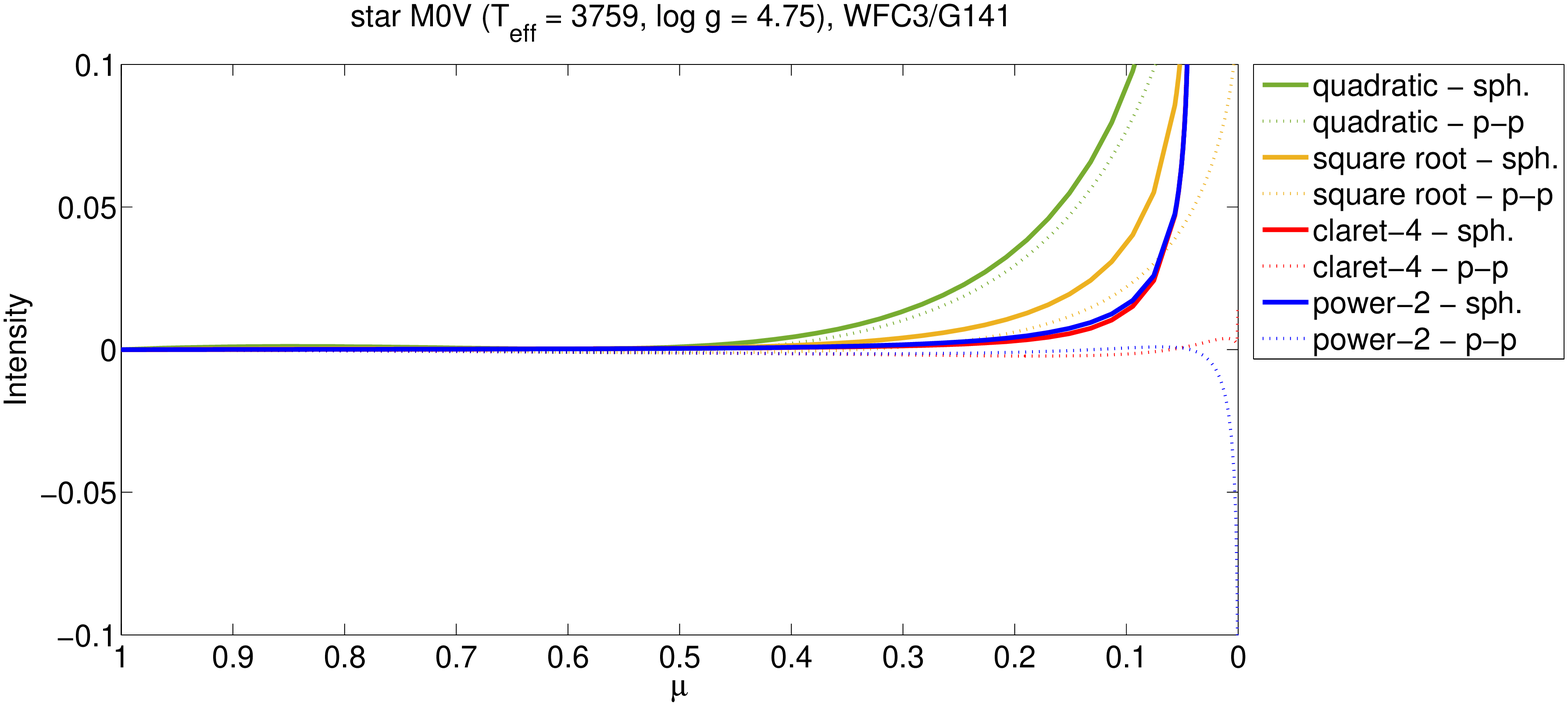}
\plotone{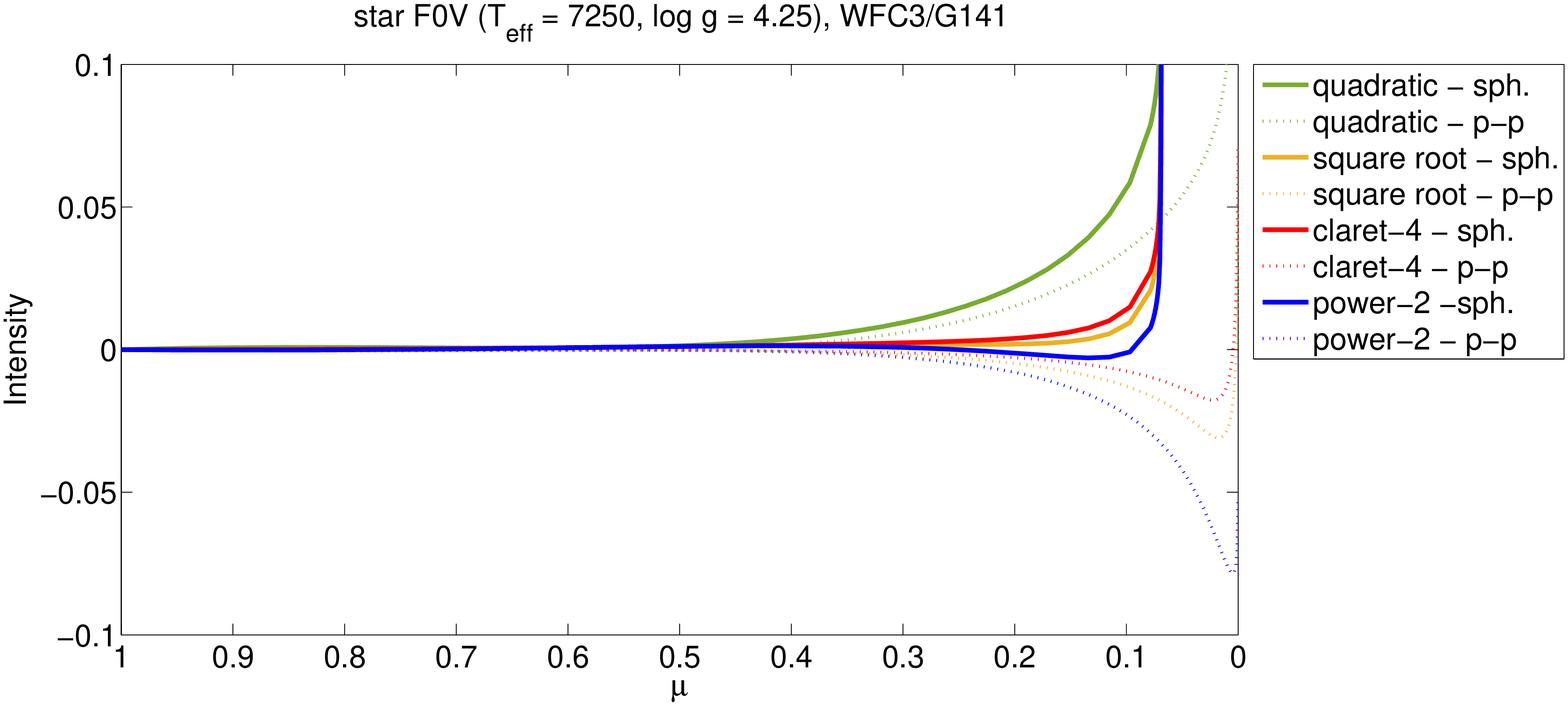}
\\
\hspace{-1cm}
\plotone{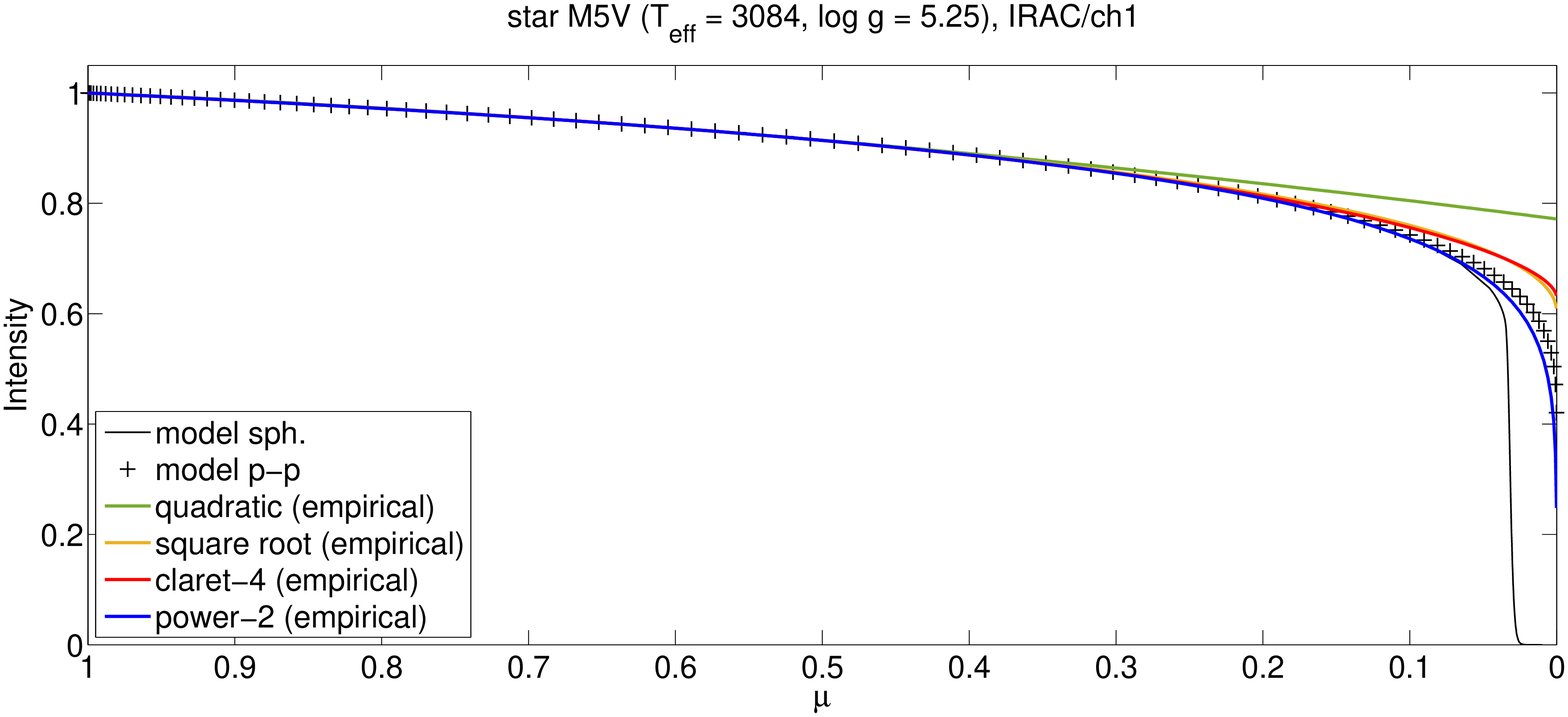}
\plotone{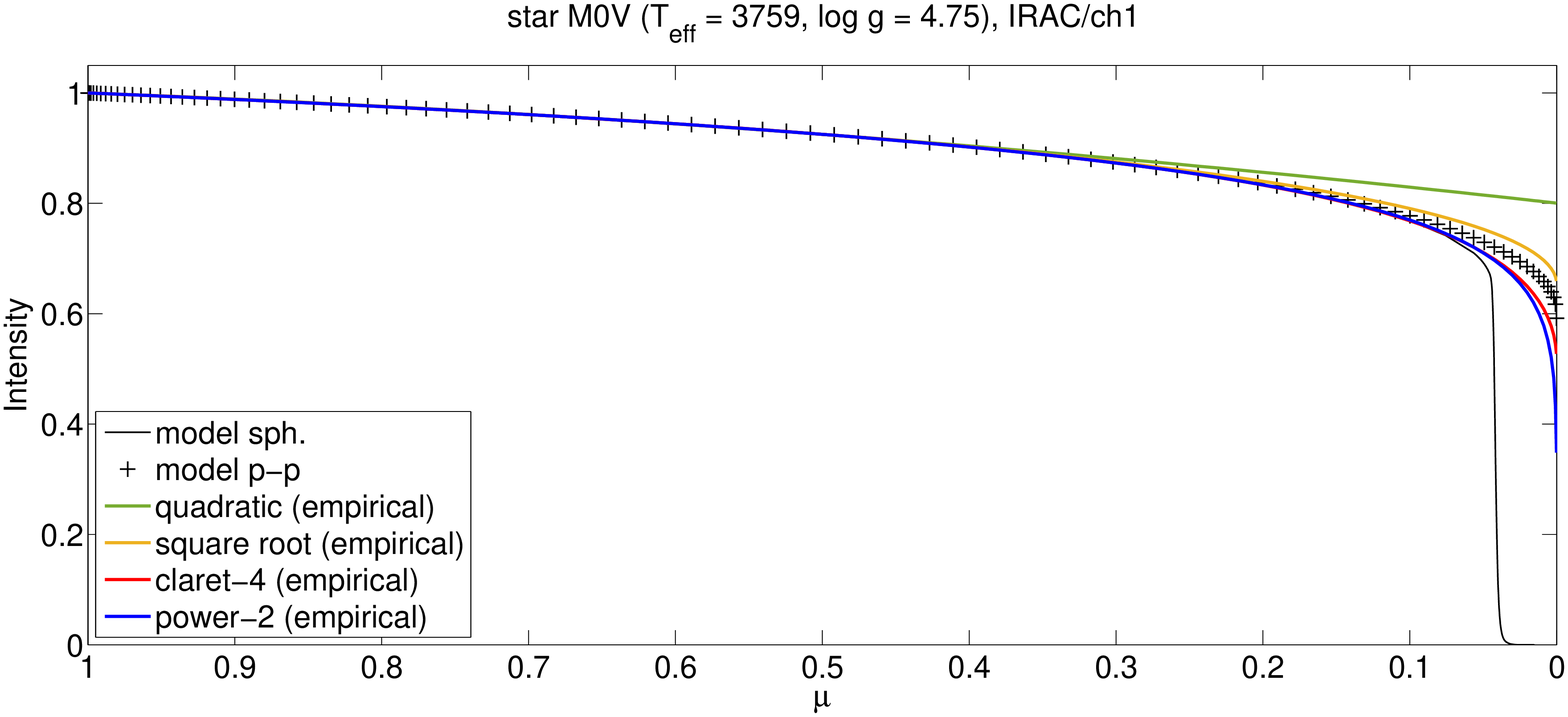}
\plotone{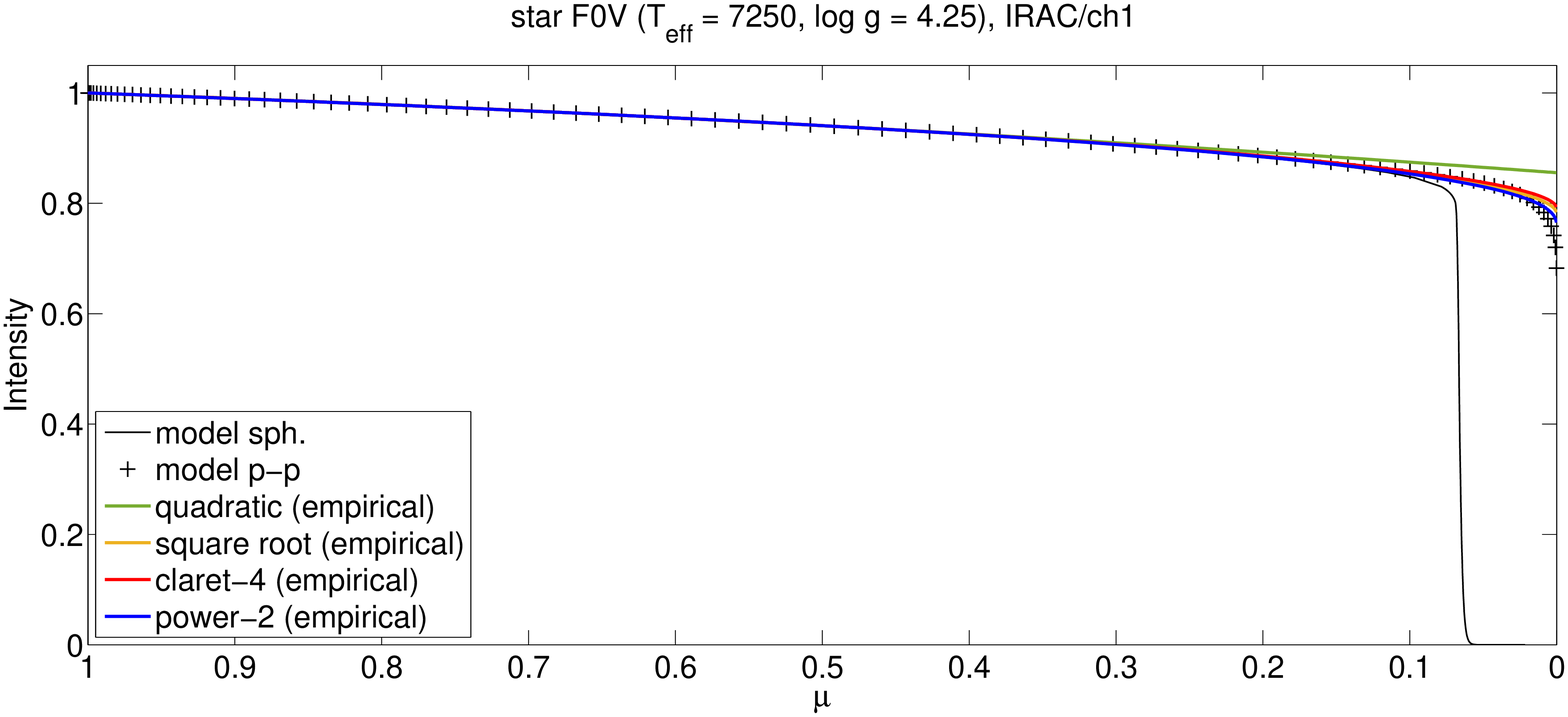}
\\
\hspace{-1cm}
\plotone{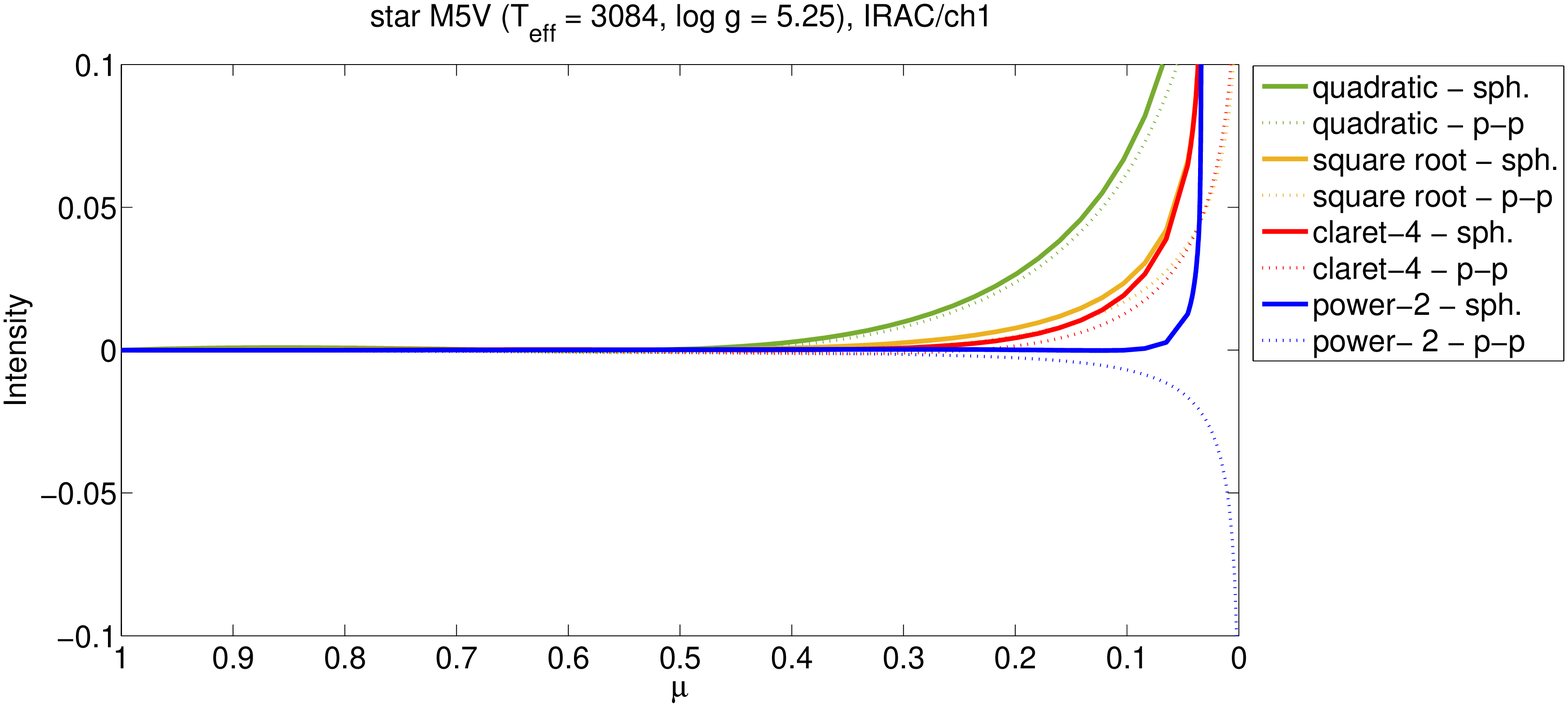}
\plotone{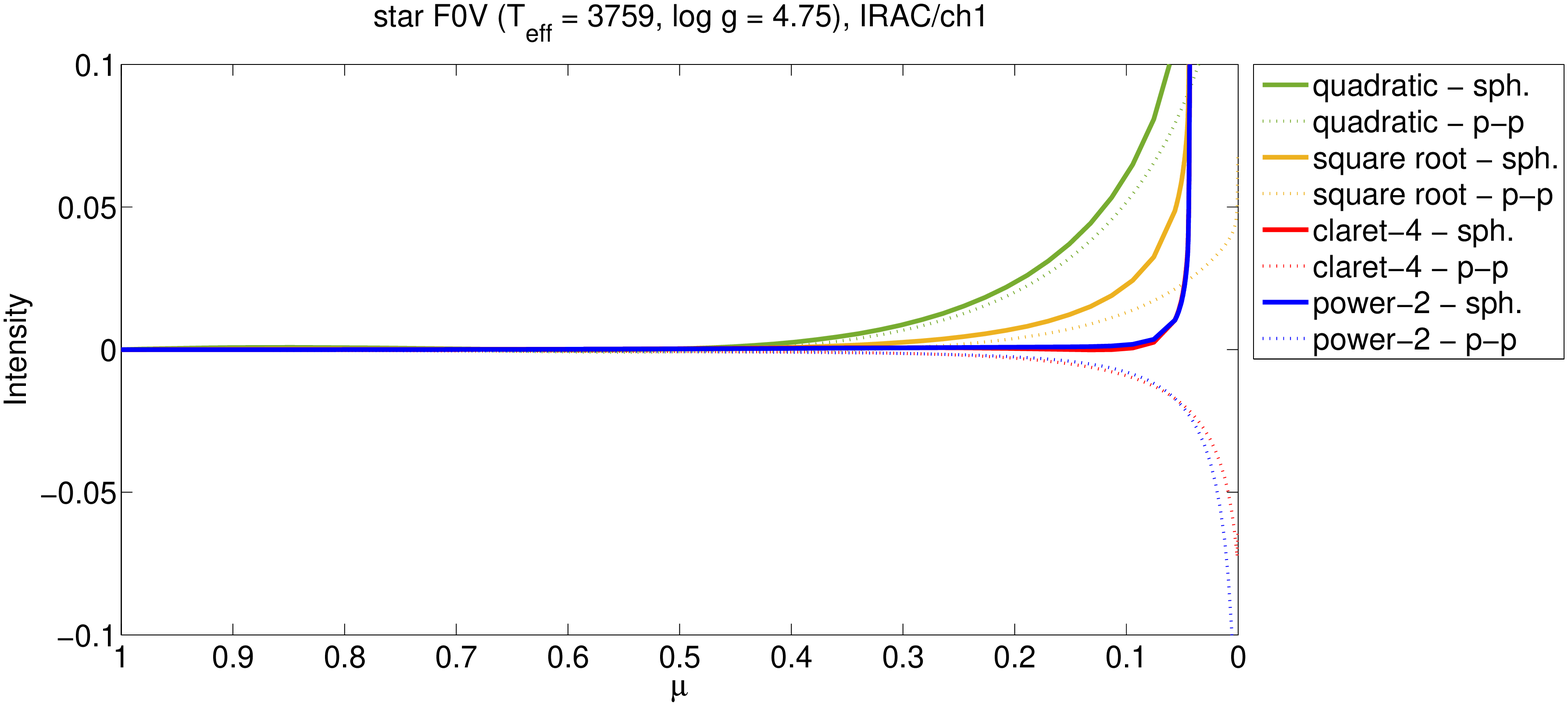}
\plotone{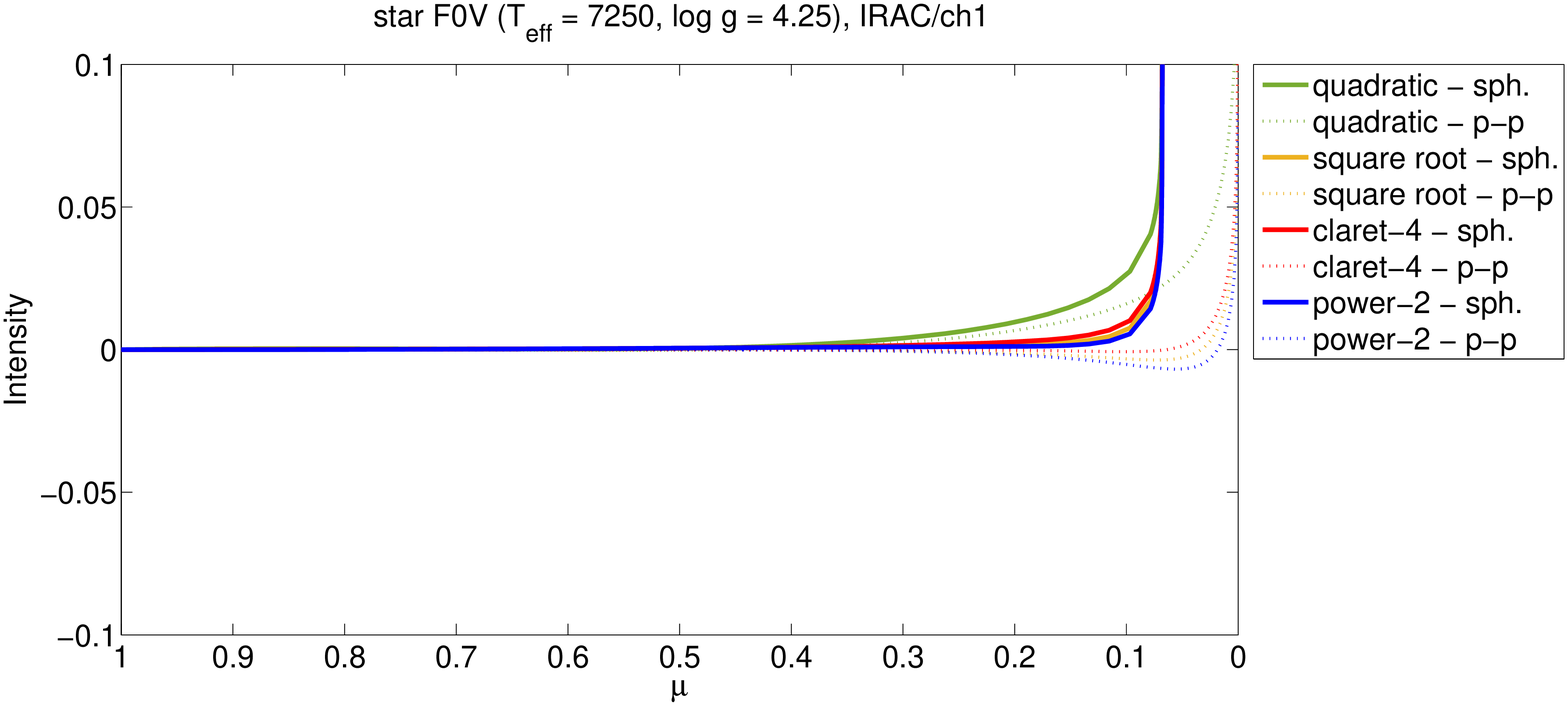}
\\
\hspace{-1cm}
\plotone{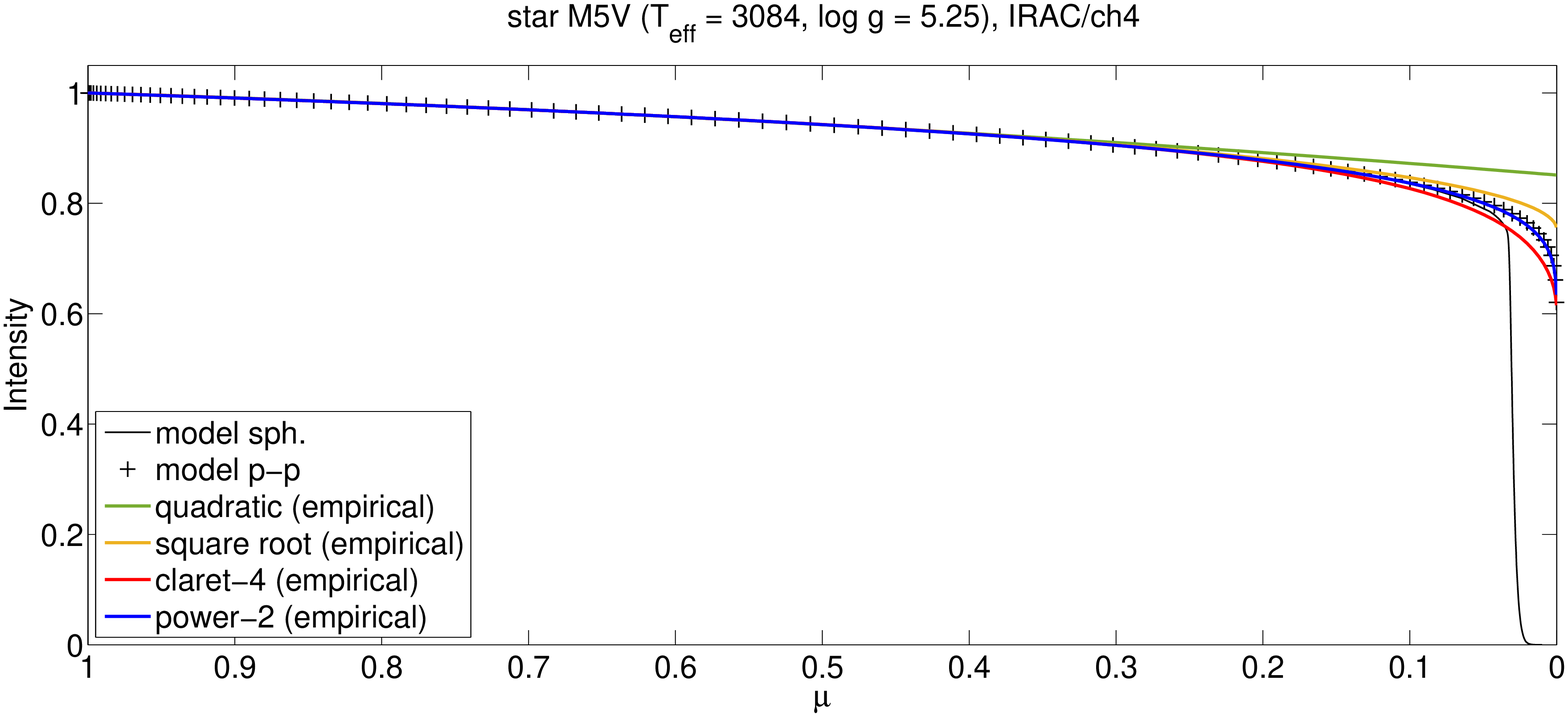}
\plotone{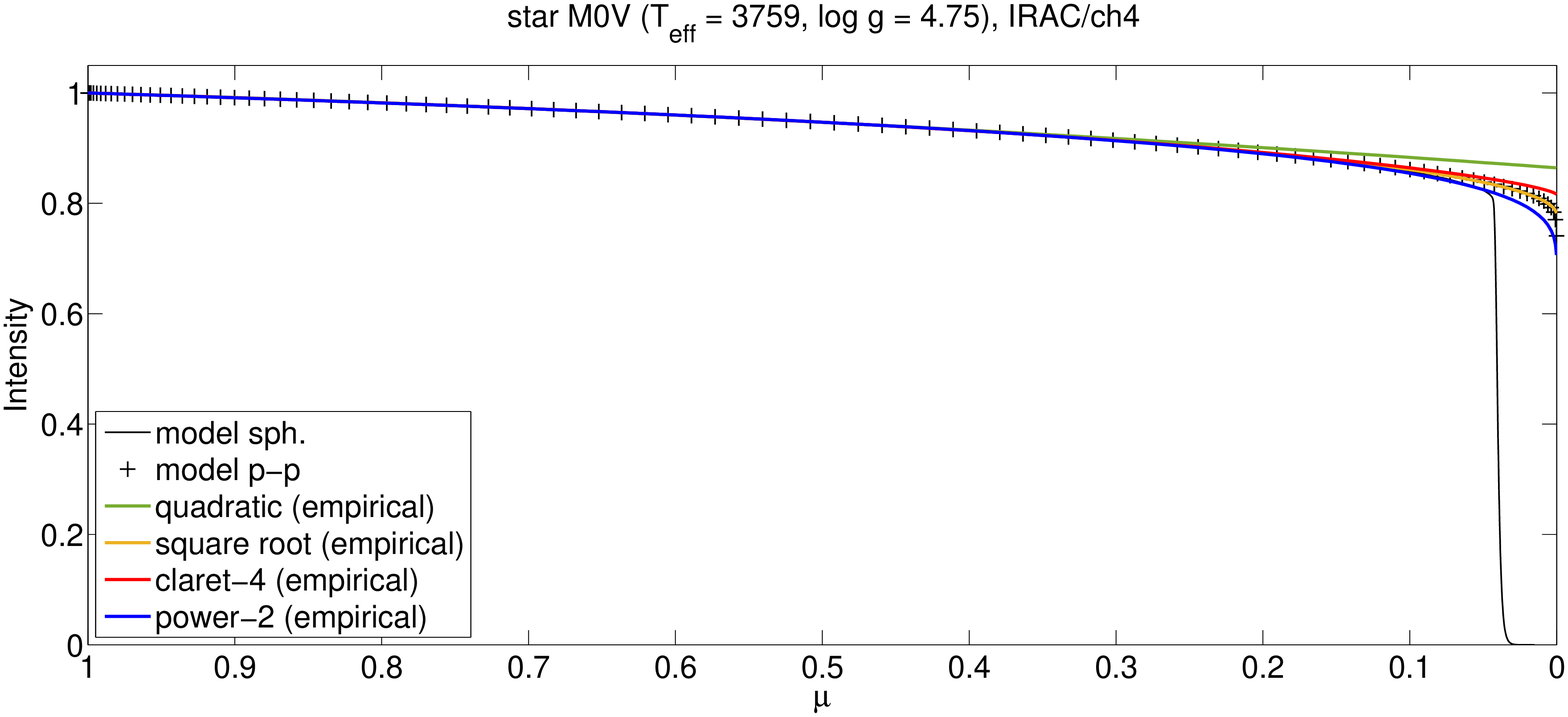}
\plotone{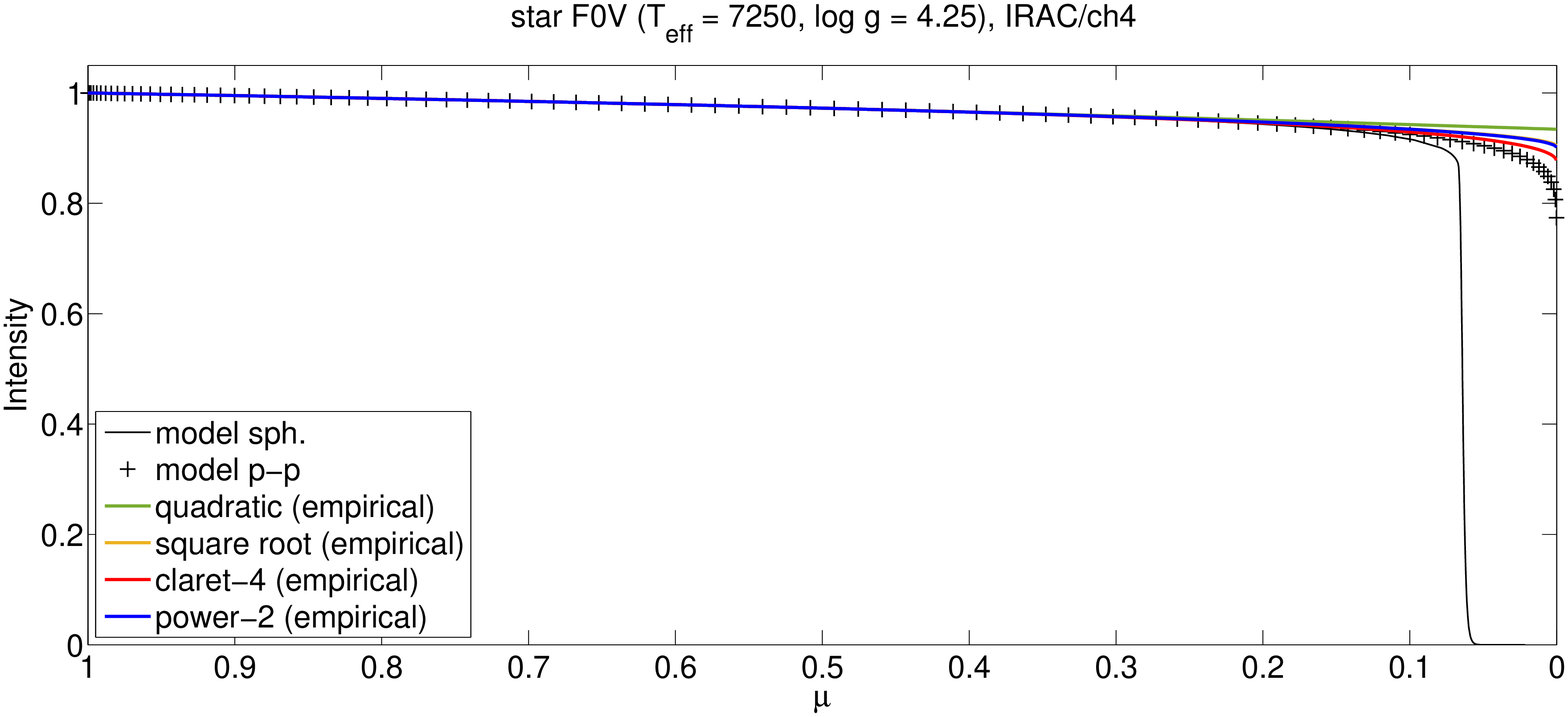}
\\
\hspace{-1cm}
\plotone{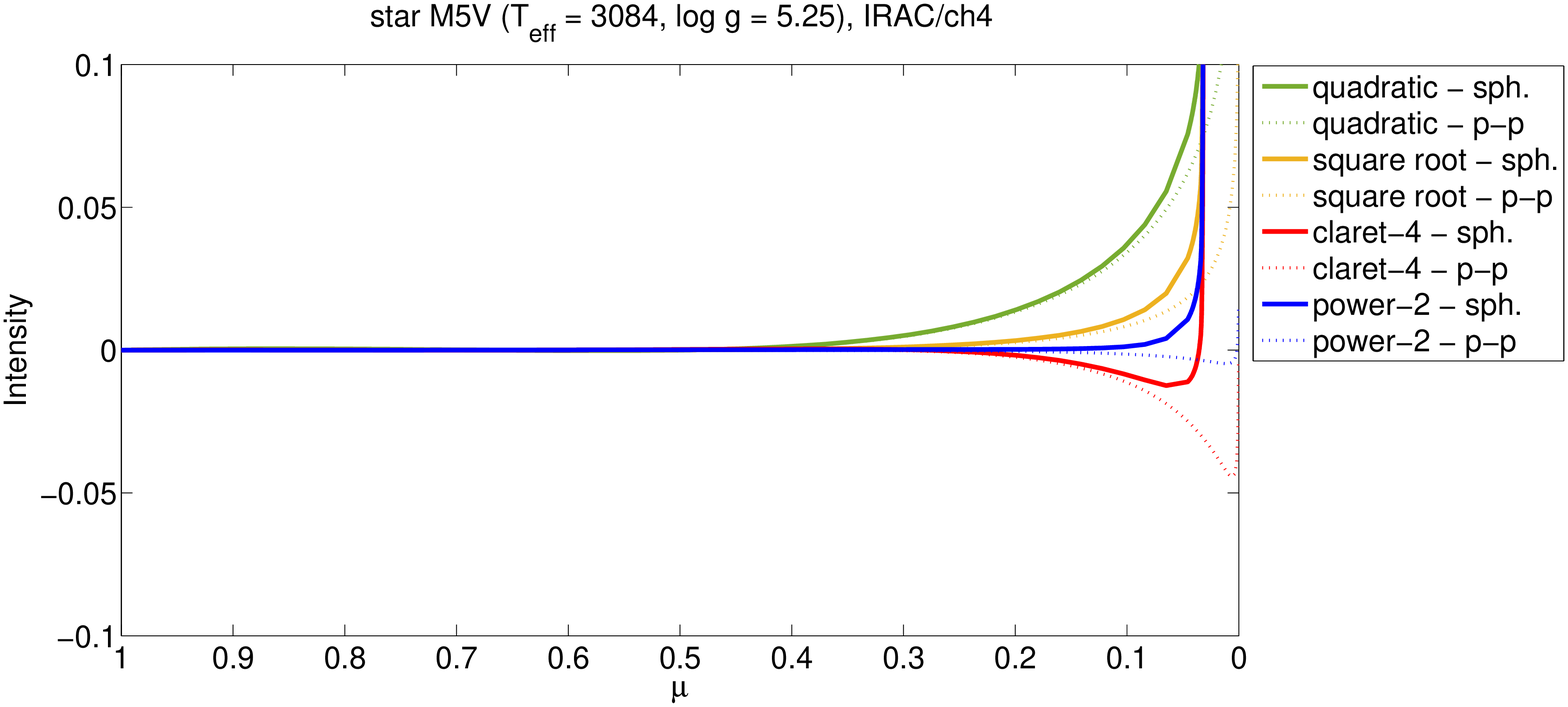}
\plotone{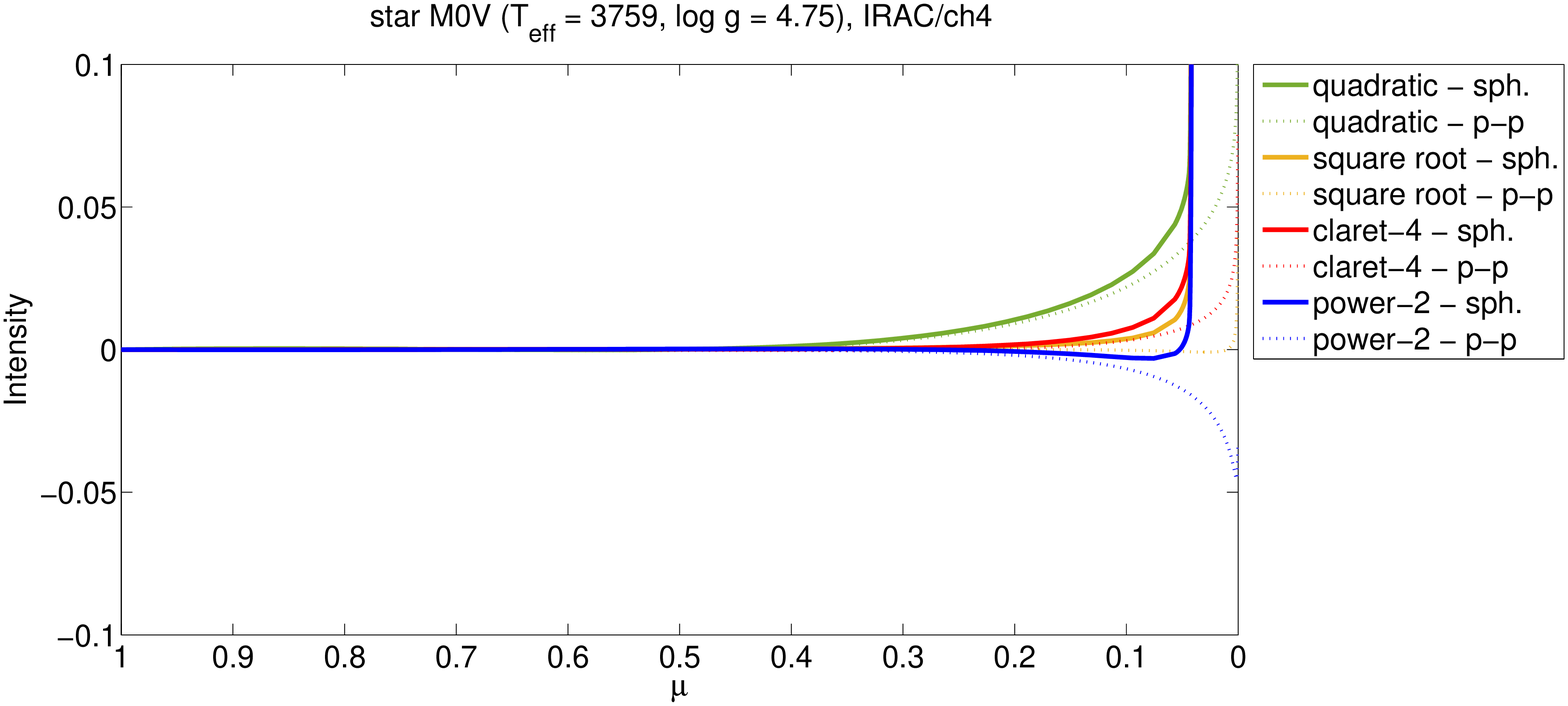}
\plotone{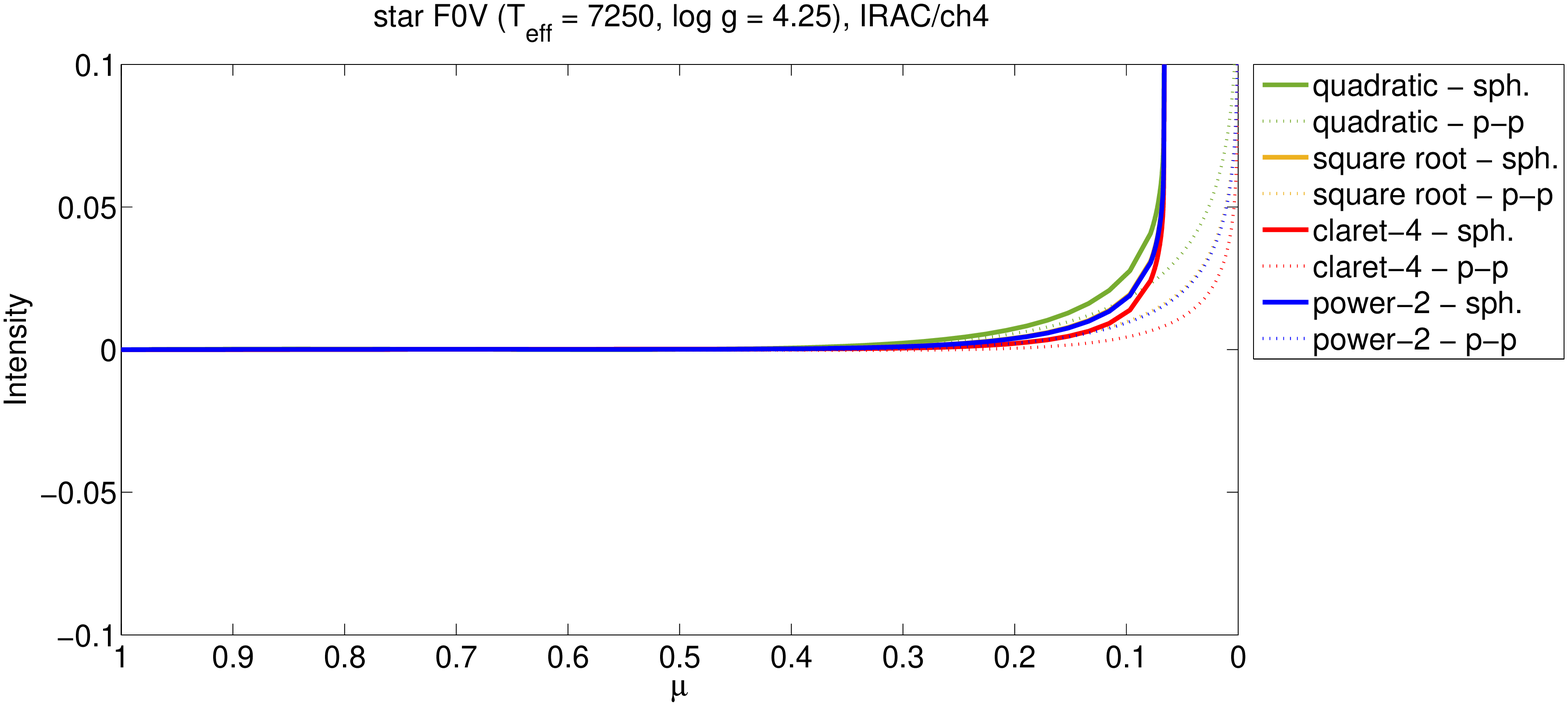}
\caption{Top panels: Plane-parallel (black `+') and spherical (black
  line) angular intensity vs.\ $\mu$. Parametric limb-darkening with
  empirical limb-darkening coefficients fitted to the transit
  light-curves with $b=0$, using quadratic (green), square-root
  (yellow), claret-4 (red), and power-2 (blue) law. Bottom panels:
  zoom of the residuals between parametric limb-darkening and
  spherical intensities (continuos lines), plane-parallel intensities
  (dashed lines). \label{fig8app}}
\end{figure}

This Appendix contains Figures~\ref{fig4app}--\ref{fig8app} showing the best-fit limb-darkening and transit models for each stellar type and passband analyzed in Sections~\ref{sec:1Dvs3D}--\ref{sec:ldc_free}.

\clearpage

\section{Supplemental figures: MCMC fitting results for inclined transits}
\label{app2}

\begin{figure}[!h]
\epsscale{1.0}
\plotone{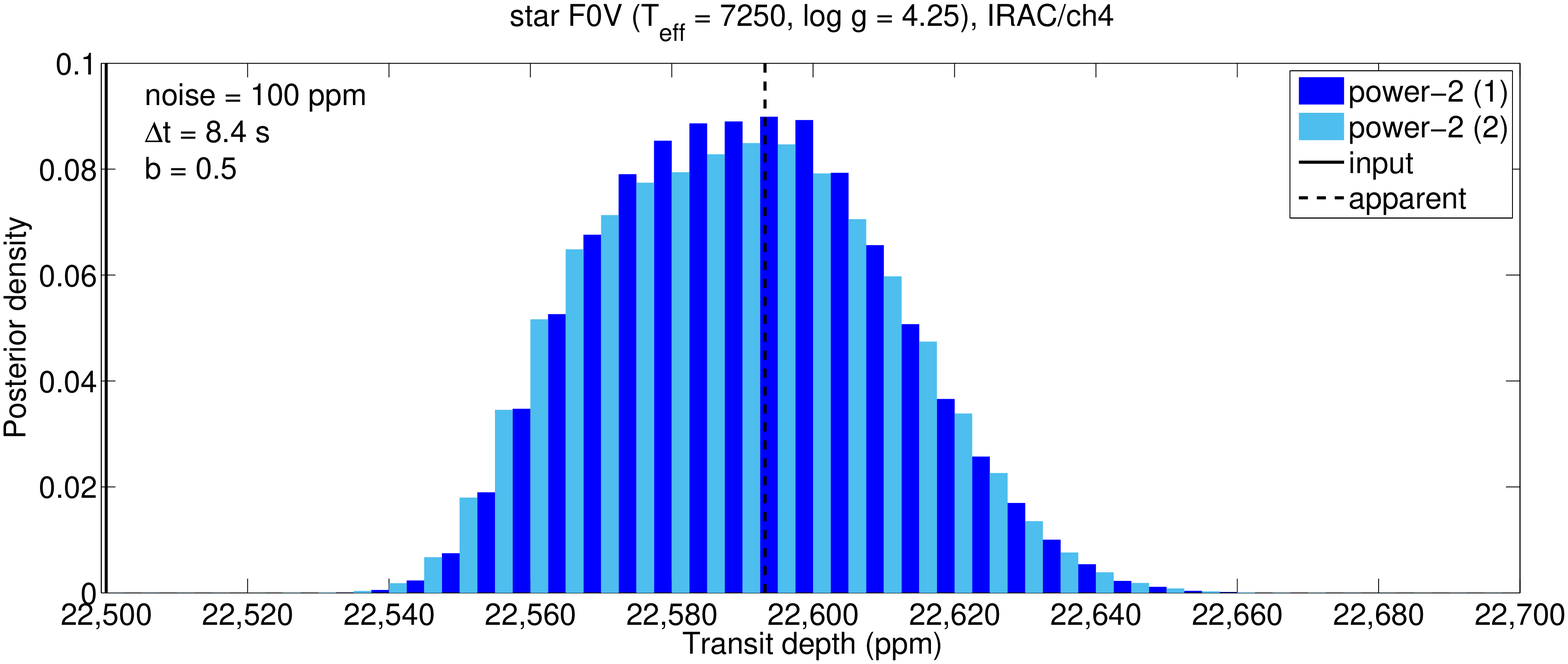}
\caption{MCMC sampled posterior distributions of the transit depth for
  the inclined transit ($b=0.5$) in front of the F0\;V model, IRAC/ch4
  passband, with 100~ppm gaussian noise; fitting $p$, $a_R$, $i$, normalization factor, and power-2
  limb-darkening coefficients. The
  histogram channels (blue and light-blue) are for two chains with
  1\,500\,000 iterations; the channels are
  half-thick and shifted to improve their visualization. The input
  (black, continuous, vertical line) and expected (black, dashed,
  vertical line) transit depths are also indicated. \label{fig15app}}
\end{figure}

\begin{figure}
\epsscale{1.0}
\plotone{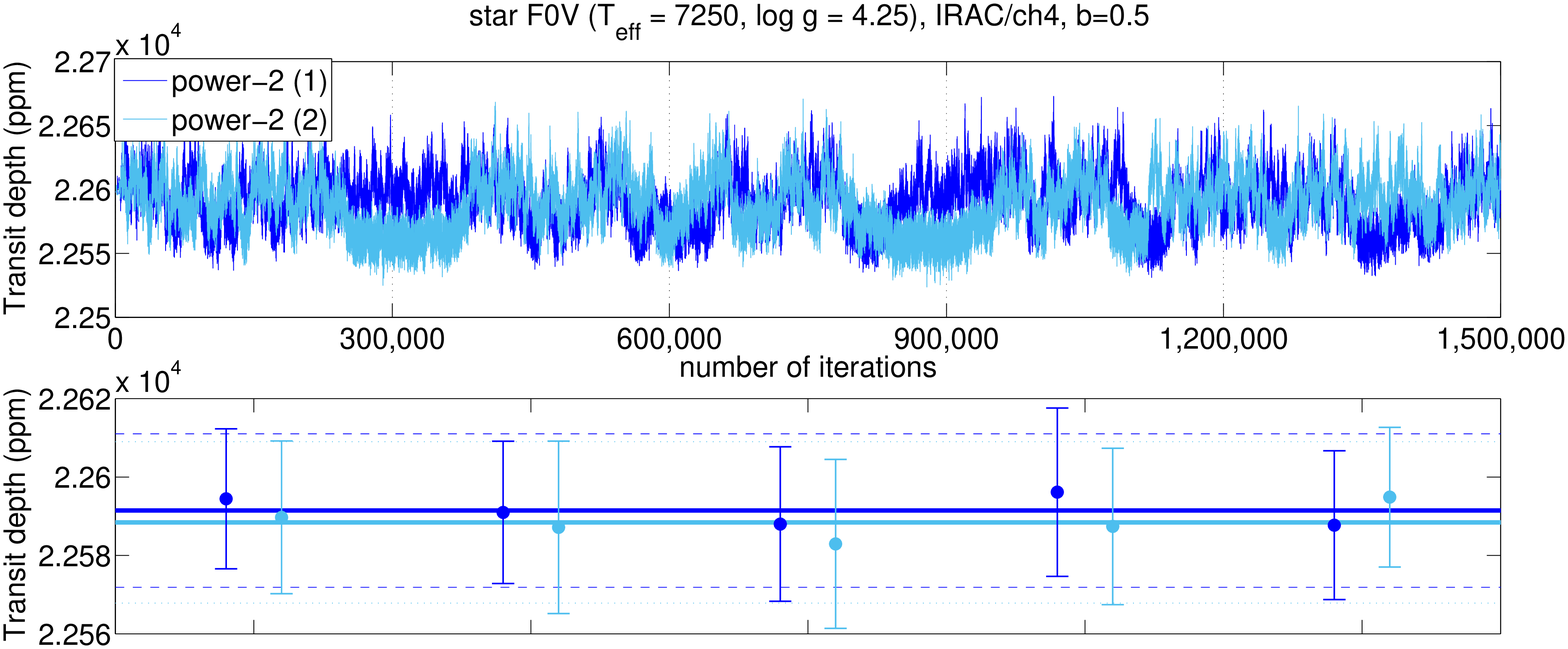}
\caption{Top panel: transit depth chains for the inclined transit ($b=0.5$) in front of the F0\;V model, IRAC/ch4 passband, with 100~ppm gaussian noise; fitting $p$, $a_R$, $i$, normalization factor and  power-2 limb-darkening coefficients. Bottom panel: mean values and standard deviations calculated over fractional chains with 300\,000 iterations; the horizontal lines indicate the mean values calculated over the whole chains (continuous lines), and the mean values plus or minus the standard deviations (dashed lines). \label{fig16app}}
\end{figure}

\begin{figure}
\epsscale{1.0}
\plotone{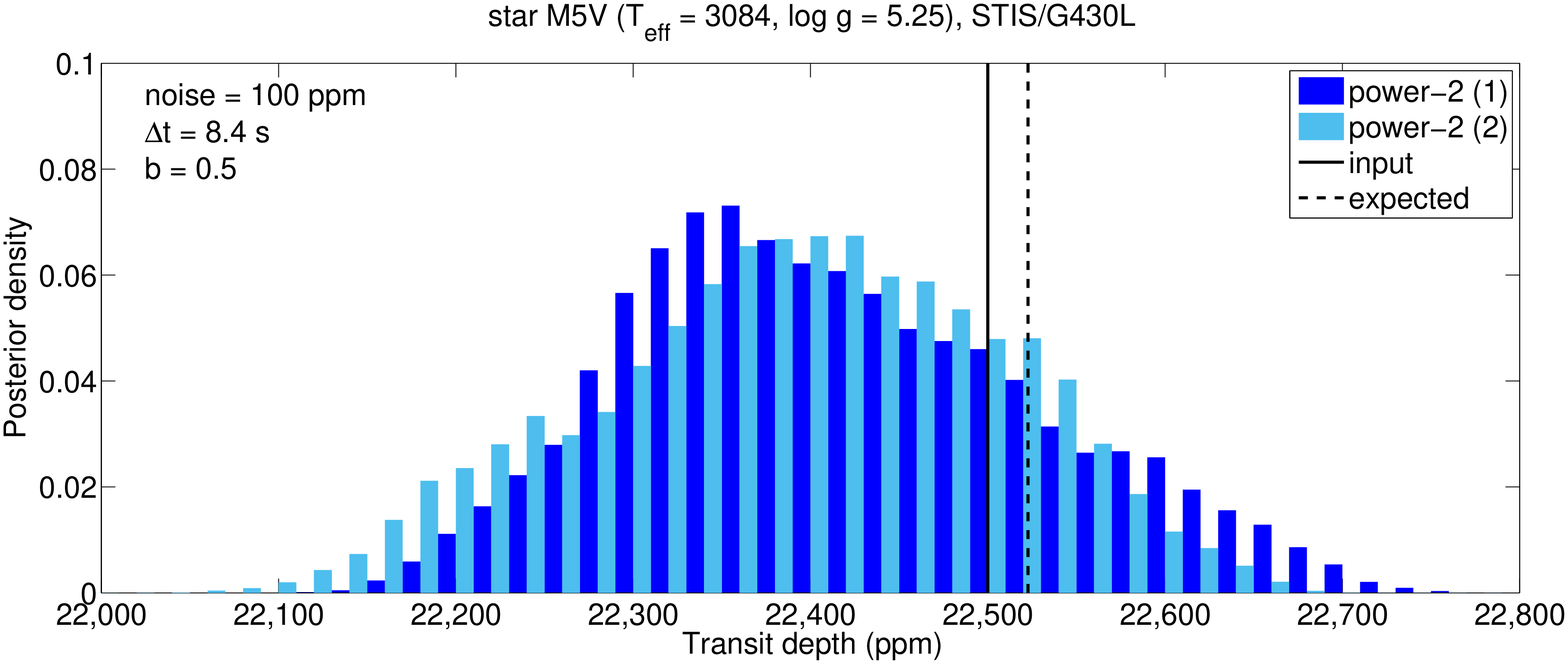}
\caption{MCMC sampled posterior distributions of the transit depth for the inclined transit ($b=0.5$) in front of the M5\;V model, STIS/G430L passband, with 100~ppm gaussian noise; fitting $p$, $a_R$, $i$, normalization factor and power-2 limb-darkening coefficients. The histogram channels (blue and light-blue) are for two chains with 1\,500\,000 iterations, using the power-2 law; the channels are half-thick and shifted to improve their visualization. The input (black, continuous, vertical line) and expected (black, dashed, vertical line) transit depths are also indicated. \label{fig22app}}
\end{figure}

\begin{figure}
\epsscale{1.0}
\plotone{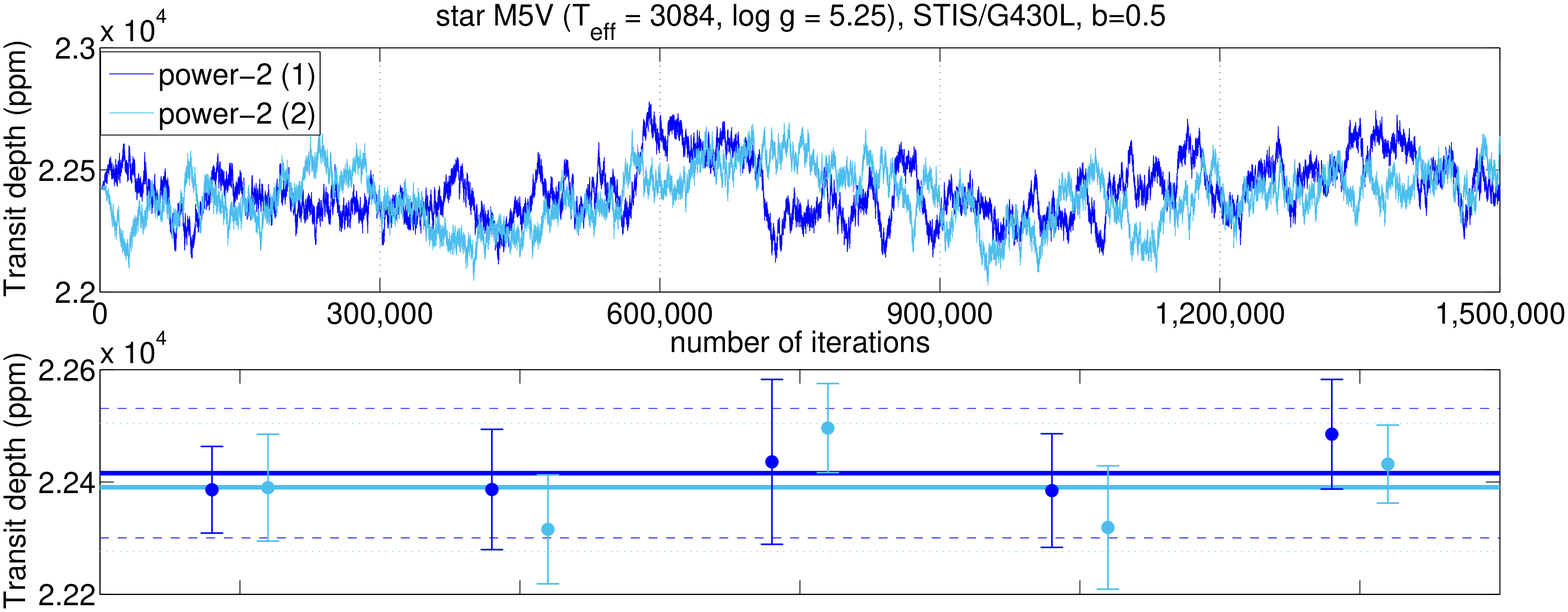}
\caption{Top panel: transit depth chains for the inclined transit ($b=0.5$) in front of the M5\;V model, STIS/G430L passband, with 100~ppm gaussian noise; fitting $p$, $a_R$, $i$, normalization factor and  power-2 limb-darkening coefficients. Bottom panel: mean values and standard deviations calculated over fractional chains with 300\,000 iterations; the horizontal lines indicate the mean values calculated over the whole chains (continuous lines), and the mean values plus or minus the standard deviations (dashed lines). \label{fig23app}}
\end{figure}

This Appendix contains Figures~\ref{fig15app}--\ref{fig23app} showing the histograms and chains for the inclined transits, analogous to those ones presented in Sections~\ref{sec:noisy_F0V_irac4}--\ref{sec:noisy_M5V_STIS_G430L} for the edge-on transits (Figures~\ref{fig12}--\ref{fig13} and \ref{fig19}--\ref{fig20}).



\clearpage

\section{Accuracy and precision of empirical limb-darkening models}
\label{app3}
In contrast to other transit parameters, the limb-darkening
coefficients do not correspond directly to some physical property of
the star-planet system. Also, for a given limb-darkening law, there exist sets of coefficients which are largely different but generate almost indistinguishable intensity profiles. Instead of studying their posterior
distributions, it is more sensitive to calculate the chains of
specific intensities at given $\mu$ values, then to compare, in the
case of simulations, with the input limb-darkening profile.

\begin{figure}[!h]
\epsscale{1.0}
\plotone{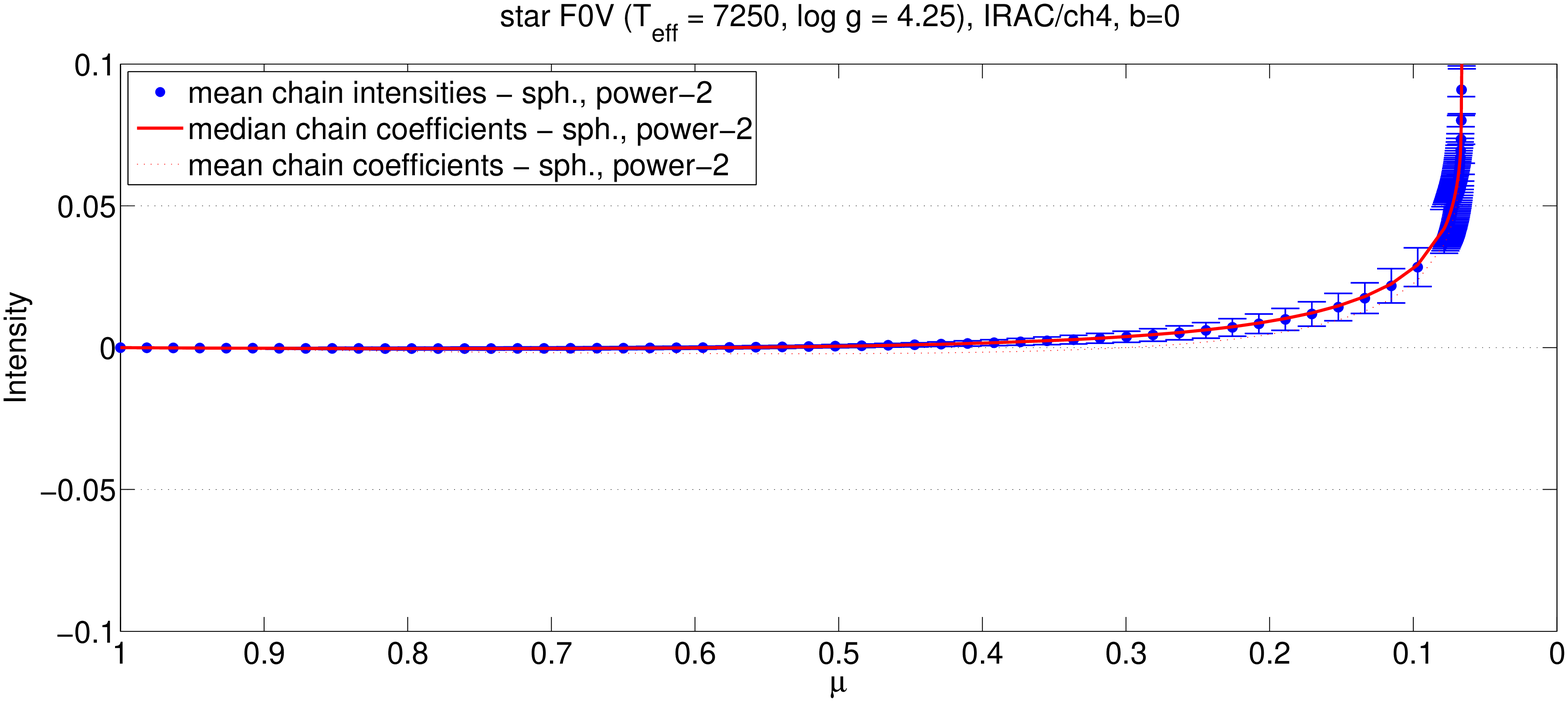}
\plotone{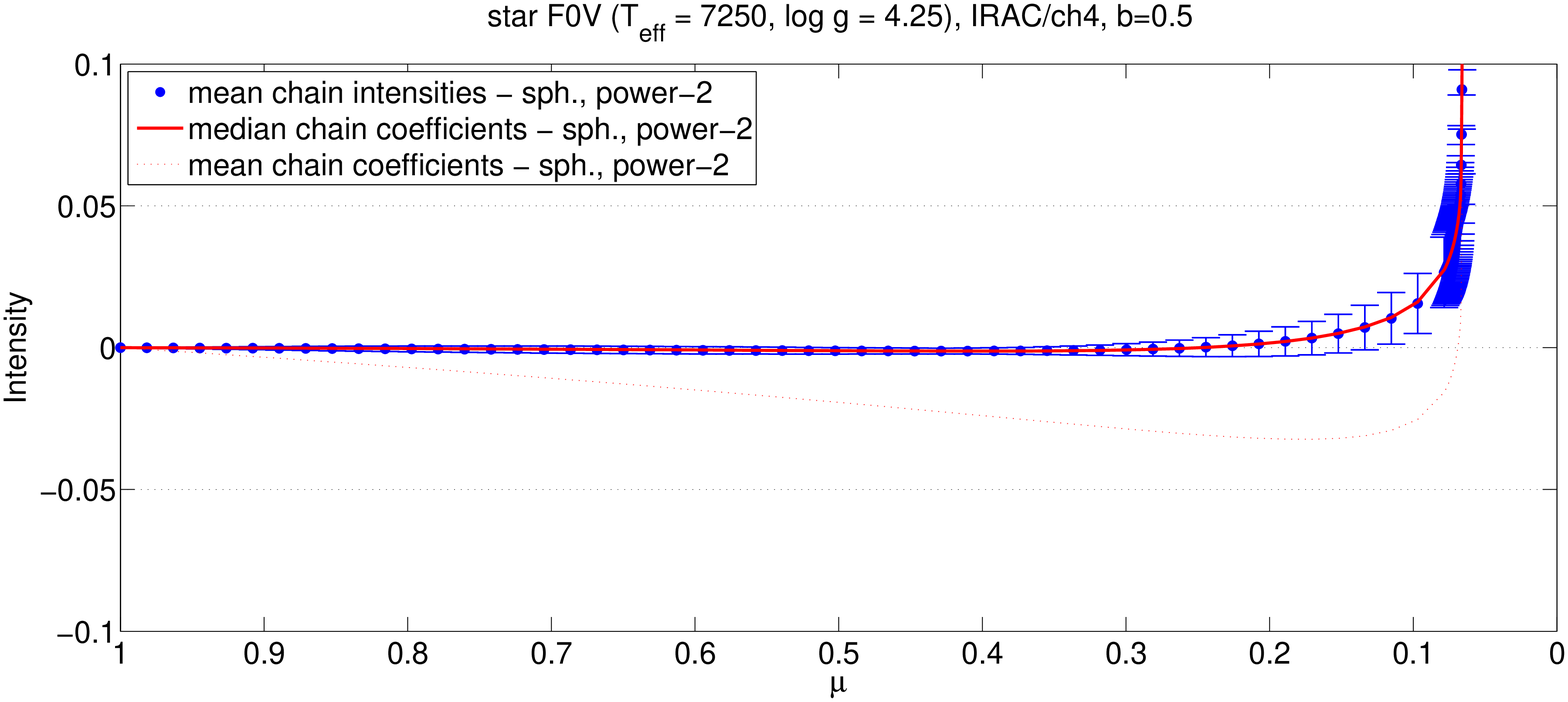}
\caption{Top panel: residuals between power-2 empirical limb-darkening profile and the spherical intensities for the F0\;V model, IRAC/ch4 passband, obtained from the edge-on transit with 100~ppm gaussian noise: estimates from the intensity chains (blue), model with median (red, continuous line) and mean (red, dotted line) chain values for the limb-darkening coefficients. Bottom panel: the same, from the inclined transit ($b=0.5$). \label{fig17app}}
\end{figure}

Figure~\ref{fig17app} shows the residuals in specific intensities obtained
from the two light-curves relative to the F0\;V model in the IRAC/ch4 passband (Section~\ref{sec:noisy_F0V_irac4}), one edge-on ($b=0$) and one inclined ($b=0.5$)
transit, using the power-2 law. The error bars (i.e., the standard
deviations of the intensity chains) are smaller than 0.2\%\ for
$\mu>0.4$, then increase up to $\gtrsim$1\%\ near the edge of the
disk. For the inclined transit, the error bars are larger by factors
1.0--2.4. The error bars of the predicted intensities along the steep
drop-off, i.e. at $\mu \lesssim0.08$, are not representative of the
true errors, as the predictions may deviate from the input values by
more than 10 $\sigma$.

\begin{figure}
\epsscale{1.0}
\plotone{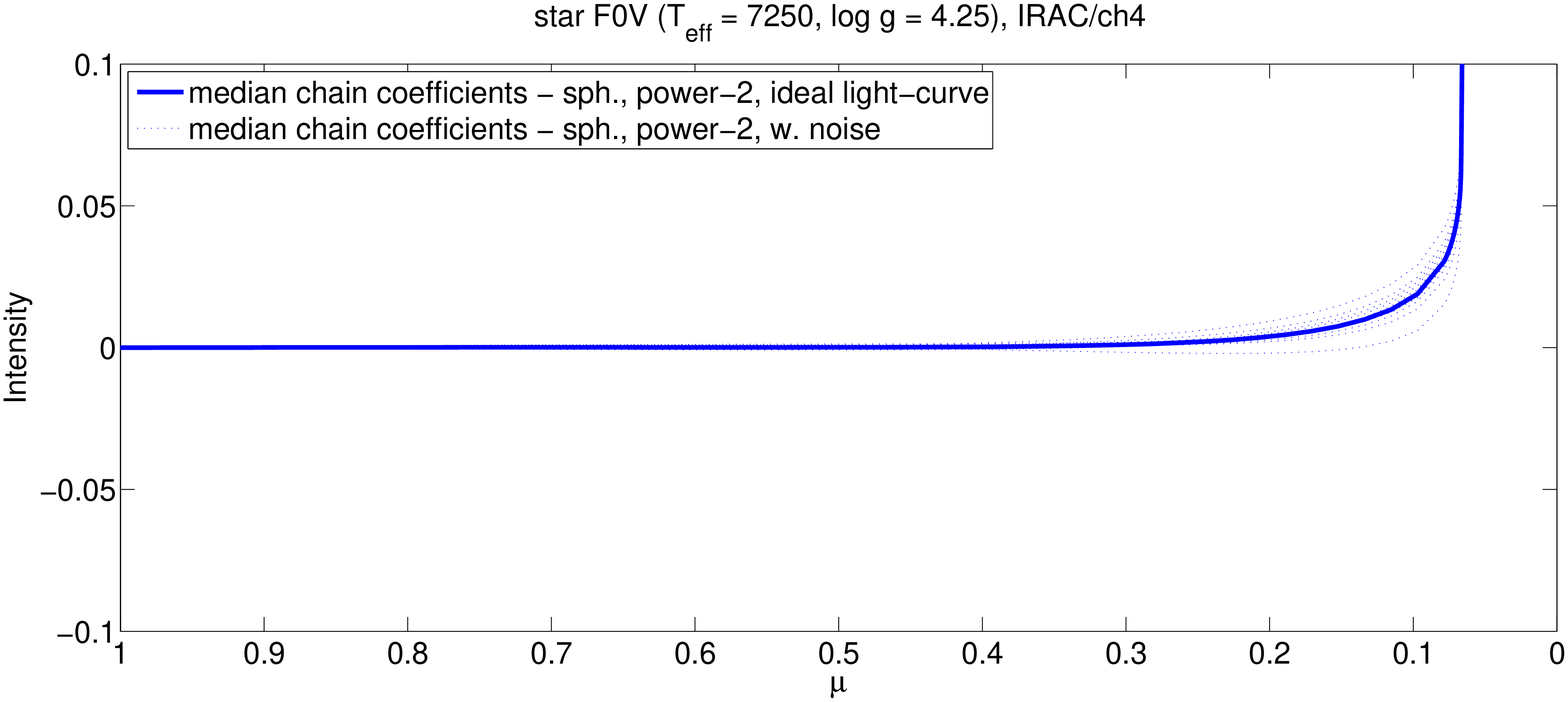}
\caption{Residuals between power-2 empirical limb-darkening profile and the spherical intensities for the F0\;V model, IRAC/ch4 passband, obtained from the edge-on transit without noise (blue, continuous line) and with 100~ppm gaussian noise (different noise time series, blue, dotted lines). Models are estimated by taking the median chain values for the limb-darkening coefficients. \label{fig18app}}
\end{figure}

We find that a good set of limb-darkening coefficients, which
reproduces intensities close to those predicted by the
intensity chains, can be obtained by taking medians of the
coefficients chains. Figure~\ref{fig18app} shows the intensity profiles
estimated in this way, from light-curves with different noise realizations.
They show, on average, the same bias obtained
for the noiseless case (see Section~\ref{sec:ldc_free}).


\begin{figure}
\epsscale{1.0}
\plotone{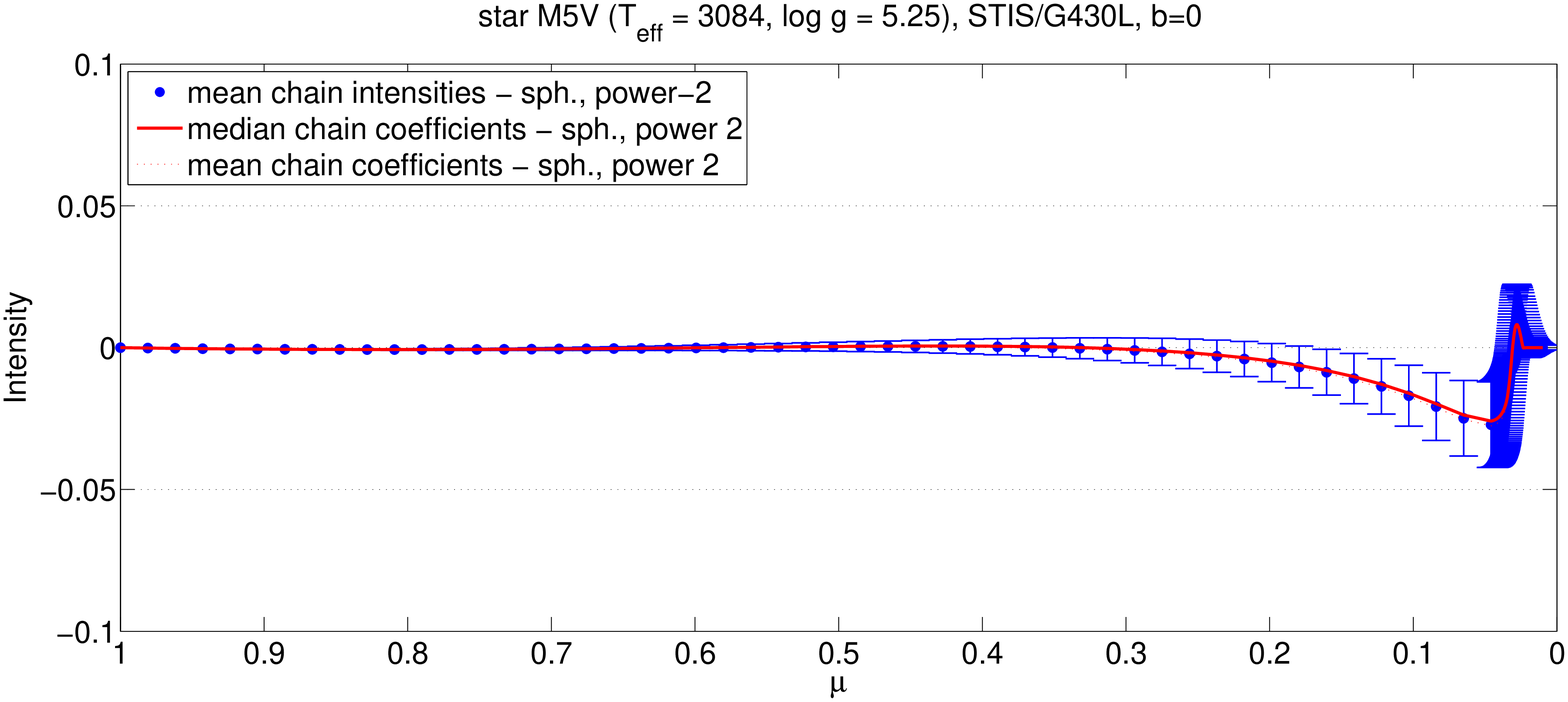}
\plotone{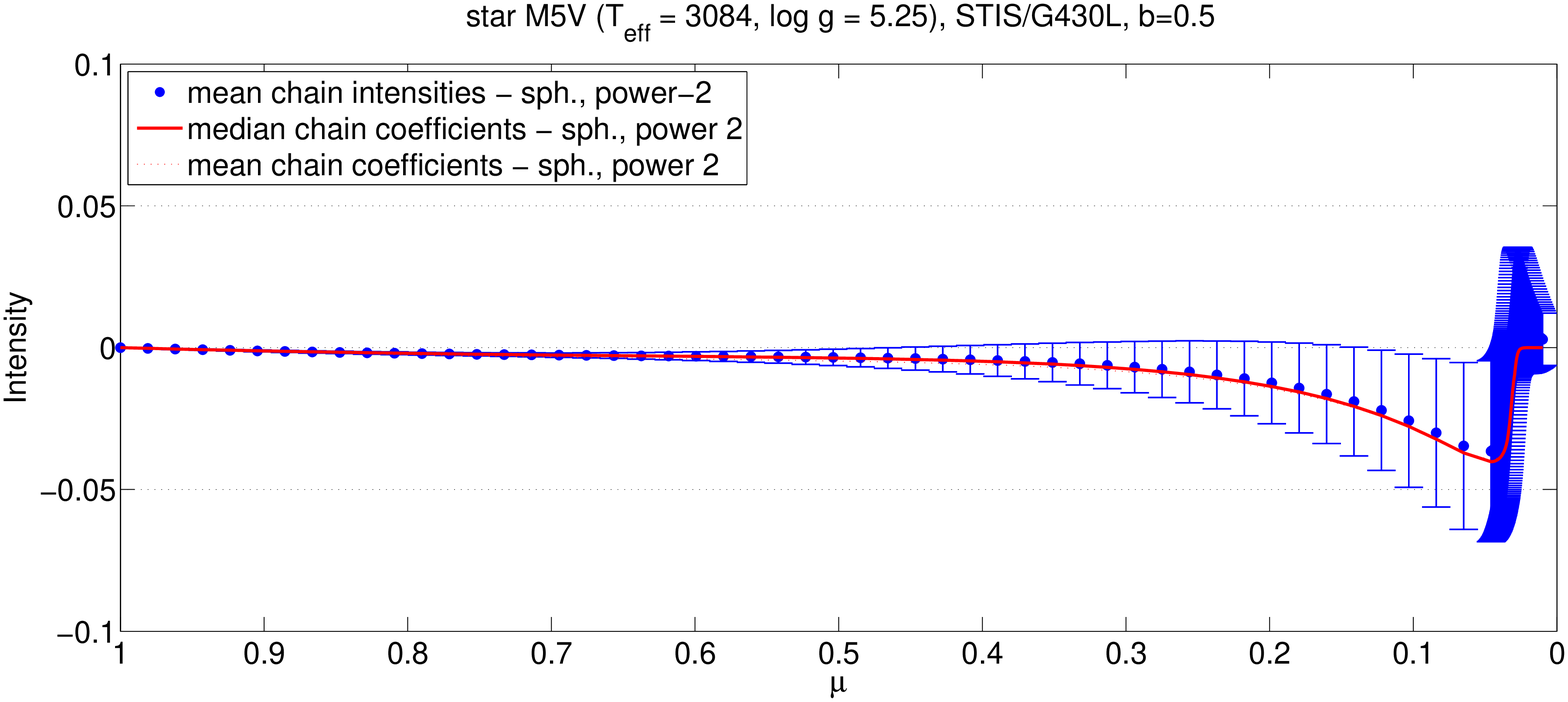}
\caption{Top panel: residuals between power-2 empirical limb-darkening profile and the spherical intensities for the M5\;V model, STIS/G430L passband, obtained from the edge-on transit with 100~ppm gaussian noise: estimates from the intensity chains (blue), model with median (red, continuous line) and mean (red, dotted line) chain values for the limb-darkening coefficients. Bottom panel: the same, from the inclined transit ($b=0.5$). \label{fig24app}}
\end{figure}

\begin{figure}
\epsscale{1.0}
\plotone{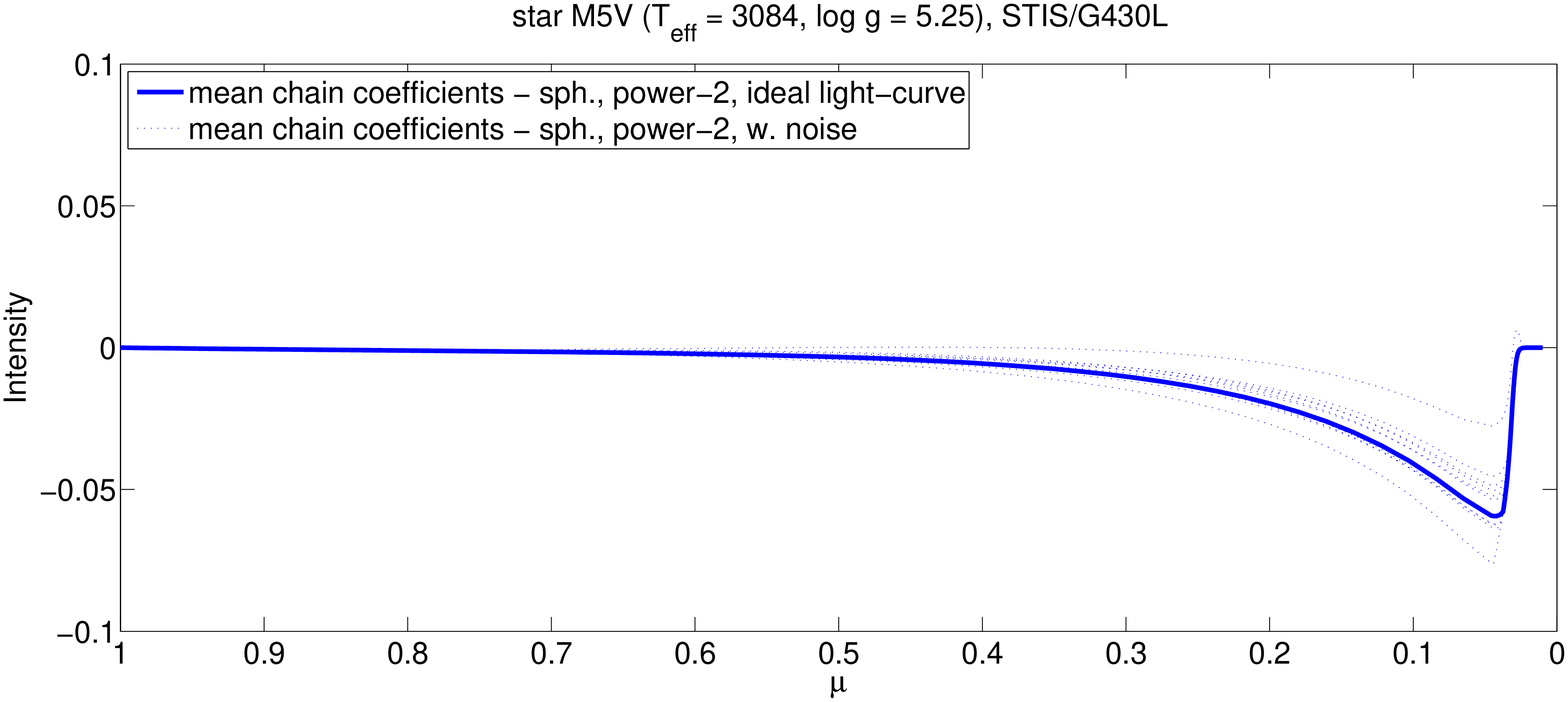}
\caption{Residuals between power-2 empirical limb-darkening profile and the spherical intensities for the M5\;V model, STIS/G430L passband, obtained from the edge-on transit without noise (blue, continuous line) and with 100~ppm gaussian noise (different noise time series, blue, dotted lines). Models are estimated by taking the mean chain values for the limb-darkening coefficients. \label{fig25app}}
\end{figure}

Figures~\ref{fig24app}--\ref{fig25app} report the analogous results obtained for the M5\;V model in the STIS/G430L passband (Section~\ref{sec:noisy_M5V_STIS_G430L}). The error bars on the specific intensities are, in average, $\gtrsim$1.5 times larger than those obtained for the less limb darkened cases. Even in this case, the bias is similar to the one obtained for the noiseless case (see Section~\ref{sec:ldc_free}).

\clearpage

\clearpage

\end{document}